\documentclass[11pt,a4paper]{article}

\pdfoutput=1 

\usepackage{jheppub}                   
\usepackage{bm}        
\usepackage{graphicx}                            
\usepackage{soul}                                
\usepackage{microtype}                           
\usepackage[sharp]{easylist}                     
\usepackage{amsmath}                             
\usepackage{amssymb}                             
\usepackage{tensor}                              
\usepackage{booktabs}                            
\usepackage{wrapfig}                             
\usepackage[T1]{fontenc}                         
\usepackage{titlesec}                            

\usepackage{diffcoeff}
\usepackage{tikz}
\usepackage{amsthm}                              
\usepackage{mathtools}
\usepackage{dsfont}
\usepackage{hyperref}   
\usepackage{derivative}

\def \dd {\mathrm{d}}
\def\d{\mathrm d}
\DeclareMathOperator{\ins}{\iota}

\usepackage[section]{placeins}

\graphicspath{{figures/}} 

\usepackage{todonotes}
\presetkeys{todonotes}{color=white, linecolor=gray, size=tiny}{}

\makeatletter
\if@todonotes@disabled
  
\else
  
\fi
\makeatother

\usepackage{color}
    \definecolor{darkgreen}{rgb}{0,0.5,0}
    \definecolor{darkred}{rgb}{0.5,0,0}
    \definecolor{darkblue}{rgb}{0,0,0.6}
    \definecolor{purple}{rgb}{0.4,.2,0.7}

\newcommand{\eg}{{\it e.g.\,}\ }
\newcommand{\ie}{{\it i.e.\,}\ }

\def\yp2{{y_+}^2}
\def\ym2{{y_-}^2}

\numberwithin{theorem}{section}

\begin{document}
\title{When AdS$_3$ Grows Hair: Boson Stars, Black Holes, and Double-Trace Deformations}
\author[a]{\'Oscar J.~C.~Dias,}
\author[a]{David Sola Gil,}
\author[b]{Jorge E. Santos}

\affiliation[a]{STAG Research Centre and Mathematical Sciences, Highfield Campus, University of Southampton, Southampton SO17 1BJ, UK}
\affiliation[b]{Department of Applied Mathematics and Theoretical Physics, University of Cambridge, Wilberforce Road, Cambridge CB3 0WA, UK} 

\emailAdd{ojcd1r13@soton.ac.uk}
\emailAdd{d.sola-gil@soton.ac.uk}
\emailAdd{jss55@cam.ac.uk}
\subheader{\today}

\abstract{
We analyse three-dimensional Einstein gravity coupled to a massive complex scalar field with double-trace boundary conditions. Using high-precision spectral methods, we construct regular  AdS$_3$ boson stars together with axisymmetric and non-axisymmetric hairy black holes. For each azimuthal number $m$, the hairy black holes bifurcate from the BTZ family at the corresponding double-trace instability onset. When the double-trace parameter satisfies $\kappa < \kappa_{\rm AdS}$, global AdS$_3$ becomes unstable and we identify its nonlinear endpoint as a zero-frequency boson star with energy below that of AdS$_3$, thereby providing the true ground state of the theory. In the microcanonical ensemble, hairy black holes always carry greater entropy than BTZ at fixed mass and angular momentum, and thus dominate whenever they exist. With notable exceptions, typically hairy black holes do not dominate the canonical nor the grand-canonical ensembles. We further show that, in the singular extremal limit, axisymmetric black holes saturate a generalised minimum‑energy theorem under double‑trace boundary conditions. These results yield the full nonlinear phase diagram of AdS$_3$ gravity with double-trace deformations.}

\maketitle
\flushbottom

\section{Introduction}\label{sec:Intro}

BTZ black holes \cite{Banados:1992wn,Banados:1992gq,Carlip:1995qv}, unlike their higher-dimensional rotating AdS counterparts, are stable against scalar field perturbations with Dirichlet or Neumann boundary conditions \cite{Hawking:1999dp,Birmingham:2001hc,Iizuka:2015vsa,Dappiaggi:2017pbe,Dias:2019ery}. This stability ultimately follows from the fact that the angular velocity of a BTZ black hole, in AdS radius units, always obeys the Hawking–Reall bound \cite{Hawking:1999dp}, $\Omega_H L \leq 1$, thereby forbidding any superradiant-type instability.

For massive scalar fields with mass $\mu$ in the window between the Breitenlohner–Freedman and unitarity bounds, $-1 < \mu^2 L^2 < 0$, one may also impose asymptotic double-trace boundary conditions, which remain normalizable and lead to finite-energy excitations \cite{Breitenlohner:1982jf,Breitenlohner:1982bm,Mezincescu:1984ev}. These mixed boundary conditions play an important role in the AdS/CFT correspondence, as they correspond to double-trace deformations of the otherwise conformal boundary theory~\cite{Klebanov:1999tb,Witten:2001ua,Amsel:2007im,Sever:2002fk,Berkooz:2002ug,Hertog:2004dr,Hertog:2004rz,Martinez:2004nb,Hertog:2004ns,Amsel:2006uf,Faulkner:2010fh,Faulkner:2010gj,Karch:2025hof}.%
\footnote{For a broader historical overview motivating double-trace deformations and their role in AdS black hole physics, see the introduction of our companion paper \cite{Dias:2025uyk}. Topics reviewed there include early AdS/CFT motivations \cite{Wald:1980jn,Klebanov:1999tb,Witten:2001ua,Amsel:2007im,Sever:2002fk,Berkooz:2002ug,Ishibashi:2003jd,Ishibashi:2004wx,Henneaux:2004zi,Henneaux:2006hk,Hertog:2004dr,Martinez:2004nb,Hertog:2004ns,Amsel:2006uf,Faulkner:2010fh,Faulkner:2010gj,Karch:2025hof}, their appearance in the holographic superconductor programme \cite{Allais:2010qq,Faulkner:2010gj,Basu:2013soa,Dias:2013bwa,Ren:2022qkr,Kinoshita:2023iad,Auzzi:2025nxx}, and studies of double-trace instabilities and black holes in higher dimensions~\cite{Katagiri:2020mvm,Harada:2023cfl,Kinoshita:2023iad}.}

In our companion paper \cite{Dias:2025uyk}, we showed that BTZ black holes can in fact be unstable under double-trace scalar perturbations. This occurs not only in the non-axisymmetric sector $-$ previously the only known unstable sector \cite{Iizuka:2015vsa,Dappiaggi:2017pbe,Ferreira:2017cta} $-$ but also, crucially, for axisymmetric modes. The origin of the instability is the possibility, under double-trace boundary conditions, of energy and angular-momentum influx through the asymptotic boundary \cite{Kinoshita:2023iad,Dias:2025uyk}. This applies to both rotating and static BTZ black holes. Moreover, the axisymmetric instability is the dominant one: whenever a non-axisymmetric instability exists, an axisymmetric instability is already present and is stronger \cite{Dias:2025uyk}.

The existence of a double-trace linear instability suggests that rotating and even static BTZ black holes should admit hairy counterparts, with scalar condensates orbiting or floating around the horizon. In a phase diagram of asymptotically AdS$_3$ solutions, such hairy black holes should bifurcate from (or merge with) the BTZ family along the curve describing the onset of the instability. Their endpoint is not a priori clear. If these hairy solutions possess higher entropy than BTZ at fixed energy and angular momentum, they could provide the natural endpoint (or at least a metastable one) of the instability evolution in the microcanonical ensemble. These expectations apply to both axisymmetric and non-axisymmetric solutions, although the axisymmetric ones are expected to be physically more relevant whenever multiple families coexist.

In this paper, we address these questions. We construct both axisymmetric (static and rotating) and non-axisymmetric (typically rotating, though in some cases static) hairy black holes with double-trace scalar condensates. These solutions indeed bifurcate from the BTZ black hole at the corresponding instability onset. We then perform a detailed study of their physical and thermodynamic properties to assess the expectations outlined above.\footnote{Several AdS$_3$ studies are relevant in the context of our study. AdS$_3$ perfect-fluid stars were constructed in \cite{Cruz:1994ar,Lubo:1998ue}. AdS$_3$ boson stars with scalar self-interaction potentials appeared in \cite{Sakamoto:1998hq,Sakamoto:1999zt,Sakamoto:1998aj}, while boson stars sourced by scalar fields were obtained in \cite{Astefanesei:2003qy,Astefanesei:2003rw}. AdS$_3$ boson stars with homogeneous Dirichlet boundary conditions and AdS$_3$ hairy black holes with Robin boundary conditions were constructed in \cite{Stotyn:2013spa,Stotyn:2012ap} and \cite{Iizuka:2015vsa}, 
respectively, providing three-dimensional analogues of the solutions in \cite{Dias:2011at,Stotyn:2011ns,Dias:2015rxy,Ishii:2018oms,Ishii:2021xmn}. 
AdS$_3$ hairy black holes with non-homogeneous Dirichlet boundary conditions were constructed in \cite{Gao:2023rqc}. 
AdS$_3$ hairy boson stars and black holes with self-interacting scalar potentials were also studied in \cite{Henneaux:2002wm,Correa:2010hf,Cardenas:2022jtz}. Finally, AdS$_4$ solitons with Robin boundary conditions were found in \cite{Bizon:2020yqs}.}

Interestingly, AdS$_3$ itself can be unstable to massive scalar fields with double-trace boundary conditions \cite{Ishibashi:2004wx,Dias:2025uyk}. Since the instability arises from boundary influx rather than near-origin effects, AdS$_3$ is linearly stable only in part of the double-trace parameter space. In the stable regime, the normal mode frequencies were computed in \cite{Dias:2025uyk}. Including backreaction at higher perturbative orders may lead to regular, fully nonlinear asymptotically AdS$_3$ boson stars with double-trace condensates. In the complementary regime where AdS$_3$ is linearly unstable, the normal-mode frequencies become purely imaginary \cite{Dias:2025uyk}, suggesting $-$ though not proving $-$ that regular boson stars might not exist. The endpoint of the AdS$_3$ double-trace instability is therefore an important open question.

We will show that regular AdS$_3$ boson stars exist for any value of the double-trace coupling. However, their qualitative properties change markedly when crossing the critical coupling at which AdS$_3$ becomes unstable. In such case, the boson stars are no longer perturbatively connected to AdS$_3$. Moreover, in addition to regular boson stars, we also find asymptotically AdS$_3$ solutions that are singular at the origin. These singular boson stars play a crucial role in completing the phase diagram, as some families of hairy black holes terminate on them.

The structure of the paper is as follows. In section~\ref{sec:Setup}, we present the physical setup: ansätze, equations of motion, boundary conditions, and thermodynamic quantities. Section~\ref{sec:NumericalSetup} describes the numerical methods used to construct both regular and singular boson stars and the hairy black holes. In section~\ref{sec:PhaseDiag-m0}, we study the $m=0$ sector: we construct boson stars perturbatively connected to AdS$_3$ when it is stable (section~\ref{sec:PhaseDiag-m0:BStar}), and non-perturbative boson stars in the unstable regime, which may represent endpoints of the Ishibashi–Wald instability \cite{Ishibashi:2004wx} (section~\ref{sec:PhaseDiag-m0:IshWald}). We also identify subfamilies of hairy black holes at constant horizon radius, scalar expectation value, or angular momentum (section~\ref{sec:PhaseDiag-m0:BHs}). Section~\ref{sec:PhaseDiag-m1} extends this analysis to $m=1$ ($m\geq 1$) solutions. Section~\ref{sec:PhaseDiag-Total} then assembles the full phase diagram of asymptotically AdS$_3$ solutions with double-trace boundary conditions, and discusses the dominant black holes in the microcanonical, canonical, and grand-canonical ensembles. Conclusions and an executive summary of our main results appear in section~\ref{sec:Conc}. The reader interested in the main findings can move immediately to sections~\ref{sec:PhaseDiag-Total}~and~\ref{sec:Conc}. Appendix~\ref{secA:superpotentials} shows that the novel zero-frequency boson stars that we find with energy below the one of AdS$_3$ agree with the ground state predictions of the superpotential analysis of~\cite{Faulkner:2010fh} and that the singular extremal hairy black holes that we find saturate a generalised minimum‑energy theorem under double‑trace boundary conditions~\cite{Faulkner:2010fh}. Appendix~\ref{secA:BStars-DN} finds the boson stars for theories whereby the double-trace boundary condition reduces to its Dirichlet and Neumann limits.
The \emph{physical} conserved mass and angular momentum of our solutions is computed from first principles using holographic renormalization~\cite{Balasubramanian:1999re,deHaro:2000vlm,Skenderis:2002wp,Papadimitriou:2005ii} in Appendix~\ref{secA:HoloRen} and, alternatively, using the covariant Noether‑charge (covariant phase‑space) formalism~\cite{Lee:1990nz,Wald:1993nt,Iyer:1994ys,Wald:1999wa} in Appendix~\ref{secA:NoetherFirstLaw}.

\section{Setup of the physical problem and thermodynamic quantities}\label{sec:Setup}

\subsection{Theory, field ans\"atze, equations of motion and symmetries}\label{sec:EoM}
We consider Einstein gravity in three-dimensional anti-de Sitter space (AdS$_3$) with a negative cosmological constant $\Lambda = -1/L^{2}$, where $L$ denotes the AdS$_3$ curvature radius. The theory is minimally coupled to a neutral, massive complex scalar field $\Phi$. Throughout, we work in natural units with $c = 1$, and $G$ denotes the three-dimensional Newton constant. The action of the theory is
\begin{align}\label{eqn:action}
    S_{\rm bulk}= \frac{1}{16\pi G} \int \mathrm{d}^{3}x\, \sqrt{-g}\,\left[\mathcal{R}+ \frac{2}{L^{2}}- 2\, \nabla_\alpha \Phi \nabla^\alpha \Phi^{\dagger}- 2 \mu^{2} \Phi \Phi^{\dagger}\right],
\end{align}
where $\mathcal{R}$ is the Ricci scalar and $\mu$ is the mass of the scalar field.

Our aim is to determine the phase diagram of asymptotically AdS$_3$ boson stars and black hole solutions of \eqref{eqn:action}. To this end, we introduce the coordinate chart $\{t,r,\phi\}$, where surfaces of constant $t$ and $r$ have the geometry of a circle $S^1$. The angular coordinate $\phi$ parametrizes this circle and is identified as $\phi \sim \phi + 2\pi$.

Working in the Schwarzschild gauge $-$ where the radial coordinate $r$ is the areal radius so that $g_{\phi\phi} = r^{2}$ $-$ the most general ansatz describing stationary, asymptotically AdS$_3$ configurations takes the form
\begin{subequations}\label{HairyAnsatz}
\begin{align}
\label{metric_ansatz}
{\rm d}s^{2}
&= -f(r) g(r)\, \mathrm{d}t^{2}
   + \frac{\mathrm{d}r^{2}}{f(r)}
   + r^{2} \left[\mathrm{d}\phi - \Omega(r)\,\mathrm{d}t \right]^{2}, \\
\label{scalar_ansatz}
\Phi(t,r,\phi)
&= e^{-i\omega t} e^{i m \phi} \psi(r)\, .
\end{align}
\end{subequations}
Here $f$, $g$, $\Omega$, and $\psi$ are real, smooth functions of the radial coordinate $r$ only. The periodicity of $\phi$ imposes that the azimuthal number $m$ must be an integer.

The metric in \eqref{HairyAnsatz} is stationary and axisymmetric, with Killing vector fields $\partial_t$ and $\partial_\phi$. However, for $m \neq 0$, the scalar field \eqref{scalar_ansatz} is invariant only under the helical Killing vector
\begin{align}\label{BS:KVF}
\xi = \partial_t + \frac{\omega}{m}\, \partial_\phi \, .
\end{align}
Ordinarily, one expects matter fields to share all the symmetries of the metric. For a complex scalar, however, this is not required: the energy--momentum tensor
\begin{equation}\label{EMtensor}
T_{\mu\nu}
  = \nabla_\mu \Phi \nabla_\nu \Phi^{\dagger}
  + \nabla_\mu \Phi^{\dagger} \nabla_\nu \Phi
  - g_{\mu\nu}\!\left(
      \nabla_c \Phi \nabla^c \Phi^{\dagger}
      + \mu^{2} |\Phi|^{2}
    \right)
\end{equation}
is invariant under the full isometry group of the metric. Thus the reduced symmetry of the scalar field does not lead to an inconsistency.

In contrast, when $m = 0$ the scalar field depends only on time, and in this case only $\partial_\phi$ is a Killing vector of the full configuration \eqref{HairyAnsatz}.

Our solutions possess two important scaling symmetries. The first is
\begin{align}\label{ScaleSym1}
    & \{t, r, \phi\} \to \{\Lambda_1 t,\, \Lambda_1 r,\, \phi\}, 
      \qquad
      \{f, g, \Omega, \psi\} \to \left\{f,\, g,\, \frac{\Omega}{\Lambda_1},\, \psi\right\}, 
      \nonumber \\
    & \{\omega, \mu, m, L\} \to 
      \left\{\frac{\omega}{\Lambda_1},\, \frac{\mu}{\Lambda_1},\, m,\, \Lambda_1 L\right\},
\end{align}
which induces the rescaling $g_{ab} \to \Lambda_1^2 g_{ab}$ while leaving the scalar field and equations of motion invariant. Under this transformation, the affine connection $\Gamma^\gamma_{\:\mu\nu}$, the curvature tensors ($\mathcal{R}^{\alpha}{}_{\beta\mu\nu}$ and $\mathcal{R}_{\mu\nu}$), and the stress tensor $T_{\mu\nu}$ remain unchanged. Consequently, the equations of motion require the AdS radius to scale as $L \to \Lambda_1 L$. 

We will use this symmetry to measure all physical quantities in units of $L$, which is equivalent to setting $L \equiv 1$. To this end, we introduce the dimensionless coordinates
\begin{align}\label{eqn:dimlessTR}
    T = \frac{t}{L},
    \qquad
    R = \frac{r}{L},
\end{align}
and denote dimensionless quantities with a hat. For example,
\begin{align}\label{eqn:dimlessParam}
    \Omega = \frac{\hat{\Omega}}{L},
    \qquad 
    \omega = \frac{\hat{\omega}}{L},
    \qquad 
    \mu = \frac{\hat{\mu}}{L}.
\end{align}

The second scaling symmetry is
\begin{align}\label{ScaleSym2}
   & \{t, r, \phi\} \to \{\Lambda_2 t,\, r,\, \phi\},  
    \qquad
    \{f, g, \Omega, \psi\} \to 
        \left\{f,\, \frac{g}{\Lambda_2},\, \frac{\Omega}{\Lambda_2},\, \psi\right\}, 
     \nonumber \\
   & 
    \{\omega, \mu, m, L\} \to 
        \left\{\frac{\omega}{\Lambda_2},\, \mu,\, m,\, L\right\},
\end{align}
which leaves the ansatz \eqref{HairyAnsatz} $-$ and therefore the equations of motion $-$ unchanged. Near the AdS boundary, the equations of motion imply that $g|_{r\to\infty} = c_0$, where $c_0$ is an arbitrary constant. We use the symmetry \eqref{ScaleSym2} to set $c_0 \equiv 1$.

Varying the action \eqref{eqn:action} yields the Einstein and scalar equations,
\begin{equation}
\mathcal{R}_{ab}-\frac{1}{2}\,\mathcal{R}\,g_{ab}
    -\frac{1}{L^{2}}\,g_{ab}
    = T_{ab},
    \qquad
    \left(\nabla^{a}\nabla_{a}-\mu^{2}\right)\Phi = 0.
\end{equation}
In dimensionless variables, these reduce to a system of four coupled second‑order ODEs for 
\(\{f, g, \hat{\Omega}, \psi\}\), together with two constraint equations \(C_1\) and \(C_2\):\footnote{One may alternatively eliminate $g$ algebraically in terms of $\{f,\hat{\Omega},\psi\}$ and their derivatives, reducing the system to three coupled second‑order ODEs.}
\begin{subequations}\label{4ODEs}
\begin{align}
& f'' + f'\!\left(\frac{3}{R} - \frac{g'}{2g}\right)
+ \frac{f g'}{g}\!\left(\frac{1}{R} - \frac{g'}{2g}\right)
+ \frac{4 \psi^{2} (\hat{\omega} - m\hat{\Omega})(\hat{\omega} - m\hat{\Omega} - 2mR\,\hat{\Omega}')}
       {fg}
\nonumber\\
& \hspace{4cm}
+ 4\psi^{2}\!\left(2\hat{\mu}^{2} + \frac{m^{2}}{R^{2}}\right)
+ \frac{8\psi\,\psi' (m^{2} + \hat{\mu}^{2}R^{2})}{R}
- 8
= 0, \label{eqn:feom} 
\\
& 
g'' + g'\!\left(\frac{2f'}{f} + \frac{1}{R}\right)
+ \frac{8\psi^{2}}{f^{2}}(\hat{\omega} - m\hat{\Omega})^{2}
  + \frac{8mR\,\psi^{2}\,\hat{\Omega}'}{f^{2}}(\hat{\omega} - m\hat{\Omega})
  - \frac{8g\psi\,\psi'}{Rf}(m^{2} + \hat{\mu}^{2}R^{2}) = 0, \label{eqn:geom}
\\
& \hat{\Omega}'' + \hat{\Omega}'\!\left(\frac{3}{R} - \frac{g'}{2g}\right)  + \frac{4m\psi^{2}}{R^{2}f}(\hat{\omega} - m\hat{\Omega}) = 0, \label{eqn:Omegaeom}
\\
& \psi'' + \psi'\!\left(\frac{1}{R} + \frac{f'}{f} + \frac{g'}{2g}\right)  + \frac{\psi}{f^{2}}
   \left(
      \frac{(\hat{\omega} - m\hat{\Omega})^{2}}{g}
      - \frac{f(m^{2} + \hat{\mu}^{2}R^{2})}{R^{2}}
   \right)
= 0. \label{eqn:chieom}
\end{align}
\end{subequations}

The constraint equations are
\begin{subequations}\label{C1C2}
\begin{align}
\label{eqn:c1eom}
C_{1} &\equiv \frac{R}{f}(f^{2} g)' 
        + 4g\!\left[\psi^{2}(m^{2} + R^{2}\hat{\mu}^{2}) - R^{2}\right]
        + R^{4}(\hat{\Omega}')^{2} = 0,
\\
\label{eqn:c2eom}
C_{2} &\equiv \frac{f'}{2Rf}
      + \frac{\psi^{2}}{f}\!\left(
            \hat{\mu}^{2}
            + \frac{m^{2}}{R^{2}}
            + \frac{(\hat{\omega} - m\hat{\Omega})^{2}}{fg}
        \right)
      + \frac{R^{2}(\hat{\Omega}')^{2}}{4fg}
      + (\psi')^{2}
      - \frac{1}{f}
      = 0.
\end{align}
\end{subequations}

Under evolution by the system \eqref{4ODEs}, the constraints satisfy
\begin{subequations}\label{C1C2eom}
\begin{align}
    C_{1}' &= \left(\frac{Rg}{Rg' - 2g}\right) C_{1},
\\
    C_{2}' &= 
    \left(\frac{(Rf g)' - 3fg}{4R^{3} f^{2} g^{2}}\right) C_{1}
    - \left(\frac{2}{R} + \frac{2f'}{f} + \frac{g'}{g}\right) C_{2}.
\end{align}
\end{subequations}
Thus, if the constraints are satisfied at one value of $R$, they remain satisfied for all $R$. In practice, we impose them at the boundary and solve the four coupled nonlinear ODEs \eqref{eqn:feom}-\eqref{eqn:Omegaeom} in the bulk.

Our system possesses additional structural properties that will be useful in our analysis. First, the action enjoys a global $U(1)$ symmetry under $\Phi \to e^{i\,\eta}\,\Phi$. Furthermore, the equations of motion are invariant under the transformation $\{m, \Omega\} \to \{-m, -\Omega\}$, allowing us to restrict without loss of generality to non‑negative azimuthal integers, $m \geq 0$.

The ansatz \eqref{HairyAnsatz} also admits a residual gauge freedom,
\begin{subequations}\label{ResidualGaugeFreedom}
\begin{align}
   & \text{For $m = 0$:} 
   && \phi \to \phi + \lambda \, t, 
   & \Omega \to \Omega + \lambda, 
   && \omega \to \omega\,; \label{ResidualGaugeFreedom:m0} \\
   & \text{For $m \neq 0$:} 
   && \phi \to \phi + \frac{\lambda}{m}\, t, 
   & \Omega \to \Omega + \frac{\lambda}{m}, 
   && \omega \to \omega + \lambda\,; \label{ResidualGaugeFreedom:m}
\end{align}
\end{subequations}
where $\lambda$ is arbitrary. This transformation leaves both \eqref{metric_ansatz} and \eqref{scalar_ansatz} invariant. In practice $-$ though not strictly required  $-$  we typically fix this gauge freedom by imposing that the spacetime has no rotation at infinity, \ie that the angular velocity at the AdS boundary vanishes: $\Omega|_{r \to \infty} = 0$.

We are interested in asymptotically AdS$_3$ boson stars and black hole solutions endowed with non‑trivial scalar hair, \ie configurations with $\psi(R) \neq 0$. For boson stars the inner boundary is the origin, $R=0$, whereas for black holes the inner boundary is the event horizon located at $R = R_+$, defined by $f(R_+) = 0$. The horizon is generated by the (dimensionless) Killing vector\footnote{\label{foot:metricTR}Under \eqref{eqn:dimlessTR}-\eqref{eqn:dimlessParam}, the metric ansatz becomes 
${\rm d}s^2 = L^2\!\left[-f(R) g(R)\, {\rm d}T^2 + \frac{{\rm d}R^2}{f(R)} + R^2\big({\rm d}\phi - \hat{\Omega}(R)\, {\rm d}T\big)^2\right]$, 
and the horizon generator is 
$K = \partial_t + \Omega_H \partial_\phi = \frac{1}{L}\big(\partial_T + \hat{\Omega}_H \partial_\phi\big)$.}
\begin{equation}\label{HorizonGenerator}
    K = \frac{1}{L}\big( \partial_T + \hat{\Omega}_H \partial_\phi \big),
\end{equation}
where $\hat{\Omega}_H = -\frac{g_{T\phi}}{g_{\phi\phi}}\big|_{R=R_+} = \hat{\Omega}(R_+)$ is the angular velocity of the horizon. Equivalently, $K^\mu K_\mu |_{R=R_+} = 0$.\footnote{If one works in a gauge where $\hat{\Omega}|_{R \to \infty} \neq 0$, then $\hat{\Omega}_H$ in \eqref{HorizonGenerator} must be interpreted as the angular velocity relative to infinity.}

The next subsection describes the boundary conditions required for physical solutions. Together with the equations of motion \eqref{4ODEs}, these conditions define the boundary value problem whose solutions correspond to hairy boson stars and hairy black holes.

Of course, the action \eqref{eqn:action} also admits solutions with vanishing scalar field. Namely, global AdS$_3$ described by
\begin{equation}\label{AdS3}
    f_{\text{\tiny AdS}} = R^2 + 1,
    \qquad g_{\text{\tiny AdS}} = 1,
    \qquad \hat{\Omega}_{\text{\tiny AdS}} = 0,
    \qquad \psi_{\text{\tiny AdS}} = 0,
\end{equation}
and the BTZ black hole \cite{Banados:1992gq, Banados:1992wn, Carlip:1995qv},
\begin{equation}\label{BTZ}
    f_{\text{\tiny BTZ}} = \frac{(R^2 - R_+^2)(R^2 - R_-^2)}{R^2},
    \qquad g_{\text{\tiny BTZ}} = 1,
    \qquad \hat{\Omega}_{\text{\tiny BTZ}} = \frac{R_+ R_-}{R^2},
    \qquad \psi_{\text{\tiny BTZ}} = 0,
\end{equation}
where $R_\pm$ (with $0\le R_- \le R_+$) are the two real roots of $f_{\text{\tiny BTZ}} = 0$. Regular BTZ black holes require $R_- \le R_+$, with equality corresponding to the extremal case.

\subsection{Boundary conditions at the asymptotic boundary}\label{sec:BCsUV}

The boundary conditions (BCs) at the asymptotic AdS$_3$ boundary, \ie as $R \to \infty$, are identical for both the boson star and black hole solutions. It is convenient to describe these BCs using Fefferman-Graham (FG) coordinates $\{T, z, \phi\}$ \cite{Fefferman:1985, Graham:1999jg, Skenderis:2002wp}.
FG coordinates are defined by imposing
\begin{equation}
    g_{zz} = \frac{L^2}{z^2}, \qquad g_{za} = 0 ,
\end{equation}
where $z$ is the radial FG coordinate, with the AdS boundary located at $z=0$, and $a = \{T,\phi\}$ denote the boundary coordinates. In Fefferman-Graham (FG) coordinates, any asymptotically AdS$_3$ spacetime admits the following near-boundary FG expansion around $z = 0$ (with $z$ dimensionless) \cite{Henningson:1998gx, deHaro:2000vlm, Skenderis:2002wp}:
\begin{align}
    \label{FGmetric}
    \mathrm{d}s^2 &= \frac{L^2}{z^2}\Big[ \mathrm{d}z^2 + g_{ab}(z,x)\,\mathrm{d}x^a \mathrm{d}x^b \Big], \nonumber \\
    g_{ab}(z,x)\big|_{z\to 0} &=
        g_{ab}^{(0)}(x) + \dots
        + z^2 g_{ab}^{(2)}(x)
        + h_{ab}^{(2)}\, z^2 \ln z
        + \dots\, .
\end{align}
Here, $g_{ab}^{(0)}(x)$ and $g_{ab}^{(2)}(x)$ are free coefficients in the UV expansion of the metric. The logarithmic term $h_{ab}^{(2)}$ corresponds to the metric variation of the gravitational conformal Weyl anomaly $\mathcal{A}$. In AdS$_3$, this term can be non-vanishing only when matter fields are present \cite{deHaro:2000vlm, Kanitscheider:2006zf, Grumiller:2008qz, Skenderis:2009nt}.  In the absence of matter, \ie in pure AdS$_3$ Einstein gravity, $\mathcal{A}$ reduces to the topological invariant $\int R_{\text{\tiny(}\partial\text{\tiny)}}$, so its variation with respect to the boundary metric vanishes; indeed, for the BTZ black hole one finds $h_{ab}^{(2)} = 0$.   However, logarithmic terms may appear in the expansions of hairy boson stars and black holes, depending on the boundary conditions imposed on the scalar field, as discussed below.

Similarly, the scalar field admits a well-known near-boundary expansion for any asymptotically AdS$_3$ background \cite{deHaro:2000vlm, Skenderis:2002wp, Dias:2015nua}:
\begin{align}\label{FGscalar}
    \psi(z)\big|_{z \to 0}
        &= \alpha(x)\, z^{\Delta_-}
        + \dots
        + \beta(x)\, z^{\Delta_+}
        + \gamma(x)\, z^{\Delta_+} \ln z
        + \mathcal{O}(z^{\Delta_+ + 1}), \nonumber \\[2mm]
    \text{where} \qquad
    \Delta_\pm &= 1 \pm \sqrt{1 + \hat{\mu}^2}
\end{align}
are the conformal dimensions of the dual operator, \ie the solutions of
\begin{equation}
    \Delta(\Delta - 2) = \mu^2 L^2 \equiv \hat{\mu}^2,
    \label{eq:Delta}
\end{equation}
with $\Delta_- = 2 - \Delta_+$. Here, $\alpha(x)$ and $\beta(x)$ are the amplitudes of the two linearly independent asymptotic solutions, while $\gamma(x)$ is related to the scalar-field conformal anomaly $\mathcal{A}_\Phi$ and is fixed in terms of $\alpha(x)$ by the equations of motion \cite{deHaro:2000vlm, Henningson:1998gx}.\footnote{At linear order, the Klein-Gordon equation on fixed global AdS$_3$ or BTZ yields $\gamma(x) = 0$ for any scalar mass \cite{deHaro:2000vlm}.  For certain masses this no longer holds once nonlinear backreaction on the metric is included.} In \eqref{FGmetric} and \eqref{FGscalar}, the ellipses denote terms that are determined as functions of $g_{ab}^{(0)}(x)$, $g_{ab}^{(2)}(x)$, $\alpha(x)$, and/or $\beta(x)$ by solving the equations of motion order by order in the $z$-expansion.

Hairy solutions of the type we construct in this paper only exist when mixed boundary conditions are imposed on the scalar field. In particular, we will be interested in double-trace boundary conditions.  
As discussed in detail in Section~2.2 of our companion paper \cite{Dias:2025uyk}, in AdS$_3$ such boundary conditions yield normalisable solutions (\ie finite-energy configurations) only when the scalar mass $\mu$ lies between the Breitenlohner-Freedman (BF) and unitarity bounds:
\begin{align}\label{2xTrace:rangeMass}
    \mu_{\text{BF}}^2 L^2
    < \mu^2 L^2
    < \mu_{\text{BF}}^2 L^2 + 1,
    \qquad \text{with} \qquad
    \mu_{\text{BF}}^2 L^2 = -1 .
\end{align}
We will always assume this mass range in the present analysis.  The upper bound is equivalent to requiring that $\Delta_-$ remains above the unitarity bound, $\Delta_-|_{\text{\tiny unit}} = 0$.

Within this range, we can impose Dirichlet boundary conditions that fix $\alpha$ (\eg $\alpha=0$, the \emph{standard quantisation}), or Neumann boundary conditions that fix $\beta$ (\eg $\beta=0$, the \emph{alternative quantisation}).  
More generally, one may impose mixed boundary conditions relating $\beta$ and $\alpha$, such as the double-trace condition
\begin{align}\label{2xTrace:BC}
    \beta = \kappa\, \alpha ,
\end{align}
where $\kappa$ is, a priori, a real constant. These boundary conditions correspond holographically to double-trace deformations of the dual CFT operator.

Considering the nonlinear backreaction of AdS$_3$ normal modes from \cite{Dias:2025uyk}, we find hairy boson stars that are perturbatively connected to AdS$_3$ for 
\begin{equation}
    \kappa \in \, ]\kappa^{\text{\tiny AdS}}_{m,\hat{\mu}^2}, +\infty[ ,
\end{equation}
where $\kappa^{\text{\tiny AdS}}_{m,\hat{\mu}^2} < 0$ marks the onset of the Wald-Ishibashi AdS$_3$ instability \cite{Ishibashi:2004wx, Dias:2025uyk}. This quantity depends on the scalar mass $\hat{\mu} = \mu L$ (restricted to the range \eqref{2xTrace:rangeMass}) and the azimuthal number $m$, and was determined in \cite{Dias:2025uyk}. For $\kappa <\kappa^{\text{\tiny AdS}}_{m,\hat{\mu}^2}$, we also find regular boson stars, but these are qualitatively different and in particular, they are \textit{not} perturbatively connected to AdS$_3$.

On the other hand, the hairy black holes we construct only exist for  
\begin{equation}
    \kappa \in \, ] -\infty ,\, \kappa^{\text{\tiny BTZ}}_{m,\hat{\mu}^2}(\hat{M},\hat{J}) [ ,
\end{equation}
where $\kappa^{\text{\tiny BTZ}}_{m,\hat{\mu}^2}(\hat{M},\hat{J}) < 0$ signals the onset of the double-trace instability of a BTZ black hole with mass $\hat{M}$ and angular momentum $\hat{J}$, as computed in \cite{Dias:2025uyk}. Importantly, the BTZ black hole is stable under both homogeneous Dirichlet boundary conditions ($\alpha=0$, equivalently $\kappa=\pm\infty$) and Neumann boundary conditions ($\beta=0$, equivalently $\kappa=0$).  Consequently, no hairy black hole solutions exist for these two specific values of~$\kappa$.

In asymptotically AdS$_3$ spacetimes, the gravitational and scalar conformal anomalies do not vanish whenever 
\begin{equation}
    \Delta_+ - \Delta_- = 2\sqrt{1+\mu^2 L^2}
\end{equation}
is an integer.  Additional cases in which conformal anomalies appear are discussed in detail in \cite{Bzowski:2015pba}.  

In what follows, we will restrict to scalar masses for which all conformal anomalies $-$ and therefore all logarithmic terms  $-$  are absent in both \eqref{FGmetric} and \eqref{FGscalar}.  We will refer to such masses as “$\hat{\mu}$'s without anomalies/logarithms.’’  This assumption is technically convenient because our numerical scheme is based on pseudospectral methods, whose exponential convergence (as the number of grid points increases) holds only for systems without logarithmic contributions.

From a physical standpoint, this restriction is not a limitation: within the double-trace mass range \eqref{2xTrace:rangeMass}, logarithmic terms arise only for a discrete set of values of the scalar mass (\eg $\mu^2 L^2 = -3/4$ and $\mu^2 L^2 = -19/100$). Since the qualitative behaviour of our hairy solutions does not change across this mass range, continuity suggests that the properties of hairy solutions at these very special values of $\hat{\mu}$ are not qualitatively different from those at generic masses, as will be argued further in the conclusions.  This expectation is also supported by the linear analysis of the double-trace instability of AdS$_3$ and BTZ performed in \cite{Dias:2025uyk}, where the case $\mu^2 L^2 = -3/4$ was examined explicitly and found to exhibit instabilities qualitatively similar to those for generic masses in the interval \eqref{2xTrace:rangeMass}.
 
Note that for our hairy solutions the asymptotic FG coefficients $g_{ab}^{(0)}$, $g_{ab}^{(2)}$, $\alpha$, and $\beta$ are constants (\ie independent of $x=\{T,\phi\}$).  The FG map $r = r(z)$ allows us to extract these coefficients from the asymptotic behaviour of the functions in \eqref{HairyAnsatz}; see Appendix~\ref{secA:HoloRen}.  

Using the scaling symmetry \eqref{ScaleSym2}, we set $\underset{R\to\infty}{\lim}\, g = 1$.  Asymptotic AdS$_3$ requires $\underset{R\to\infty}{\lim}\, f = R^2$, and in a frame that does not rotate at infinity one must further impose $\underset{R\to\infty}{\lim}\, \hat{\Omega} = 0$.  With these conditions, and for any $\hat{\mu} \in\, ]-1,0[$ “without anomalies/logarithms,’’ a Frobenius expansion around the asymptotic boundary yields
\begin{align}
\label{UVexpansion}
f \big|_{R\to\infty} &\simeq R^2 + \cdots - C_f + \cdots , \qquad
g \big|_{R\to\infty} \simeq 1 + \cdots , \qquad
\Omega \big|_{R\to\infty} \simeq \frac{\hat{C}_{\Omega}}{R^2} + \cdots , \nonumber \\
\Phi \big|_{R\to\infty} &\simeq \frac{\alpha}{R^{\Delta_-}} + \cdots + \frac{\beta}{R^{\Delta_+}} + \cdots
\end{align}
where $C_f$, $\hat{C}_{\Omega}$, $\alpha$, and $\beta$ are free ultraviolet (UV) parameters, and the ellipses denote terms depending on these UV parameters and, in some cases, also on $m$ and $\hat{\omega}$.  
We further fix $\beta$ by imposing the double-trace boundary condition \eqref{2xTrace:BC}.\footnote{\label{foot:UVexpm15o16}
We will use $\mu^2 L^2 = -15/16$ (\ie $\Delta_- = 3/4$, $\Delta_+ = 5/4$) to illustrate our main results.  
In this case, the expansion \eqref{UVexpansion} takes the explicit form:
\begin{align} \label{UVexpansion:m15o16}
    f \big|_{R \to \infty} \:&= R^2+\frac{3 \alpha ^2}{2}R^{1/2}-C_f+\frac{5 \beta ^2}{2 }R^{-1/2}+\frac{45 \alpha ^4}{16}R^{-1} + \mathcal{O}(R^{-3/2}), 
    \nonumber \\
    g \big|_{R \to \infty} \:&= 1 -\frac{3 \alpha ^2}{2}R^{-3/2}-\frac{15 \alpha  \beta }{4}R^{-2}-\frac{5 \beta ^2}{2}R^{-5/2}-\frac{9 \alpha ^4}{16}R^{-3} + \mathcal{O}(R^{-7/2}),
    \nonumber \\
    \hat{\Omega} \big|_{R \to \infty} \:&= \hat{C}_{\Omega} R^{-2} -\frac{\alpha^2}{21}\big(9\hat{C}_{\Omega} +16 m \hat{\omega}\big)R^{-7/2}
    -\frac{\alpha \beta}{16}\big(15\hat{C}_{\Omega} +16 m \hat{\omega}\big)R^{-4}+ \mathcal{O}(R^{-9/2})\,, 
    \\
\psi \big|_{R \to \infty} \:&= \alpha R^{-3/4}+\beta R^{-5/4} +\frac{3 \alpha ^3}{8}R^{-9/4} +\frac{\alpha  \left(45 \alpha  \beta +9 C_f+16 m^2-16 \hat{\omega}^2\right)}{48}R^{-11/4} + \mathcal{O}(R^{-13/4}).\nonumber
\end{align}
}

The FG coordinate transformation $R = R(z)$ is a function of the UV parameters appearing in \eqref{UVexpansion}.  Up to the order needed to compute the holographic energy, angular momentum, and scalar expectation value, and for any $\hat{\mu} \in\, ]-1,0[$ “without anomalies/logarithms,’’ it is given by
\begin{align}\label{HoloRen:FGgeneral}
T = \frac{t}{L}\,, \qquad
R = \frac{r}{L} = \frac{1}{z} - \frac{1}{2}\,\alpha^2\, z^{1-(\Delta_+ - \Delta_-)} + \frac{1}{4} C_f\, z + \cdots
\end{align}
where $\alpha$ and $C_f$ were introduced in \eqref{UVexpansion}, and $\Delta_\pm$ are defined in \eqref{FGscalar}.
 
\subsection{Boundary conditions at the inner boundary (boson stars and black holes)}\label{sec:BCsIR}

Having discussed the asymptotic boundary conditions (BCs), we now turn to the BCs at the inner boundary. This inner boundary corresponds to the origin of the spacetime, $R=0$, in the case of hairy boson stars, or to the event horizon at $R = R_+$ when considering hairy black holes. Naturally, the analysis of the inner BCs differs between these two situations. We discuss the boson star case in Section~\ref{sec:BCsIRorigin}, followed by the black hole case in Section~\ref{sec:BCsIRhorizon}.

\subsubsection{Regularity of boson stars at the origin}\label{sec:BCsIRorigin}

For boson stars, the inner boundary lies at the origin, $R = 0$.  We are (primarily) interested in regular boson stars $-$ although the system also admits singular yet physically relevant solutions, discussed later $-$ which must be smooth everywhere, including at the origin.  
Under these conditions, we assume a Taylor expansion in powers of $R$ around $R = 0$.  
At leading order, the metric behaves as
\begin{equation}
L^{-2} {\rm d}s^2\big|_{R=0} \sim - f_0\, g_0\, {\rm d}T^2 
    + f_0^{-1}\big( {\rm d}R^2 + f_0 R^2 {\rm d}\phi^2 \big),
\end{equation}
where $f_0 \equiv f(R=0)$ and $g_0 \equiv g(R=0)$. To avoid a conical singularity, we must impose $f_0 = 1$ \cite{Deser:1983tn,Deser:1983nh,Astefanesei:2003qy,Astefanesei:2003rw}.

The discussion of boson star boundary conditions at the origin requires separating the cases $m = 0$ and $m > 0$ (recall that the system is symmetric under $\{m,\hat{\Omega}\} \to \{-m,-\hat{\Omega}\}$).

We begin with the $m = 0$ case. Setting $m = 0$ in \eqref{eqn:Omegaeom} gives $\hat{\Omega}' = c_1 \sqrt{g(R)}/R^3$, with $c_1$ an integration constant. A Taylor expansion of the constraint \eqref{eqn:c1eom} around $R = 0$ forces $c_1 = 0$, implying that $\hat{\Omega}$ is constant. Working in a frame that is non-rotating at infinity then sets $\hat{\Omega} = 0$. Thus, as expected, $m=0$ boson stars (BS) must be static with $SO(2)$ symmetry. A Frobenius analysis of the equations of motion around $R = 0$ yields
\begin{equation}\label{BCs:OriginBSm0}
\begin{cases}
     f\big|_{R \to 0}  = 1+ \left[1-\left(\hat{\mu}^2+\frac{\hat{\omega}^2}{g_0}\right)\psi_0^2\right] R^2  + \mathcal{O}(R^4),  \\ 
     g\big|_{R \to 0} = g_0 + 2 \psi_0^2\, \hat{\omega}^2\, R^2 + \mathcal{O}(R^4),  \\ 
     \psi\big|_{R \to 0} = \psi_0 + \frac{\psi_0}{4}\left(\hat{\mu}^2 - \frac{\hat{\omega}^2}{g_0}\right) R^2+\mathcal{O}(R^4), 
\end{cases}
\qquad \hbox{for regular $m=0$ BS.}
\end{equation}
Here $g_0$ and $\psi_0$ are infrared (IR) free parameters, and $\hat{\omega} = \omega L$ is the boson star frequency. Using these expansions, one finds that the bulk energy-momentum tensor violates the dominant energy condition at $R=0$ when $\hat{\omega}=0$.

We now turn to the $m>0$ case, for which $\hat{\Omega} \neq 0$. A Frobenius analysis of the equations of motion near $R=0$ yields
\begin{equation}\label{BCs:OriginBSm}
\begin{cases}
     f\big|_{R \to 0} = 1 + R^2 - 2m\,\psi_{m}^2 R^{2m} + \mathcal{O}(R^{2m+2}),  \\
     g\big|_{R \to 0} = \tilde{g}_0 + 2 m\, \tilde{g}_0\, \psi_{m}^2 R^{2m} + \mathcal{O}(R^{2m+2}),  \\
     \hat{\Omega}\big|_{R \to 0} = \tilde{\Omega}_0 + \frac{\psi_{m}^2}{1+m}\big(m\tilde{\Omega}_0 - \hat{\omega}\big)R^{2m} + \mathcal{O}(R^{2m+2}),  \\
     \psi\big|_{R \to 0} = \psi_{m} R^{m} + \mathcal{O}(R^{m+2}), 
\end{cases}
\qquad \hbox{for regular $m>0$ BS.}
\end{equation}
Here $\tilde{g}_0$, $\tilde{\Omega}_0$, and $\psi_{m}$ are arbitrary free parameters.  As explained in Section~\ref{sec:NumericalSetup:RegBS}, we use $\psi_{m}$ to label our regular boson star solutions for fixed $m > 0$.
\subsubsection{Boundary conditions at the black hole event horizon}\label{sec:BCsIRhorizon}

In the hairy black hole case, the inner boundary is located at the event horizon, $R = R_+$, whose position is determined by the largest root of the condition $f(R_+) = 0$. Regularity of the remaining functions $g$, $\hat{\Omega}$, and $\psi$ on the horizon hypersurface requires
\begin{align}\label{BCs:horizonBH}
    f \big|_{R \to R_+} &= \mathcal{O}(R - R_+), 
    \qquad g \big|_{R \to R_+} = \mathcal{O}(1), 
    \qquad \Omega \big|_{R \to R_+} = \mathcal{O}(1), 
    \qquad \psi\big|_{R \to R_+} = \mathcal{O}(1).
\end{align}

The condition $f(R_+) = 0$ imposes an additional constraint. Multiplying \eqref{eqn:chieom} by $f^2$ and evaluating it at $R = R_+$ yields the requirement that the scalar field frequency $\hat{\omega}$ must equal $m$ times the angular velocity of the horizon, $\hat{\Omega}_H \equiv \hat{\Omega}(R_+)$:
\begin{align}\label{EoMhorizon:wmOcond}
    \hat{\omega} = m\, \hat{\Omega}(R_+) \equiv m\, \hat{\Omega}_H\,,
\end{align}
a relation that holds for any hairy black hole with $m \geq 0$.

The Killing horizon generator $K$ of the hairy black holes is given by \eqref{HorizonGenerator}. For $m>0$, this helical Killing vector is the \emph{only} Killing vector of the solution (see the discussion around \eqref{BS:KVF}). Thus, our hairy AdS$_3$ black holes with $m>0$ are the three-dimensional analogues of the AdS black resonators constructed in \cite{Dias:2011at,Stotyn:2011ns,Dias:2015rxy,Ishii:2018oms,Ishii:2021xmn}. In particular, $m>0$ AdS$_3$ black resonators are neither time-independent nor axisymmetric; instead, they are time-periodic solutions that can coexist with the stationary BTZ black hole.

On the other hand, \eqref{EoMhorizon:wmOcond} implies that $m=0$ hairy black holes must satisfy $\hat{\omega}=0$. Consequently, $m=0$ hairy black holes remain time-independent and axisymmetric (\ie they are stationary), so both $\partial_T$ and $\partial_\phi$ are Killing vector fields of the ansatz \eqref{HairyAnsatz}. We will see that such $m=0$ hairy black holes also exist when the angular momentum vanishes, \ie there exist static $m=0$ hairy black holes that coexist with the static BTZ black hole.

The asymptotic boundary conditions \eqref{UVexpansion} imply that the norm of the horizon generator behaves as $|K|^2 \to R^2 (\hat{\Omega}_H^2 - 1)$ as $R \to \infty$. We will find that all our hairy black holes satisfy $\Omega_H L < 1$, so they always possess a Killing vector field (for $m>0$ the unique one) that is timelike everywhere. This contrasts sharply with higher-dimensional black resonators, which always satisfy $\Omega_HL > 1$ (at least under Dirichlet boundary conditions) \cite{Dias:2011at,Stotyn:2011ns,Dias:2015rxy,Ishii:2018oms,Ishii:2021xmn}.

\subsection{Thermodynamic quantities and conserved charges}\label{sect:thermodynamics}

Naturally, to discuss the physical properties of our hairy boson stars and black holes, we rely on invariant thermodynamic quantities. In this section we present these quantities.

We begin by recalling that the action \eqref{eqn:action} has a global $U(1)$ symmetry, $\Phi \to e^{i\,\eta}\,\Phi$. The associated conserved current is
\begin{align}
j^\mu = -2 i g^{\mu\nu} \Big(\Phi \nabla_\nu \Phi^{\dagger} - \Phi^{\dagger}\nabla_\nu \Phi \Big).
\end{align}
Using the equations of motion, one verifies that it is conserved, $\nabla_{\mu} j^{\mu}=0$. The corresponding $U(1)$ charge is therefore
\begin{align}\label{N:Def}
  \hat{N}\equiv \frac{8G}{L} \, N 
  = \frac{1}{2}\int_{\Sigma_T} {\rm d}V_\Sigma \, j_\mu n^\mu
  = \int_{R_0}^\infty {\rm d}R \, \frac{4 R \, \psi(R)^2 \big[\hat{\omega} - m\,\hat{\Omega}(R)\big]}{f(R)\sqrt{g(R)}},
\end{align}
where $\Sigma_T$ is a constant-$T$ hypersurface with normal $n^\mu$, and $dV_\Sigma$ is its induced volume element. For hairy black holes one has $R_0 = R_+$, whereas for boson stars $R_0 = 0$. Following the literature \cite{Ruffini:1969qy,Schunck:1996he,Schunck:1999pm}, this quantity $N$ is often interpreted as the “number of scalar particles.’’

For hairy black holes described by the ansatz \eqref{HairyAnsatz}, the horizon is generated by the helical Killing vector field $K$ defined in \eqref{HorizonGenerator}. Their temperature equals the surface gravity divided by $2\pi$, \ie $T_H = \sqrt{- |\nabla K|^2_{R_+}}/(2\sqrt{2}\pi)$, and their Bekenstein-Hawking entropy $S_H$ is the horizon area divided by $4G$:
\begin{align}\label{TS:Def}
 \hat{T}_H \equiv T_H L = \frac{f'(R_+)\sqrt{g(R_+)}}{4\pi},
 \qquad 
 \hat{S}_H \equiv \frac{8G}{L}\,S_H = 4\pi R_+\,.
\end{align}
Boson stars, being horizonless, have zero entropy and no well-defined temperature.

Although $\partial_T$ and $\partial_\phi$ are not Killing fields of the full hairy solutions when $m\neq0$, they remain asymptotic Killing fields because the scalar field decays to zero at large $R$. Thus, we may define conserved charges associated to these asymptotic symmetries. To compute the mass $M$ and angular momentum $J$ of the hairy solutions, we employ either holographic renormalization \cite{Balasubramanian:1999re,deHaro:2000vlm,Skenderis:2002wp,Papadimitriou:2005ii} or the covariant Noether charge formalism \cite{Lee:1990nz,Wald:1993nt,Iyer:1994ys,Wald:1999wa}. Because the literature sometimes contains incorrect expressions, we recompute the charges from first principles in Appendix~\ref{secA:HoloRen} (holographic renormalization) and Appendix~\ref{secA:NoetherFirstLaw} (Noether charges). The two methods agree, and our derivations follow closely the references cited above. We also indicate how these computations generalize to AdS$_{d+1}$ with $d>2$. Here we simply quote the final results.

The mass and angular momentum of the hairy AdS$_3$ solutions are (see \eqref{HoloRen:EnergyAngMomFinal} or \eqref{HoloRen:EnergyAngMomAlternative}):
\begin{align}
   \hat{M} &\equiv 8G\, M = C_f + 4 \alpha \beta, \label{Mass:Def} \\
   \hat{J} &\equiv \frac{8G}{L}\, J = 2 \hat{C}_\Omega\,, \label{AngMom:Def}
\end{align}
where $C_f$, $\hat{C}_\Omega$, $\alpha$, and $\beta$ are the UV coefficients introduced in \eqref{UVexpansion}, and we impose the double-trace condition \eqref{2xTrace:BC}, $\beta=\kappa \alpha$.  
The expectation value of the operator dual to the scalar is (see \eqref{HoloRen:ScalarVEVs})
\begin{align}\label{ScalarVEV}
    \big\langle  \hat{\mathcal{O}}_{\Phi}^{\hbox{\tiny $(\Delta_-)$}}  \big\rangle 
    \equiv 8 G \big\langle  \mathcal{O}_{\Phi}^{\hbox{\tiny $(\Delta_-)$}}  \big\rangle 
    = \frac{1}{2\pi} (\Delta_+ - \Delta_-) \alpha\,.
\end{align}

Any family of solutions with a Killing vector satisfies a first law involving variations of \emph{conserved} charges. Using the covariant Noether formalism, Appendix~\ref{secA:NoetherFirstLaw} shows that our families obey
\begin{subequations}\label{FirstLaw}
\begin{align}
\text{Black hole:} \qquad 
\mathrm{d}\hat{M} &= \hat{T}_H\,\mathrm{d}\hat{S}_H + \hat{\Omega}_H\,\mathrm{d}\hat{J}\,, 
\label{FirstLawBH} \\
\text{Boson star:} \qquad 
\mathrm{d}\hat{M} &= \hat{\omega}\,\mathrm{d}\hat{N}\,. 
\label{FirstLawBStar}
\end{align}
\end{subequations}

Two remarks are in order.

First, for $m\neq0$ the boson star first law \eqref{FirstLawBStar} may be rewritten in the perhaps more familiar form $\mathrm{d}\hat{M} = \frac{\hat{\omega}}{m}\,\mathrm{d}\hat{J}$, using the equation of motion \eqref{eqn:Omegaeom} (see also \eqref{eqn:Omegaeom3}) to show that $\hat{J} = m\,\hat{N}$. However, \eqref{FirstLawBStar} also applies to axisymmetric boson stars with $m=0$.

Second, computing the mass via holographic renormalization requires not only divergent counterterms to make the renormalized action $S_{\text{\tiny ren}}$ finite, but also specific \emph{finite} counterterms needed for a well-posed variational problem under double-trace boundary conditions. Without these finite terms, the resulting “holographic” energy and angular momentum ($\widetilde{E}$, $\widetilde{J}$) are \emph{not} conserved; see Appendices~\ref{secA:HoloRen} and~\ref{secA:NoetherFirstLaw}, especially the discussion below \eqref{HoloRen:conservedCharges}. In that case, the first law would acquire an extra term proportional to $\langle \mathcal{O}_\Phi^{(\Delta_-)} \rangle d\alpha$. Obtaining correct conserved charges is essential for meaningful thermodynamic comparisons (\eg between hairy and BTZ black holes at fixed $(\hat{M},\hat{J})$).

We will compare the thermodynamics of our hairy black holes with that of AdS$_3$ and BTZ. Pure AdS$_3$ satisfies $8GM^{\text{\tiny AdS}} = -1$ and carries no angular momentum or entropy. The BTZ black hole thermodynamics can be described via the quantities 
\begin{align}\label{BTZ:thermo}
   8 G M^{\hbox{\tiny BTZ}} &= R_+^2 + R_-^2\,, 
   \qquad 
   \frac{8 G}{L}\, J^{\hbox{\tiny BTZ}} = 2 R_+ R_-\,,  \\
   \Omega_H^{\hbox{\tiny BTZ}} L &= \frac{R_-}{R_+}\,,\qquad
   T_H^{\hbox{\tiny BTZ}} L = \frac{1}{2\pi}\frac{R_+^2 - R_-^2}{R_+}\,,\qquad
   \frac{8G}{L}\, S_H^{\hbox{\tiny BTZ}} = 4\pi R_+\,, \nonumber
\end{align}
where $R_\pm$ are the two real roots of $f_{\text{\tiny BTZ}}=0$ defined in \eqref{BTZ}, which may be written as
\begin{equation}
R_\pm = \frac{1}{2}\left(\sqrt{\hat{M}+\hat{J}} \pm \sqrt{\hat{M}-\hat{J}}\right)\Big|_{\text{\tiny BTZ}},
\end{equation}
with $\hat{M}^{\text{\tiny BTZ}} = 8GM^{\text{\tiny BTZ}}$ and $\hat{J}^{\text{\tiny BTZ}} = \frac{8G}{L}J^{\text{\tiny BTZ}}$.  
Using this, the BTZ entropy becomes
\begin{align}\label{BTZ:thermoS}
\frac{8G}{L}\, S_H^{\hbox{\tiny BTZ}}
  = 2\pi \sqrt{2}\,\sqrt{\hat{M} + \sqrt{\hat{M}^2 - \hat{J}^2}}\Big|_{\text{\tiny BTZ}},
\end{align}
a relation we will employ later.
\section{Numerical strategy to find hairy \texorpdfstring{AdS$_3$}{AdS3} boson stars and black holes}\label{sec:NumericalSetup}

To find the double-trace hairy AdS$_3$ boson stars and black holes, we must solve the system of coupled nonlinear ODEs \eqref{eqn:feom}-\eqref{eqn:chieom} for $\{\mathfrak{f}\} \equiv \{f,g,\Omega,\psi\}$, subject to the asymptotic boundary conditions discussed in Section~\ref{sec:BCsUV} and the inner boundary conditions described in Section~\ref{sec:BCsIR}. To improve numerical convergence, it is advantageous to perform field redefinitions that factor out the leading behaviour required by the boundary conditions. Concretely, it is often convenient to redefine
\begin{equation}
\mathfrak{f}(R) = \chi_{\mathfrak{f}}(R)\, p_{\mathfrak{f}}(R),
\end{equation}
where $\chi_{\mathfrak{f}}(R)$ encodes the known boundary behaviour and the auxiliary function $p_{\mathfrak{f}}(R)$ is smooth and is the object solved for numerically. In addition, rather than working with the original non-compact radial coordinate $R$, it is convenient to introduce a compact coordinate $y = y(R)$ such that the integration domain is mapped to $y \in [0,1]$.

As explained in Sections~\ref{sec:BCsUV} and~\ref{sec:BCsIR}, the boundary conditions depend on whether we search for boson stars or black holes, on the scalar field mass $\hat{\mu}$, and even on the azimuthal number $m$. Accordingly, we will use different auxiliary functions $p_{\mathfrak{f}}(R)$ depending on the specific type of solution we seek; thus several cases must be treated separately.\footnote{Note, however, that we use the same symbols for the auxiliary functions across these cases for the sake of notational simplicity.}

The analysis depends sensitively on the scalar mass, or equivalently on the conformal dimensions $\Delta_\pm$. For concreteness, throughout this section we present the construction for the scalar mass $\mu^2 L^2 = -15/16$ (\ie $\Delta_- = 3/4,\ \Delta_+ = 5/4$), for which the asymptotic behaviour of $\{f,g,\Omega,\psi\}$ is given in \eqref{UVexpansion:m15o16} of footnote~\ref{foot:UVexpm15o16}. All physical results in the next sections also refer to this mass. Nevertheless, the qualitative physical properties of the system should hold for other masses in the range $-1<\hat{\mu}<0$, as argued in the conclusion section~\ref{sec:Conc}.

To construct the hairy boson stars and black holes, we employ a Newton-Raphson relaxation algorithm combined with spectral collocation methods on a Gauss-Lobatto-Chebyshev grid (reviewed in detail in \cite{Dias:2015nua}). We discretise equations \eqref{eqn:feom}-\eqref{eqn:chieom} and solve the resulting algebraic system. Recall that we have already used the constraint equations \eqref{C1C2} to extract the correct boundary conditions at infinity and at the inner boundary (origin for boson stars, horizon for black holes). Once these constraints are enforced at the boundaries, the evolution of the equations of motion into the bulk ensures that they are satisfied everywhere. We also monitor the residuals of \eqref{C1C2} to assess the accuracy of our numerical solutions.

\subsection{Numerical setup to find regular \texorpdfstring{AdS$_3$}{AdS3} boson stars}\label{sec:NumericalSetup:RegBS}
Boson stars constitute a one–parameter family of solutions. A convenient way to analyse them is to introduce the auxiliary functions
\begin{equation}
f(r)=\bar{g}(r), 
\qquad 
g(r)=\frac{\bar{f}(r)}{\bar{g}(r)}\, .
\end{equation}
We then compactify the radial coordinate $R\in[0,\infty)$ via
\begin{align}\label{regBS:compactCoord}
y=\frac{1}{(1+R^{2})^{1/4}},
\end{align}
so that $y\in[0,1]$, with the origin at $y=1$ and the AdS boundary at $y=0$. As discussed in Section~\ref{sec:BCsIRorigin}, the cases $m=0$ and $m>0$ are qualitatively distinct, and we therefore treat them separately.

For $m=0$ we search for solutions that are axisymmetric in both the scalar and metric sectors. In this case the stress–energy component $T_{t\phi}$ vanishes, so the $t\phi$ component of the Einstein equation can be integrated explicitly to yield
\begin{equation}
\hat{\Omega}
   = C_{0}
     + C_{1}\int \frac{\sqrt{\bar{f}(r)}}{\sqrt{\bar{g}(r)}\, r^{3}}\, \mathrm{d}r,
\end{equation}
with $C_{0}$ and $C_{1}$ integration constants. Regularity at the origin $r=0$ requires $f$ and $g$ to be regular there, which in turn implies that the only way to keep $\hat{\Omega}$ finite is to set $C_{1}=0$. As we shall see later, $C_{1}$ parametrises the angular momentum, so regularity enforces vanishing angular momentum in the $m=0$ sector. The remaining constant $C_{0}$ can be removed using the residual gauge freedom~\eqref{ResidualGaugeFreedom:m0}, allowing us to set
\begin{equation}
\hat{\Omega}=0 \qquad (m=0).
\end{equation}

In this sector it is convenient to introduce the field redefinitions
\begin{align}
\bar{g}(y) &= \frac{1}{y^{4}} + \frac{\tilde{g}(y)}{y}, &
\bar{f}(y) &= \frac{1}{y^{4}} + \frac{\tilde{f}(y)}{y}, &
\psi = y^{3/2}\, \tilde{\psi},
\end{align}
after which we solve for the functions $\tilde{g}(y)$, $\tilde{f}(y)$ and $\tilde{\psi}(y)$, together with the frequency $\hat{\omega}$, which must be nonzero. The boundary conditions \eqref{BCs:OriginBSm0} imply that the scalar field remains finite and non-vanishing at the origin, so the value $\tilde{\psi}(1)\equiv \psi_{0}$ of the scalar at $y=1$ provides a convenient parameter for labelling regular $m=0$ boson stars. Each regular $m=0$ boson star constructed in this way is characterised by a corresponding non-vanishing frequency $\hat{\omega}$.

Regular boson stars with $m \geq 1$ admit a Taylor expansion around the origin $r = 0$, as detailed in \eqref{BCs:OriginBSm}. Unlike the case $m = 0$, the scalar field is forced to vanish at the origin as $R^{m}$, and for this reason the corresponding ansatz must be modified. Moreover, $\hat{\Omega}$ is necessarily non‑vanishing since $T_{t\phi} \neq 0$. In this sector, it is convenient to introduce the field redefinitions
\begin{align}
\bar{g}(y) &= \frac{1}{y^{4}} + \frac{\tilde{g}(y)}{y}, &
\bar{f}(y) &= \frac{1}{y^{4}} + \frac{\tilde{f}(y)}{y}, &
\psi &= y^{3/2}(1-y^4)^{m/2}\tilde{\psi}, &
\hat{\Omega} &= y^4 \tilde{\Omega},
\end{align}
after which we solve for the functions $\tilde{g}(y)$, $\tilde{f}(y)$, $\tilde{\psi}(y)$ and $\tilde{\Omega}(y)$, together with the frequency $\hat{\omega}$. To parametrise our $m\neq 0$ regular boson stars (\eg $m=1$) we use $\tilde{\psi}(1)\equiv \psi_{m}$. Note that here we are working in a gauge where $\tilde{\Omega}$ vanishes at the conformal boundary. The boundary conditions at the origin, $y=1$, are then obtained via \eqref{BCs:OriginBSm}, while at the conformal boundary we use \eqref{UVexpansion}.

\subsection{Numerical setup to find regular \texorpdfstring{AdS$_3$}{AdS3} hairy black holes}\label{sec:NumericalSetup:BHs}

Hairy black holes form a two‑parameter family of solutions, which may conveniently be parametrized by the horizon radius $R_{+}$ and the amplitude $\alpha$ of the leading asymptotic mode of the scalar field. Equivalently, one may use other pairs of parameters, such as the mass and angular momentum $\{8G M,\,8GJ/L\}$, or $\{\alpha,\,8GJ/L\}$. 

For the purpose of numerically exploring the phase space of solutions, we will typically work with the parameter pairs $\{R_{+},\alpha\}$ or $\{\hat{J},\alpha\}$. In the following, we describe in detail how the numerical construction of hairy black hole solutions is implemented.
\subsubsection{Numerical setup to find hairy black holes: parametrization \texorpdfstring{$\{ R_+,\alpha \}$}{\{R+, alpha\}}}\label{sec:NumericalSetup:BHs1}

We begin by compactifying the radial coordinate $R\in[R_{+},\infty)$ according to
\begin{align}
y = 1 - \frac{\sqrt{R_{+}}}{\sqrt{R}},
\label{eq:compactBH}
\end{align}
so that $y\in[0,1]$, with the event horizon located at $y=0$ and the AdS boundary at $y=1$.
Hairy AdS$_3$ black holes must satisfy the horizon boundary conditions~\eqref{BCs:horizonBH} and the asymptotic boundary conditions~\eqref{UVexpansion} (see also the discussion for $\mu^{2}L^{2}=-15/16$ in~\eqref{UVexpansion:m15o16}). Moreover, evaluating the equations of motion at the horizon requires the condition~\eqref{EoMhorizon:wmOcond} to hold, namely that the scalar field frequency and the horizon angular velocity satisfy
\begin{equation}
\hat{\omega} = m\,\hat{\Omega}_{H},
\end{equation}
for any value of $m$.

We make use of the residual gauge freedom~\eqref{ResidualGaugeFreedom} to set $\hat{\omega}=0$, which in turn enforces
\begin{equation}
\hat{\Omega}\big|_{y=1} = -\hat{\omega} = -m\,\hat{\Omega}_{H}.
\end{equation}
In practice, for a given value of $m$ we extract $\hat{\omega}$ and $\hat{\Omega}_{H}$ from the asymptotic value of the function $\hat{\Omega}(y)$ obtained in our numerical solutions. This procedure applies to all $m\geq0$. Note, however, that for $m=0$ one necessarily has $\hat{\omega}=0$, while $\hat{\Omega}_{H}$ may still be nonzero. In addition, typically, only in the $m=0$ case can one obtain static hairy black holes with $\hat{\Omega}(y)=0$, which we will also encounter. There are, however, notable exceptions; see the discussion of the $m=1$ static black holes in the bottom panel of Fig.~\ref{fig:m1TotalPhaseDiag-3k}.

For any $m\geq0$, we implement the boundary conditions~\eqref{UVexpansion:m15o16} and~\eqref{BCs:horizonBH} by introducing the auxiliary fields $\{q_f, q_g, q_\Omega, q_\psi\}$ via
\begin{align}
f(R) &= (1+R^{2})\left(1-\frac{\sqrt{R_{+}}}{\sqrt{R}}\right) q_f(R),
&
g(R) &= 1 + \frac{\sqrt{R_{+}}}{\sqrt{R}}\, q_g(R), \nonumber\\
\hat{\Omega}(R) &= \left(1-\frac{\sqrt{R_{+}}}{\sqrt{R}}\right) q_\Omega(R),
&
\psi(R) &= \frac{\alpha\, R}{(R+1)^{7/4}}
          + \frac{R}{(R+1)^{9/4}} q_\psi(R).
\end{align}
Substituting these field redefinitions, together with the compact radial coordinate, into the equations of motion~\eqref{eqn:feom}–\eqref{eqn:chieom} yields a system of equations for $\{q_f(y), q_g(y), q_\Omega(y), q_\psi(y)\}$. The associated horizon and asymptotic boundary conditions are now simply the conditions obtained by evaluating the equations of motion at $y=0$ (the horizon) and $y=1$ (the AdS$_3$ boundary), after imposing the double‑trace boundary condition $\beta=\kappa\alpha$ for a given value of~$\kappa$.

Fixing $\kappa$ and $m$, once an initial hairy black hole solution has been obtained, it is straightforward to generate families of solutions at constant $R_{+}$ or constant $\alpha$ by marching in $\alpha$ and/or $R_{+}$ using a Newton–Raphson algorithm. To obtain the initial solution, we recall that in the phase diagram hairy black holes are expected to merge with (or bifurcate from) BTZ black holes at the onset of the double‑trace $m$‑instability. This onset curve for BTZ black holes, expressed as $\hat{J}=\hat{J}(\hat{M})$ or equivalently $\hat{\Omega}_{H}(R_{+})$, was determined for each choice of $\hat{\mu}$, $\kappa$, and $m$ in our companion paper~\cite{Dias:2025uyk}. Starting from one of these BTZ solutions and introducing a small but finite scalar field provides a suitable seed for the Newton–Raphson algorithm to converge to a hairy black hole with small $\alpha$ (and correspondingly small $\beta=\kappa\alpha$). From this initial configuration, we can then efficiently march in $\alpha$ or $R_{+}$ to construct the full two‑parameter family $\{\hat{J},\hat{M}\}$ of hairy black holes for fixed $\hat{\mu}$, $\kappa$, and $m$.

\subsubsection{Numerical setup to find hairy black holes: parametrization \texorpdfstring{$\{ \hat{J}, \alpha \}$}{\{Jhat, alpha\}}}\label{sec:NumericalSetup:BHs2}

To study the properties of hairy black holes in the zero‑size limit ($R_{+}\to 0$), it is convenient to construct a numerical scheme that searches for solutions at fixed angular momentum. For this purpose, instead of the formulation used above, we introduce the following field redefinitions:
\begin{equation}
\bar{f} = (R^{2}-R_{+}^{2})\,\tilde{f},
\qquad
\bar{g} = (R^{2}-R_{+}^{2})\,\tilde{g},
\qquad
\psi = \left(\frac{R_{+}}{R}\right)^{3/4}\tilde{\psi},
\qquad
\hat{\Omega} = \left(\frac{R_{+}}{R}\right)^{2}\tilde{\Omega}.
\end{equation}
As before, we define
\begin{equation}
f(R)=\bar{g}(R),
\qquad
g(R)=\frac{\bar{f}(R)}{\bar{g}(R)} \, .
\end{equation}
We again employ the compact radial coordinate introduced in \eqref{eq:compactBH}. In terms of the variables above, the boundary conditions at the asymptotic AdS$_3$ boundary ($y=1$) read
\begin{equation}
\tilde{f}(1)=\tilde{g}(1)=1,
\qquad
\tilde{\psi}'(1) + \frac{\kappa}{\sqrt{R_{+}}}\tilde{\psi}(1)=0,
\qquad
\tilde{\Omega}'(1)=0,
\end{equation}
while at the black hole event horizon ($y=0$) we obtain three Robin boundary conditions for $\tilde{f}$, $\tilde{g}$, and $\tilde{\psi}$, together with
\begin{equation}
\tilde{\Omega}(0)= m\,\hat{\omega} \quad\text{for}\quad |m|>0.
\label{eq:conomega}
\end{equation}

To explore the moduli space of solutions for $|m|>0$ at fixed $\kappa$, we use $\hat{J}=2R_{+}^{2}\tilde{\Omega}(1)$ together with $R_{+}$ as control parameters. For fixed values of $\hat{J}$, $\kappa$, and $R_{+}$, we then solve for the functions $\tilde{f}$, $\tilde{g}$, $\tilde{\psi}$, and $\tilde{\Omega}$, as well as for the frequency $\hat{\omega}$.

For $m=0$, we instead fix $\hat{\omega}=0$ and search for the value of $\tilde{\Omega}(0)$ for which a smooth solution exists. Note that in this case the constraint appearing in \eqref{eq:conomega} is absent.

\subsection{\label{sec:NumericalSetup:SingBS} Finding singular black holes at the boundary of moduli space}
The moduli space of regular black holes minimally coupled to a massive complex scalar field with AdS$_3$ asymptotics forms an open subset of the $(\hat M,\hat J)$ plane and were discussed in the previous subsection~\ref{sec:NumericalSetup:BHs}. The (topological) boundary of this set is given by singular black hole solutions, which we describe in this subsection. In what follows, three distinct classes of singular black hole solutions will be relevant: solutions with $m=0$ and $\hat J=0$ (subsection~\ref{sec:NumericalSetup:SingBHm0J0}); solutions with $m=0$ and $\hat J\neq 0$ (subsection~\ref{sec:NumericalSetup:singBHm0J}); and solutions with $|m|\geq 1$ and $\hat J\neq 0$ (subsection~\ref{sec:NumericalSetup:singBHmJ}).

\subsubsection{\label{sec:NumericalSetup:SingBHm0J0} Numerical setup to find singular black holes with \texorpdfstring{$m=\hat{J}=0$}{m=8GJ/L=0}}
This section is devoted to the construction of singular static hairy black holes that realise the minimum-energy configurations among black holes with scalar hair and zero angular momentum, and that saturate the positivity of energy theorem~\eqref{GlobalMin} of Appendix~\ref{secA:superpotentials} when $\hat J=0$. These solutions are characterised by $\hat J = m = 0$.

Since $\hat J = m = 0$, it follows that $\hat\Omega = 0$ (see discussion around~\eqref{eq:integrateOmega} for a detailed justification). Numerical evidence indicates that the corresponding singular solution is boost invariant in the $(t,\phi)$ directions, with
\begin{equation}
g_{\phi\phi} = - g_{tt} = r^{2},
\end{equation}
in close analogy with the boost‑invariant solutions discussed in section 4.2 of \cite{Horowitz:2009ij}. Motivated by this observation, we adopt the metric ansatz
\begin{equation}
\mathrm{d}s^{2}
   = r^{2}\left(-\mathrm{d}t^{2} + \mathrm{d}\phi^{2}\right)
     + \frac{\mathrm{d}r^{2}}{g(r)} \, .
\end{equation}
The Einstein equations then fix the metric function to be
\begin{equation}
g(r)
   = \frac{r^{2}}{L^{2}}
     \frac{1 - L^{2}\mu^{2}\psi(r)^{2}}
          {1 - r^{2}\psi'(r)^{2}} \, .
\end{equation}
The system therefore reduces to a single nonlinear second‑order equation for the scalar field profile $\psi(r)$, which takes the form
\begin{multline}
L^{2}\mu^{2}\psi(r)
   - 3 r \left[1 - L^{2}\mu^{2}\psi(r)^{2}\right]\psi'(r)
   - L^{2} r^{2} \mu^{2} \psi(r)\psi'(r)^{2}
\\
   + 2 r^{3} \left[1 - L^{2}\mu^{2}\psi(r)^{2}\right]\psi'(r)^{3}
   - r^{2} \left[1 - L^{2}\mu^{2}\psi(r)^{2}\right]\psi''(r)
   = 0 \, .
   \label{eq:psir}
\end{multline}
To proceed, we introduce the auxiliary variable
\begin{equation}
X \equiv -\log r,
\end{equation}
which simplifies the equation of motion for $\psi$ to
\begin{multline}
2 \psi'(X) - 2 \psi'(X)^{3}
+ (\Delta - 2)\Delta\, \psi(X)\left[1 - \psi'(X)^{2}\right]
- \psi''(X)
\\
- (\Delta - 2)\Delta\, \psi(X)^{2}
\left[2 \psi'(X) - 2 \psi'(X)^{3} - \psi''(X)\right]
= 0 \, ,
\label{eq:psiX}
\end{multline}
where we have used \eqref{eq:Delta}. We are interested in the behaviour of solutions in the limit $X \to +\infty$. In this regime, the equation admits two linearly independent asymptotic behaviours, $\psi \sim X$ and $\psi \sim X^{1/2}$. The former corresponds to a highly singular geometry and does not appear to play a role in the moduli space of physically relevant solutions. We therefore focus on the latter case.

Assuming $\psi \sim X^{1/2}$ at large $X$, we find a consistent asymptotic expansion of the form
\begin{equation}
\psi(X)
   = X^{1/2}
     \left(
        1
        + \sum_{i=1}^{\infty}
          \sum_{j=0}^{i}
            \frac{a_{ij}}{X^{i}}
            \log^{j} X
     \right).
\end{equation}
Since we are solving a second‑order differential equation and have already fixed one boundary condition by eliminating the $\psi\sim X$ branch, the expansion is expected to contain a single free parameter. Indeed, this remaining freedom is parametrised precisely by the coefficient $a_{10}$. The first few coefficients in the expansion are given by
\begin{subequations}
\begin{align}
& a_{11}= \frac{1}{2}\left[\frac{1}{(\Delta - 2)\Delta}- \frac{1}{4}\right],\\
& a_{22}= -\frac{1}{8}\left[\frac{1}{4}+ \frac{1}{(2 - \Delta)\Delta}\right]^{2},
\\
& a_{21}= \frac{1}{2}\left[\frac{1}{4}+ \frac{1}{(2 - \Delta)\Delta}\right]\left[\frac{1}{4}+ \frac{1}{(2 - \Delta)\Delta}+ a_{10}\right],
\\
& a_{20}= -\frac{1}{16}\left[\frac{4 + 16a_{10}}{(2 - \Delta)\Delta}+ 1+ 4a_{10}+ 8a_{10}^{2}\right].
\end{align}
\end{subequations}
Using the asymptotic expansion derived above, we can determine the behaviour of the metric function $g(r)$ near the origin. We find
\begin{equation}
g(r) \sim \frac{r^{2}}{L^{2}} (2-\Delta)\Delta\,(-\log r),
\end{equation}
which corresponds to a very mild null singularity~\cite{Horowitz:2009ij}.

The strategy is now clear. Solutions of \eqref{eq:psiX} satisfying our chosen boundary conditions as $X\to+\infty$ form a one‑parameter family, labeled by the coefficient $a_{10}$. For each value of $a_{10}$, we integrate \eqref{eq:psir} outward toward large $r$. In the asymptotic region $r\to+\infty$, the scalar field takes the Fefferman-Graham form given in \eqref{FGscalar}, with coefficients $\alpha$ and $\beta$ determined by the chosen value of $a_{10}$. Taking the ratio $\beta/\alpha$ then fixes the corresponding value of the coupling $\kappa$. Hence, for a given $\kappa$, the desired solution is obtained by tuning $a_{10}$ appropriately.

The singular $m=0$ static extremal hairy black hole of this subsection will always be represented by a red diamond in subsequent plots (\eg Figs.~\ref{fig:BS-m0},~\ref{fig:m0_J0-BH},~\ref{fig:dMJ:3families},~\ref{fig:m0TotalPhaseDiag-k04},~\ref{fig:m0_kappa_m4o10_DeltaF_DeltaG}, and~\ref{fig:m0TotalPhaseDiag-3k}). 
For reference, for $\hat{\mu}^2=-15/16$ and $\kappa=-4/10$, the thermodynamic quantities of this solution are given in \eqref{sBHm0J0:Thermo}.
\subsubsection{Numerical setup for singular, extremal hairy black holes with \texorpdfstring{$m=0$ and $\hat{J}\neq 0$}{m=0}}\label{sec:NumericalSetup:singBHm0J}

The singular solutions discussed in this section saturate the positivity-of-energy theorem \eqref{GlobalMin} derived in Appendix~\ref{secA:superpotentials} for $\hat{J}\neq 0$. They still have $m=0$, and, in a phase diagram, these solutions should reduce to those of the previous subsection~\ref{sec:NumericalSetup:SingBHm0J0} in the limit $\hat J \to 0$. To construct these solutions, we begin by introducing an alternative method for generating configurations with  $m=0$. In this case, the $t\phi$ component of the Einstein equations can be integrated directly, yielding
\begin{equation}
\hat{\Omega}(R)
   = C_{0}\int_{+\infty}^{R}
        \frac{e^{-\chi(\tilde{R})/2}}{\tilde{R}^{3}}
        \,\mathrm{d}\tilde{R}
     + C_{1},
     \label{eq:integrateOmega}
\end{equation}
where $C_{0}$ and $C_{1}$ are integration constants, and we have defined $g \equiv e^{-\chi}$. As before, the constant $C_{1}$ can be removed using the residual gauge freedom~\eqref{ResidualGaugeFreedom:m0}. The remaining constant $C_{0}$ coincides with the conserved angular momentum $\hat{J}$.

Since $m=0$, the Einstein–scalar system depends only on $\hat{\Omega}'(R)$. Substituting the expression above into the remaining field equations therefore reduces the system to a first‑order equation for $f$ and $\chi$, together with a second‑order equation for the scalar field $\psi$. Moreover, $\chi$ does not enter explicitly in the equations governing $f$ and $\psi$, so these two functions can be solved for independently, after which $\chi$ may be obtained a posteriori. We are thus left with a coupled nonlinear system for $f(R)$ and $\psi(R)$ of the form
\begin{subequations}
\begin{align}
& f'(R)
   + \frac{C_{0}^{2}}{2R^{3}}
   - 2R
   + 2R(\Delta-2)\Delta\,\psi(R)^{2}
   + 2R f(R)\psi'(R)^{2}
   = 0,
\\
& \frac{(2-\Delta)\Delta\,\psi(R)}{f(R)}
  + \left[\frac{1}{R} + \frac{2R}{f(R)}\right]\psi'(R)
  - \frac{C_{0}^{2}\psi'(R)}{2R^{3}f(R)}
  + \frac{2R(2-\Delta)\Delta\,\psi(R)^{2}\psi'(R)}{f(R)}
  + \psi''(R)
  = 0,
\end{align}
where for the moment we keep $\Delta$ arbitrary, and will later specialise to $\Delta=3/4$. For completeness, the equation determining $\chi$ is
\begin{equation}
\chi'(R) = -4R\,\psi'(R)^{2}.
\end{equation}
\end{subequations}
One may use the equation for $\psi''(R)$ to solve algebraically for $f(R)$ as a function of $\psi(R)$, $\psi'(R)$, and $\psi''(R)$. Substituting this expression into the equation for $f'(R)$ then yields a third‑order differential equation for $\psi(R)$, whose behaviour we wish to analyse in the limit $R\to0$. Being a third‑order equation, the resulting equation for $\psi(R)$ admits three independent singular behaviours as $R\to0$. We find that $\psi(R)$ admits an expansion of the form
\begin{multline}
\psi(R)
   = \frac{C_{0}}{2\sqrt{\Delta(2-\Delta)}\,R^{2}}
     - \frac{4-(2-\Delta)\Delta}{4C_{0}\sqrt{(2-\Delta)\Delta}}\,R^{2}
     - \frac{16 + (2-\Delta)\Delta\,[8 - 7(2-\Delta)\Delta]}
            {16C_{0}^{3}\sqrt{(2-\Delta)\Delta}}\,R^{6}
\\
     - \frac{\bigl[4-(2-\Delta)\Delta\bigr]
              \bigl\{16+(2-\Delta)\Delta\,[8+33(2-\Delta)\Delta]\bigr\}}
            {32C_{0}^{5}\sqrt{(2-\Delta)\Delta}}\,R^{10}
     + \mathcal{O}(R^{14})
     + \delta\psi(R),
\end{multline}
where
\begin{equation}
\delta\psi(R)
   = A_{0}\exp\!\left[-\frac{C_{0}^{2}}
      {2(2-\Delta)\Delta\,R^{4}}\right] r^{\frac{2 [4-(2-\Delta ) \Delta ]}{(2-\Delta ) \Delta }},
\end{equation}
and $A_{0}$ is a constant. In deriving this expansion, we have fixed \emph{two} integration constants corresponding to branches that would otherwise lead to exponentially growing behaviour of $\delta\psi(R)$ as $R\to0$. The integration procedure is now straightforward. We fix the parameter $C_{0}$ (equivalently, the angular momentum $\hat{J}$) and choose a value for $A_{0}$. We then integrate the equations outward to large $R$, where both asymptotic coefficients $\alpha$ and $\beta$ can be extracted. Taking the ratio $\beta/\alpha$ subsequently determines the corresponding value of the coupling $\kappa$. It remains only to specify a convenient set of field redefinitions that implement this procedure in practice. From the near‑origin expansion derived above, we find in particular that
\begin{equation}
f(R)
   = \frac{(2-\Delta)^{2}\Delta^{2}}{4C_{0}^{2}}\,R^{6}
     + \mathcal{O}(R^{10}) \, .
\end{equation}
Motivated by this behaviour, we introduce the redefinitions
\begin{equation}
f(R)
   = \frac{R^{6}}{(1+R^{2})^{2}}
     \left[1+\frac{1}{(1+R^{2})^{3/4}}\tilde{f}(R)\right],
\qquad
\psi(R)
   = \left(1+\frac{1}{R^{2}}\right)
     \left(\frac{1}{1+R^{2}}\right)^{\Delta/2}
     \tilde{\psi}(R) \, .
\end{equation}
We now specialize to $\Delta=3/4$. As before, we employ the compact radial coordinate defined in \eqref{regBS:compactCoord}. The boundary conditions at $y=1$ (corresponding to $R=0$) are
\begin{equation}
\tilde{f}(1)=\frac{225-1024C_{0}^{2}}{1024C_{0}^{2}},
\qquad
\tilde{\psi}(1)=\frac{2C_{0}}{\sqrt{15}} \, .
\end{equation}
At the conformal boundary, located at $y=0$, we impose
\begin{equation}
\tilde{f}(0)=\frac{3}{2}\,\tilde{\psi}(0)^{2},
\qquad
\tilde{\psi}'(0)-\kappa\,\tilde{\psi}(0)=0 \, .
\end{equation}
Finally, we note that the singularity at $R=0$ is again null, see \eg~\cite{Horowitz:2009ij}, with the associated Kretschmann scalar diverging as $R^{-8}$ in the limit of small $R$.

The singular $m=0$ rotating extremal hairy black holes of this subsection will always be represented by dark-blue triangles (which emanate from the static red diamond of the previous subsection) in subsequent plots (\eg Figs.~\ref{fig:dMJ:3families},~\ref{fig:m0-hBTZ-J3},~\ref{fig:m0TotalPhaseDiag-k04},~\ref{fig:m0_kappa_m4o10_DeltaF_DeltaG},~\ref{fig:m0TotalPhaseDiag-3k}, and Fig.~\ref{fig:m0m1FinalPhaseDiag-3k}). 
For reference, for $\hat{\mu}^2=-15/16$ and $\kappa=-4/10$, the thermodynamic quantities of this solution are given in \eqref{sBHm0J:Thermo}, and it reduces to \eqref{sBHm0J0:Thermo} in the $\hat J \to 0$ limit.

\subsubsection{Numerical setup for singular hairy black holes with \texorpdfstring{$m\neq0$}{m!=0} and \texorpdfstring{$\hat{J}\neq0$}{8GJ/L!=0}}\label{sec:NumericalSetup:singBHmJ}

For $|m|\geq 1$, we must necessarily have $\hat{\Omega}\neq0$. Guided by the finite‑temperature analysis  discussed in subsection~\ref{sec:NumericalSetup:BHs}, we find that near the origin the spacetime approaches that of an $M=J=0$ BTZ black hole, namely
\begin{equation}
\mathrm{d}s^{2}
   = - a_{0}\,\frac{r^{2}}{L^{2}}\,\mathrm{d}t^{2}
     + \frac{L^{2}}{r^{2}}\,\mathrm{d}r^{2}
     + r^{2}\left(\mathrm{d}\phi + \Omega_{0}\,\mathrm{d}t\right)^{2},
\end{equation}
where $a_{0}>0$ and $\Omega_{0}$ are real constants. Although this geometry is locally AdS$_3$, it is globally singular: the angular identification introduces a fixed point at $r=0$, resulting in an orbifold (conical) singularity.

On this background, the scalar field equation can be solved exactly. One finds
\begin{equation}
\psi(\tilde{y})
   = \tilde{y}
     \left[
        C_{0}\,K_{\Delta-1}(\tilde{y})
        + C_{1}\,I_{\Delta-1}(\tilde{y})
     \right],
\end{equation}
where $C_{0}$ and $C_{1}$ are integration constants, and $I_{\nu}$ and $K_{\nu}$ denote the modified Bessel functions of the first and second kind, respectively. The argument $\tilde{y}$ is defined as
\begin{equation}
\tilde{y}
   \equiv \frac{L}{r}\,\sqrt{\gamma_{0}},
\qquad
\gamma_{0}
   \equiv m^{2}
        - \frac{L^{2}\left(\omega - m\Omega_{0}\right)^{2}}{a_{0}} \, .
\end{equation}

Regularity of the scalar field at the origin, corresponding to $\tilde{y}\to+\infty$, singles out a unique branch of solutions. Since $I_{\Delta-1}(\tilde{y})$ grows exponentially at large $\tilde{y}$, regularity forces $C_{1}=0$. The remaining solution involving $K_{\Delta-1}(\tilde{y})$ decays exponentially and therefore yields the physically relevant branch.

Having established the appropriate boundary conditions at $r=0$, we now turn to the numerical construction of the corresponding solutions. We once again employ the compact radial coordinate defined in \eqref{regBS:compactCoord}, together with the following field redefinitions:
\begin{align}
\bar{g}(y) &= (1-y^{4})\left[\frac{1}{y^{4}} + \frac{\tilde{g}(y)}{y}\right], &
\bar{f}(y) &= (1-y^{4})\left[\frac{1}{y^{4}} + \frac{\tilde{f}(y)}{y}\right], &
\psi &= y^{3/2}\tilde{\psi}, &
\hat{\Omega} &= y^{4}\tilde{\Omega}.
\end{align}
With these definitions, the boundary conditions at the origin, located at $y=1$, take the simple form
\begin{equation}
\tilde{f}'(1) + 3\tilde{f}(1) = 0,
\qquad
\tilde{g}(1) = 0,
\qquad
\tilde{\psi}(1) = 0,
\qquad
\tilde{\Omega}'(1) + 4\tilde{\Omega}(1) = 0.
\end{equation}
At the conformal boundary, located at $y=0$, we impose
\begin{equation}
\tilde{f}(0) = 0,
\qquad
\tilde{g}(0) = \frac{3}{2}\tilde{\psi}(0)^{2},
\qquad
\tilde{\psi}'(0) - \kappa\,\tilde{\psi}(0) = 0,
\qquad
\tilde{\Omega}'(0) = 0.
\end{equation}
To generate solutions, we use $\tilde{\psi}(0)$ as the parameter at fixed $\kappa$, thereby obtaining a one‑parameter family of solutions for each value of $\kappa$. The angular momentum is then determined uniquely as a function of $\tilde{\psi}(0)$.

The singular $m=1$ rotating extremal hairy black holes of this subsection will always be represented by dark-red triangles in subsequent plots (\eg Figs.~\ref{fig:BS-m1},~\ref{fig:m1:dMJ:3families}~and~\ref{fig:m1TotalPhaseDiag-3k}). 

\section{Exploratory study of \texorpdfstring{AdS$_3$}{AdS3} hairy solutions with \texorpdfstring{$m=0$}{m=0}}\label{sec:PhaseDiag-m0}
In this section, we study AdS$_3$ Einstein–scalar field theory with double‑trace boundary conditions and analyse the phase diagram of asymptotically AdS$_3$ stationary solutions with $m=0$. Well‑known solutions of this theory include static and rotating BTZ black holes with dimensionless mass $\hat{M}>0$ and angular momentum $\hat{J}\geq 0$, satisfying $\hat{M}\geq \hat{J}L$, as well as global AdS$_3$, which corresponds to $\hat{M}=-1$ and $\hat{J}=0$. Recall that AdS$_3$ exhibits a mass gap relative to the singular vacuum BTZ geometry with $\hat{M}=0=\hat{J}$.

All of these solutions have a vanishing scalar field. Nonetheless, the theory also admits boson stars and black holes endowed with non‑trivial scalar hair obeying double‑trace boundary conditions, which we will present below. The main qualitative features of the resulting phase diagram are largely insensitive to the precise value of the scalar mass $\hat{\mu}=\mu L$. For definiteness, we focus on the case $\mu^{2}L^{2}=-15/16$, which lies within the range
\begin{equation}
-1 < \mu^{2}L^{2} < 0,
\end{equation}
where double‑trace boundary conditions of the form~\eqref{2xTrace:BC} are allowed.

Recall that the relation $\beta=\kappa\,\alpha$, with $\kappa\in\mathbb{R}$, interpolates between Neumann boundary conditions ($\kappa=0$) and Dirichlet boundary conditions ($\kappa\to\pm\infty$). While boson stars exist for both Neumann and Dirichlet boundary conditions, hairy black hole solutions do not. A discussion of boson stars with Dirichlet and Neumann boundary conditions is deferred to Appendix~\ref{secA:BStars-DNm0}.

\subsection{Hairy \texorpdfstring{AdS$_3$}{AdS3} boson stars with \texorpdfstring{$m=0$}{m=0}}\label{sec:PhaseDiag-m0:BStar}
Before embarking on the construction of (regular) hairy boson stars, it is useful to recall some background that will be important both for guiding our search and for interpreting the resulting solutions. Ishibashi and Wald~\cite{Ishibashi:2004wx} famously showed that, for any $d\geq3$, AdS$_d$ is stable under scalar perturbations with Dirichlet or Neumann boundary conditions, but can become unstable under certain mixed (Robin) boundary conditions when the scalar mass lies between the Breitenlohner–Freedman and unitarity bounds. This is precisely the situation relevant for scalar fields with double‑trace boundary conditions in the mass range~\eqref{2xTrace:rangeMass}, which includes the AdS$_3$ setup studied here.

Ishibashi and Wald~\cite{Ishibashi:2004wx} established the existence and onset conditions for this instability, while in our companion paper~\cite{Dias:2025uyk} we determined its characteristic time scale. In particular, AdS$_3$ is unstable whenever $\kappa < \kappa^{\rm AdS}_{m,\hat{\mu}^2}$ and linearly stable otherwise. For a given scalar mass $\hat{\mu}$ and azimuthal number $m$, the critical onset value $\kappa^{\rm AdS}_{m,\hat{\mu}^2}$ is negative and given by
\begin{align}\label{kAdS-onset}
\kappa^{\rm AdS}_{m,\hat{\mu}^2}
 = \frac{\Gamma\!\left(-\sqrt{1+\mu^2L^2}\right)}
        {\Gamma\!\left(\sqrt{1+\mu^2L^2}\right)}
   \frac{\Gamma\!\left(\tfrac{1}{2}(\Delta_{+}+m)\right)^{2}}
        {\Gamma\!\left(\tfrac{1}{2}(\Delta_{-}+m)\right)^{2}} .
\end{align}
For instance, for $\mu^2L^2=-15/16$ and $m=0$, this yields
$\kappa^{\rm AdS}_{0,-15/16}\simeq -0.4951294$.

A natural question we wish to address is what happens when $\kappa$ crosses $\kappa^{\rm AdS}_{m,\hat{\mu}^2}$, and in particular whether unstable AdS$_3$ evolves toward a new stationary configuration with scalar hair.

With this motivation in mind, we begin by discussing $m=0$ hairy boson stars that are everywhere regular when $\kappa > \kappa^{\rm AdS}_{m,\hat{\mu}^2}$. The numerical strategy used to construct these solutions was described in Section~\ref{sec:NumericalSetup:RegBS}. For $\kappa>\kappa^{\rm AdS}_{m,\hat{\mu}^2}$, the double‑trace normal modes of AdS$_3$ were computed in~\cite{Dias:2025uyk} by solving the linear Klein–Gordon equation with the boundary condition~\eqref{2xTrace:BC}. These modes have purely real frequencies (for example, in the case $\mu^2L^2=-15/16$ their dependence on $\kappa$ is shown in Fig.~1 of~\cite{Dias:2025uyk}). Beyond linear order, once backreaction is included, one expects these perturbative modes to continue into fully nonlinear, horizonless, and everywhere regular solutions - namely, boson stars.

\begin{figure}
    \centering
    \includegraphics[width=0.45\linewidth]{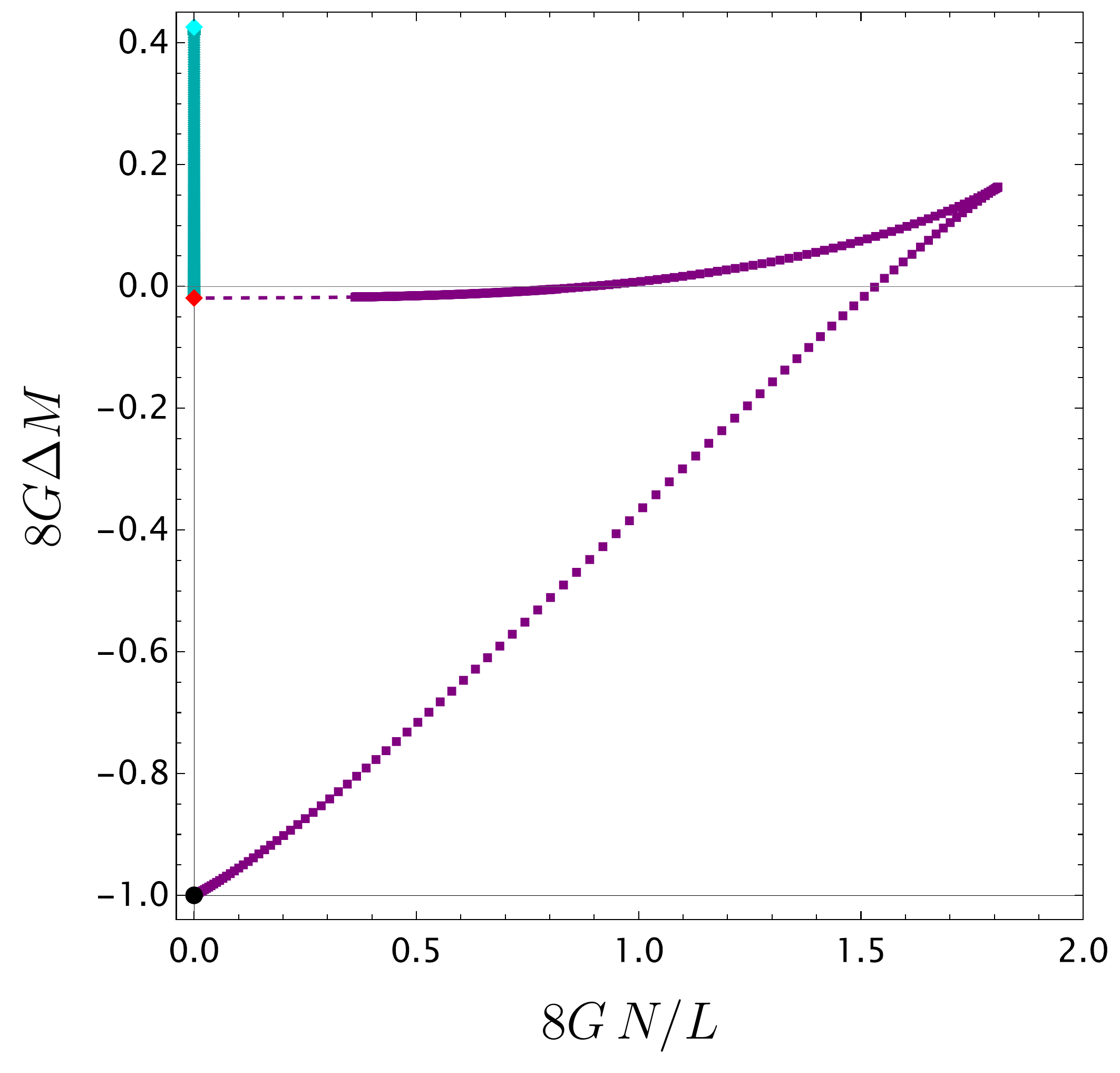}
    \includegraphics[width=0.45\linewidth]{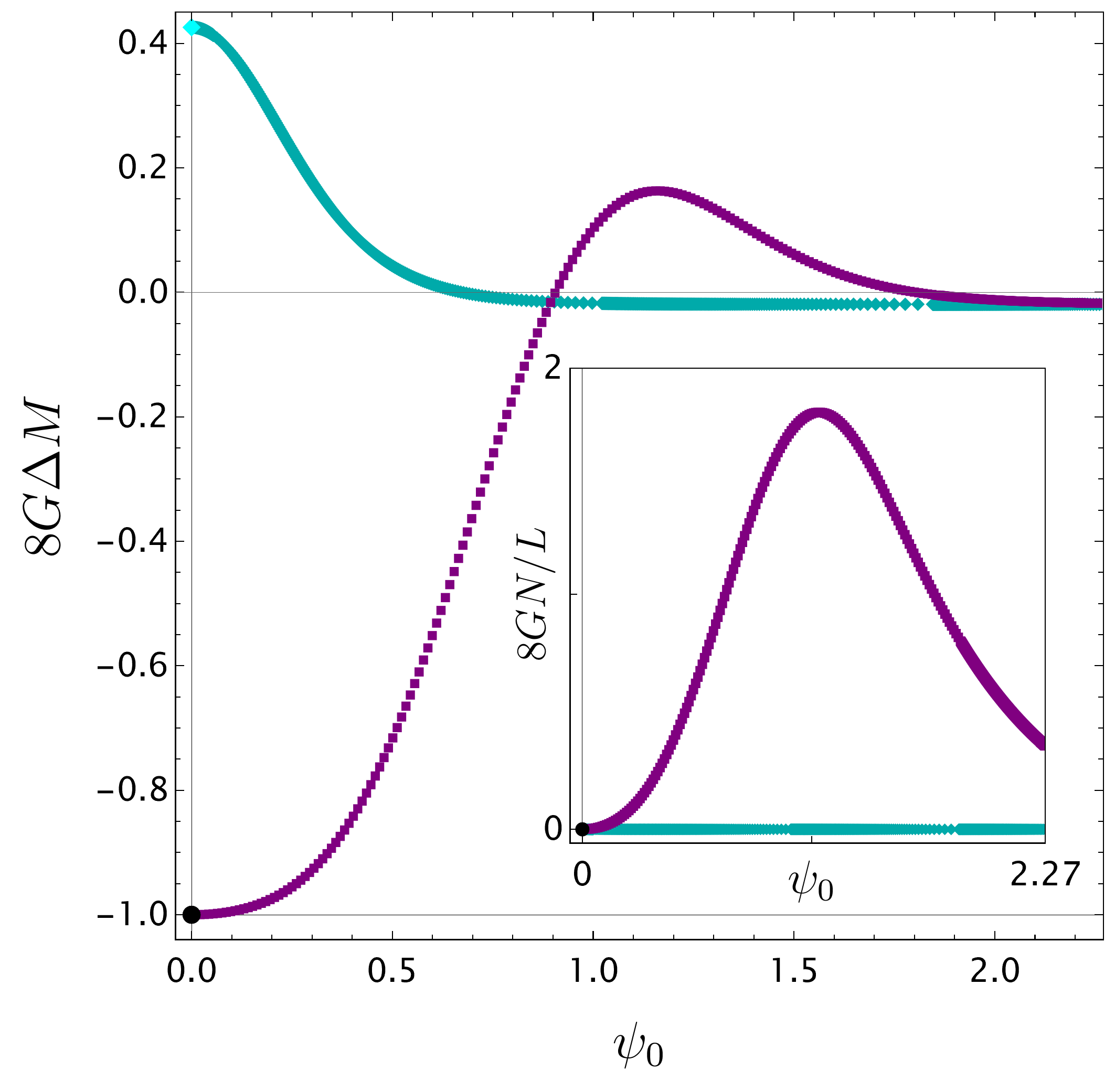}
    \includegraphics[width=0.45\linewidth]{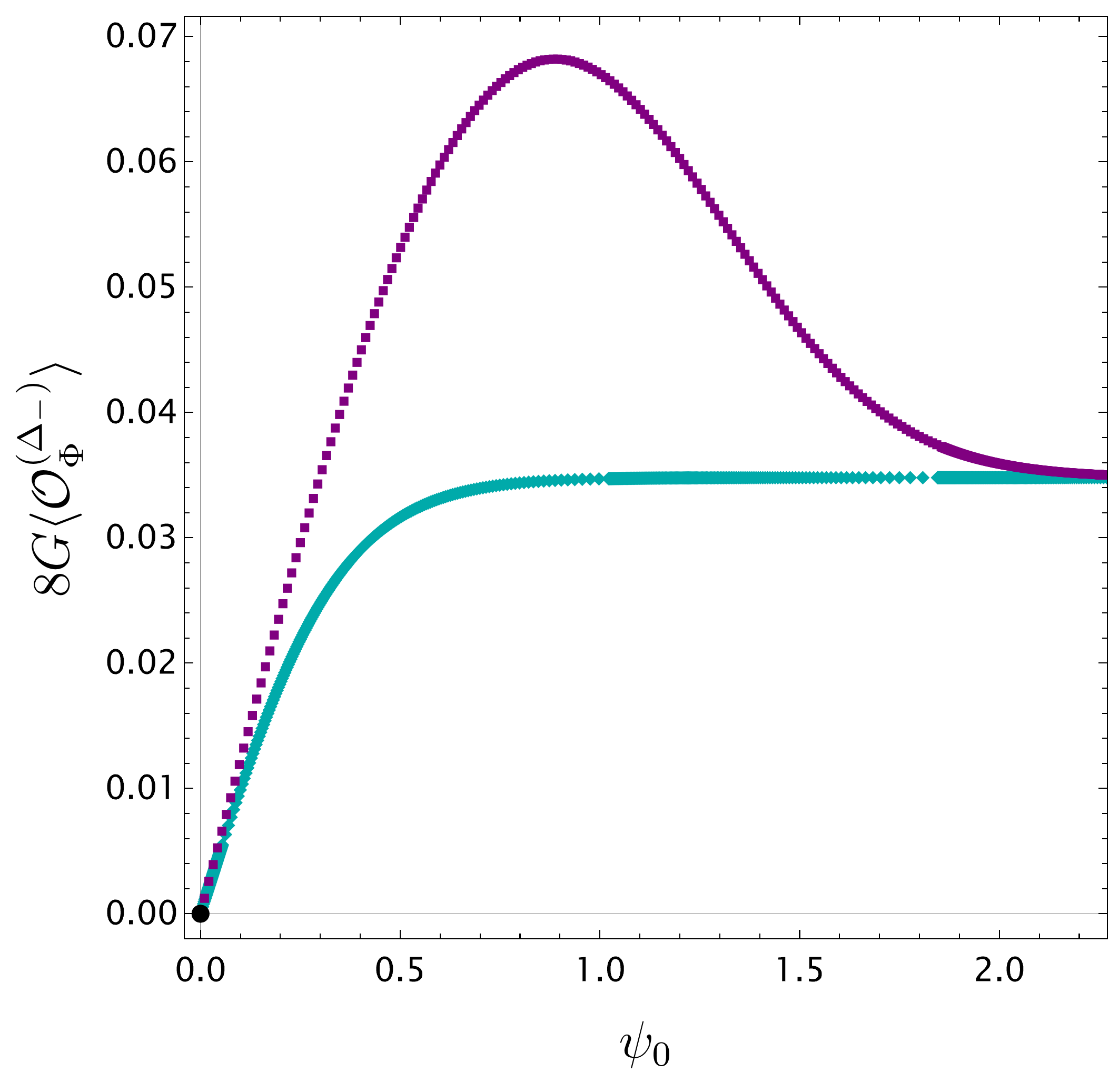}
    \includegraphics[width=0.45\linewidth]{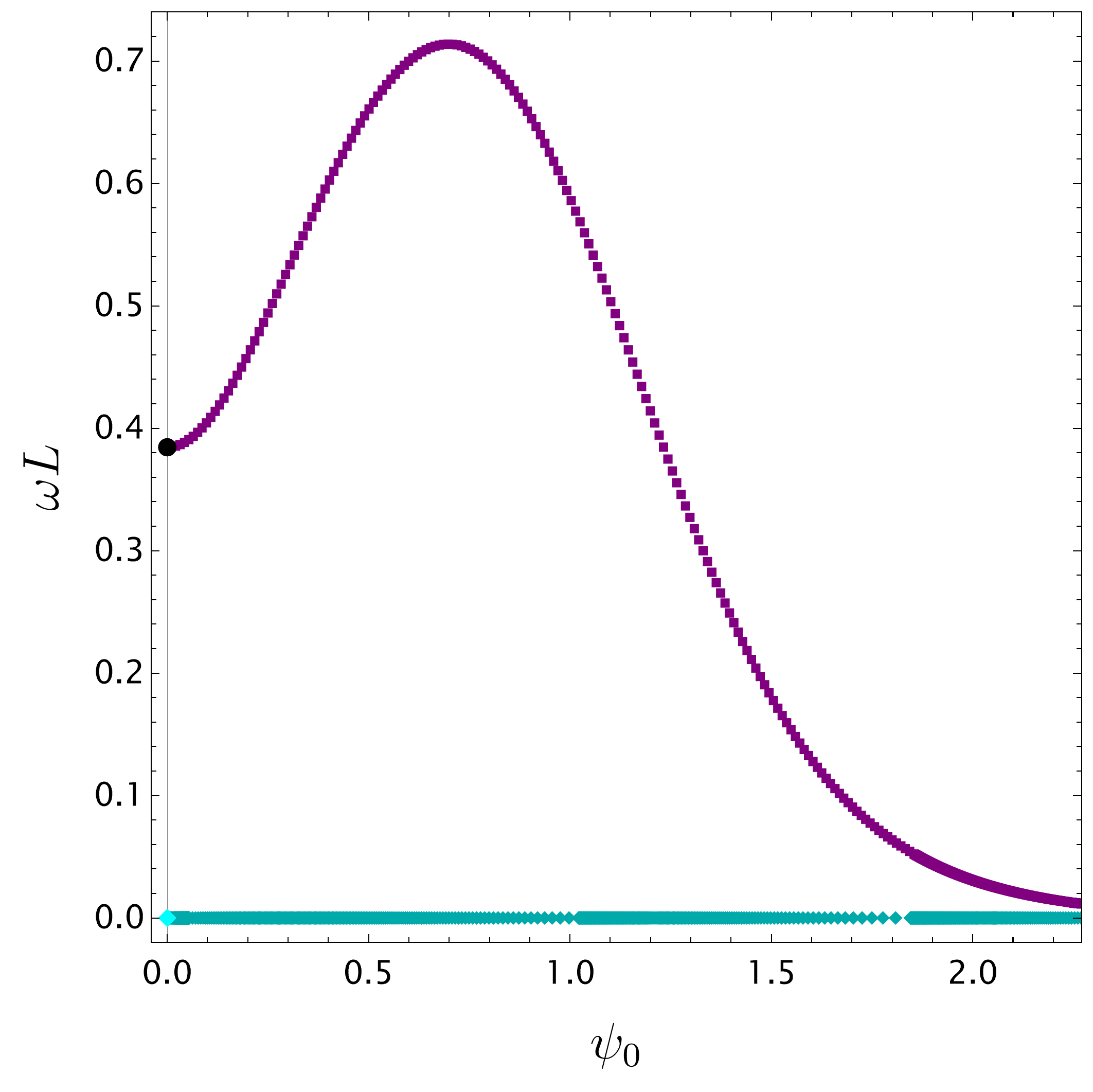}
    \includegraphics[width=0.46\linewidth]{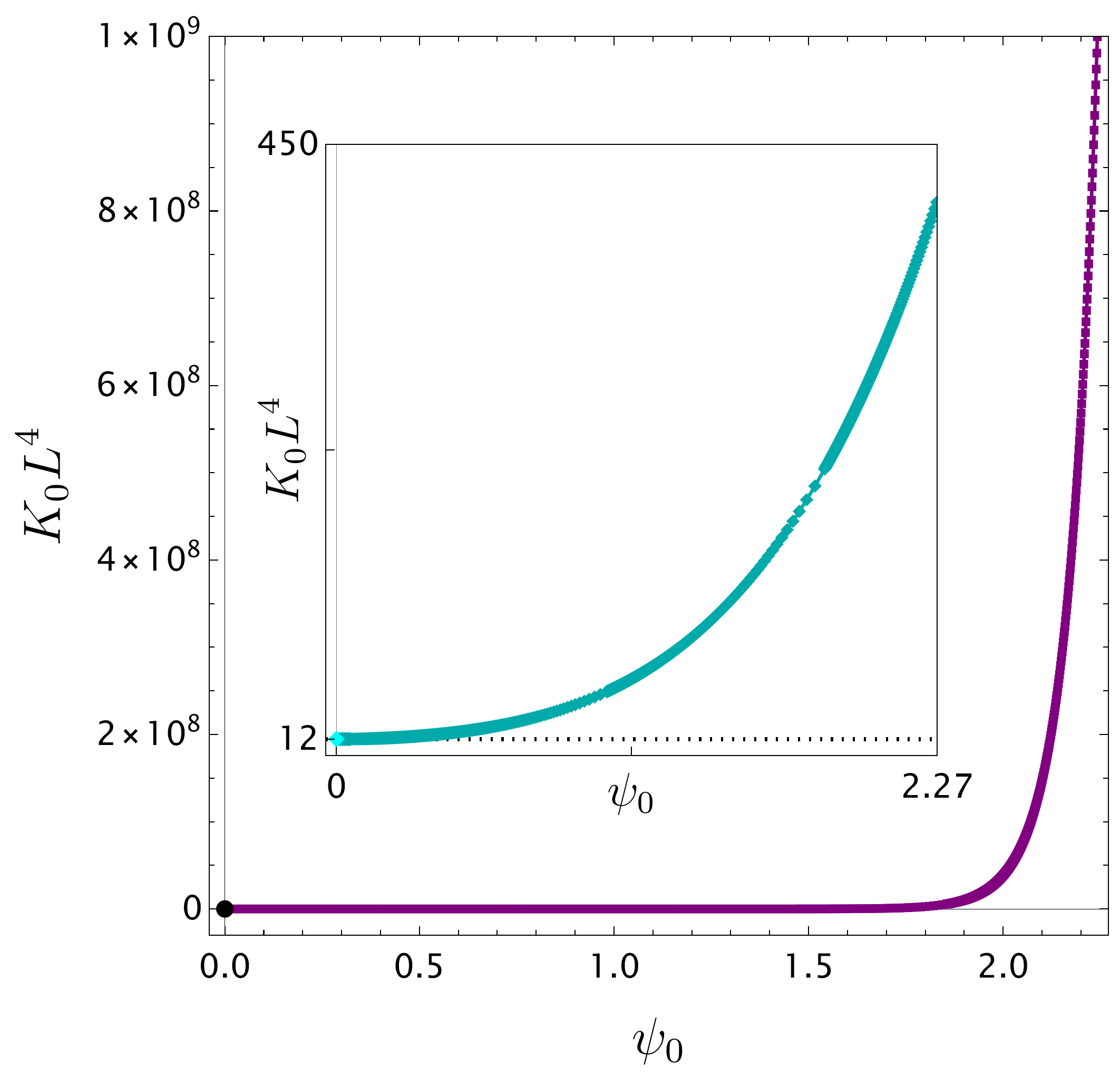}
    \includegraphics[width=0.45\linewidth]{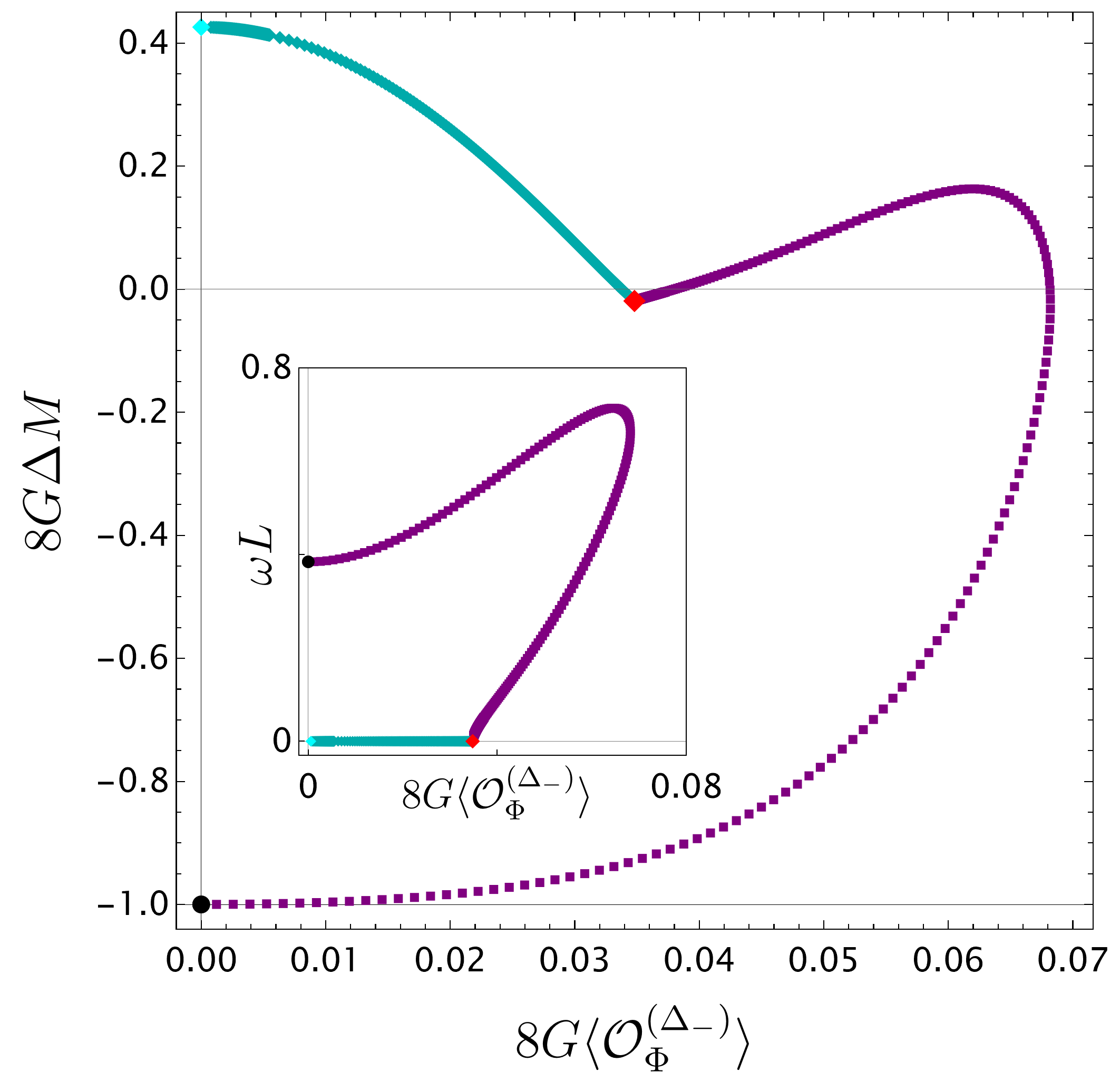}
    \caption{Properties of $\bm{m=0}$ boson stars (purple squares; discussed in section~\ref{sec:PhaseDiag-m0:BStar}),  and regular $\bm{m=0}$ static hairy black holes (petrol green diamonds; discussed in section~\ref{sec:PhaseDiag-m0:BHsStatic}) with $\kappa = -4/10$, $\mu^2L^2 = -15/16$. The black disk describes global AdS$_3$ and the red diamond describes the singular $m=0$ static extremal hairy black hole \eqref{sBHm0J0:Thermo} described in section~\ref{sec:NumericalSetup:SingBHm0J0}. In the boson star case, $\psi_0$ is the value of the scalar field at the origin while in the black hole case, $\psi_0\equiv \psi_H$ describes the value of the scalar field at the horizon.
    }
    \label{fig:BS-m0}
\end{figure}

Indeed, regular $m=0$ hairy boson stars exist for $\kappa>\kappa^{\rm AdS}_{0,\hat{\mu}^2}$ and $-1<\hat{\mu}<0$. This expectation is fully borne out by our numerical results. As an illustration, in Fig.~\ref{fig:BS-m0} we set $\mu^2L^2=-15/16$ and $\kappa=-4/10>\kappa^{\rm AdS}_{0,-15/16}$, and display various properties of the resulting regular boson stars (purple squares), which are perturbatively connected to AdS$_3$.

These solutions are static, \ie with vanishing angular momentum $\hat{J}=0$, but carry finite mass $\hat{M}$, defined in~\eqref{Mass:Def}, and a finite conserved $U(1)$ charge $\hat{N}$, defined in~\eqref{N:Def} (with $R_0\equiv0$). The top‑left panel of Fig.~\ref{fig:BS-m0} shows their location in the $(\hat{N},\hat{M})$ phase diagram. In practice, as explained in Section~\ref{sec:NumericalSetup:RegBS}, these solutions are most conveniently parametrized by the value $\psi_0$ of the scalar field at the center of the boson star. For each $\psi_0$ there exists a unique boson star, as illustrated in the top‑right panel of Fig.~\ref{fig:BS-m0}, where $\hat{M}$ and $\hat{N}$ are plotted as functions of $\psi_0$.

Another important diagnostic is the vacuum expectation value
$\langle\hat{\mathcal{O}}_{\Phi}^{(\Delta_-)}\rangle$, given in~\eqref{ScalarVEV}, which characterizes the dual double‑trace operator. This quantity is shown in the middle‑left panel of Fig.~\ref{fig:BS-m0}. It vanishes in the AdS$_3$ limit, grows to a maximum as $\psi_0$ increases, and then decreases again toward a finite value. The scalar field frequency $\hat{\omega}$ is displayed in the middle‑right panel and likewise varies smoothly along the family.

In the limit $\psi_0\to0$, the boson star solutions reduce continuously to global AdS$_3$ with $\hat{M}=-1$ and $\hat{N}=0$, represented by the black disk in Fig.~\ref{fig:BS-m0}. In this limit, $\alpha\to0$, $\langle\hat{\mathcal{O}}_{\Phi}^{(\Delta_-)}\rangle\to0$, and $\hat{\omega}$ approaches the linear normal‑mode frequency $\hat{\omega}\simeq0.384480$ computed in~\cite{Dias:2025uyk}. At the opposite endpoint, the family terminates at a Chandrasekhar‑like limit, shown as the red diamond in Fig.~\ref{fig:BS-m0}. As this limit is approached, the Kretschmann scalar at the center of the boson star,
$\hat{K}_0=\hat{R}^{\mu\nu\alpha\beta}\hat{R}_{\mu\nu\alpha\beta}|_{R=0}$,
grows without bound, signaling the onset of a singular geometry (bottom‑left panel).

This limiting boson star solution coincides with the singular $m=0$ static extremal hairy black hole (red diamond in all plots) discussed in Section~\ref{sec:NumericalSetup:SingBHm0J0}. That is, in this limit all thermodynamic quantities of the $m=0$ boson star
approach the singular values of the
\begin{align}\label{sBHm0J0:Thermo}
&\hbox{Singular $m=0$ static extremal hairy black hole } (\hat{\mu}^2=-15/16,\kappa=-4/10):\nonumber \\
 & \hspace{0.5cm}
\{\hat{M},\hat{N},\alpha,\langle\hat{\mathcal{O}}_{\Phi}^{(\Delta_-)}\rangle,\hat{\omega},\hat{S}_H, \hat{T}_H,\hat{\Omega}_H\}=\{-0.019,\,0,\,0.437,\,0.034,\,0,0,0,0\},
\end{align}
which depend on $\hat{\mu}$ and $\kappa$ (for this extremal black hole, $\psi_0$ in Fig.~\ref{fig:BS-m0} denotes the value of the scalar field at the horizon). A very important finding is that the singular $m=0$ static extremal hairy black hole saturates the positivity-of-energy theorem~\eqref{GlobalMin}–\eqref{Super:Emin} derived in Appendix~\ref{secA:superpotentials} for $\hat J=0$ and $p=0$.

As a consistency check, we verify that our numerical solutions satisfy the first law of boson star thermodynamics~\eqref{FirstLawBStar},
\begin{equation}
100\left[1-\frac{\hat{\omega}\,\hat{N}'(\psi_0)}{\hat{M}'(\psi_0)}\right]=0,
\end{equation}
with relative errors typically below $10^{-3}\%$.

Finally, note that in Fig.~\ref{fig:BS-m0} the boson star family is usually plotted as a function of $\psi_0$, since this parameter labels solutions uniquely. This is not the case when parametrizing by the VEV: as illustrated in the bottom‑right panel, there exists a range where two distinct boson stars share the same VEV but have different masses and frequencies. For later reference, we also indicate in Fig.~\ref{fig:BS-m0} a petrol‑green curve terminating on the singular $m=0$ extremal hairy black hole (red diamond). This curve represents a one‑parameter family of regular $m=0$ static hairy black holes, whose detailed discussion is deferred to Section~\ref{sec:PhaseDiag-m0:BHsStatic}.

\subsection{Hairy \texorpdfstring{AdS$_3$}{AdS3} boson stars with \texorpdfstring{$m=0$}{m=0} and  Ishibashi-Wald instability of \texorpdfstring{AdS$_3$}{AdS3}}\label{sec:PhaseDiag-m0:IshWald}
\begin{figure}
    \centering
    \vskip -0.5cm
    \includegraphics[width=0.50\linewidth]{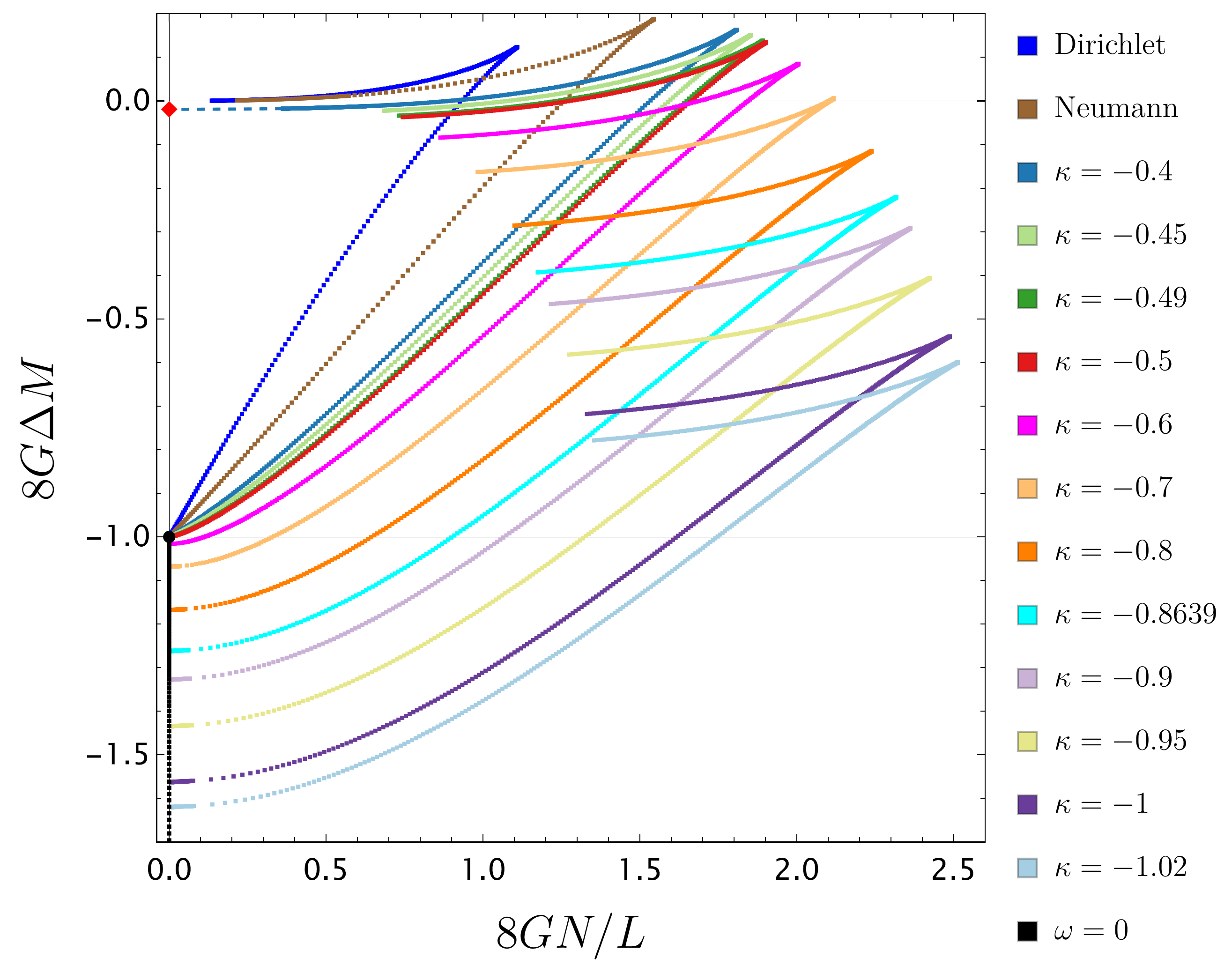}
    \includegraphics[width=0.45\linewidth]{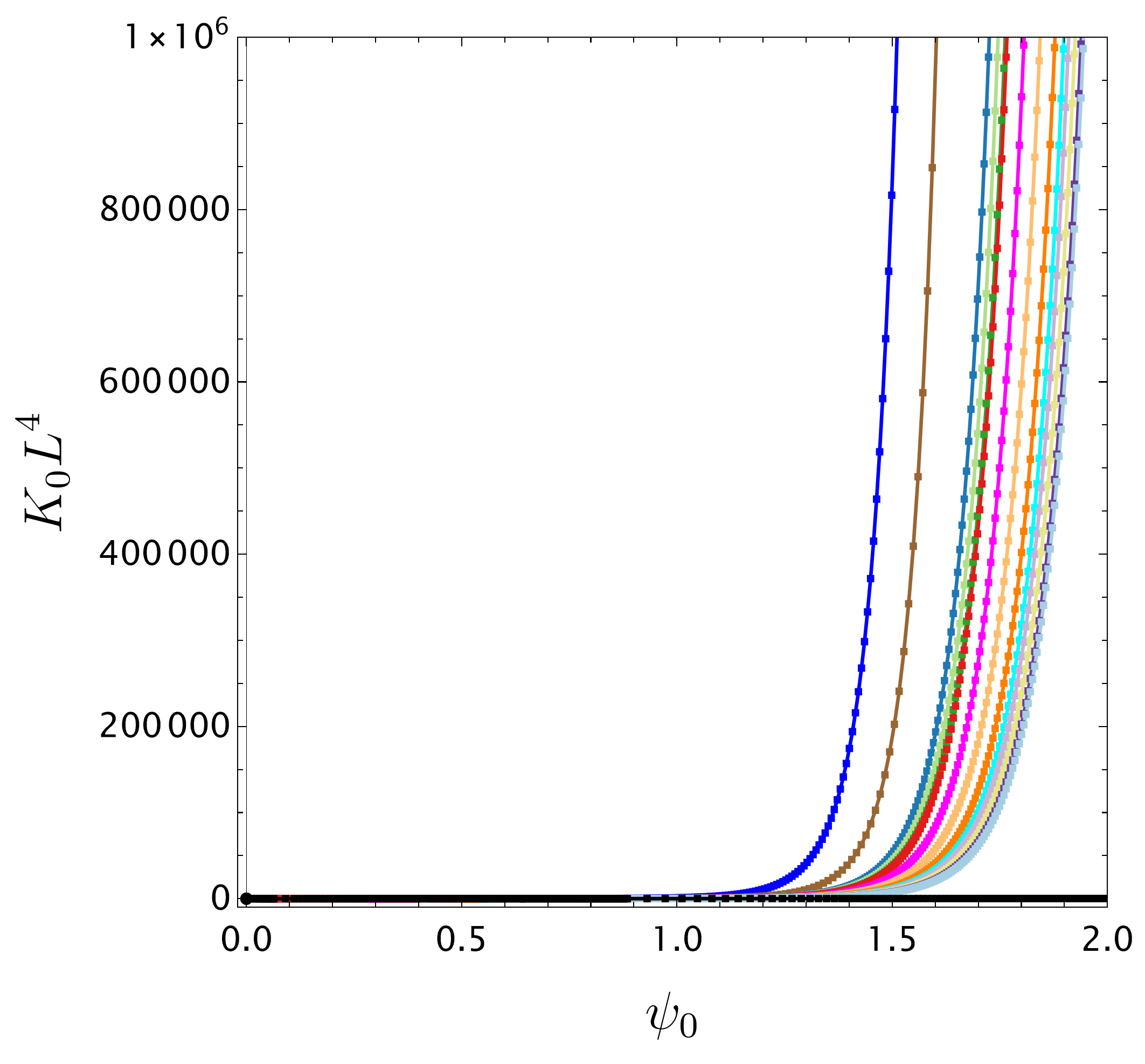}
    \includegraphics[width=0.47\linewidth]{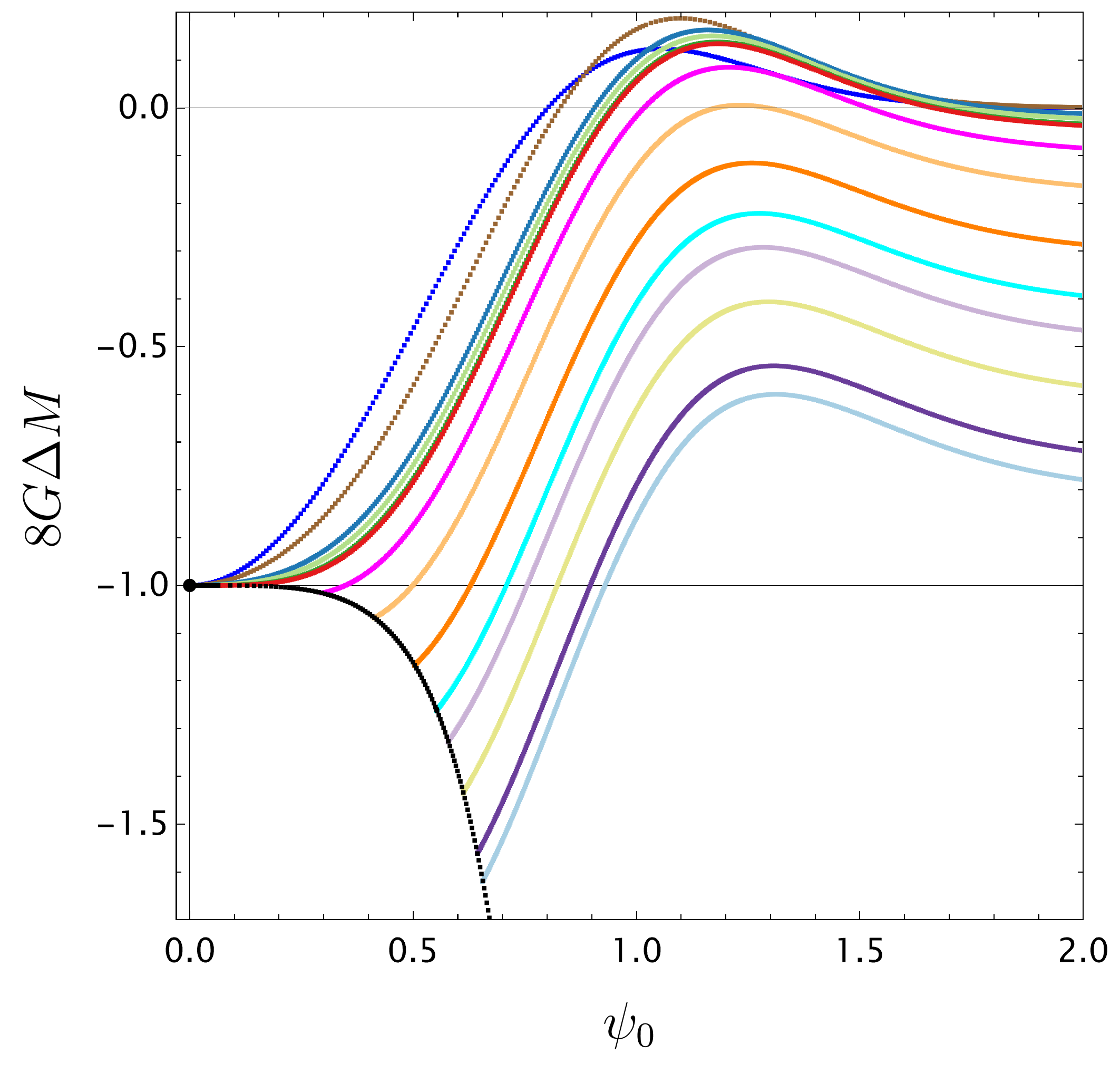}
    \includegraphics[width=0.47\linewidth]{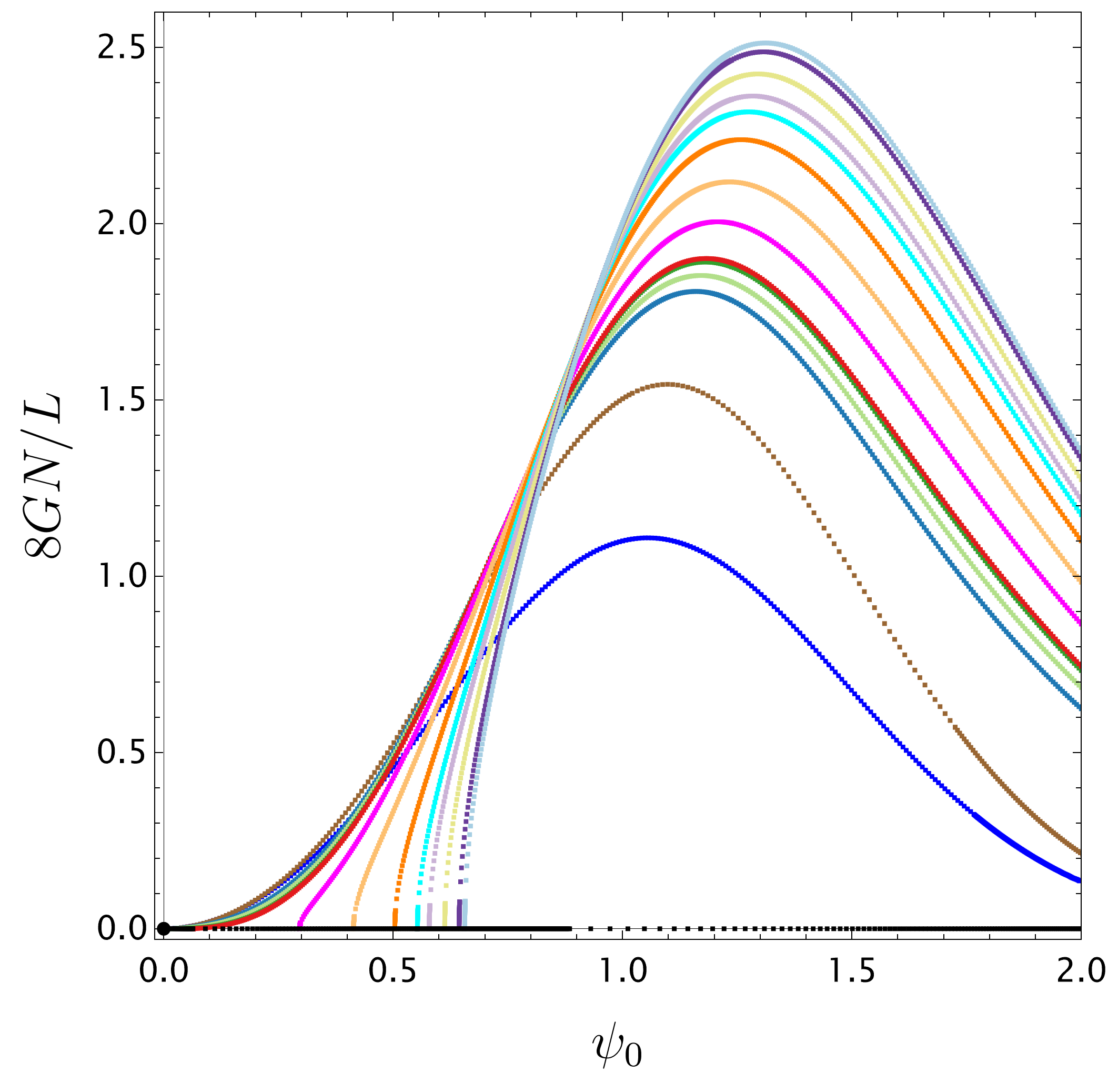}
    \includegraphics[width=0.47\linewidth]{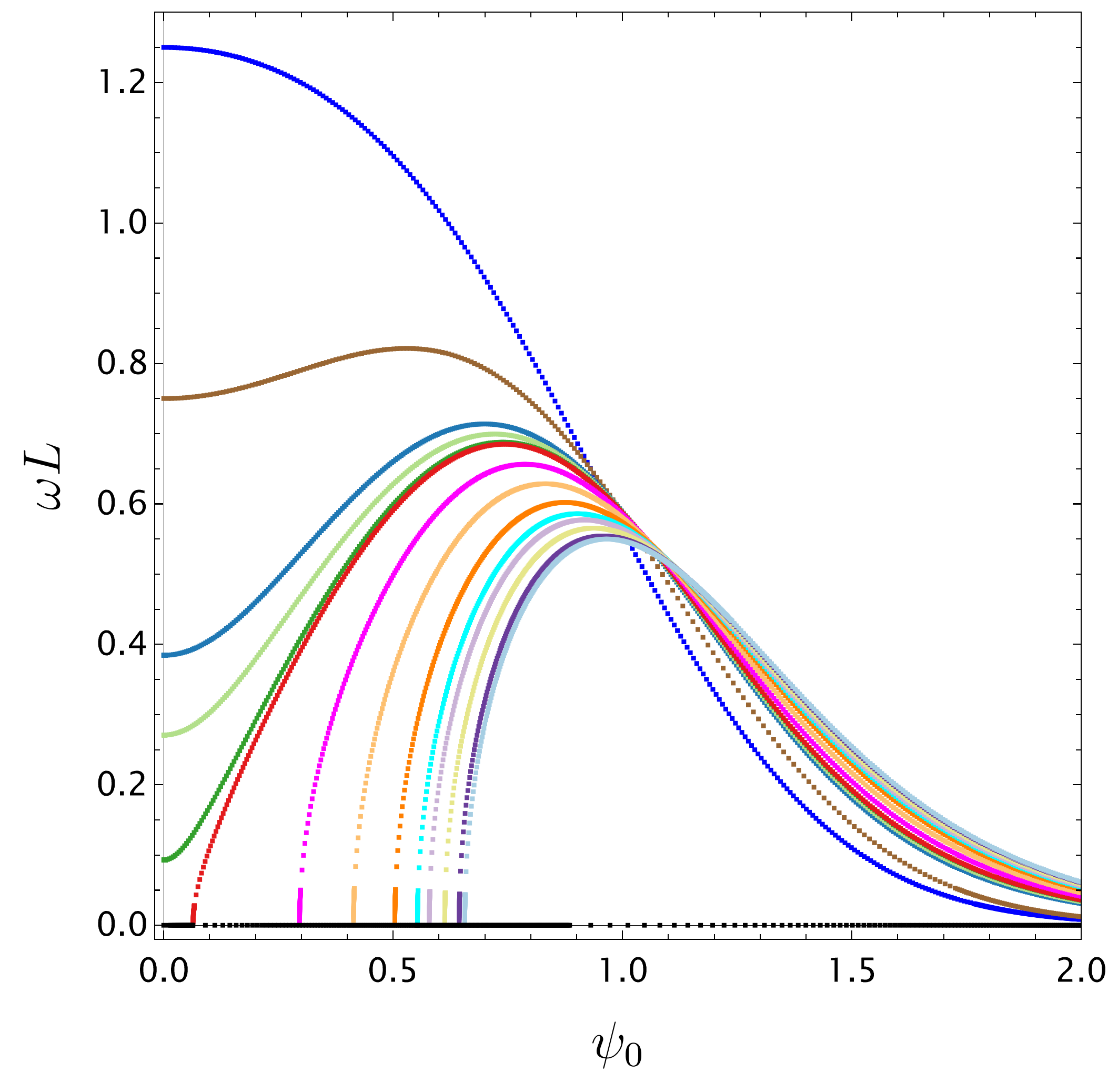}
    \includegraphics[width=0.47\linewidth]{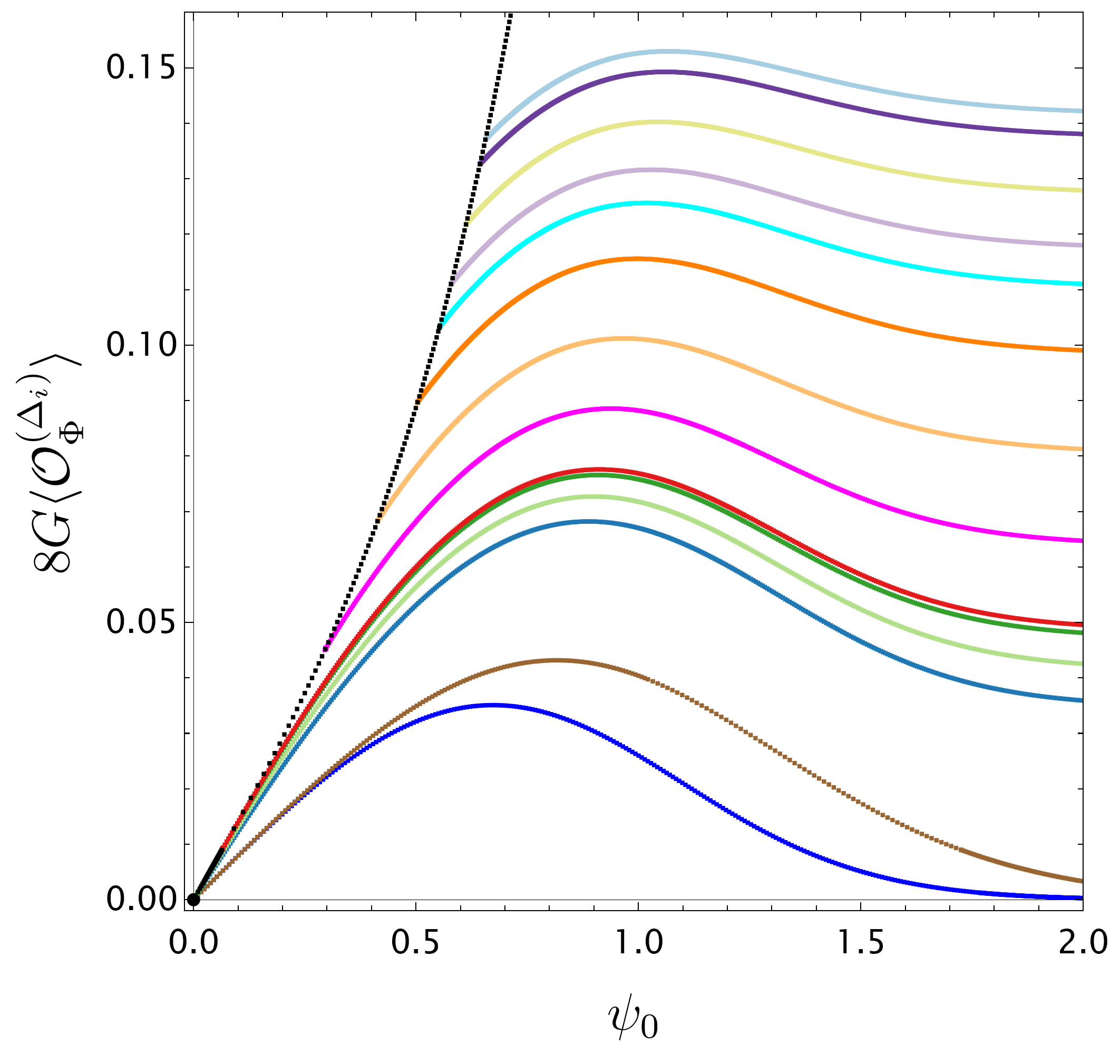}
    \caption{Regular $\bm{m=0}$ boson stars with $\mu^{2}L^{2}=-15/16$ for several values of $\kappa$ (see legend), displayed in the range $\psi_{0}\in[0,2]$. 
    For $\kappa>\kappa^{\rm AdS}_{m,\hat{\mu}^{2}} \simeq-0.4951294$, the boson star families originate at global AdS$_3$, with $\{\hat{M},\hat{N}\}=\{-1,0\}$, $\psi_{0}=0$, $\langle\mathcal{O}_{\Phi}^{(\Delta_-)}\rangle=0$, and $\hat{\omega}(\kappa)\neq0$. 
    For $\kappa<\kappa^{\rm AdS}_{m,\hat{\mu}^{2}}$, where AdS$_3$ is unstable, the families instead start at a regular zero-frequency boson star (black squares) with $\hat{M}(\kappa)<-1$, $\hat{N}=0$, $\psi_{0}(\kappa)>0$, $\langle\mathcal{O}_{\Phi}^{(\Delta_-)}\rangle(\kappa)>0$, and $\hat{\omega}=0$. 
    When present, this $m=0$ zero-frequency boson star is the ground state of the theory and provides the endpoint of the AdS$_3$ instability (see Fig.~\ref{fig:m0m1omega0BS} and  discussions of \eqref{GlobalMin}-\eqref{Super:Emin}, when $p=1$, of Appendix~\ref{secA:superpotentials}).}
    \label{fig:m0BSevoK}
\end{figure}

Figure~\ref{fig:BS-m0} describes the properties of $m=0$ boson stars for $\kappa>\kappa^{\rm AdS}_{m=0,\hat{\mu}^{2}}$, where AdS$_3$ is linearly stable against double-trace perturbations~\cite{Ishibashi:2004wx,Dias:2025uyk}. We now examine how the boson star phase diagram evolves as $\kappa$ is decreased through the critical value $\kappa^{\rm AdS}_{m=0,\hat{\mu}^{2}}$.

In Fig.~\ref{fig:m0BSevoK} we show regular $m=0$ boson stars with $\mu^{2}L^{2}=-15/16$ for several values of $\kappa$ above and below $\kappa^{\rm AdS}_{0,-15/16}\simeq-0.4951294$, as determined from~\eqref{kAdS-onset}. Specifically, we display solutions for
\begin{subequations}
\begin{equation}
\kappa=\{-0.4,-0.45,-0.49\}>\kappa^{\rm AdS}_{0,-15/16}
\end{equation}
and
\begin{equation}
\kappa=\{-0.5,-0.6,-0.7,-0.8,-0.8639,-0.9,-0.95,-1,-1.02\}<\kappa^{\rm AdS}_{0,-15/16}\,.
\end{equation}
\end{subequations}
For each value of $\kappa$, we plot $\hat{N}$ versus $\hat{M}$ (top-left), $\psi_{0}$ versus the central Kretschmann invariant $\hat{K}_{0}$ (top-right), $\psi_{0}$ versus $\hat{M}$ (middle-left), $\psi_{0}$ versus $\hat{N}$ (middle-right), $\psi_{0}$ versus $\hat{\omega}$ (bottom-left), and $\psi_{0}$ versus $\langle\mathcal{O}_{\Phi}^{(\Delta_-)}\rangle$ (bottom-right).

Two qualitatively distinct behaviours emerge. For $\kappa>\kappa^{\rm AdS}_{m,\hat{\mu}^{2}}$, the regular boson star family is perturbatively connected to AdS$_3$: the solutions originate at $\hat{M}=-1$, $\hat{N}=0$, and extend toward a Chandrasekhar-like limit at which the curvature diverges, while $\hat{N}\to0$ and $\hat{\omega}\to0$. These solutions can be interpreted as the fully nonlinear backreaction of the double-trace normal modes of AdS$_3$ computed in~\cite{Dias:2025uyk}.

Figure~\ref{fig:m0BSevoK} also displays boson stars with Dirichlet ($\kappa=\pm\infty$) and Neumann ($\kappa=0$) boundary conditions. These families behave similarly to those with $\kappa>\kappa^{\rm AdS}_{m,\hat{\mu}^{2}}$: they are perturbatively connected to global AdS$_3$, possess a Chandrasekhar limit, and admit a perturbative construction near AdS$_3$. A detailed comparison between perturbative and numerical solutions is presented in Appendix~\ref{secA:BStars-DNm0}; see Figs.~\ref{fig:m0_DirBS_numerics_VS_perturbation} and~\ref{fig:m0_NeuBS_numerics_VS_perturbation} (the agreement between these two analyses for small mass further testifies our numerics).

A qualitatively new behaviour appears for $\kappa<\kappa^{\rm AdS}_{m,\hat{\mu}^{2}}$, where AdS$_3$ is unstable to double-trace perturbations~\cite{Ishibashi:2004wx,Dias:2025uyk}. In this regime, regular boson stars still exist but are {\it no longer} perturbatively connected to AdS$_3$. Instead, the boson star family begins at a regular {\it zero-frequency} boson star solution with $\hat{M}(\kappa)<-1$, $\hat{N}=0$, $\psi_{0}(\kappa)>0$, $\langle\mathcal{O}_{\Phi}^{(\Delta_-)}\rangle(\kappa)>0$, and $\hat{\omega}=0$ (shown by the black squares in Fig.~\ref{fig:m0BSevoK}). These solutions are obtained numerically by imposing $\hat{\omega}=0$.

As $\psi_{0}$ increases, the mass grows monotonically until a cusp at $\hat{N}=\hat{N}_{\mathrm{max}}(\kappa)$ is reached, beyond which a second branch with higher mass appears. The family ultimately terminates at the singular $m=0$ static extremal hairy black hole  discussed in Section~\ref{sec:NumericalSetup:SingBHm0J0}, where the curvature diverges and again $\hat{N}=0$ and $\hat{\omega}=0$. For all $\kappa$, the scalar amplitude $\psi_{0}$ continues to label the boson star solutions uniquely.

We emphasize the central result of this analysis. For $\kappa<\kappa^{\rm AdS}_{m,\hat{\mu}^{2}}$, where AdS$_3$ is unstable, the $m=0$ zero-frequency boson star has a mass strictly below that of global AdS$_3$ and constitutes the ground state of the theory. Its energy decreases monotonically as $\kappa$ is reduced further below $\kappa^{\rm AdS}_{m,\hat{\mu}^{2}}$ (this will be better seen in Fig.~\ref{fig:m0m1omega0BS}). This configuration provides the natural endpoint of the Ishibashi-Wald instability of AdS$_3$ with double-trace boundary conditions and realizes explicitly the minimum-energy solution predicted by the superpotential analysis of~\cite{Faulkner:2010fh}, which we revisit in Appendix~\ref{secA:superpotentials}: see discussion of \eqref{GlobalMin}-\eqref{Super:Emin} when $p=1$.

\subsection{Static and rotating hairy \texorpdfstring{AdS$_3$}{AdS3} black holes with \texorpdfstring{$m = 0$}{m=0}}\label{sec:PhaseDiag-m0:BHs}

In this section, we present the hairy black hole solutions that merge with (or bifurcate from) the BTZ black hole along the one‑parameter curve $\hat{M}(\hat{J})\big|_{\text{BTZ onset}}$, which describes the onset of the $m=0$ double‑trace instability of the BTZ geometry. This BTZ onset curve was determined in our companion paper~\cite{Dias:2025uyk}. As in the boson star analysis, we take $\mu^{2}L^{2}=-15/16$ to illustrate the properties of the phase diagram within the mass range~\eqref{2xTrace:rangeMass}, $-1<\mu^{2}L^{2}<0$, where the double‑trace boundary condition~\eqref{2xTrace:BC} is allowed. Recall that the relation $\beta=\kappa\,\alpha$, with $\kappa\in\mathbb{R}$, interpolates between Neumann ($\kappa=0$) and Dirichlet ($\kappa\to\pm\infty$) boundary conditions, although hairy black hole solutions do not exist for either Dirichlet or Neumann boundary conditions.

We focus on the theory with $\kappa=-4/10>\kappa^{\rm AdS}_{m,\hat{\mu}^{2}}$, which is the same value used in the boson star discussion of Section~\ref{sec:PhaseDiag-m0:BStar} and in Fig.~\ref{fig:BS-m0}, in order to illustrate our results. We will later also present $m=0$ hairy black holes for $\kappa<\kappa^{\rm AdS}_{m,\hat{\mu}^{2}}$. Based on our analysis, the main qualitative features of the $m=0$ hairy black hole solutions are insensitive to the precise values of $\hat{\mu}\in(-1,0)$ and $\kappa$, as we will argue in the conclusion section~\ref{sec:Conc}.

\subsubsection{Static hairy black holes with \texorpdfstring{$m = 0$}{m=0}}\label{sec:PhaseDiag-m0:BHsStatic}

The $m=0$ hairy black holes exist for $\hat{J}\geq 0$. We therefore begin by describing the family of static $m=0$ hairy black holes. All these solutions have $\hat{J}=0$, $\hat{\Omega}_{H}=0$ (in fact, $\hat{\Omega}=0$ everywhere), and furthermore $\hat{\omega}=0$ (see Section~\ref{sec:NumericalSetup:BHs}). As a result, the conserved $U(1)$ charge~\eqref{N:Def}, evaluated with $R_{0}\equiv R_{+}$, vanishes identically, $\hat{N}\propto(\hat{\omega}-m\hat{\Omega})=0$. Nevertheless, these black holes have a finite mass $\hat{M}$, as defined in~\eqref{Mass:Def}. This constitutes a one‑parameter family of solutions, which we may parametrise by the horizon radius $R_{+}$. Equivalently, one could use the value of the scalar field at the horizon, $\psi_{H}$, or the scalar condensate amplitude $\alpha$ (or the associated VEV).

To illustrate how the family of static $m=0$ hairy black holes sits relative to the $m=0$ boson stars (recall that $m=0$ boson stars necessarily have $\hat{J}=0$ but generically $\hat{N}\neq0$), we already displayed this black hole branch $-$ shown by petrol‑green diamonds $-$ in the plots of Fig.~\ref{fig:BS-m0}. In that figure, $\psi_{0}\equiv\psi_{H}$ denotes the value of the scalar field at the horizon. In the top‑left panel of Fig.~\ref{fig:BS-m0}, we see that the static hairy black hole family, with $\hat{N}=0=\hat{J}$, emerges from static BTZ at the instability onset point (cyan disk with $\hat{M}\simeq0.413$ and $\hat{N}=0$) and extends all the way down to the singular $m=0$ extremal hairy black hole (red diamond with $\hat{M}\simeq-0.019$, \ie \eqref{sBHm0J0:Thermo}). Remarkably, the singular $m=0$ static extremal hairy black hole identified in Section~\ref{sec:NumericalSetup:SingBHm0J0}  therefore constitutes both the endpoint of the regular $m=0$ boson stars (purple squares) and of the static $m=0$ non-extremal hairy black holes (petrol‑green diamonds).

This conclusion is further supported by the remaining panels of Fig.~\ref{fig:BS-m0}, particularly the bottom‑right panel, where the singular $m=0$ extremal hairy black hole (red diamond given by \eqref{sBHm0J0:Thermo}) is seen to be the endpoint $-$ across all observables $-$ of both the static non-extremal hairy black hole branch and the regular boson star family.

\begin{figure}
    \centering
    \includegraphics[width=0.45\linewidth]{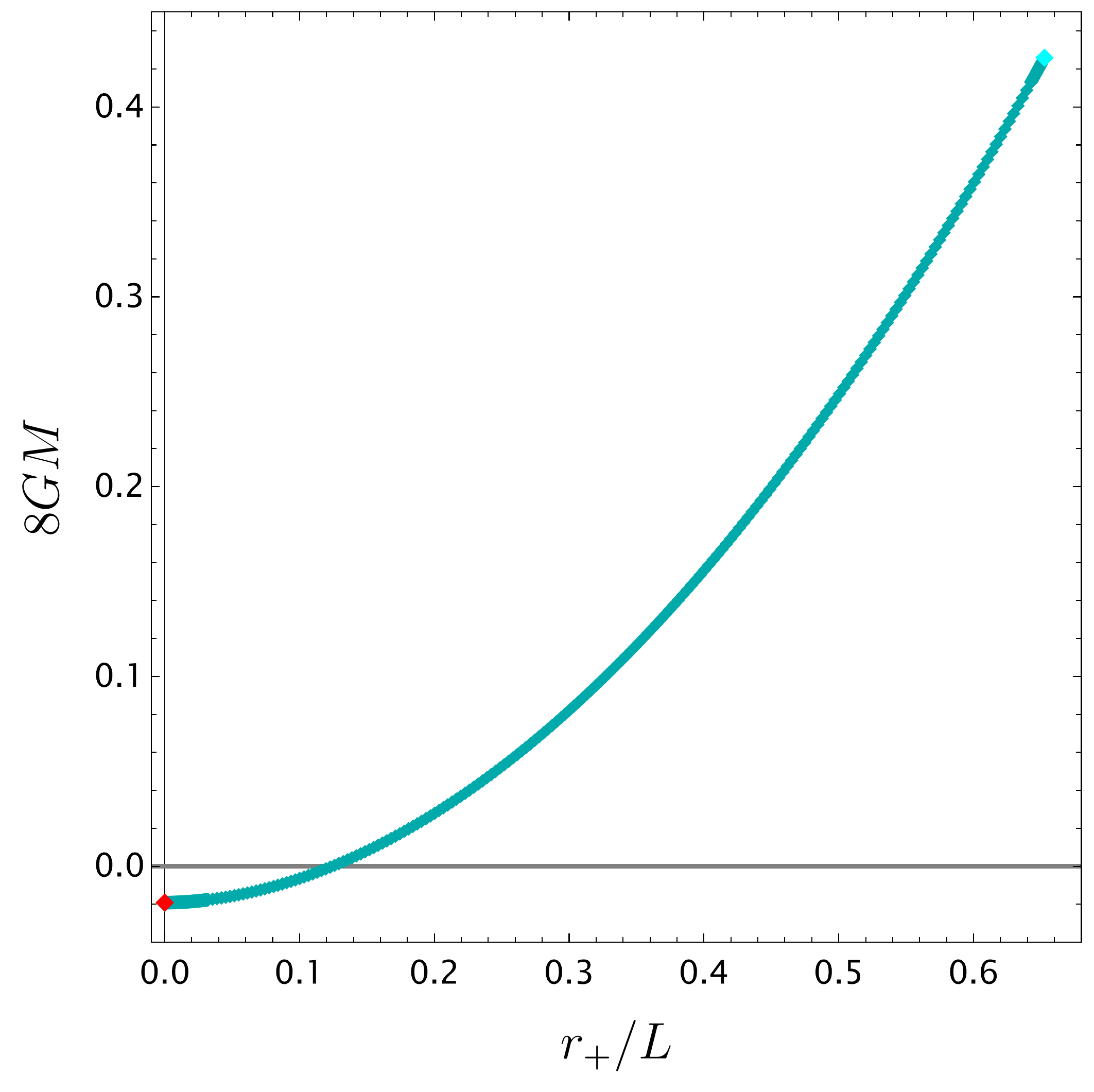}
    \includegraphics[width=0.45\linewidth]{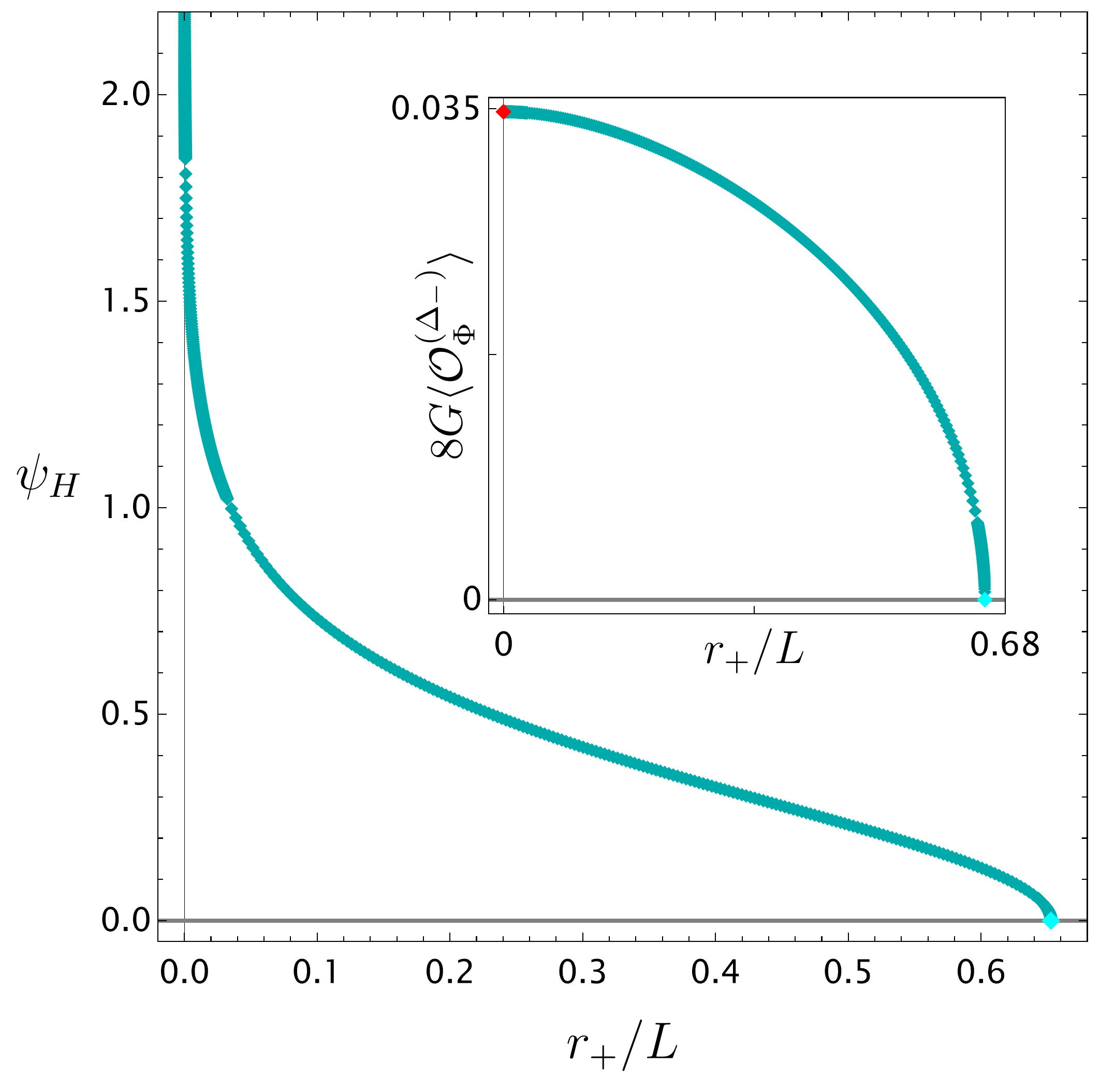}
    \includegraphics[width=0.45\linewidth]{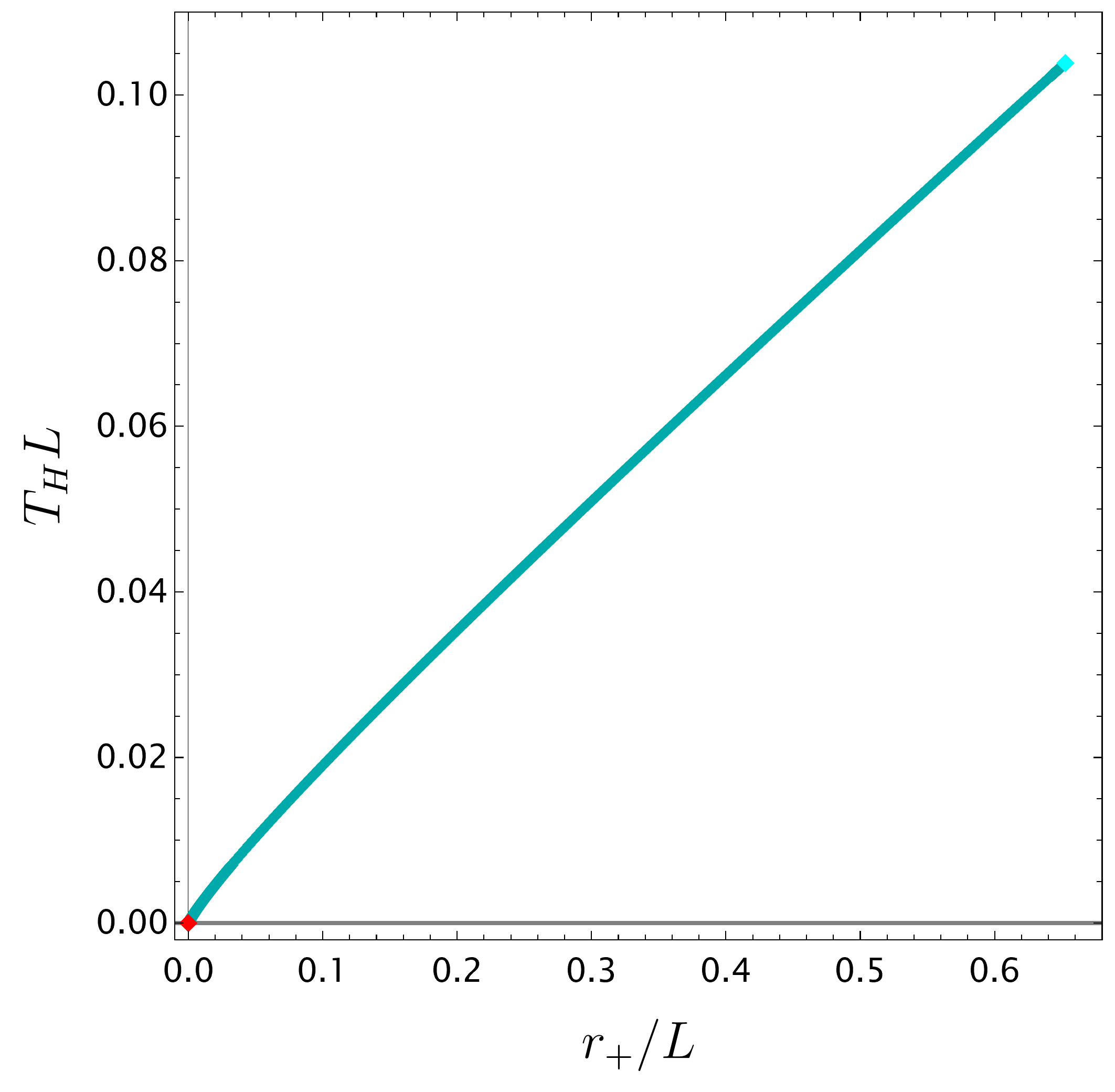}
    \includegraphics[width=0.45\linewidth]{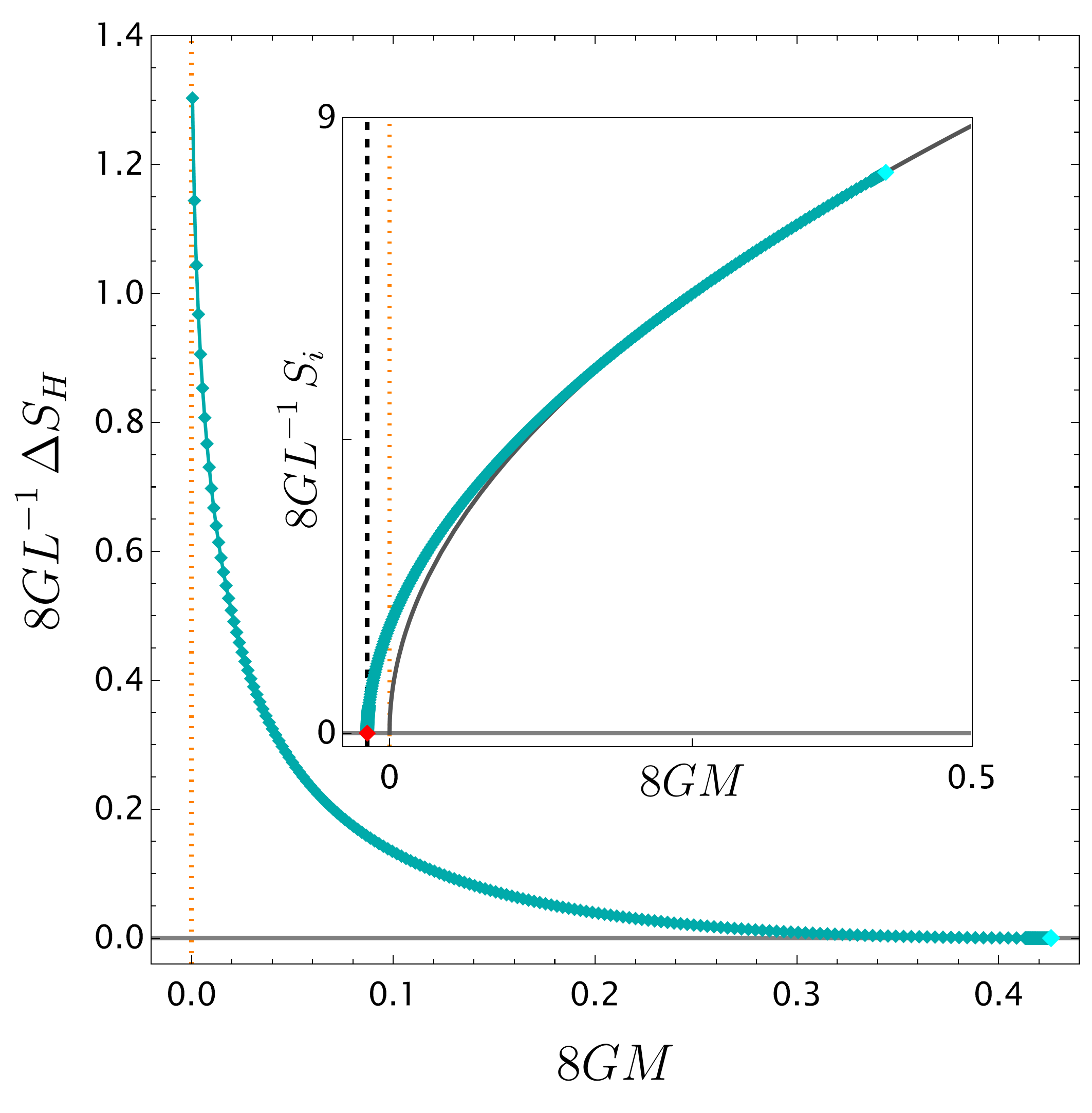}
    \caption{Mass, scalar condensate (the horizon curvature scales as $\psi_{H}^{2}$), temperature, and entropy of static ($\hat{J}=0$) $\bm{m=0}$ hairy black holes (petrol‑green diamonds) with $\kappa=-4/10$, $\mu^{2}L^{2}=-15/16$, plotted as functions of the horizon radius $R_{+}$. The cyan disk marks the merger with static BTZ at the onset of the double‑trace instability. The red diamond denotes the singular $m=0$ extremal hairy black hole~\eqref{sBHm0J0:Thermo}. This figure complements the information shown in Fig.~\ref{fig:BS-m0}.}
    \label{fig:m0_J0-BH}
\end{figure}

Figure~\ref{fig:m0_J0-BH} presents further properties of the static $m=0$ hairy black holes, now typically displayed as functions of the horizon radius $R_{+}$. Specifically, we plot the mass $\hat{M}$ (top‑left), the value of the scalar field at the horizon $\psi_{H}$ together with the VEV $\langle\hat{\mathcal{O}}_{\Phi}^{(\Delta_-)}\rangle$ (top‑right, including inset), the Hawking temperature $\hat{T}_{H}$ (bottom‑left), and the entropy difference $\Delta\hat{S}_{H}$ along with the entropy $\hat{S}_{H}$ (bottom‑right, including inset). In the inset of the last panel, we compare the entropy of the hairy black hole, $\hat{S}_{H}$ (petrol‑green diamonds), with that of the BTZ solution, $\hat{S}_{H}^{\text{\tiny BTZ}}$ (solid black curve). The main panel shows $\Delta\hat{S}_{H}\equiv \hat{S}_{H}-\hat{S}_{H}^{\text{\tiny BTZ}}$ for black holes with the same mass. We have verified that the first law of thermodynamics~\eqref{FirstLawBH},
\begin{equation}
100\!\left[1-\hat{T}_{H}\frac{\hat{S}_{H}'(\psi_{0})}{\hat{M}'(\psi_{0})}\right]=0,
\end{equation}
is satisfied with a relative error smaller than $10^{-3}\%$.

From Fig.~\ref{fig:m0_J0-BH} (and Fig.~\ref{fig:BS-m0}) we see that the static $m=0$ hairy black hole branch bifurcates from static BTZ at the cyan disk. As anticipated, this point coincides precisely with the onset of the double‑trace instability of static BTZ found in the independent linear analysis of~\cite{Dias:2025uyk}, providing a strong consistency check of both the linear and nonlinear numerical calculations. The bifurcation corresponds to a second‑order phase transition, with $\Delta\hat{S}_{H}=0$ and vanishing scalar condensate $\psi_{H}=\alpha=\langle\hat{\mathcal{O}}_{\Phi}^{(\Delta_-)}\rangle=0$.

Moving along the hairy branch, both $\psi_{H}$ and $\langle\hat{\mathcal{O}}_{\Phi}^{(\Delta_-)}\rangle$ increase monotonically as $R_{+}$ decreases, until the red‑diamond endpoint at $R_{+}=0$ is reached. Near this endpoint, the horizon Kretschmann scalar and $|\psi_{H}|^{2}$ diverge, while the temperature and entropy $\hat{S}_{H}=4\pi R_{+}$ both decrease to zero. This confirms that the red diamond \eqref{sBHm0J0:Thermo} corresponds to the zero‑horizon‑radius limit of the static hairy black hole and is described by the singular $m=0$ extremal hairy black hole identified in Section~\ref{sec:NumericalSetup:SingBHm0J0}.

Whenever static hairy and BTZ black holes coexist, the former always possesses a higher entropy at fixed mass, \emph{\ie}\ $\Delta\hat{S}_{H}\geq0$. Moreover, static hairy black holes exist in a small window of negative mass $\hat{M}<0$, while regular static BTZ black holes only exist for $\hat{M}>0$ (recall that vacuum BTZ with $\hat{M}=0=\hat{J}$ is singular). This is possible because double‑trace boundary conditions break supersymmetry, allowing violations of the usual BPS bound $M\geq |J|/L$.\footnote{\label{footBPS}
Here we are referring to the BPS bound associated with the spin structure for which the BTZ black hole admits supersymmetric configurations, namely periodic boundary conditions for fermions along the angular direction. This spin structure differs from that of global AdS$_3$, which admits supersymmetry only for anti\-periodic (NS) boundary conditions and has the supersymmetric ground state at $\hat{M}=-1$. Throughout this work, we therefore compare masses with respect to the BTZ‑supersymmetric spin structure rather than the global AdS$_3$ one. See Appendix~\ref{secA:superpotentials} for further discussion.}

\subsubsection{Rotating hairy black holes with \texorpdfstring{$m = 0$}{m=0}}\label{sec:PhaseDiag-m0:BHsRotating}

In this section, we begin the discussion of rotating hairy black holes that merge with (or bifurcate from) the spinning BTZ black hole along the one‑parameter curve $\hat{M}(\hat{J})^{\text{\tiny BTZ}}_{\text{\tiny onset}}$, which describes the onset of the $m=0$ double‑trace instability of BTZ. This onset curve was computed in the linear analysis of Ref.~\cite{Dias:2025uyk}. The corresponding hairy black hole solutions form a two‑parameter family which, as discussed in Section~\ref{sec:NumericalSetup:BHs}, may be conveniently parametrized for instance by the horizon radius and scalar condensate amplitude, $\{R_{+},\alpha\}$, or equivalently by the mass and angular momentum, $\{\hat{M},\hat{J}\}$.

To gain insight into the structure of this two‑dimensional moduli space, we perform an exploratory analysis by focusing on three representative one‑parameter sub‑families of solutions:  
(i) a family of hairy black holes at fixed horizon radius $R_{+}=3/4$ (equivalently fixed entropy $\hat{S}_{H}=3\pi$, shown as pink diamonds in our plots);  
(ii) a family at fixed scalar condensate amplitude $\alpha=0.206$ (blue diamonds); and  
(iii) a family at fixed angular momentum $\hat{J}=3$ (brown diamonds; see also the static $\hat{J}=0$ case in Fig.~\ref{fig:m0_J0-BH}).  
These preliminary investigations prove useful for identifying the boundaries, as well as the qualitative nature, of the two‑dimensional region in parameter space where hairy black holes exist. The specific numerical values chosen here are not special, but rather illustrate generic qualitative features shared by other families at constant $R_{+}$, $\alpha$, or $\hat{J}$. A complete scan of the full two‑dimensional parameter space of $m=0$ static and rotating hairy black holes is presented in Section~\ref{sec:PhaseDiag-Total}.

Recall that BTZ black holes possess no scalar condensate and that this two‑parameter family exists for $\hat{M}\geq\hat{J}$, with the line $M=J/L$ corresponding to the one‑parameter family of extremal BTZ black holes of zero temperature. As is well known, extremal BTZ black holes saturate the BPS bound $M\geq|J|/L$, in contrast with extremal Kerr–AdS black holes in higher dimensions. When presenting our results, instead of plotting the mass $\hat{M}$ directly, it is useful to introduce
\begin{equation}
\Delta\hat{M}=\hat{M}-\hat{M}^{\text{\tiny BTZ}}_{\text{\tiny ext}},
\qquad
\hat{M}^{\text{\tiny BTZ}}_{\text{\tiny ext}}=\hat{J},
\end{equation}
which measures the mass relative to extremal BTZ and enhances the visualization of the region between extremality and the instability onset. Similarly, when discussing the entropy, we present
\begin{equation}
\Delta\hat{S}_{H}=\hat{S}_{H}-\hat{S}_{H}^{\text{\tiny BTZ}},
\end{equation}
where $\hat{S}_{H}^{\text{\tiny BTZ}}$ is the entropy of the BTZ black hole with the same mass $\hat{M}$ and angular momentum $\hat{J}$ as the hairy black hole under consideration, see \eqref{BTZ:thermoS}, whenever the two solutions coexist.

\begin{figure}[t]
    \centering
    \includegraphics[width=0.55\linewidth]{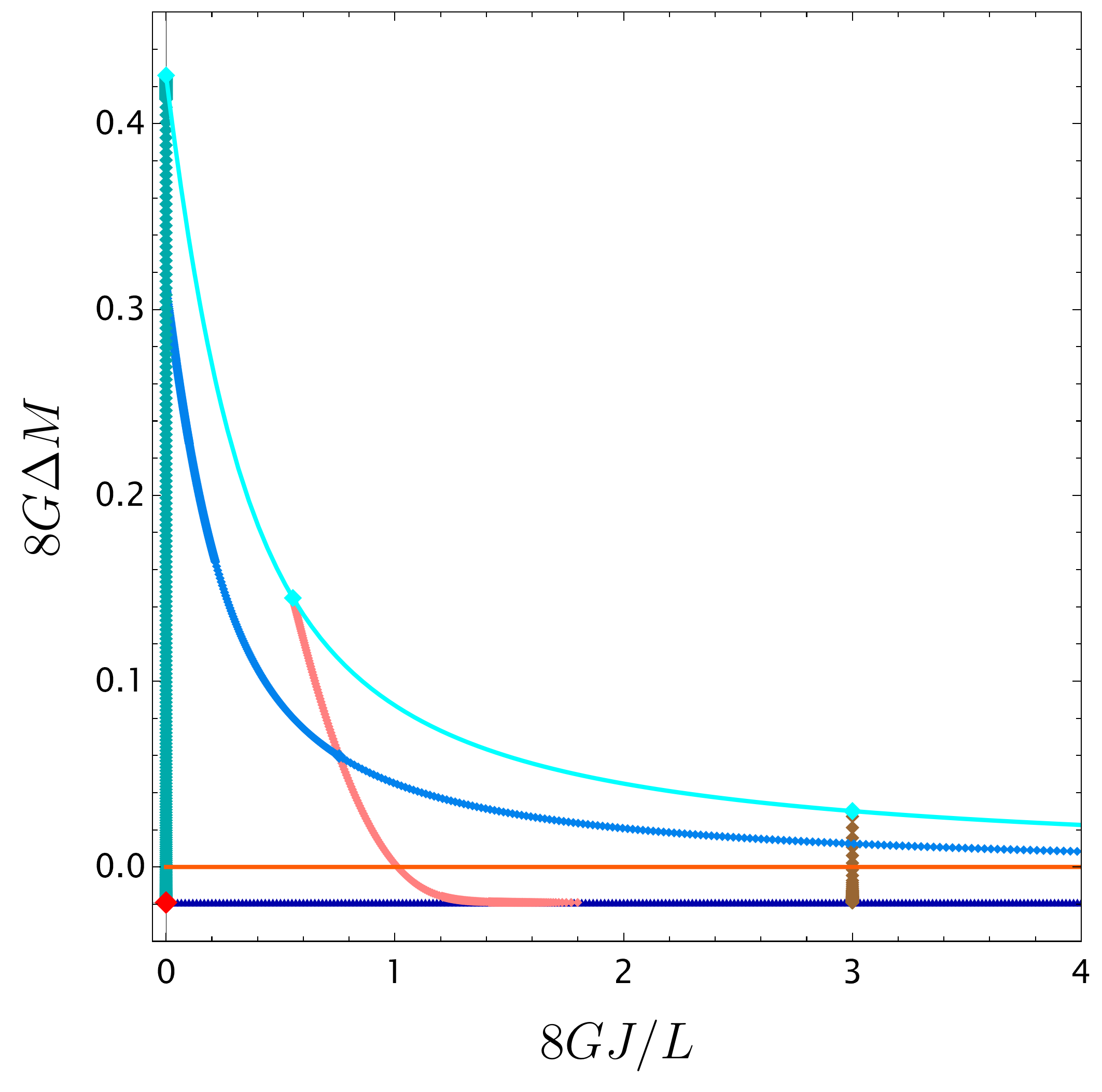}
    \caption{The three 1-parameter sub-families of $\bm{m=0}$ hairy black holes studied in section~\ref{sec:PhaseDiag-m0:BHsRotating}, namely the family with constant: $R_+=3/4$ (pink diamonds), $\alpha=0.206$ (blue diamonds), and $\hat{J}=3$ (brown diamonds). We also show the static ($\hat{J}=0$) hairy black hole family (petrol green), that ends on the singular $m=0$ static extremal hairy black hole (red diamond) described by \eqref{sBHm0J0:Thermo} and section~\ref{sec:NumericalSetup:SingBHm0J0}, both already displayed in Figs.~\ref{fig:BS-m0}~and~\ref{fig:m0_J0-BH}. The dark-blue triangles (that start at the red diamond and have $\Delta{M}(\hat{J})\simeq -0.019$) describe the singular $m=0$ rotating extremal hairy black hole identified in section~\ref{sec:NumericalSetup:singBHm0J} and \eqref{sBHm0J:Thermo}.
    This for a theory with $\mu^2L^2 = -15/16$ and $\kappa = -4/10$. We plot $\Delta \hat{M}$ vs  $\hat{J}$ where $\Delta \hat{M} = \hat{M} - \hat{M}^{\hbox{\tiny BTZ}}_{\hbox{\tiny ext}} $ is the mass difference of a given solution with respect to the extremal BTZ mass, $\hat{M}^{\hbox{\tiny BTZ}}_{\hbox{\tiny ext}} =\hat{J}$. So, the extremal BTZ family is the horizontal orange line with $\Delta\hat{M}=0$ and non-extremal BTZ black holes exist above this line for arbitrarily large $\Delta\hat{M}$. The cyan curve describes the onset curve $\hat{M}(\hat{J})|^{\hbox{\tiny BTZ}}_{\hbox{\tiny onset}}$  of the double-trace instability of BTZ black holes as found in \cite{Dias:2025uyk}.  Thermodynamic properties of these three sub-families will be presented in Fig.~\ref{fig:m0-hBTZ-Rp075} (family with  $R_+=3/4$), Fig.~\ref{fig:m0-hBTZ-alpha0206} (family with  $\alpha=0.206$), and Fig.~\ref{fig:m0-hBTZ-J3} (family with  $\hat{J}=3$) using the same colour code. }
    \label{fig:dMJ:3families}
\end{figure}

\begin{figure}
    \centering
    \includegraphics[width=0.45\linewidth]{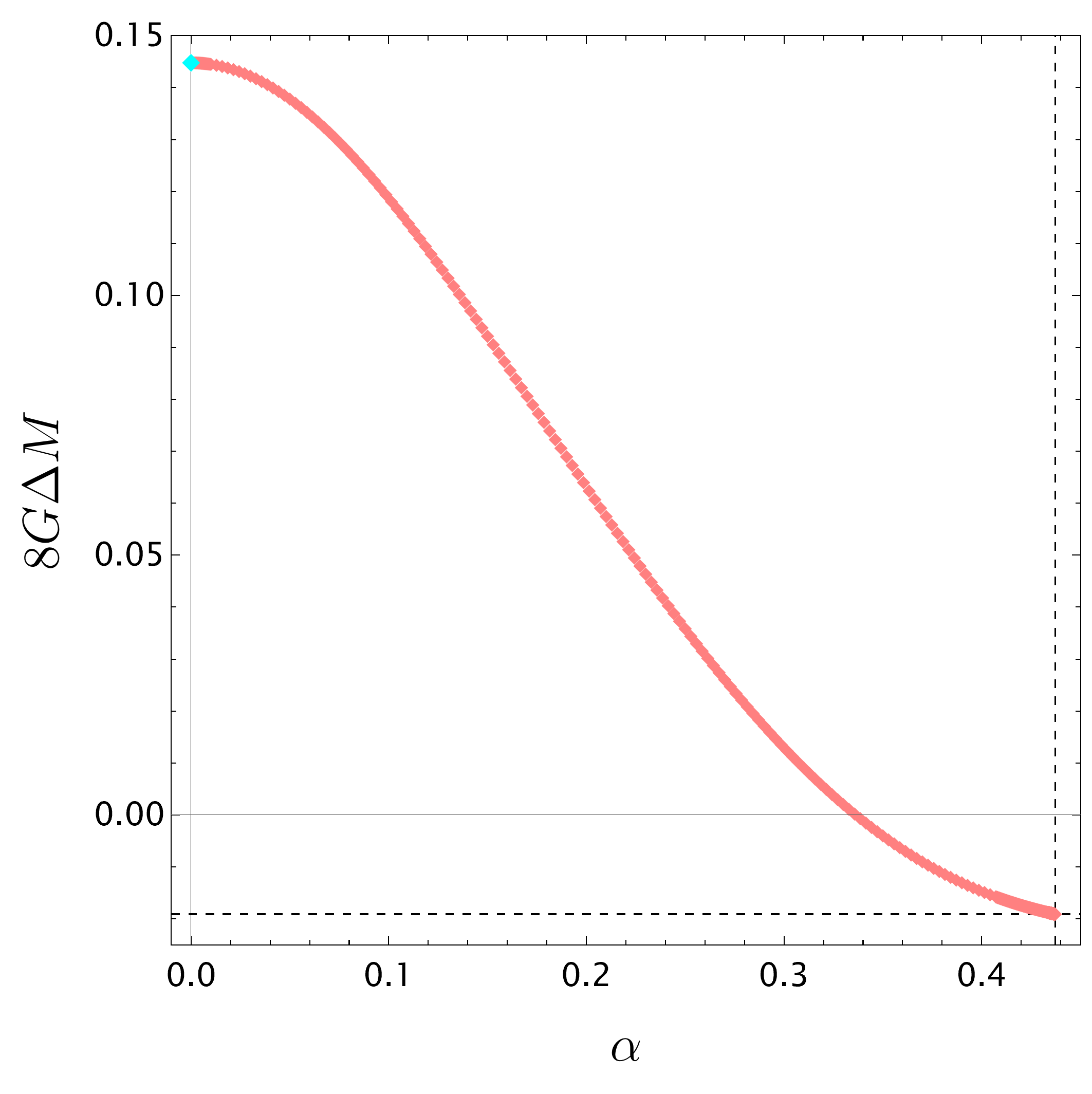}
    \includegraphics[width=0.45\linewidth]{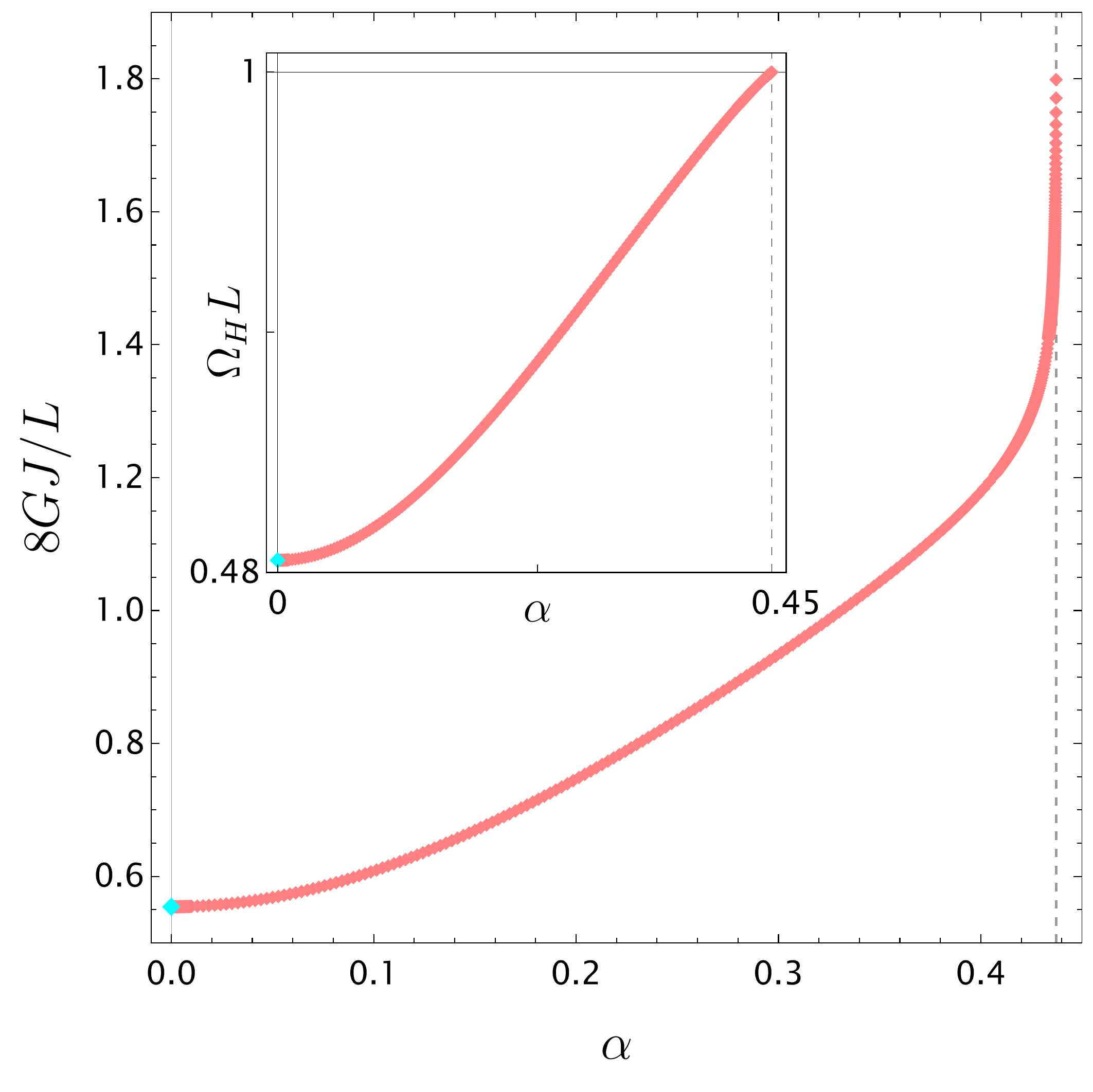}
    \includegraphics[width=0.45\linewidth]{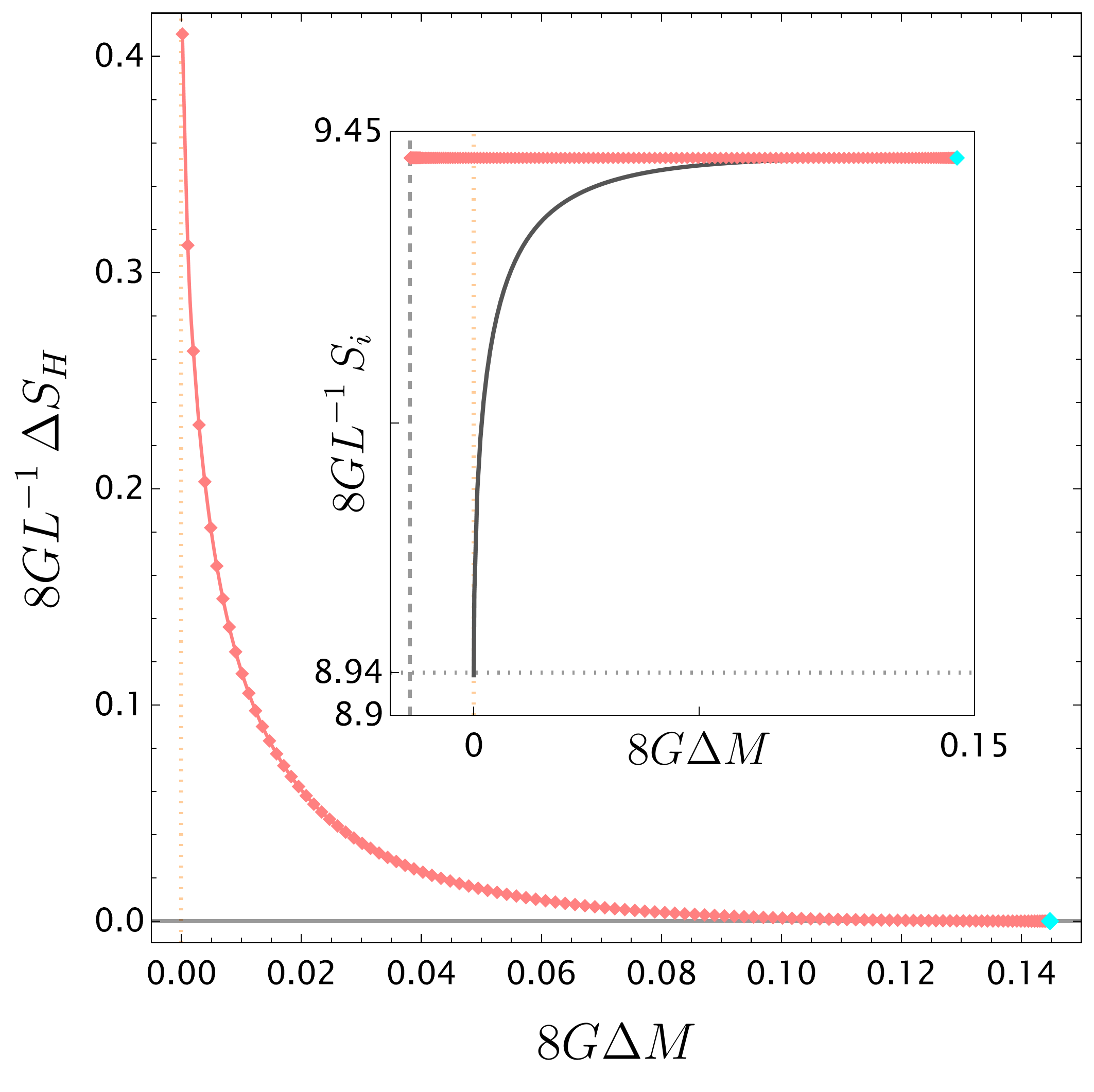}
     \includegraphics[width=0.45\linewidth]{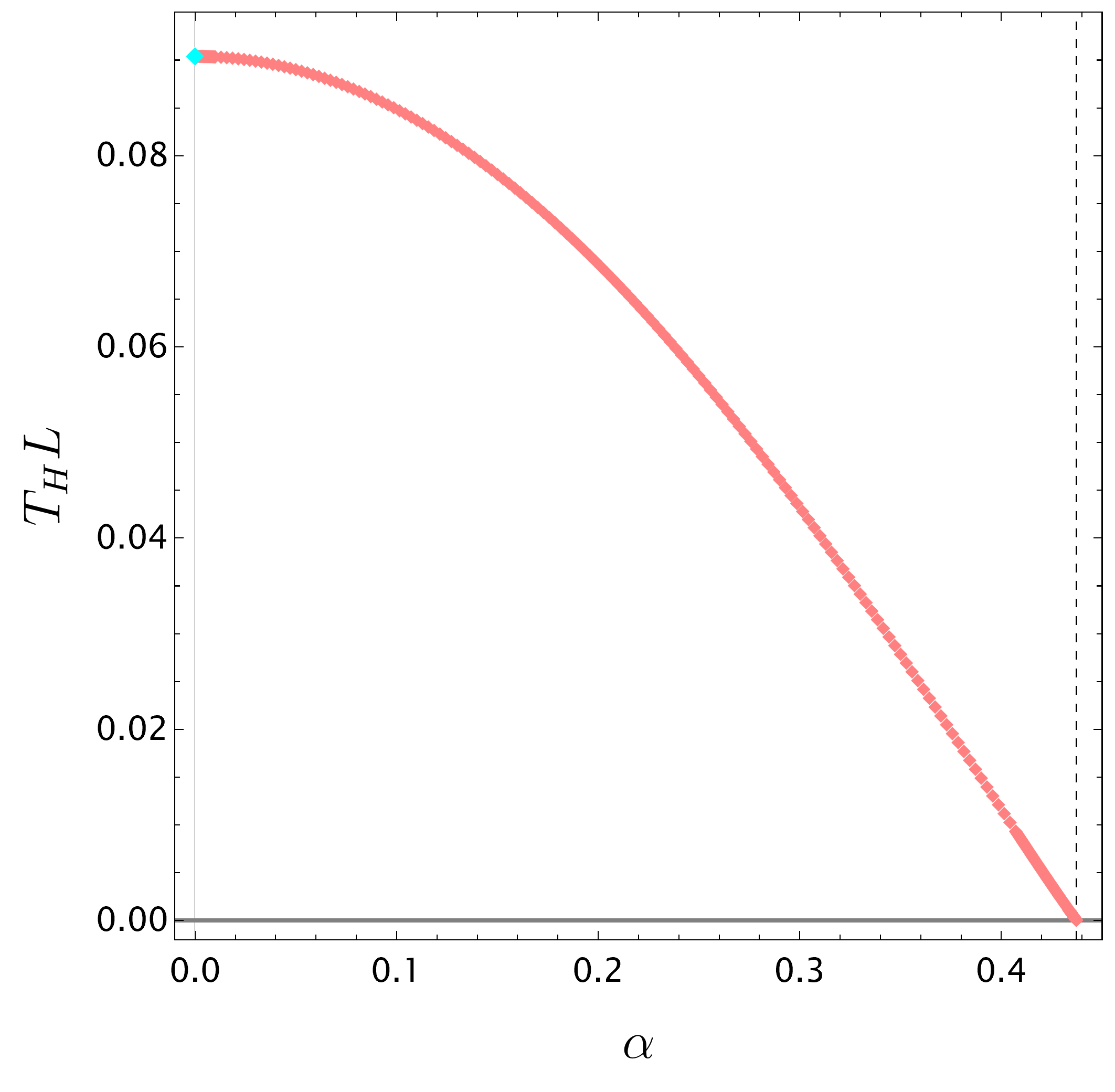}
     \includegraphics[width=0.45\linewidth]{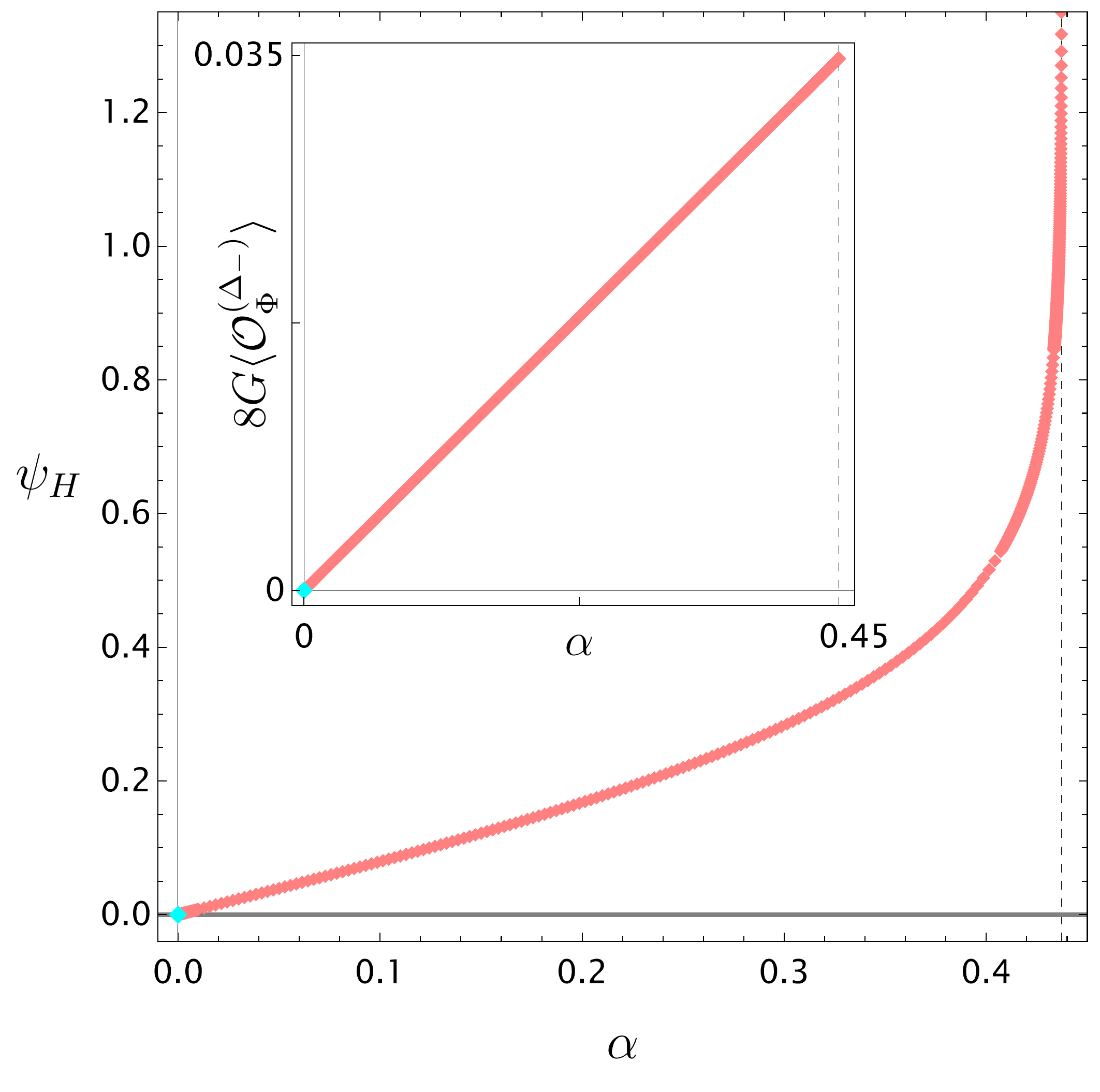}
    \includegraphics[width=0.45\linewidth]{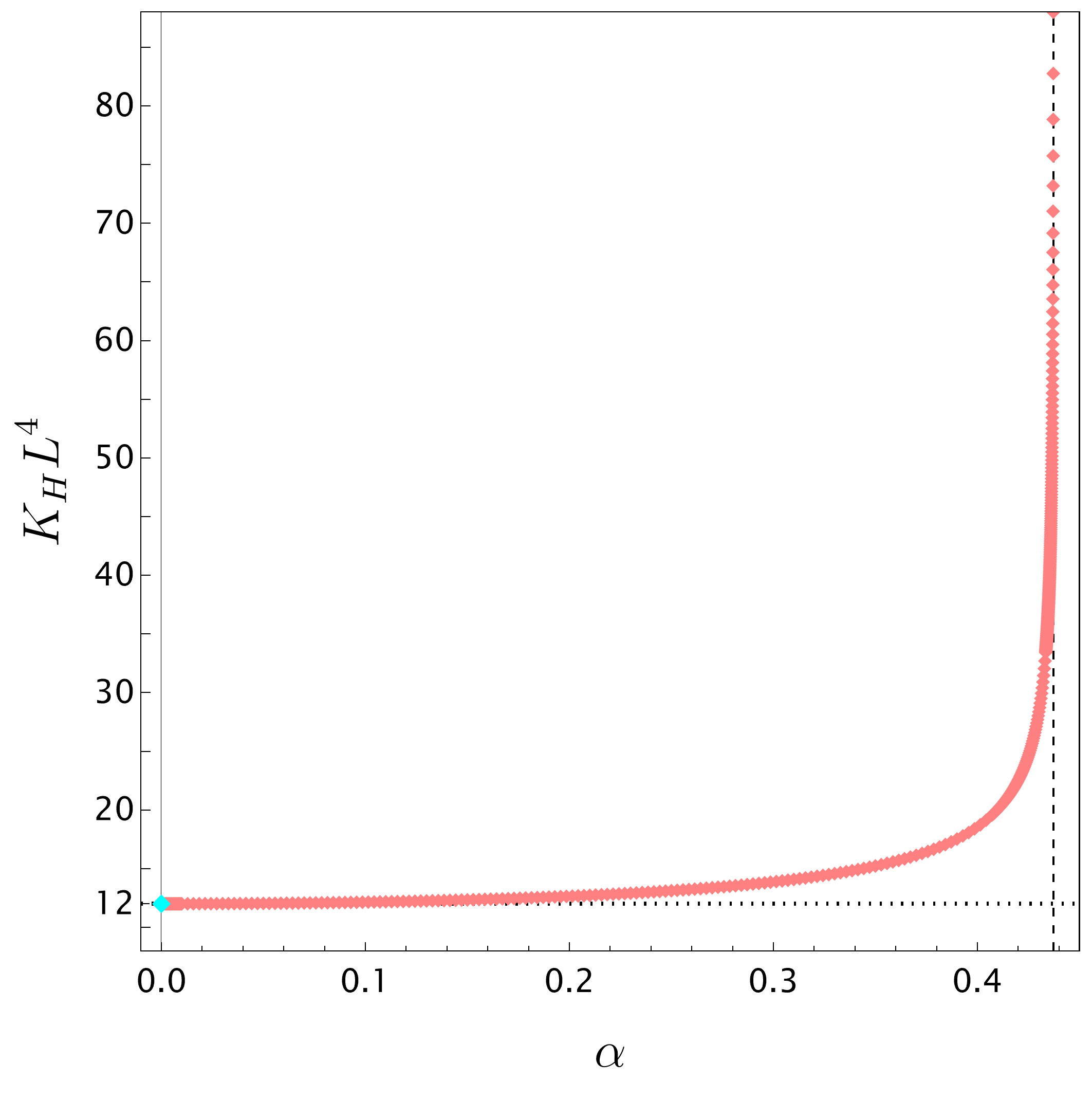}
    \caption{Thermodynamic properties of $\bm{m=0}$ hairy black hole sub-family with $\bm{R_+=0.75}$ (\ie $\bm{\hat{S}_H=3\pi}$) for $\kappa = -4/10$, $\mu^2L^2 = -15/16$.  The cyan diamond is the BTZ instability onset point.  Auxiliary orange dotted lines describe extremal BTZ ($\Delta \hat{M}=0$). $\Delta \hat{M}=\hat{M}- \hat{M}^{\hbox{\tiny BTZ}}_{\hbox{\tiny ext}} =\hat{M} -\hat{J}$ describes how far off the mass of a given hairy black hole is from the extremal BTZ mass.
    $\Delta \hat{S}_H=\hat{S}_H-\hat{S}_H^{\hbox{\tiny BTZ}}$ is the entropy difference between a given hairy black hole and BTZ that has the same  $\hat{M}$ and $\hat{J}$  (when they co-exist).
    }
    \label{fig:m0-hBTZ-Rp075}
\end{figure}

\begin{figure}
    \centering
    \includegraphics[width=0.46\linewidth]{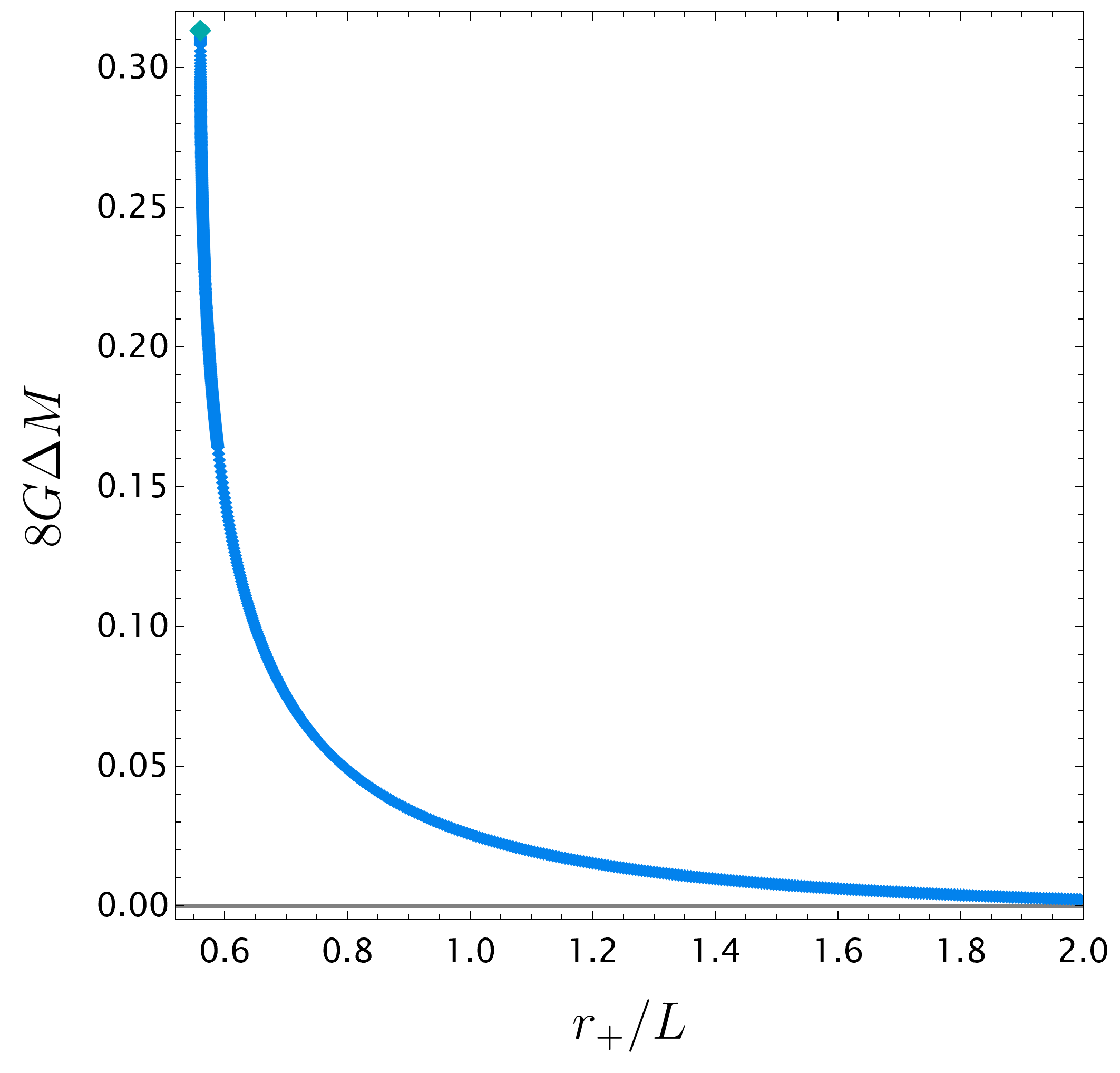}
    \includegraphics[width=0.45\linewidth]{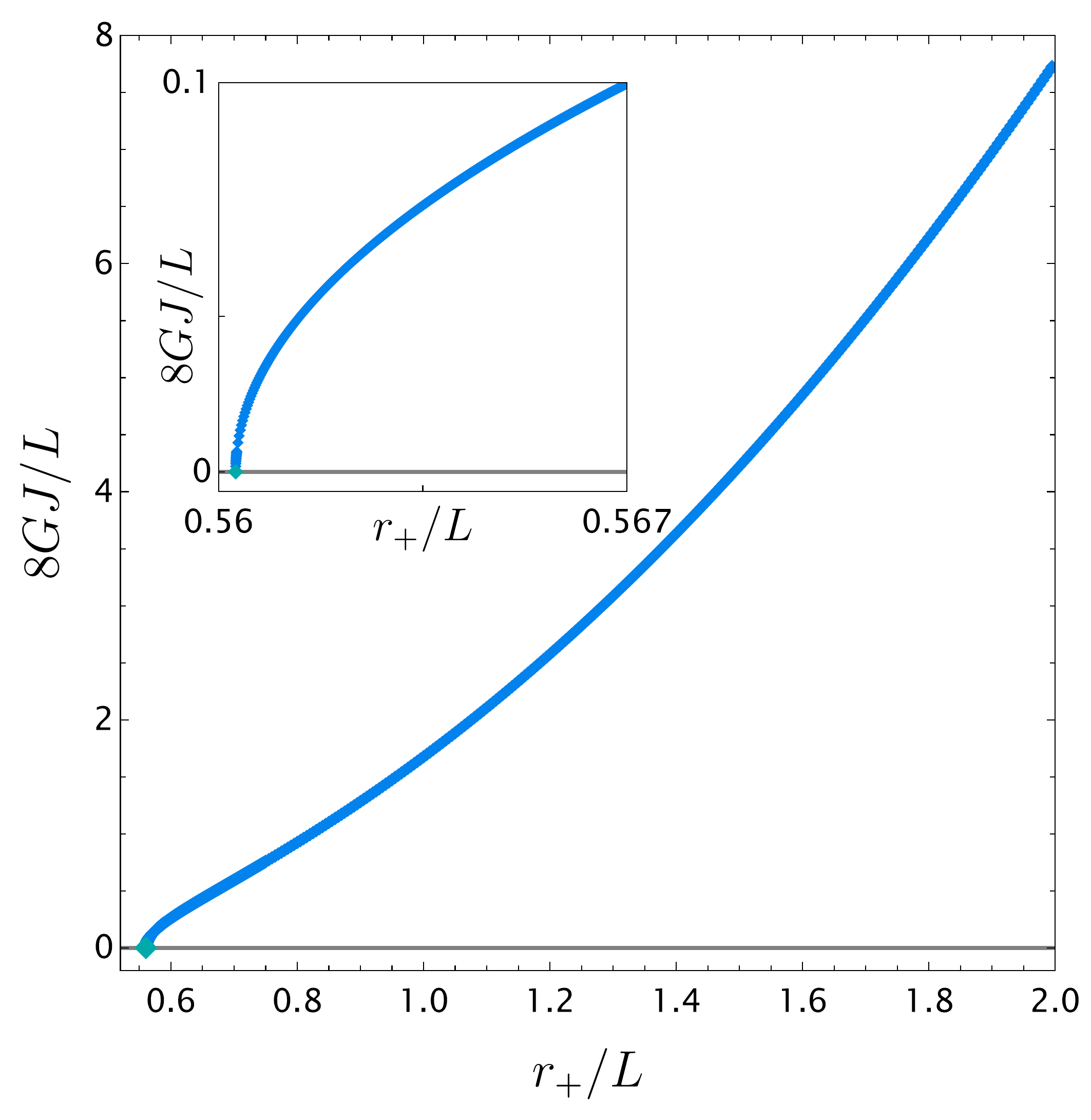}
    \includegraphics[width=0.45\linewidth]{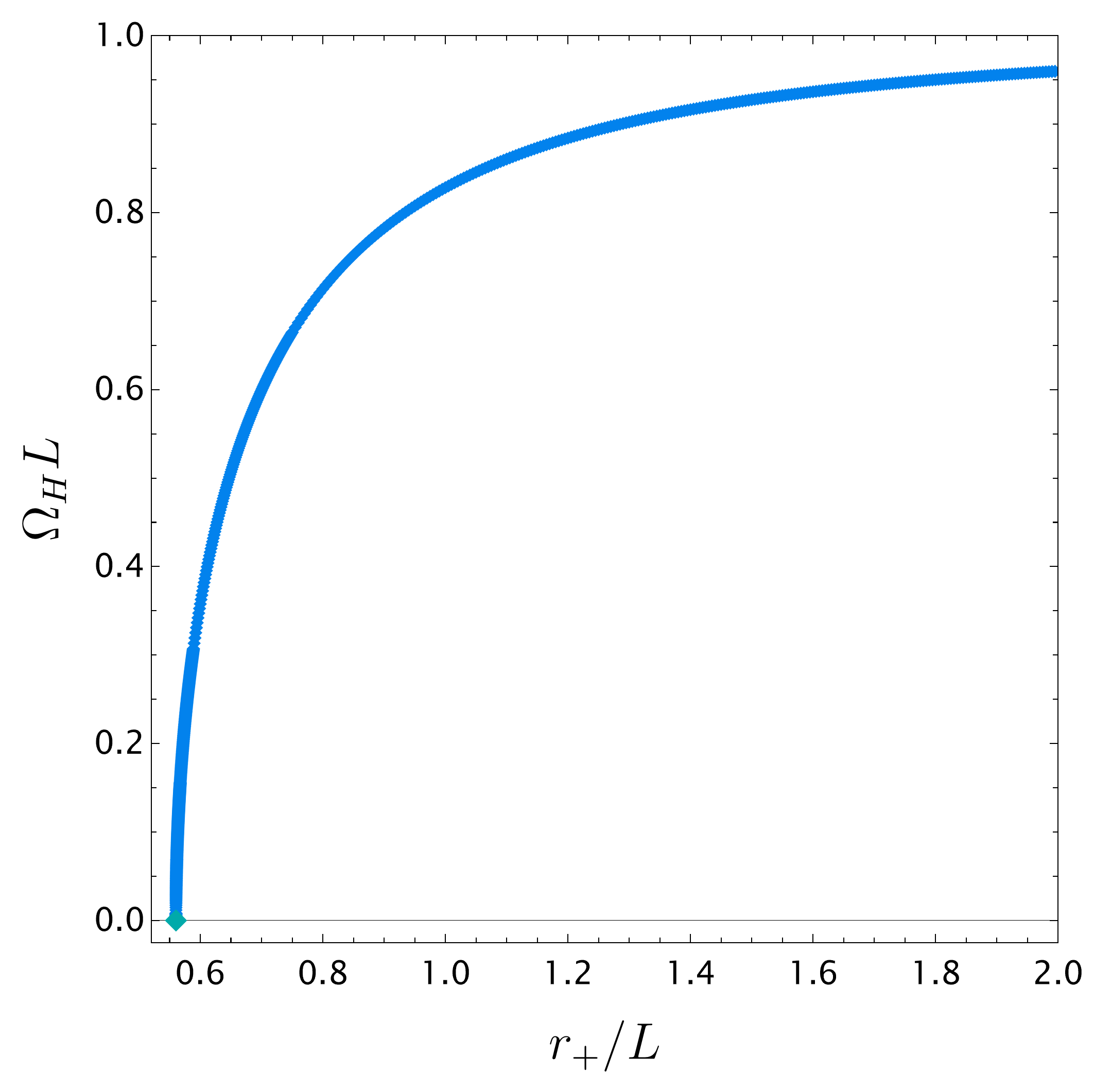}
    \includegraphics[width=0.45\linewidth]{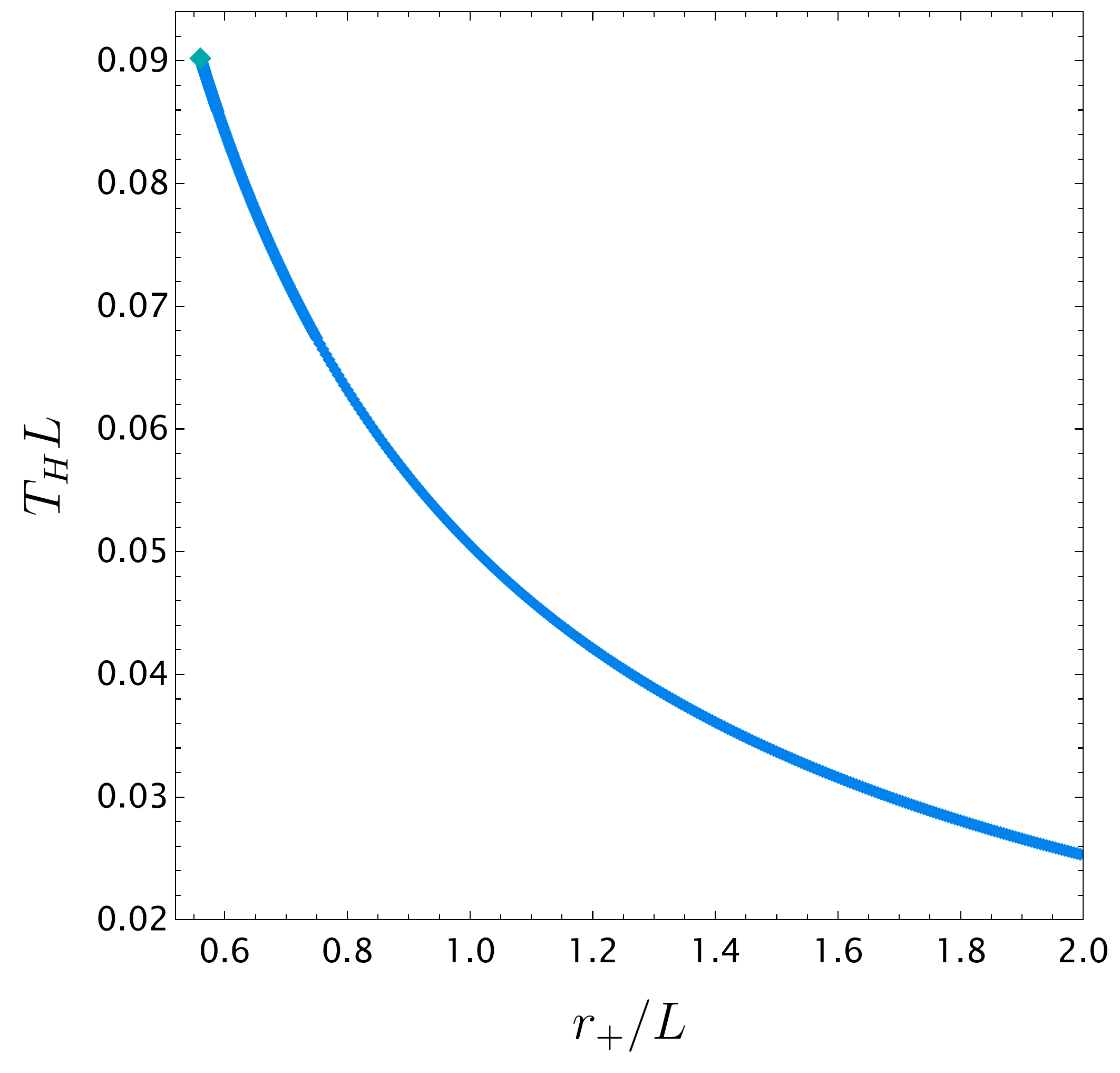}
    \includegraphics[width=0.47\linewidth]{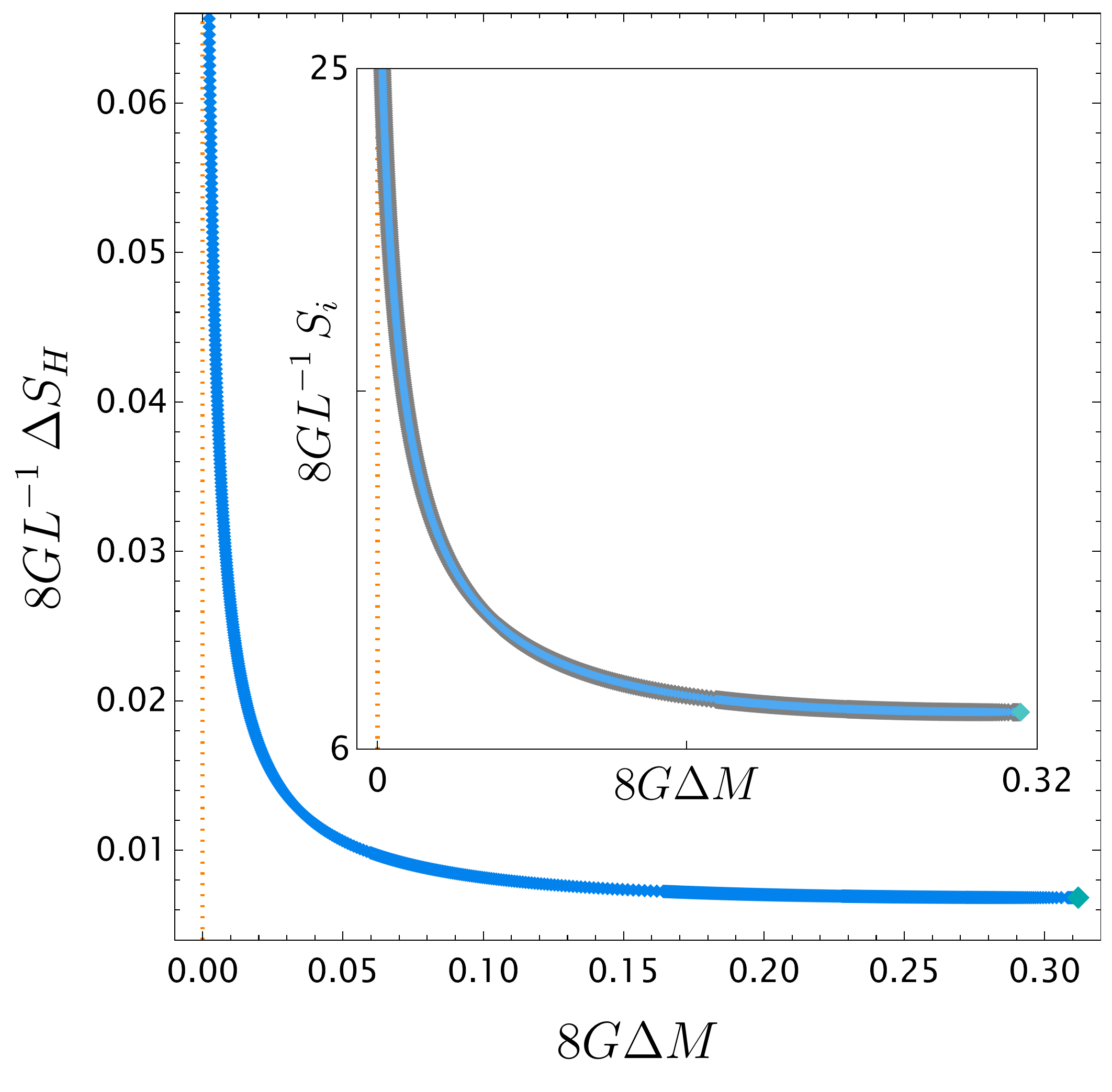}
    \caption{ 
Thermodynamic properties of $\bm{m=0}$ hairy black hole sub-family with $\bm{\alpha =0.206}$ for $\kappa = -4/10$, $\mu^2L^2 = -15/16$.  The green petrol diamond is the static $m=0$ hairy black hole.   $\Delta \hat{M}=\hat{M}- \hat{M}^{\hbox{\tiny BTZ}}_{\hbox{\tiny ext}} =\hat{M} -\hat{J}$ describes how far off the mass of a given hairy black hole is from the extremal BTZ mass. $\Delta \hat{S}_H=\hat{S}_H-\hat{S}_H^{\hbox{\tiny BTZ}}$ is the entropy difference between a given hairy black hole and BTZ that has the same  $\hat{M}$ and $\hat{J}$  (when they co-exist).
Auxiliary orange dotted line describes extremal BTZ ($\Delta \hat{M}=0$).
    }
    \label{fig:m0-hBTZ-alpha0206}
\end{figure}

\begin{figure}
    \centering
    \vskip -0.5cm
    \includegraphics[width=0.45\linewidth]{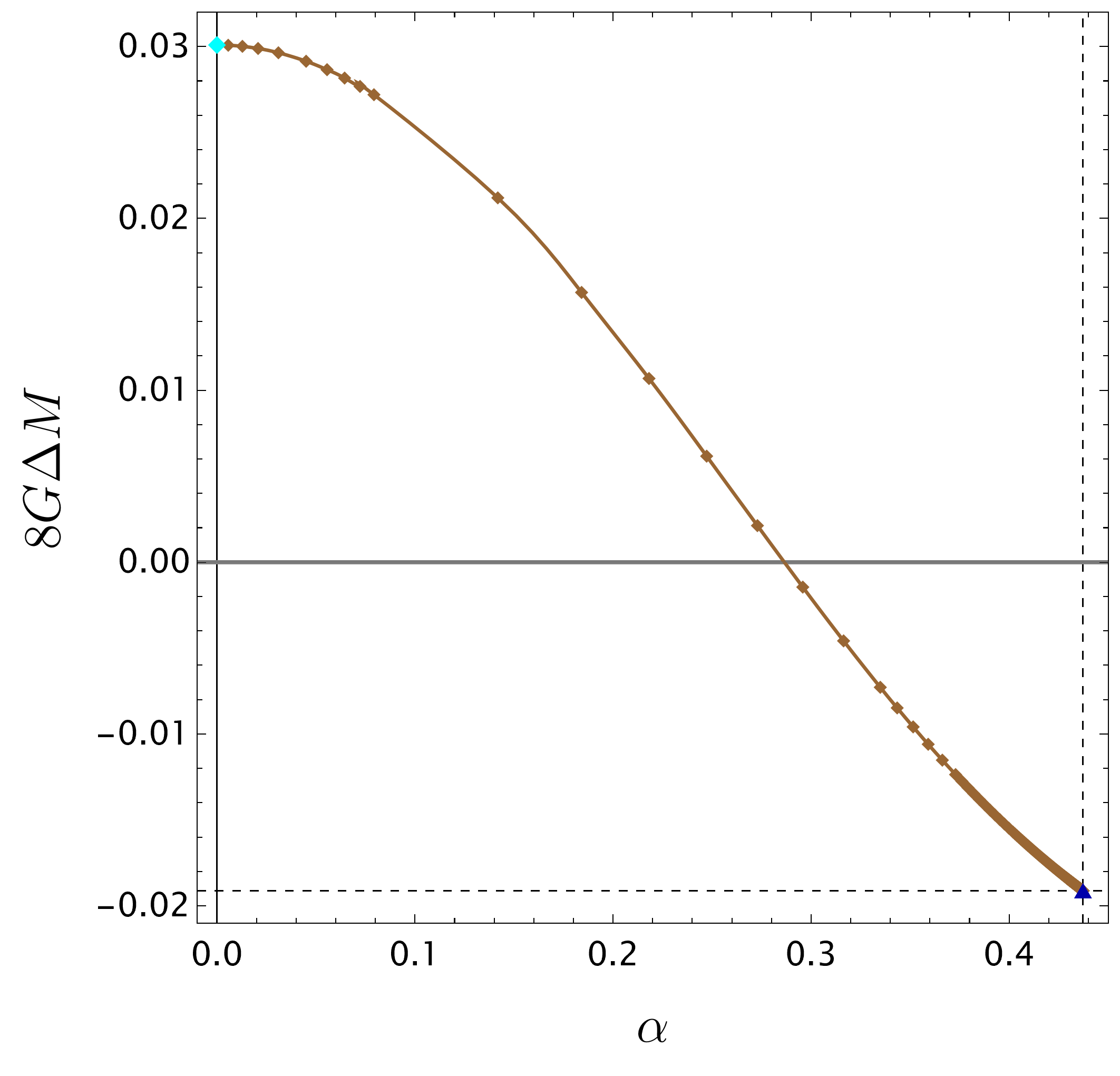}
      \includegraphics[width=0.45\linewidth]{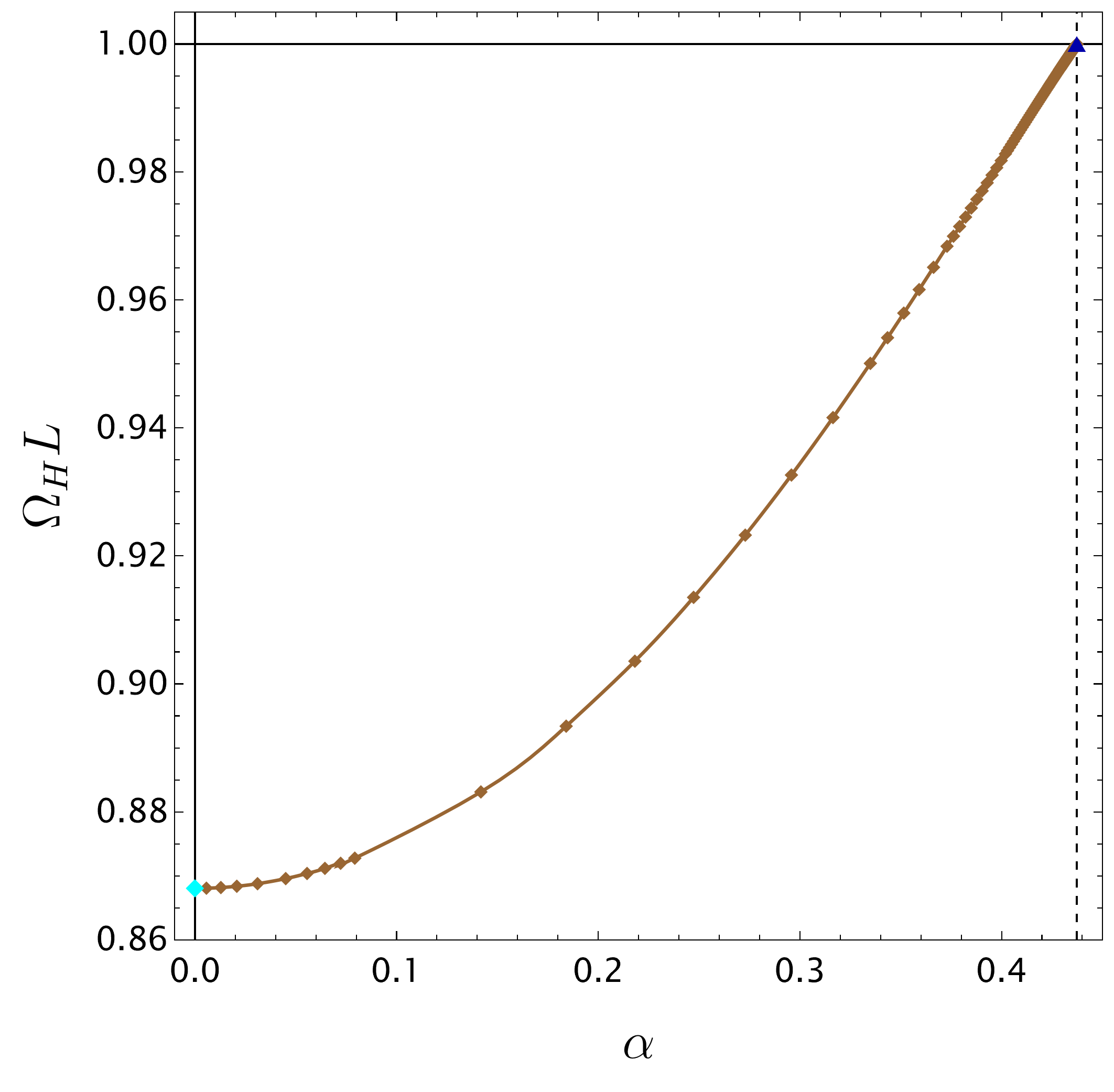}
     \includegraphics[width=0.45\linewidth]{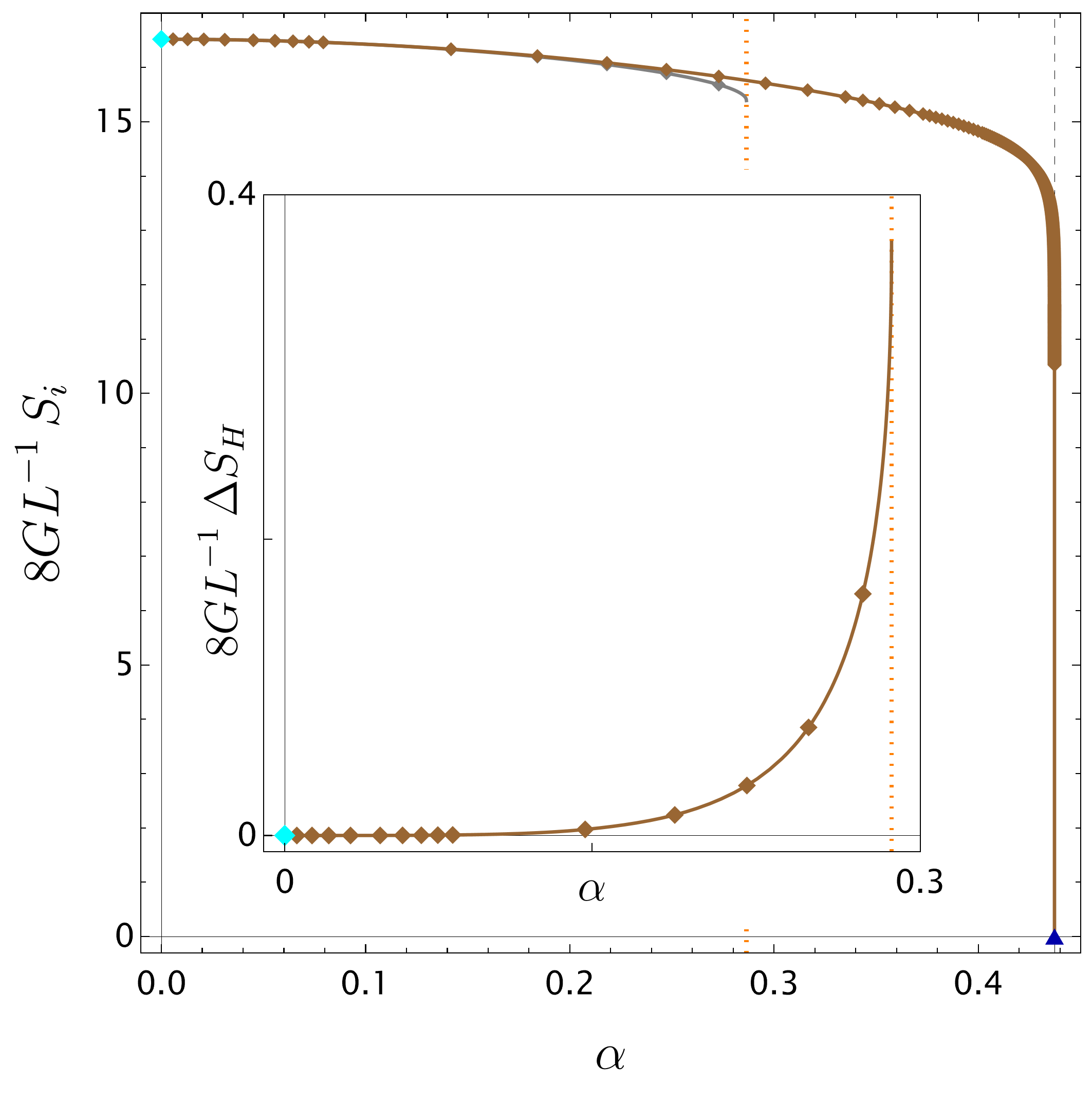}
    \includegraphics[width=0.45\linewidth]{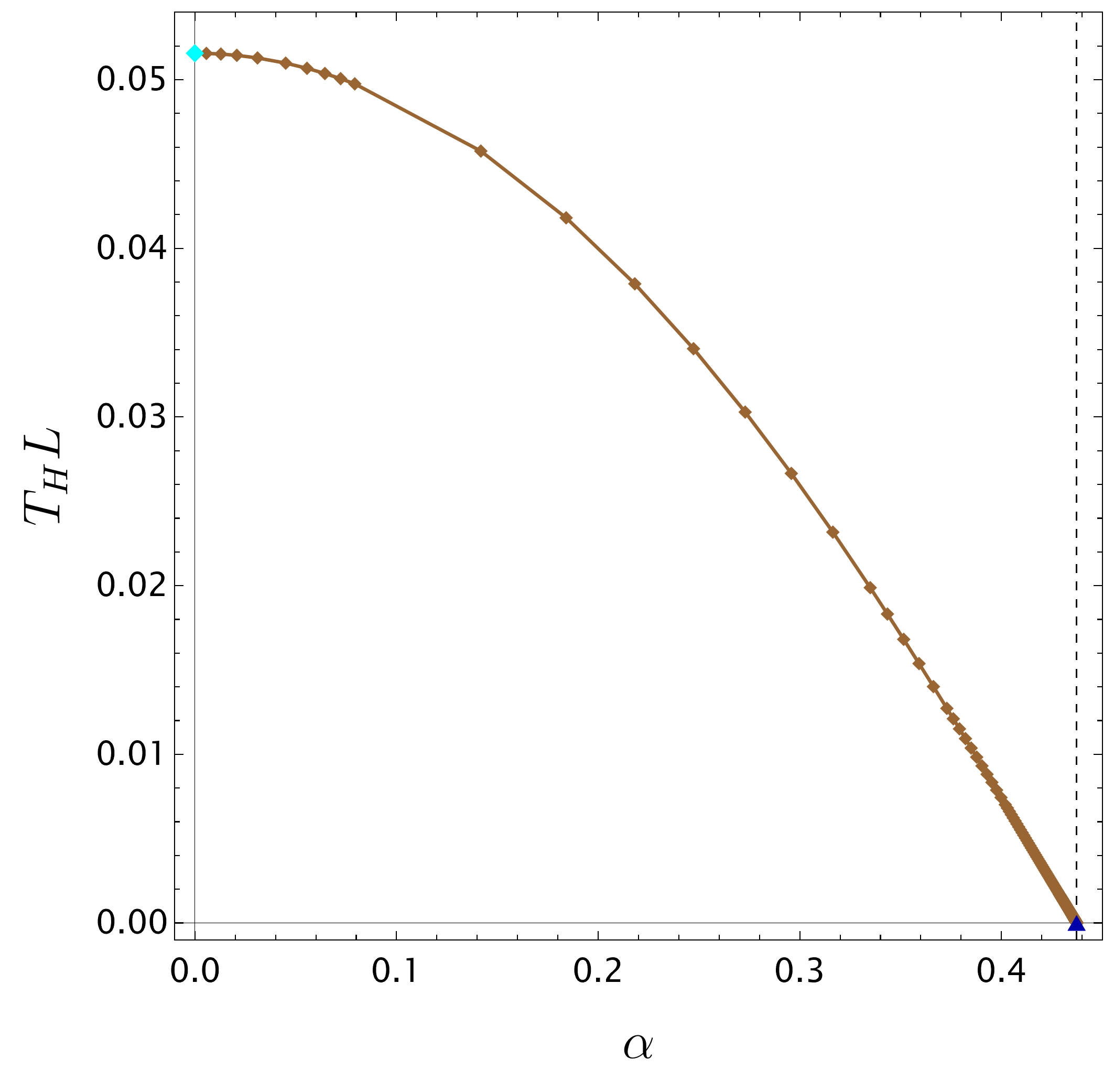}
     \includegraphics[width=0.44\linewidth]{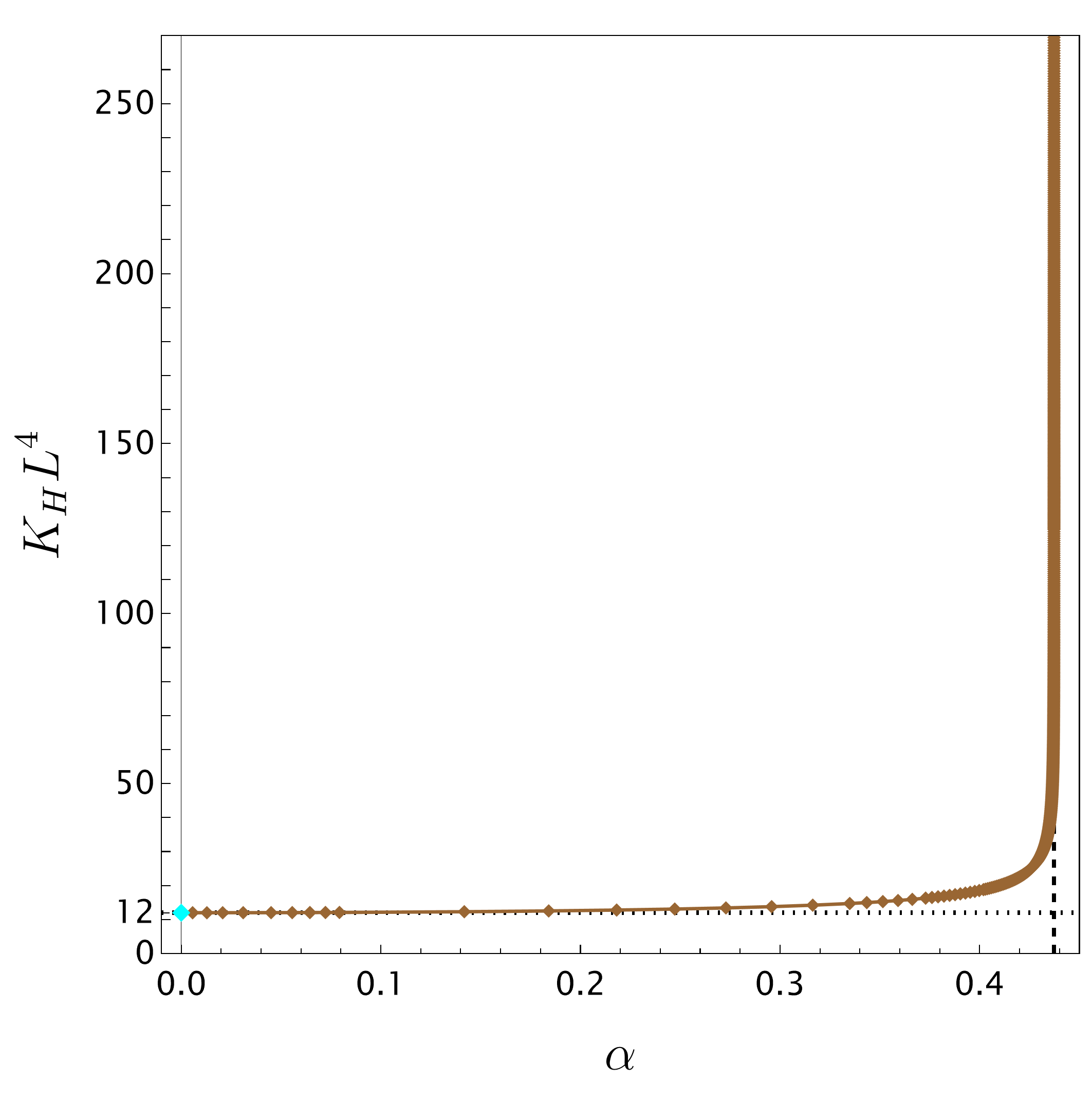}
     \includegraphics[width=0.43\linewidth]{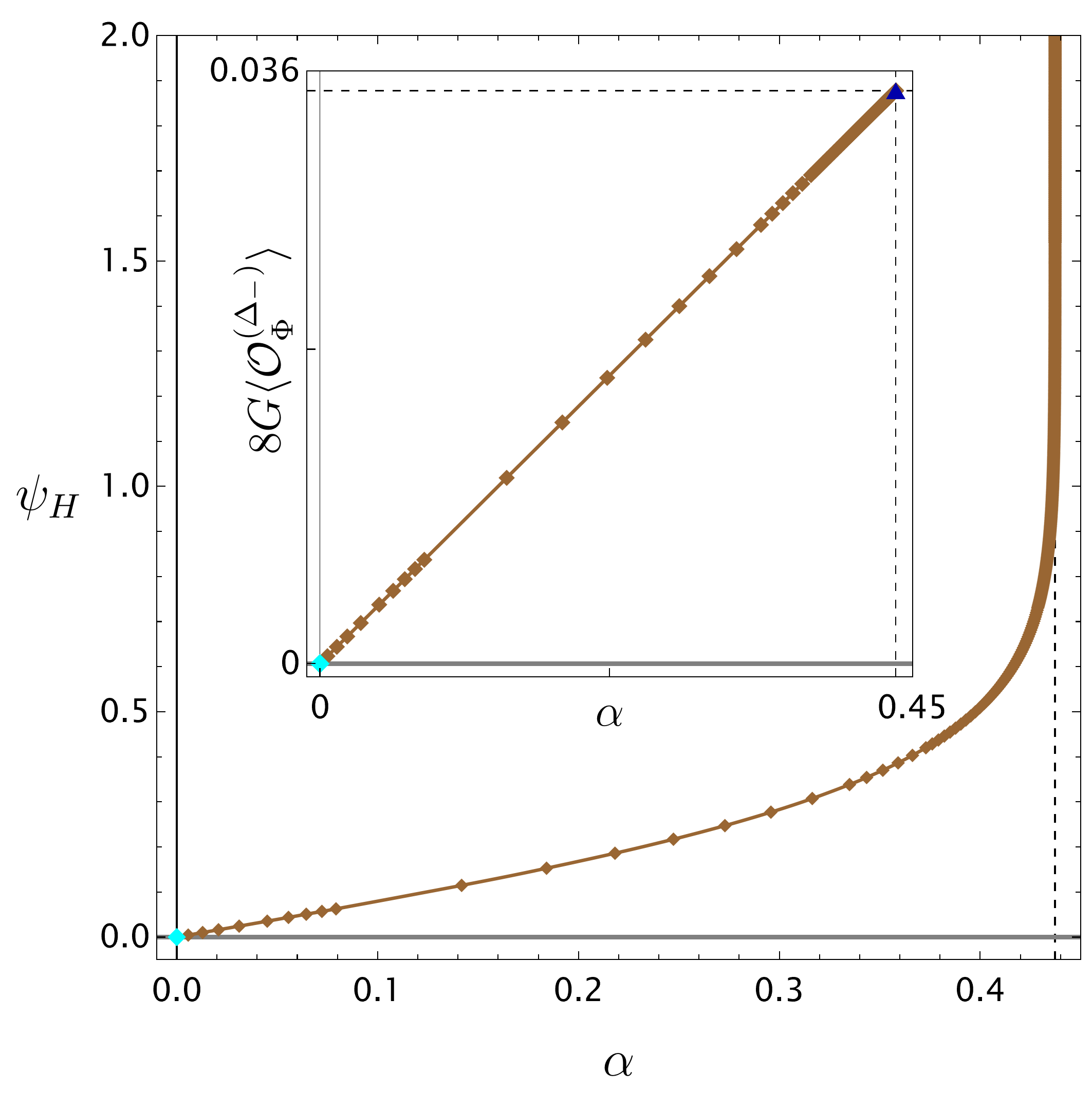}
    \caption{
Thermodynamic properties of $\bm{m=0}$ hairy black hole sub-family with $\bm{\hat{J}=3}$ for $\kappa = -4/10$, $\mu^2L^2 = -15/16$.  The green petrol diamond is the static $m=0$ hairy black hole. The dark-blue triangle with $\Delta{M}\simeq -0.019, \,\hat{\Omega}_H=1,\, \hat{S}_H=0, \, \hat{T}_H=0, \alpha \simeq 0.437,\, \langle\hat{\mathcal{O}}_{\Phi}^{(\Delta_-)}\rangle \simeq 0.034$
describes the singular $m=0$ rotating extremal hairy black hole identified in section~\ref{sec:NumericalSetup:singBHm0J} and \eqref{sBHm0J:Thermo}.
$\Delta \hat{M}=\hat{M}- \hat{M}^{\hbox{\tiny BTZ}}_{\hbox{\tiny ext}} =\hat{M} -\hat{J}$ describes how far off the mass of a given hairy black hole is from the extremal BTZ mass. $\Delta \hat{S}_H=\hat{S}_H-\hat{S}_H^{\hbox{\tiny BTZ}}$ is the entropy difference between a given hairy black hole and BTZ that has the same  $\hat{M}$ and $\hat{J}$  (when they co-exist).
Auxiliary orange dotted line describes extremal BTZ ($\Delta \hat{M}=0$).
}\label{fig:m0-hBTZ-J3}
\end{figure}

The three one‑parameter sub‑families of $m=0$ hairy black holes introduced above are displayed in Fig.~\ref{fig:dMJ:3families}. Their thermodynamic properties are analysed in detail in Fig.~\ref{fig:m0-hBTZ-Rp075} for the family with $R_{+}=3/4$, in Fig.~\ref{fig:m0-hBTZ-alpha0206} for the family with $\alpha=0.206$, and in Fig.~\ref{fig:m0-hBTZ-J3} for the family with fixed $\hat{J}=3$.

In more detail, Fig.~\ref{fig:m0-hBTZ-Rp075} shows several physical quantities for the sub‑family with fixed horizon radius $R_{+}=3/4$, plotted as functions of the asymptotic scalar amplitude $\alpha$. Specifically, we display: $\Delta\hat{M}$ (top‑left panel); $\hat{J}$ and $\hat{\Omega}_{H}$ (top‑right panel, including inset); $\Delta\hat{S}_{H}$ and $\hat{S}_{H}$ (middle‑left panel, including inset); the Hawking temperature $\hat{T}_{H}$ (middle‑right panel); the value of the scalar field at the horizon $\psi_{H}$ together with the VEV $\big\langle\hat{\mathcal{O}}_{\Phi}^{(\Delta_-)}\big\rangle$ (bottom‑left panel and inset); and the Kretschmann scalar at the horizon $\hat{K}_{H}$ (bottom‑right panel). We have verified that the first law of thermodynamics~\eqref{FirstLawBH} is satisfied, with relative errors below $10^{-3}\%$.

The main qualitative features of Fig.~\ref{fig:m0-hBTZ-Rp075} are straightforward to identify. Families of hairy black holes at fixed entropy - such as the one shown - always bifurcate from the BTZ solution at the cyan diamond marking the onset of the double‑trace instability of BTZ, as determined in~\cite{Dias:2025uyk}. This bifurcation corresponds to a second‑order phase transition at which $\Delta\hat{S}_{H}=0$ and the scalar condensate vanishes, \ie
$\alpha=0=\big\langle\hat{\mathcal{O}}_{\Phi}^{(\Delta_-)}\big\rangle=\psi_{H}$. The constant‑entropy branch then extends to arbitrarily large angular momentum, much as BTZ itself exists for arbitrarily large $\hat{J}$. In fact, a very small increase in $\alpha$ for $\alpha\gtrsim0.43$ leads to a rapid growth in $\hat{J}$. Along this sub‑family, the scalar condensate increases while the temperature decreases (equivalently, the horizon curvature grows), eventually reaching $\hat{T}_{H}\to0$ and $\hat{K}_{H}\to\infty$ as $\hat{J}\to\infty$, with $\hat{\Omega}_{H}\to1$.

Turning next to Fig.~\ref{fig:m0-hBTZ-alpha0206}, which corresponds to the sub‑family with fixed scalar amplitude $\alpha=0.206$, we plot several quantities as functions of the horizon radius $R_{+}$: $\Delta\hat{M}$ (top‑left panel), $\hat{J}$ (top‑right panel), $\hat{\Omega}_{H}$ (middle‑left panel), $\hat{T}_{H}$ (middle‑right panel), and $\Delta\hat{S}_{H}$ (bottom panel). Once again, the first law of thermodynamics~\eqref{FirstLawBH} is satisfied within relative errors below $10^{-3}\%$. Along this family, the VEV $\big\langle\hat{\mathcal{O}}_{\Phi}^{(\Delta_-)}\big\rangle$ is fixed by construction, while both $\psi_{H}$ and the horizon Kretschmann scalar $\hat{K}_{H}$ remain constant.

The plots of Fig.~\ref{fig:m0-hBTZ-alpha0206} further show that families of hairy black holes at fixed $\alpha$ always originate at $\hat{J}=0$ from the static hairy black hole (petrol‑green diamond), previously discussed in Figs.~\ref{fig:BS-m0} and~\ref{fig:m0_J0-BH}, and then extend to arbitrarily large $R_{+}$ or $\hat{J}$. Importantly, whenever a hairy black hole and a BTZ black hole coexist with the same $(\hat{M},\hat{J})$, the $m=0$ hairy black hole is always entropically favored, \ie
$\Delta\hat{S}_{H}>0$.

The analyses of Figs.~\ref{fig:m0-hBTZ-Rp075} and~\ref{fig:m0-hBTZ-alpha0206} suggest that, in the $\hat{J}$–$\Delta\hat{M}$ plane of Fig.~\ref{fig:dMJ:3families}, the two‑dimensional region populated by hairy black holes is bounded from above by the double‑trace instability onset curve (cyan) and from the left by the static hairy black hole branch at $\hat{J}=0$ (the petrol‑green curve), which extends from the cyan onset diamond down to the red diamond corresponding to the singular boson star. To complete the characterization of this region, it remains to identify its lower boundary. For this purpose, it is convenient to follow a one‑parameter sub‑family of hairy black holes at fixed angular momentum $\hat{J}$, corresponding to a vertical line in Fig.~\ref{fig:dMJ:3families}.

This strategy is implemented in Fig.~\ref{fig:m0-hBTZ-J3}, where we analyse a representative sub‑family with $\hat{J}=3$. We plot several quantities as functions of the asymptotic scalar amplitude $\alpha$, namely: $\Delta\hat{M}$ (top‑left panel), the horizon angular velocity $\hat{\Omega}_{H}$ (top‑right panel), $\Delta\hat{S}_{H}$ together with $\hat{S}_{H}$ (middle‑left panel, including inset), the Hawking temperature $\hat{T}_{H}$ (middle‑right panel), the Kretschmann curvature scalar at the horizon $K_{H}$ (bottom‑left panel), and the value of the scalar field at the horizon $\psi_{H}$ together with the VEV $\big\langle\hat{\mathcal{O}}_{\Phi}^{(\Delta_-)}\big\rangle$ (bottom‑right panel, including inset). We have verified that the first law of thermodynamics~\eqref{FirstLawBH} is satisfied, with relative errors below $10^{-3}\%$.

The main features of Fig.~\ref{fig:m0-hBTZ-J3} are transparent. Nevertheless, it is worth emphasizing that families of hairy black holes at fixed angular momentum $-$ such as the one shown $-$ always bifurcate from the BTZ solution at the cyan diamond marking the onset of the double‑trace instability, as computed in~\cite{Dias:2025uyk}. This bifurcation is a second‑order phase transition, characterized by $\Delta\hat{S}_{H}=0$ and vanishing scalar condensate, \ie $\alpha=0=\big\langle\hat{\mathcal{O}}_{\Phi}^{(\Delta_-)}\big\rangle=\psi_{H}$. Away from the onset, the constant‑$\hat{J}$ branch extends while both the entropy and the temperature decrease, eventually reaching a singular endpoint where $\hat{S}_{H}\to 0$, $\hat{T}_{H}\to 0$, $\hat{\Omega}_{H}\to 1$ and the horizon curvature diverges (see dark-blue triangle in Fig.~\ref{fig:m0-hBTZ-J3}). These features were not apparent from the analyses of Figs.~\ref{fig:m0-hBTZ-Rp075} and~\ref{fig:m0-hBTZ-alpha0206}. Remarkably, as this singular limit is approached, the entropy and temperature drop sharply to zero even though the scalar amplitude $\alpha$ changes only mildly.

This singular endpoint turns out to be  the singular $m=0$ rotating extremal hairy black hole identified in section~\ref{sec:NumericalSetup:singBHm0J} (dark-blue triangle in Fig.~\ref{fig:m0-hBTZ-J3}).  
Its thermodynamic observables are
given by 
\begin{align}\label{sBHm0J:Thermo}
&\hbox{Singular $m=0$ rotating extremal hairy black hole}\: (\hat{\mu}^2=-15/16,\kappa=-4/10): \nonumber\\
&
\Delta{M}(\hat{J})\simeq -0.019, \,\, \hat{S}_H=0, \,\, \hat{T}_H=0, \,\, \hat{\Omega}_H=1,\,\, \alpha (\hat{J}) \simeq 0.437,\,\, \langle\hat{\mathcal{O}}_{\Phi}^{(\Delta_-)}\rangle (\hat{J})\simeq 0.034\,,
\end{align}
and describe the horizontal dark-blue triangle line in Fig~\ref{fig:dMJ:3families}.
At $\hat{J}=0$, this singular zero‑radius limit reduces to the  singular $m=0$ static extremal hairy black hole \eqref{sBHm0J0:Thermo} studied in section
~\ref{sec:NumericalSetup:SingBHm0J0} (red diamond).

A key result is that the singular $m=0$ rotating extremal hairy black hole family-identified in \eqref{sBHm0J:Thermo} and Fig.~\ref{fig:dMJ:3families} for the given $\hat{\mu}$ and $\kappa$—saturates the positivity-of-energy theorem~\eqref{GlobalMin}-\eqref{Super:Emin} derived in Appendix~\ref{secA:superpotentials} for $\hat J\neq 0$ and $p=0$.

\section{Exploratory study of \texorpdfstring{AdS$_3$}{AdS3} hairy solutions with \texorpdfstring{$m=1$ ($m\geq 1$)}{m=1 (m >= 1)}}\label{sec:PhaseDiag-m1}
In our companion paper~\cite{Dias:2025uyk}, we studied linear double‑trace perturbations of AdS$_3$ and BTZ black holes, not only in the axisymmetric sector ($m=0$), but also in the non‑axisymmetric sectors with $m\geq1$. We found that the $m=0$ sector is the most relevant one, in the sense that for any fixed value of the double‑trace parameter $\kappa$, any BTZ black hole that is unstable to modes with $m\geq1$ is already unstable to the $m=0$ mode, with the latter typically exhibiting a stronger instability. A similar conclusion applies to perturbations of global AdS$_3$ and to the associated Ishibashi–Wald instability~\cite{Ishibashi:2004wx,Dias:2025uyk}.

Despite this, in the $\hat{J}$–$\hat{M}$ phase diagram there exists a distinct one‑parameter onset curve
\begin{equation}
\hat{M}(\hat{J})\big|^{m\geq1}_{\text{\tiny BTZ onset}}
   < \hat{M}(\hat{J})\big|^{m=0}_{\text{\tiny BTZ onset}},
\end{equation}
below which BTZ black holes are unstable to perturbations with $m\geq1$. As in the $m=0$ case, one therefore expects the existence of $m\geq1$ hairy black holes bifurcating from the onset curve $\hat{M}(\hat{J})\big|^{m\geq1}_{\text{\tiny BTZ onset}}$ and extending into the unstable region below it. These non‑axisymmetric ($m\geq1$) hairy black holes are expected to exhibit qualitative features distinct from those of the axisymmetric $m=0$ hairy BTZ solutions. Similarly, again in close analogy with the $m=0$ sector, one anticipates that the $m\geq1$ Ishibashi–Wald instability of AdS$_3$ signals a transition between $m\geq1$ boson stars that are perturbatively connected to AdS$_3$ and those that are not.

In this section, we test these expectations explicitly by constructing and analysing $m\geq1$ hairy boson stars and black holes, and by comparing their properties with those of their $m=0$ counterparts.

As in the $m=0$ case, the main qualitative features of the $m\geq1$ solutions are expected to be largely insensitive to the precise value of the scalar mass $\hat{\mu}=\mu L$ (see discussion in the conclusion section~\ref{sec:Conc}). Moreover, solutions with $m\geq2$ are qualitatively similar to those with $m=1$, although they differ significantly from the axisymmetric $m=0$ solutions. For concreteness, and to allow for direct comparison with earlier sections, we therefore focus on $m=1$ and take $\mu^{2}L^{2}=-15/16$ throughout. This value lies within the range~\eqref{2xTrace:rangeMass}, $-1<\mu^{2}L^{2}<0$, where the double‑trace boundary condition~\eqref{2xTrace:BC} is allowed.

Finally, recall that the relation $\beta=\kappa\,\alpha$, with $\kappa\in\mathbb{R}$, interpolates between Neumann ($\kappa=0$) and Dirichlet ($\kappa\to\pm\infty$) boundary conditions. As in the $m=0$ case, $m\geq1$ hairy black holes do not exist for either Dirichlet or Neumann boundary conditions. Hairy boson stars with $m\geq1$, however, do exist for Dirichlet and Neumann boundary conditions, and a discussion of these solutions is deferred to Appendix~\ref{secA:BStars-DNm1}.

\subsection{Hairy \texorpdfstring{AdS$_3$}{AdS3} boson stars and singular extremal black holes with \texorpdfstring{$m=1$}{m=1}}\label{sec:PhaseDiag-m1:BStar}
In Ref.~\cite{Ishibashi:2004wx}, Ishibashi and Wald established that AdS$_{d}$, with $d\geq3$, can be unstable under mixed (Robin) boundary conditions not only in the axisymmetric sector ($m=0$), but in fact for scalar perturbations with arbitrary azimuthal number $m\geq0$. By contrast, AdS$_{d}$ is linearly mode stable for all $m\geq0$ perturbations obeying Dirichlet or Neumann boundary conditions. Of direct relevance for the present discussion, this instability arises for scalar fields with double‑trace boundary conditions in the mass range~\eqref{2xTrace:rangeMass}, propagating in AdS$_3$, including modes with $m\geq1$. Ishibashi and Wald~\cite{Ishibashi:2004wx} established the existence and onset conditions for this instability, while in our companion paper~\cite{Dias:2025uyk} we determined its characteristic time scale. In particular, AdS$_3$ is unstable whenever $\kappa<\kappa^{\rm AdS}_{m,\hat{\mu}^{2}}$ and linearly mode stable otherwise, where for a given scalar mass $\hat{\mu}$ and azimuthal number $m$ the critical onset value of $\kappa$ is given by \eqref{kAdS-onset}. As an explicit example, for $\mu^{2}L^{2}=-15/16$ and $m=1$, \eqref{kAdS-onset} yields $\kappa^{\rm AdS}_{1,-15/16}\simeq-1.009837$.

In this subsection, we focus on the regime $\kappa>\kappa^{\rm AdS}_{m\geq1,\hat{\mu}^{2}}$, where AdS$_3$ is linearly stable against $m\geq1$ perturbations. The complementary unstable regime, $\kappa<\kappa^{\rm AdS}_{m\geq1,\hat{\mu}^{2}}$, will be discussed separately in Section~\ref{sec:PhaseDiag-m1:IshWald}.

When $\kappa>\kappa^{\rm AdS}_{m\geq1,\hat{\mu}^{2}}$, the corresponding normal modes of AdS$_3$ have purely real frequencies. For instance, in the case $\mu^{2}L^{2}=-15/16$ and $m=1$, these frequencies were computed and plotted as functions of $\kappa$ in Fig.~2 of Ref.~\cite{Dias:2025uyk}. Going beyond linear perturbation theory, \ie once the back‑reaction of the scalar field on the metric is taken into account, one expects these normal modes to extend to fully nonlinear, horizonless, and everywhere regular solutions. In other words, in close analogy with the $m=0$ case, regular rotating ($m\geq1$) hairy AdS$_3$ boson stars should exist for all $\kappa>\kappa^{\rm AdS}_{m\geq1,\hat{\mu}^{2}}$ and $-1<\hat{\mu}<0$.

We confirm that this expectation is indeed realized. More precisely, for $m\geq1$ we find regular hairy boson stars following the numerical strategy outlined in section~\ref{sec:NumericalSetup:RegBS}. They necessarily rotate since the self‑gravitating scalar field carries angular momentum in the non‑axisymmetric sector.  For concreteness, we illustrate the structure of the $m\geq1$ phase diagram using the representative case $m=1$ with $\kappa=-0.8>\kappa^{\rm AdS}_{1,-15/16}\simeq-1.009837$, and scalar mass $\mu^{2}L^{2}=-15/16$.

\begin{figure}
    \centering
    \includegraphics[width=0.45\linewidth]{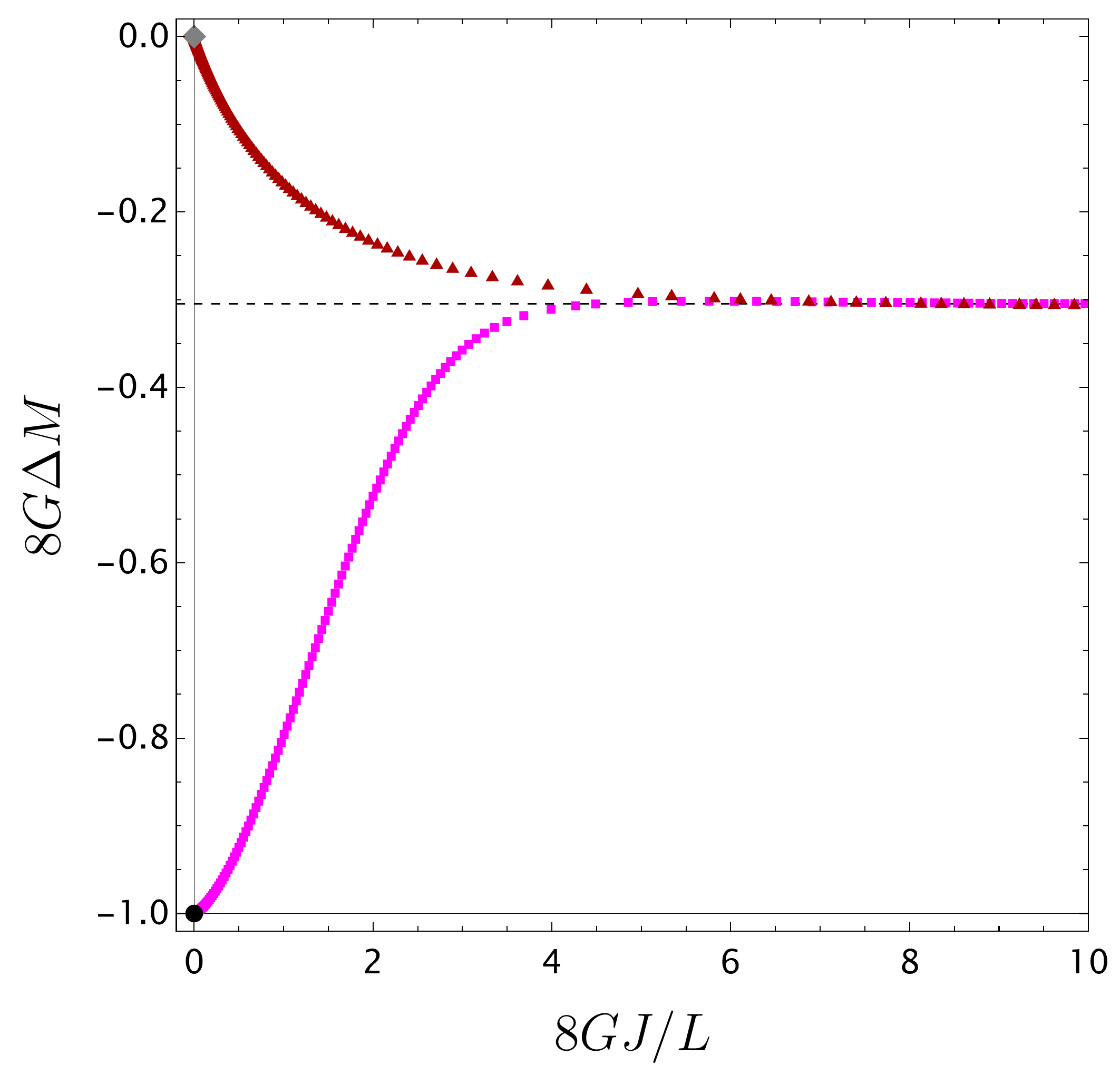}
     \includegraphics[width=0.45\linewidth]{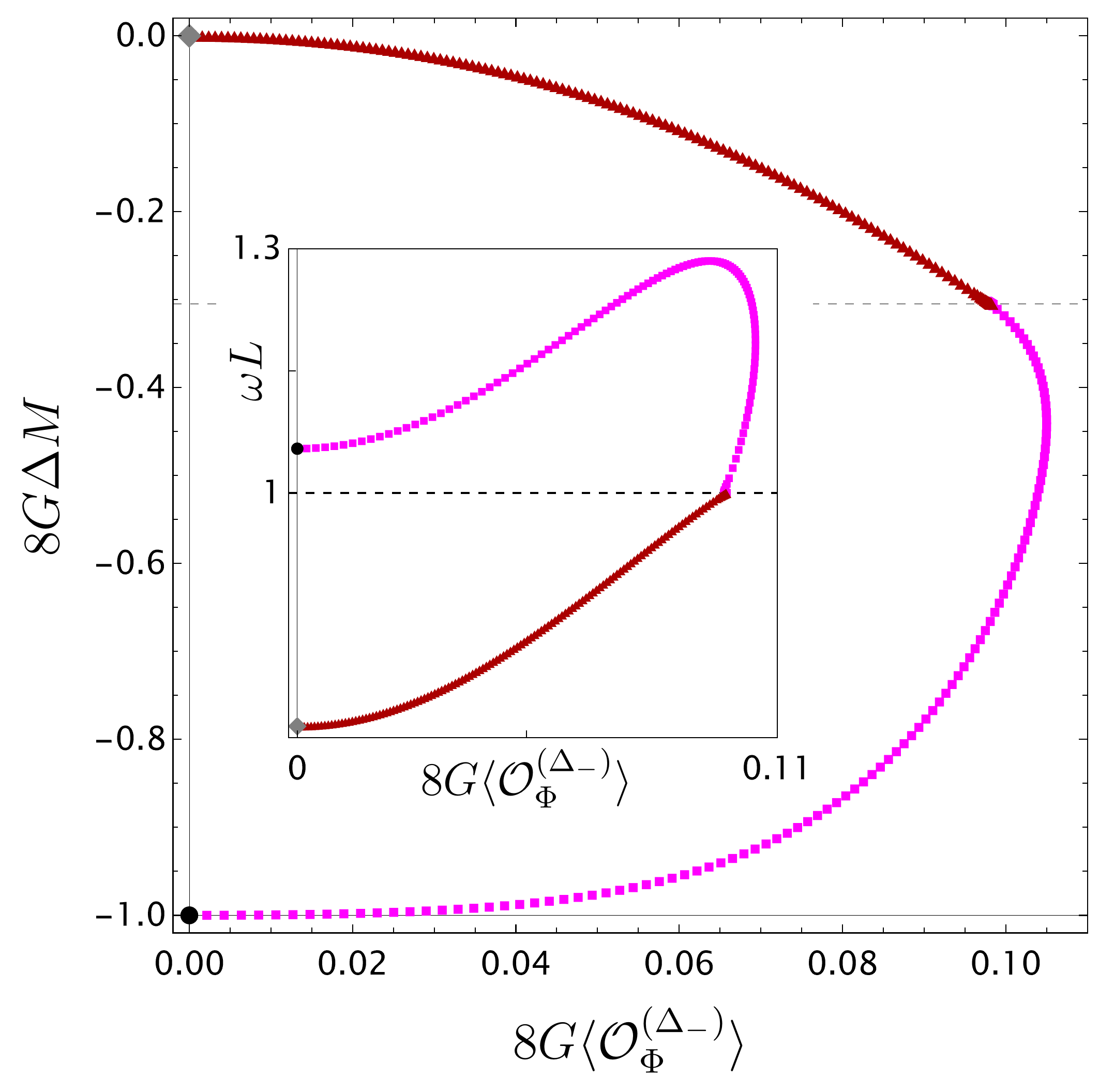}
    \includegraphics[width=0.45\linewidth]{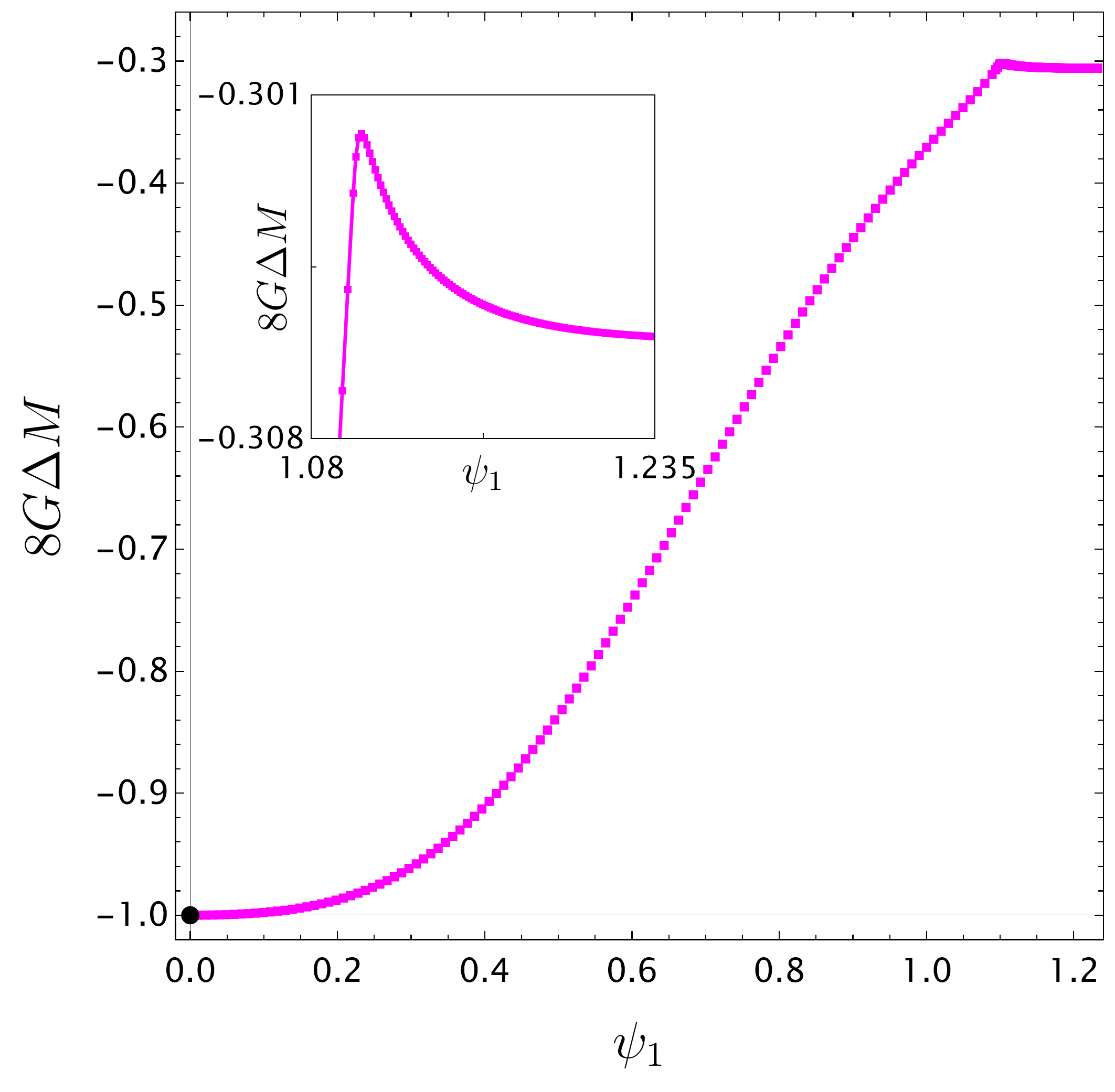}
    \includegraphics[width=0.45\linewidth]{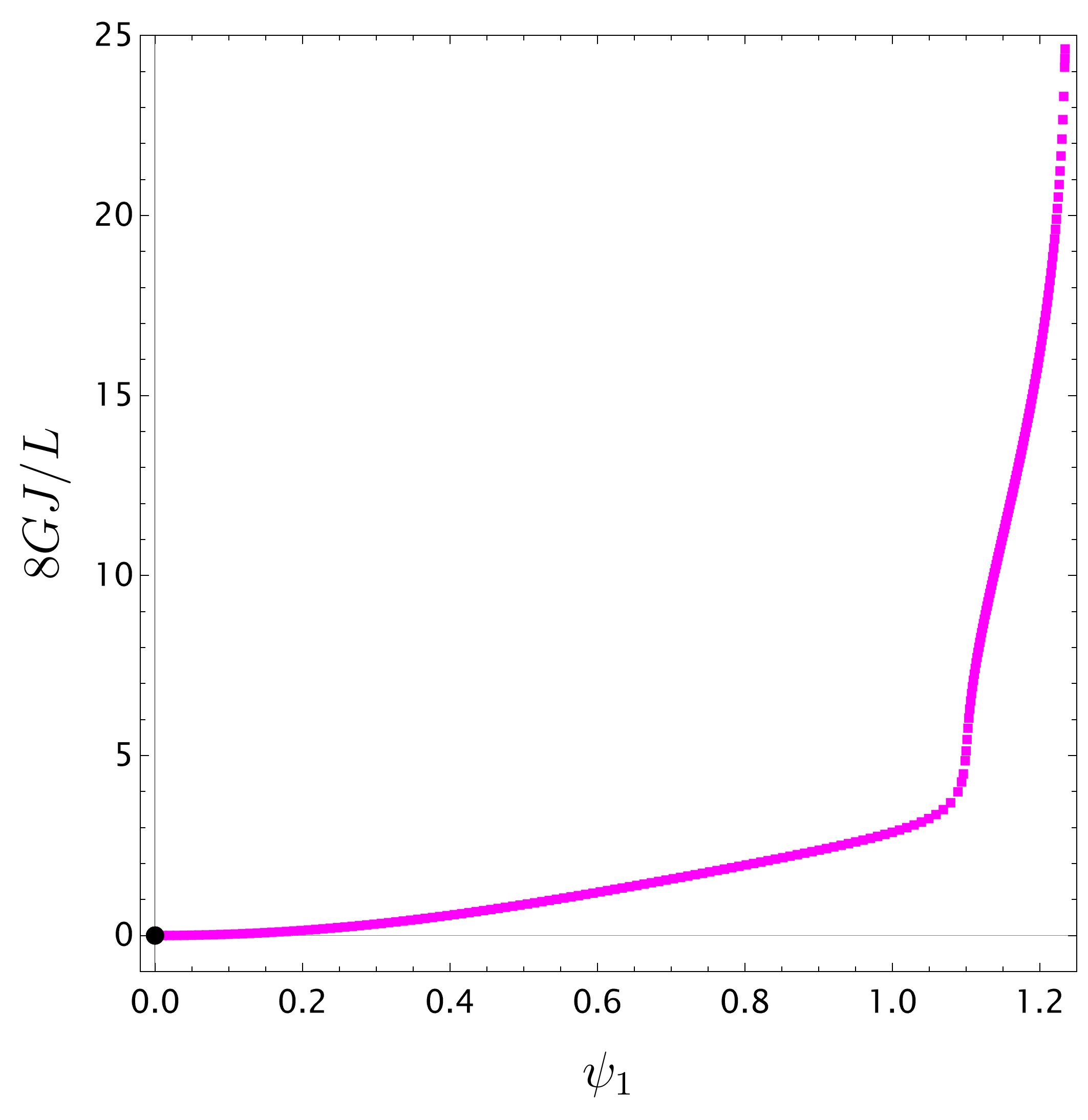}
    \includegraphics[width=0.45\linewidth]{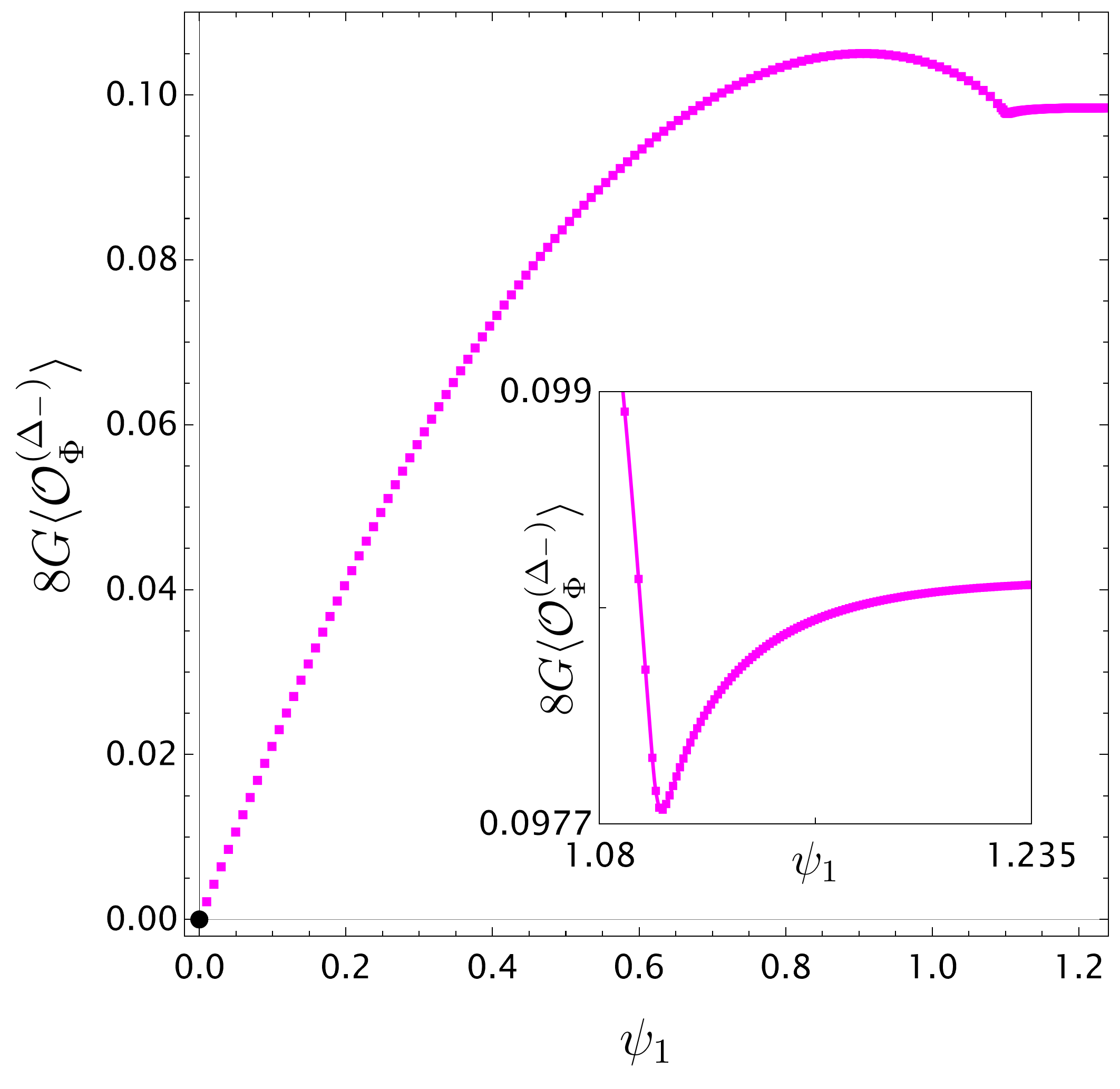}
    \includegraphics[width=0.45\linewidth]{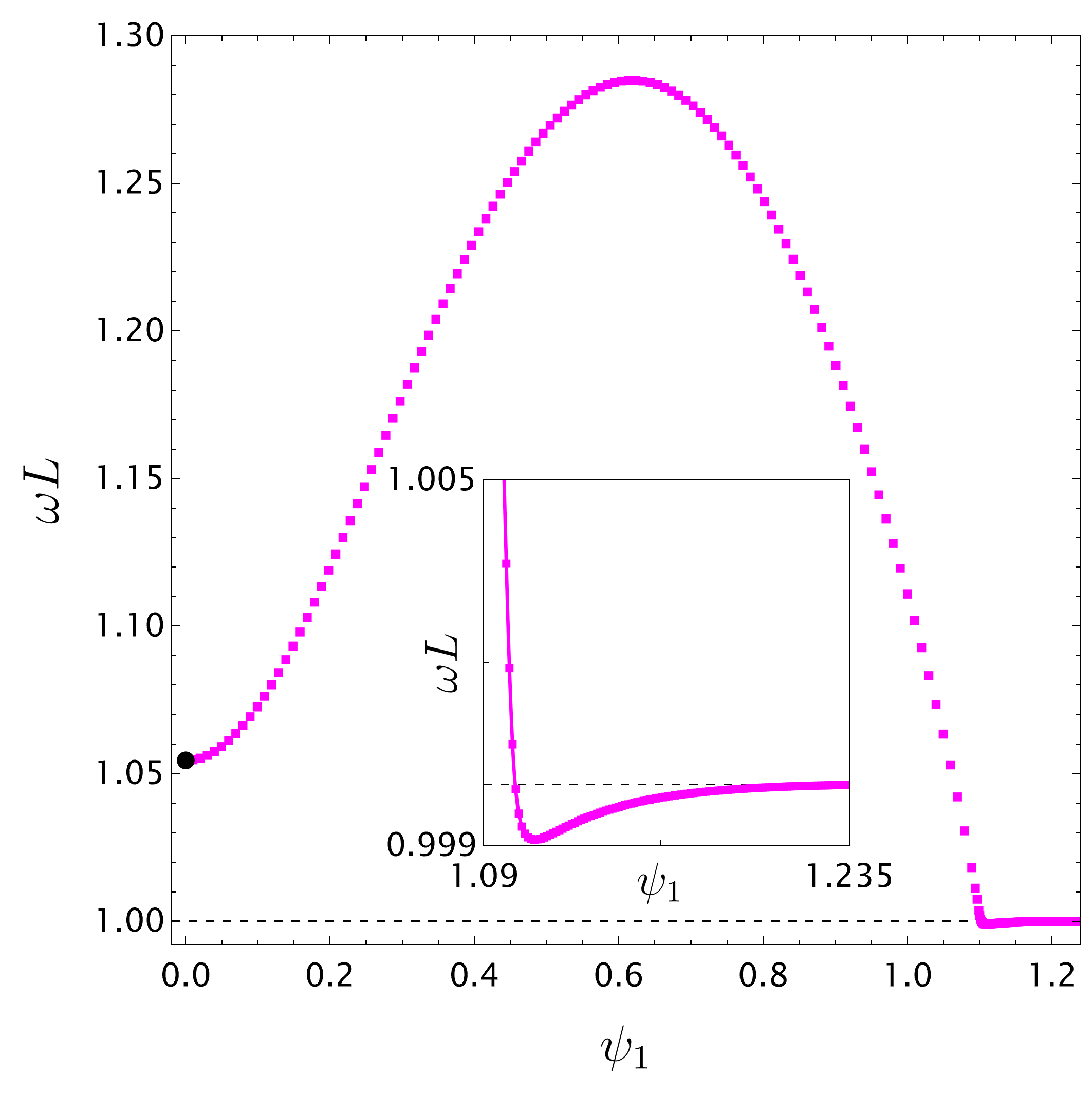}
    \caption{Physical properties and phase diagram of regular boson stars (magenta squares) and singular extremal hairy black holes (dark-red triangles) with $\bm{m=1}$ for $\mu^2L^2=-15/16$ and $\kappa=-8/10 > \kappa^{\rm AdS}_{m=1, -15/16}$. The black disk describes AdS$_3$ and the grey diamond is the singular vaccum BTZ. The value of the scalar field at the centre of the boson star, $\psi_1$, was introduced in the last equation of \eqref{BCs:OriginBSm}.
    At large $\hat{J}$, the regular and singular extremal black hole curves approach each other at \eqref{mergeBH-BSm1} and $\hat{\omega} \sim 1$ (in some plots these values are represented by the dashed black lines).
    }
    \label{fig:BS-m1}
\end{figure}

In Fig.~\ref{fig:BS-m1} we display the properties of regular $m=1$ hairy boson stars, shown as magenta squares. Unlike the $m=0$ boson stars, the $m=1$ solutions necessarily rotate and therefore carry non‑vanishing angular momentum, $\hat{J}>0$. They possess a finite mass $\hat{M}$, angular momentum $\hat{J}$, and a conserved $U(1)$ charge $\hat{N}$ (interpreted as the number of scalar particles), defined in \eqref{Mass:Def}, \eqref{AngMom:Def}, and \eqref{N:Def} with $R_{0}\equiv0$, respectively.

In the top‑left panel of Fig.~\ref{fig:BS-m1} we plot $\Delta\hat{M}\equiv\hat{M}-\hat{M}^{\hbox{\tiny BTZ}}_{\hbox{\tiny ext}}$ as a function of $\hat{J}$, while the top‑right panel shows $\Delta\hat{M}$ as a function of the expectation value $\big\langle\hat{\mathcal{O}}_{\Phi}^{(\Delta_-)}\big\rangle$, defined in~\eqref{ScalarVEV}. In practice, as described in Section~\ref{sec:NumericalSetup:RegBS}, these solutions are obtained by marching along the value $\psi_{1}$ of the scalar field at the centre of the boson star, introduced in the final equation of~\eqref{BCs:OriginBSm}. Accordingly, the remaining panels of Fig.~\ref{fig:BS-m1} display, in lexicographic order, $\Delta\hat{M}$, $\hat{J}$, $\big\langle\hat{\mathcal{O}}_{\Phi}^{(\Delta_-)}\big\rangle$, and the scalar field frequency $\hat{\omega}$ as functions of $\psi_{1}$.

For $m=1$, the Kretschmann scalar evaluated at the origin is given by the polynomial expression
\begin{equation}
\hat{K}_{0}=4\bigl(3-2\psi_{1}^{2}+3\psi_{1}^{4}\bigr),
\end{equation}
and the conserved charges satisfy $\hat{N}=\hat{J}$. We therefore do not include separate plots for these quantities in Fig.~\ref{fig:BS-m1}.

The regular $m=1$ boson stars form a one‑parameter family of solutions, uniquely specified by the central scalar amplitude $\psi_{1}$. The family originates at global AdS$_3$ (black disk) with $\psi_{1}=0$, $\alpha=0=\beta$, $\{\hat{J},\hat{M}\}=\{0,0\}$, and frequency $\hat{\omega}\simeq1.05452346$, which coincides with the linear normal‑mode frequency of AdS$_3$ computed in Fig.~2 of Ref.~\cite{Dias:2025uyk} for $\kappa=-0.8$ and $m=1$. As $\psi_{1}$ increases, the solutions evolve toward $\Delta\hat{M}\simeq-0.304832$, $\big\langle\hat{\mathcal{O}}_{\Phi}^{(\Delta_-)}\big\rangle\simeq0.098244$, and $\hat{\omega}\to1$, while the angular momentum $\hat{J}$ grows without bound. Thus, as for the $m=0$ case shown in Fig.~\ref{fig:BS-m0}, the regular $m=1$ boson stars with $\kappa>\kappa^{\rm AdS}_{1,-15/16}$ are perturbatively connected to AdS$_3$.

Although these solutions appear to exist for arbitrarily large angular momentum (we have tested this up to $\hat{J}=10$), they may still be said to exhibit a Chandrasekhar‑like limit in the sense that they occupy only a finite range of the parameters $\psi_{1}$ and $\big\langle\hat{\mathcal{O}}_{\Phi}^{(\Delta_-)}\big\rangle$. The boson star first law~\eqref{FirstLawBStar},
\begin{equation}
100\left(1-\hat{\omega}\frac{\hat{N}'(\psi_{1})}{\hat{M}'(\psi_{1})}\right)=0,
\end{equation}
is satisfied with relative errors below $10^{-3}\%$ (recall that $\hat{J}=m\,\hat{N}$ for boson stars).

In Fig.~\ref{fig:BS-m1}, we also took the opportunity to display the 
singular $m\geq 1$ rotating extremal hairy black holes for reasons that will become clear after \eqref{mergeBH-BSm1}. The numerical strategy to construct these solutions was detailed in Section~\ref{sec:NumericalSetup:singBHmJ}. For $m=1$, this 1-parameter family of solutions is shown as dark‑red triangles in the top panels of Fig.~\ref{fig:BS-m1}. In contrast to the $m=0$ case, these singular solutions do not have constant $\Delta\hat{M}(\hat{J})$. This singular $m=1$ family of rotating extremal black holes  starts at the singular vacuum BTZ geometry with $\hat{M}=\hat{J}=\big\langle\hat{\mathcal{O}}_{\Phi}^{(\Delta_-)}\big\rangle=0$ (grey diamond) and evolves, at arbitrarily large $\hat{J}$, toward
\begin{equation}\label{mergeBH-BSm1}
\hat{M}\simeq-0.30483190,
\qquad
\big\langle\hat{\mathcal{O}}_{\Phi}^{(\Delta_-)}\big\rangle\simeq0.098244.
\end{equation}
In this limit, the singular $m=1$ extremal black hole and  $m=1$ boson star families merge, as illustrated in the top panels of Fig.~\ref{fig:BS-m1}.

\subsection{Hairy \texorpdfstring{AdS$_3$}{AdS3} boson stars with \texorpdfstring{$m=1$}{m=1} and  Ishibashi-Wald instability of \texorpdfstring{AdS$_3$}{AdS3}}\label{sec:PhaseDiag-m1:IshWald}

Figure~\ref{fig:BS-m1} illustrates the properties of $m=1$ boson stars in the regime where the double‑trace parameter satisfies $\kappa>\kappa^{\rm AdS}_{m=1,\hat{\mu}^{2}}$, so that AdS$_3$ is linearly stable against double‑trace perturbations. It is then natural to ask how the boson star phase diagram evolves as $\kappa$ is decreased toward, and eventually below, the critical value $\kappa^{\rm AdS}_{m=1,\hat{\mu}^{2}}$, where AdS$_3$ becomes unstable to double‑trace perturbations~\cite{Ishibashi:2004wx,Dias:2025uyk}. Understanding this behaviour provides insight into the possible endpoint or metastable states of the $m=1$ AdS$_3$ instability.

\begin{figure}
    \centering
    \vskip -0.5cm
    \includegraphics[width=0.55\linewidth]{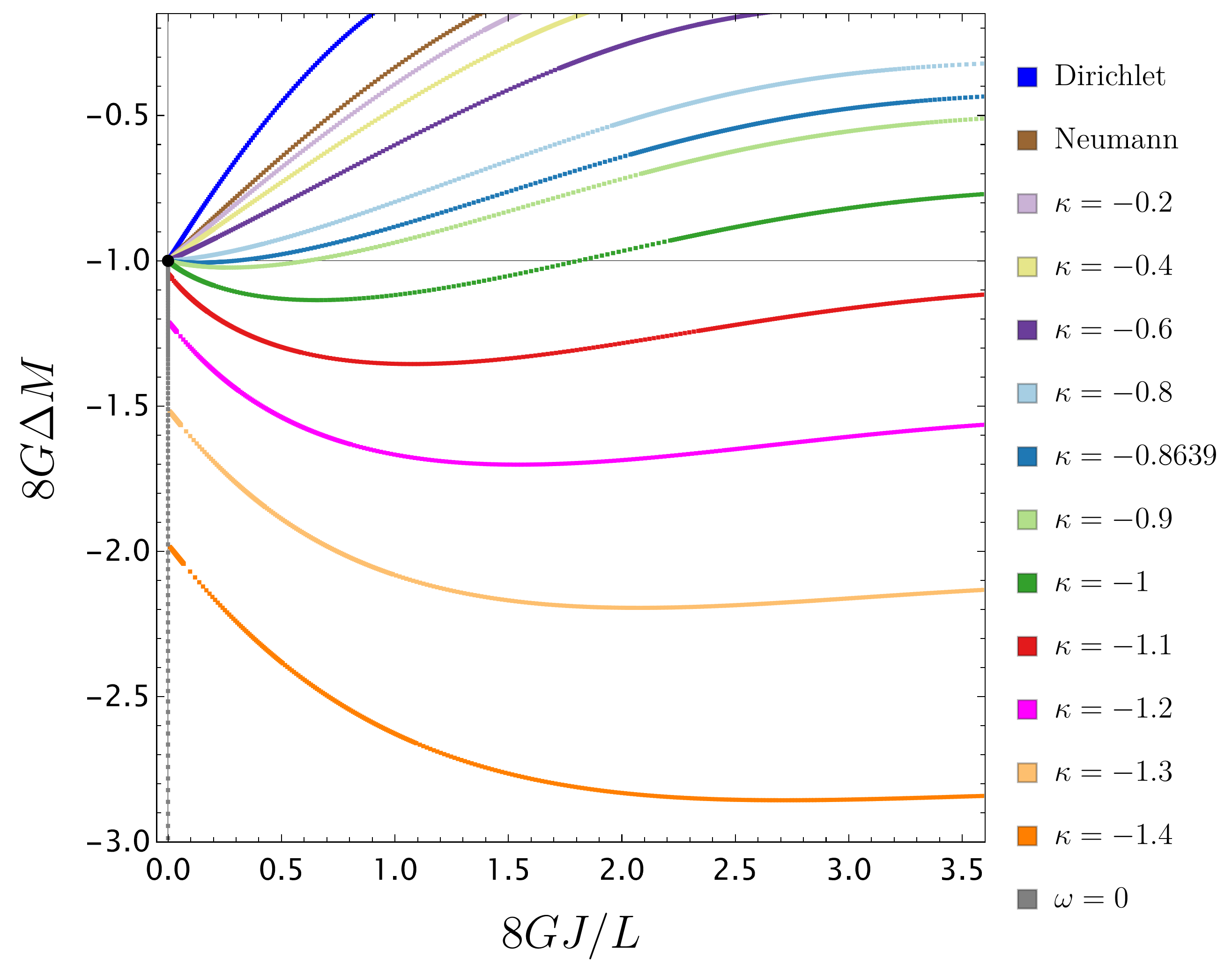}
    \includegraphics[width=0.44\linewidth]{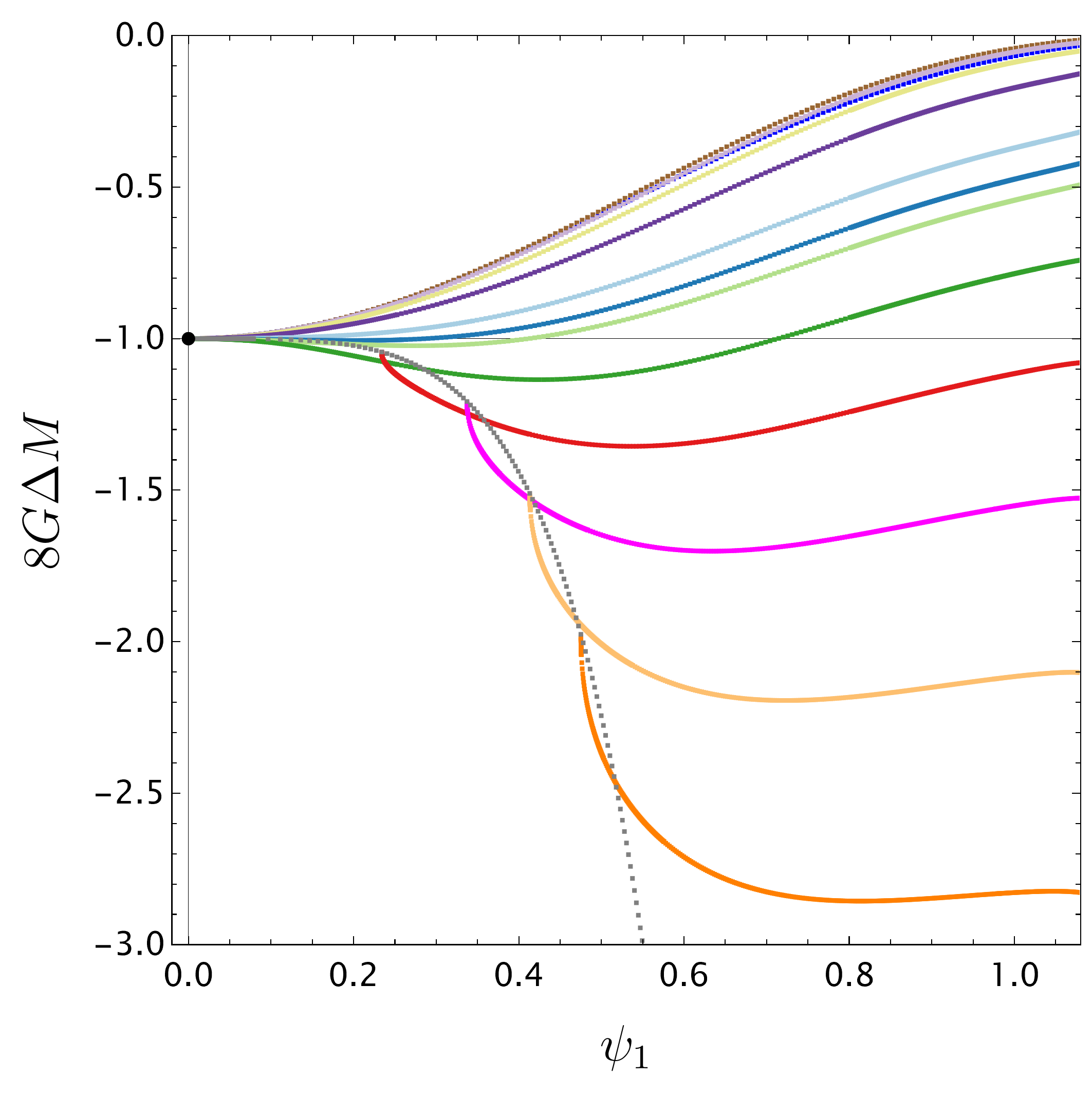}
    \includegraphics[width=0.47\linewidth]{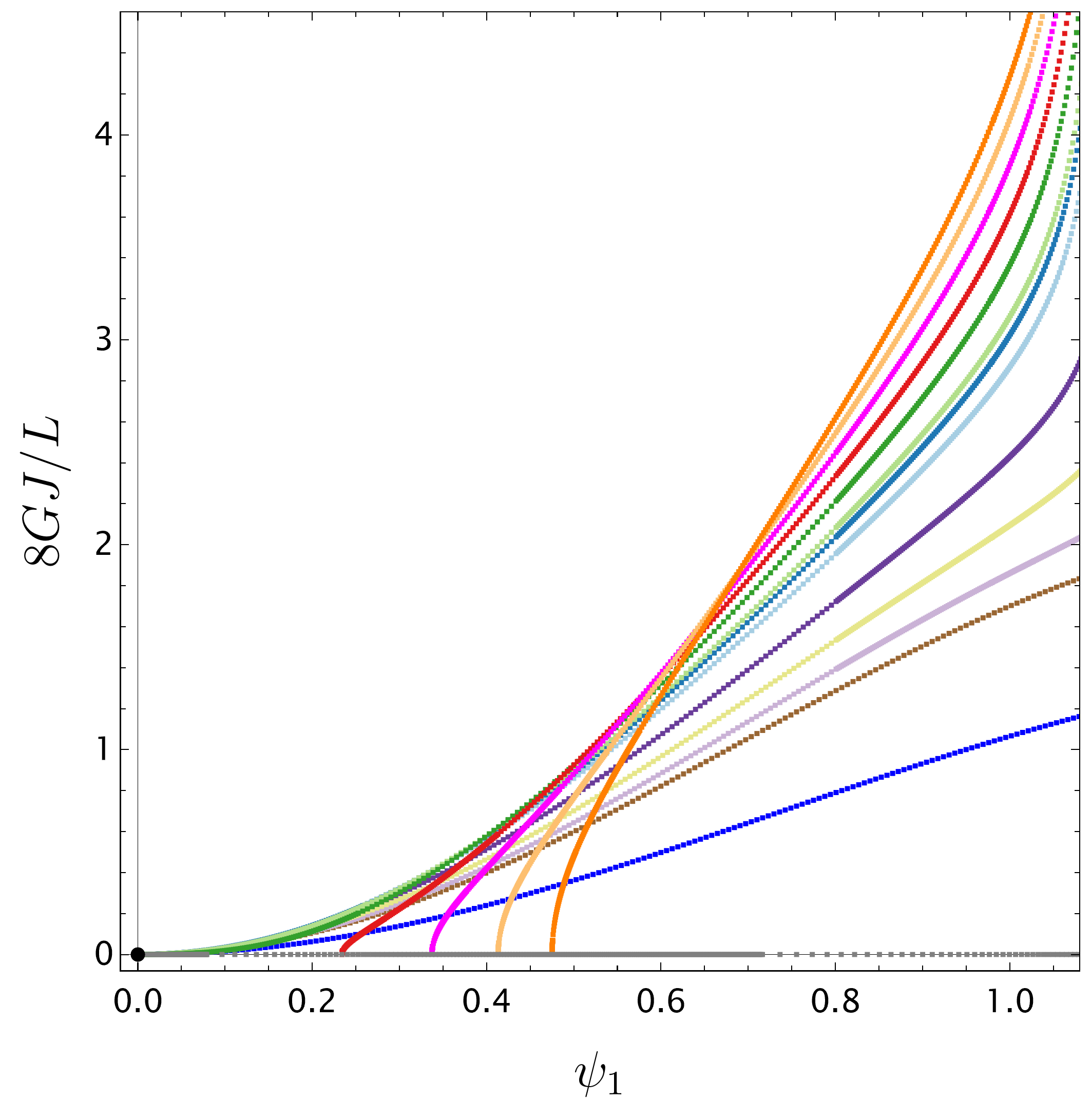}
    \includegraphics[width=0.48\linewidth]{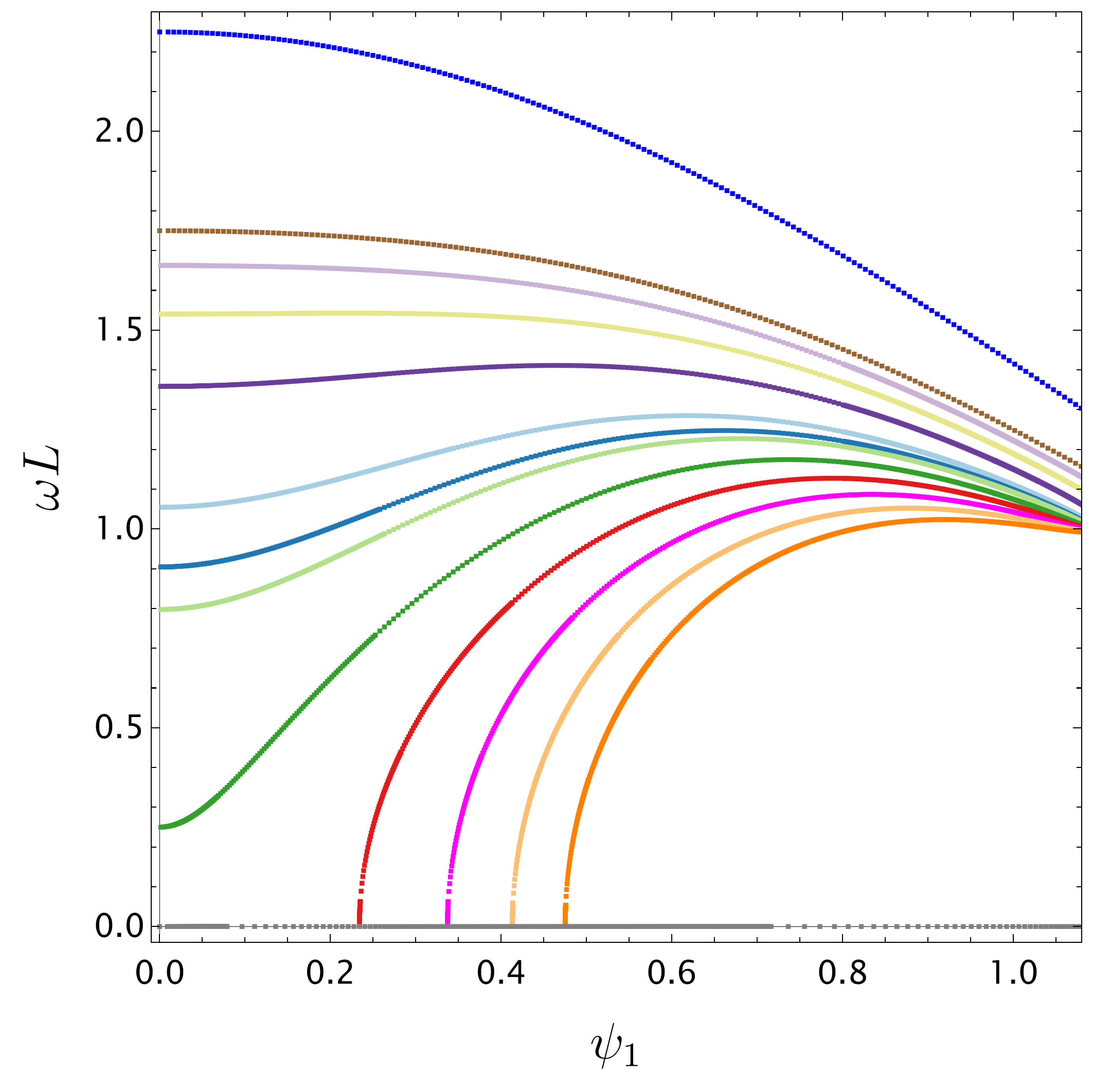}
    \includegraphics[width=0.47\linewidth]{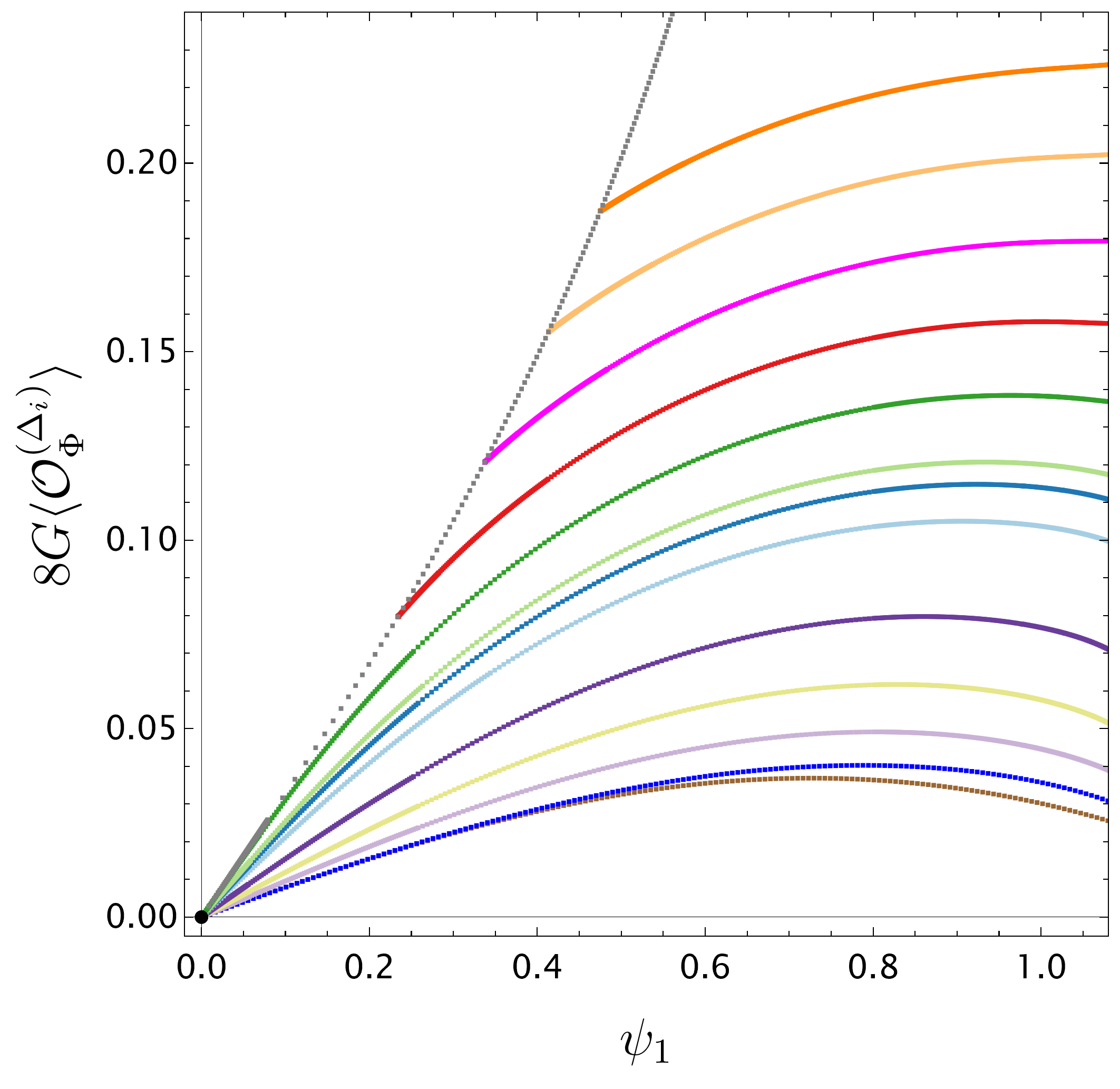}
    \caption{\footnotesize Regular boson stars with $\bm{m=1}$ and $\mu^2L^2 = -15/16$ for different values of $\kappa$ (see legend caption) in the window $\psi_1 \in [0,1.08]$ (singular extremal black holes are {\it not} shown). When $\kappa>\kappa^{\rm AdS}_{m=1, \hat{\mu}^2}\simeq -1.009837$ these boson stars start at global AdS$_3$, $\{\Delta\hat{M},\hat{J}\}=\{-1.0\}$, $\psi_1=0=\langle \mathcal{O}_{\Phi}^{(\Delta_i)} \rangle$ and with $\hat{\omega}(\kappa) \neq 0$. But for $\kappa<\kappa^{\rm AdS}_{m=1, \hat{\mu}^2}$, \ie when AdS$_3$ is unstable, these boson stars  start along the zero-frequency ($\hat{\omega}=0$) grey square curve, with $\Delta \hat{M}(\kappa)<-1$, $\hat{J}(\kappa)>0$, $\psi_1(\kappa)>0$ and $\langle \mathcal{O}_{\Phi}^{(\Delta_-)} \rangle(\kappa)>0$.}
    \label{fig:m1BSevoK}
\end{figure}

To address this question, in Fig.~\ref{fig:m1BSevoK} we again fix $\mu^{2}L^{2}=-15/16$ and consider $m=1$, for which \eqref{kAdS-onset} gives $\kappa^{\rm AdS}_{m=1,\hat{\mu}^{2}}\simeq-1.009837$. We then display several families of regular $m=1$ boson stars as $\kappa$ is varied from values above the critical $\kappa^{\rm AdS}_{m=1,\hat{\mu}^{2}}$ to values below it. Specifically, we show solutions for
\begin{subequations}
\begin{equation}
\kappa=\left\{-\tfrac{2}{10},-\tfrac{4}{10},-\tfrac{6}{10},-\tfrac{8}{10},-0.8639,-\tfrac{9}{10},-1\right\}>\kappa^{\rm AdS}_{1,-15/16}
\end{equation}
and for
\begin{equation}
\kappa=\left\{-\tfrac{11}{10},-\tfrac{12}{10},-\tfrac{13}{10},-\tfrac{14}{10}\right\}<\kappa^{\rm AdS}_{1,-15/16},
\end{equation}
\end{subequations}
with the corresponding colour coding indicated in the legend. For each value of $\kappa$ we plot the following quantities: $\hat{J}$ versus $\Delta\hat{M}$ (top panel); $\psi_{1}$ versus $\Delta\hat{M}=\hat{M}-\hat{M}^{\hbox{\tiny BTZ}}_{\hbox{\tiny ext}}$ (middle‑left panel); $\psi_{1}$ versus $\hat{J}$ (middle‑right panel); $\psi_{1}$ versus $\hat{\omega}$ (bottom‑left panel); and $\psi_{1}$ versus $\langle\mathcal{O}_{\Phi}\rangle\equiv\langle\mathcal{O}_{\Phi}^{(\Delta_i)}\rangle$ (bottom‑right panel), with $\Delta_i = \Delta_+$ for Dirichlet BCs and $\Delta_i= \Delta_-$ for Neumann and/or double-trace BCs. This presentation closely parallels that of Fig.~\ref{fig:BS-m1} for the representative case $\kappa=-0.8$, which is also included in Fig.~\ref{fig:m1BSevoK} for comparison. In this figure, we restrict attention to solutions with $\psi_{1}\in[0,1.08]$, although the families extend to larger values of $\psi_{1}$. For clarity, we do not display the singular $m=1$ extremal black holes that also exist for each $\kappa$, as these are distinct for different $\kappa$.

A key observation from Fig.~\ref{fig:m1BSevoK} is that for any $\kappa>\kappa^{\rm AdS}_{m=1,\hat{\mu}^{2}}$ (including the case $\kappa=-8/10$ already shown in Fig.~\ref{fig:BS-m1}), the corresponding family of regular $m=1$ boson stars originates at AdS$_3$. At this endpoint one has $\Delta\hat{M}=-1$, $\hat{J}=0$, $\langle\mathcal{O}_{\Phi}\rangle=0$, and a frequency $\hat{\omega}$ equal to the linear normal‑mode frequency of AdS$_3$ computed in~\cite{Dias:2025uyk} (black disk). These solutions are therefore perturbatively connected to AdS$_3$ and may be interpreted as the fully nonlinear back‑reaction of the double‑trace normal modes found in~\cite{Dias:2025uyk}.

Figure~\ref{fig:m1BSevoK} also includes the families of boson stars obeying Dirichlet ($\kappa=\pm\infty$) and Neumann ($\kappa=0$) boundary conditions, shown respectively by the blue and brown curves. These Dirichlet and Neumann boson star families are qualitatively similar to those with $\kappa>\kappa^{\rm AdS}_{m=1,\hat{\mu}^{2}}$. In particular, they are perturbatively connected to global AdS$_3$ with $\Delta\hat{M}=-1$ and $\hat{J}=0$, which is linearly stable under Dirichlet and Neumann perturbations. Near the AdS$_3$ point, the $m=1$ Dirichlet and Neumann boson stars can be constructed analytically within perturbation theory. This analysis is presented in Appendix~\ref{secA:BStars-DNm1}, where we also compare the perturbative results with the full numerical solutions to validate both the numerical implementation and the regime of applicability of perturbation theory (see Figs.~\ref{fig:m1_DirBS_numerics_VS_perturbation} and~\ref{fig:m1_NeuBS_numerics_VS_perturbation}). Incidentally, unlike the Neumann and double‑trace solitons, Dirichlet boson stars also exist for scalar field masses $\mu\geq0$. In this regime $-$ above the AdS$_3$ unitarity bound $-$ only Dirichlet modes are normalizable, and the corresponding boson stars can again be constructed perturbatively as discussed in Appendix~\ref{secA:BStars-DNm1}.

To discuss a second main feature of Fig.~\ref{fig:m1BSevoK}, let us first recall that AdS$_3$ is unstable to double‑trace boundary conditions whenever
$\kappa<\kappa^{\rm AdS}_{m=1,\hat{\mu}^{2}}$~\cite{Ishibashi:2004wx,Dias:2025uyk}. This Ishibashi–Wald instability leaves a clear imprint on the phase diagram of solutions. Indeed, for $\kappa<\kappa^{\rm AdS}_{m=1,\hat{\mu}^{2}}$, regular $m=1$ boson stars continue to exist, but they are \emph{no longer perturbatively connected} to AdS$_3$. Instead, the corresponding boson star families now originate at
$\Delta\hat{M}(\kappa)<-1$, \ie at energies below that of global AdS$_3$, with $\hat{J}=0$, $\hat{\omega}=0$, and finite values of $\psi_{1}(\kappa)$ and
$\langle\mathcal{O}_{\Phi}\rangle(\kappa)$, as shown by the grey square curves in Fig.~\ref{fig:m1BSevoK}. These starting points are obtained numerically by imposing $\omega=0$ (with $m=1$), which is precisely how the grey square curves in Fig.~\ref{fig:m1BSevoK} were generated.

For any value of $\kappa$, all regular $m=1$ boson stars approach, in the large‑$\hat{J}$ limit, the singular $m=1$ extremal black hole solution identified in Section~\ref{sec:NumericalSetup:SingBS}. This singular branch is illustrated for $\kappa=-0.8$ by the dark‑red curve in Fig.~\ref{fig:BS-m1}, although it is not shown explicitly in Fig.~\ref{fig:m1BSevoK}.

We emphasize that for $\kappa<\kappa^{\rm AdS}_{m=1,\hat{\mu}^{2}}$ the lowest‑energy regular $m=1$ boson star has zero frequency and a mass strictly below that of AdS$_3$, \ie $\Delta\hat{M}<-1$. Moreover, this minimum mass decreases monotonically as $\kappa$ is lowered further below $\kappa^{\rm AdS}_{m=1,\hat{\mu}^{2}}$. These solutions correspond to the grey dashed $\omega=0$ curves in Fig.~\ref{fig:m1BSevoK}.

\begin{figure}[b]
    \centering
    \includegraphics[width=0.47\linewidth]{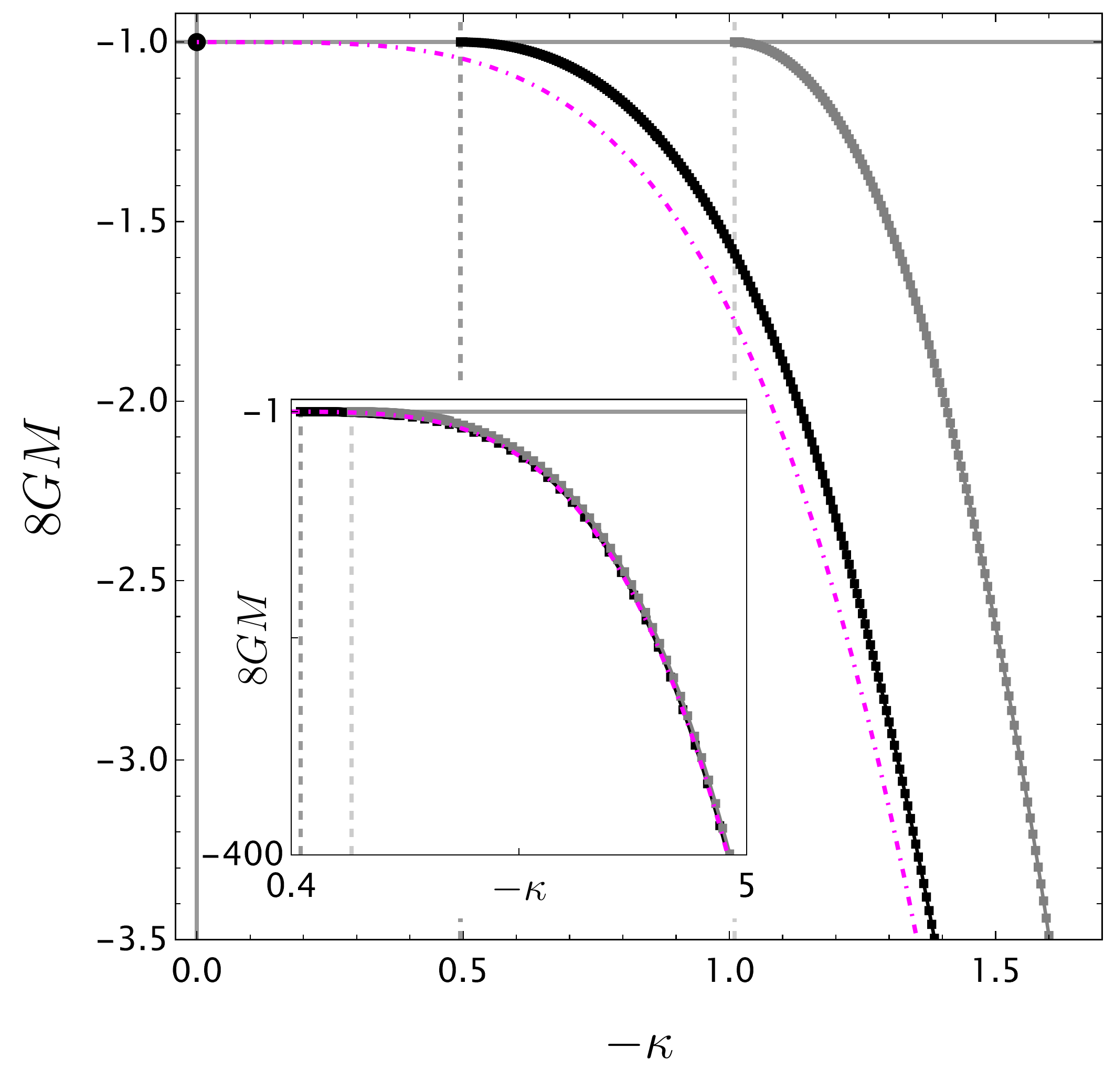}\hspace{0.5cm}
    \includegraphics[width=0.47\linewidth]{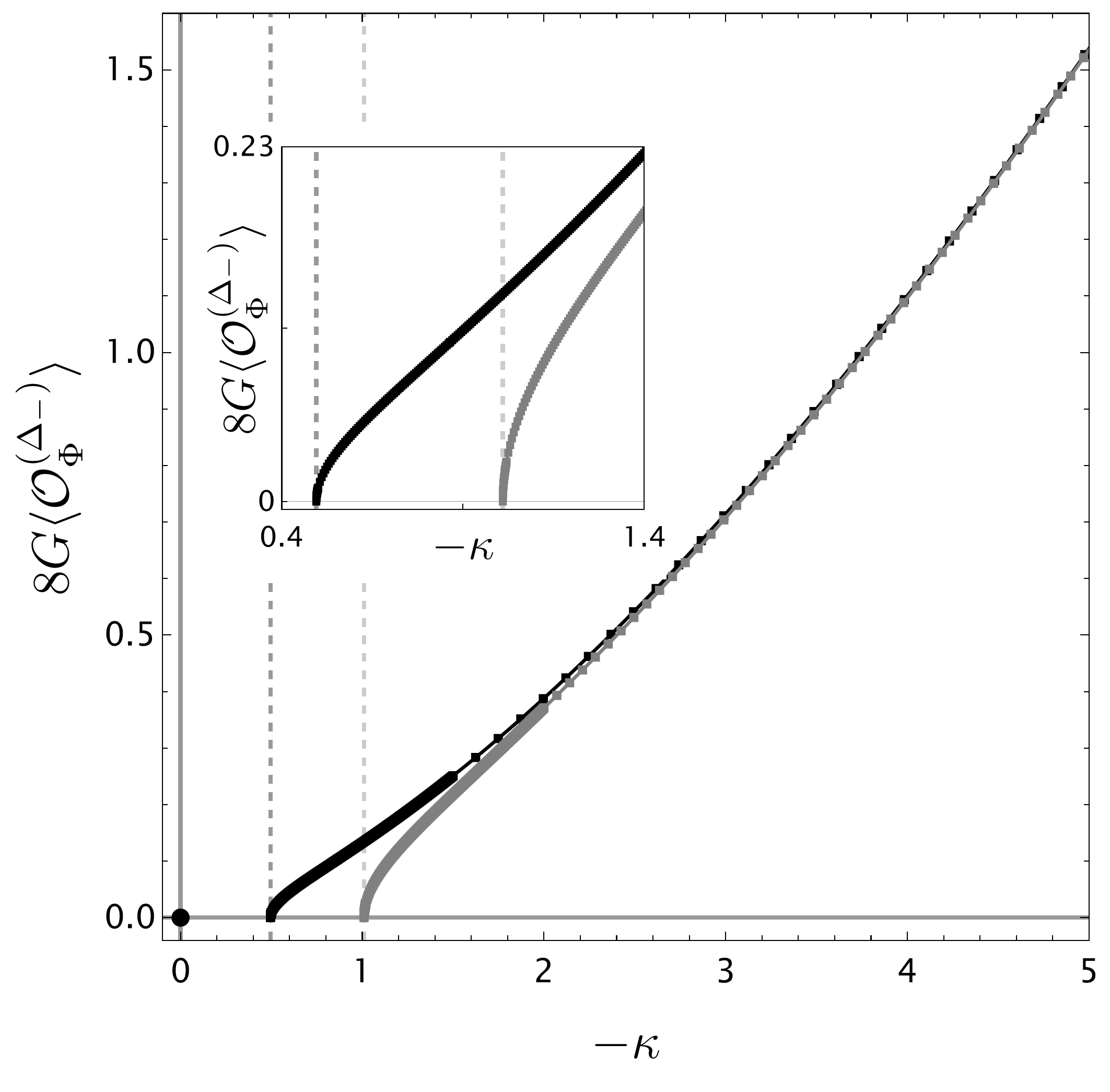}
    
    \caption{Ground state solutions as a function of the double-trace parameter $-\kappa$ (which specifies the theory).
    For $\kappa > \kappa^{\rm AdS}_{m=0, \hat{\mu}^2}\simeq -0.495129$, the solution with minimum energy is AdS$_3$ (continuous black line with $\hat{M}=-1$ for any $\kappa$). But for $\kappa < \kappa^{\rm AdS}_{m=0, \hat{\mu}^2}$, the minimum energy solution is now the $\bm{m=0}$ zero-frequency regular boson star with $\hat{M}(\kappa)<-1$ (black square curve). We also show the $\bm{m=1}$ zero-frequency regular boson star (grey squares) which exists for $\kappa < \kappa^{\rm AdS}_{m=1, \hat{\mu}^2}\simeq -1.00984$; it also has $\hat{M}(\kappa)<-1$ but higher than the $m=0$ solution (when they coexist). AdS$_3$ is unstable for $\kappa < \kappa^{\rm AdS}_{m=0, \hat{\mu}^2}$ and can decay into the $m=0$, zero-frequency boson star. The magenta dot-dashed curve represents the lower bound implied by the $p=1$ positivity-of-energy theorem~\eqref{GlobalMin}–\eqref{Super:Emin} derived in Appendix~\ref{secA:superpotentials}.
    }
    \label{fig:m0m1omega0BS}
\end{figure}

One of the main findings of this work is that the $m=1$ zero‑frequency regular boson star with $\hat{M}(\kappa)<-1$ provides a natural endpoint for the Ishibashi–Wald instability of AdS$_3$ triggered by $m=1$ double‑trace perturbations with $\kappa<\kappa^{\rm AdS}_{m=1,\hat{\mu}^{2}}$~\cite{Ishibashi:2004wx,Dias:2025uyk}. The endpoint of this AdS$_3$ instability had not been identified previously. However, for those values of $\kappa$ where both solutions coexist, the $m=1$ zero‑frequency boson star is itself unstable to $m=0$ double‑trace perturbations and is therefore expected to decay into the corresponding $m=0$ zero‑frequency boson star shown in Fig.~\ref{fig:m0BSevoK}, at the same value of $\kappa$.

This observation deserves particular emphasis. For a given theory, \ie for fixed $\kappa$ (and $\hat{\mu}$), when $m=0$ and $m=1$ zero‑frequency regular boson stars coexist, the $m=1$ solution has both a higher mass and a larger VEV $\langle\mathcal{O}_{\Phi}^{(\Delta_-)}\rangle$ than the $m=0$ solution. More generally, for any $m$, our numerical studies suggest that the zero‑frequency regular boson star with azimuthal number $m+1$ has larger mass and VEV than the corresponding solution with azimuthal number $m$, whenever both exist at the same $\kappa$. This ordering is displayed explicitly in Fig.~\ref{fig:m0m1omega0BS} for $m =0$ and $m =1$. In particular, for $\kappa>\kappa^{\rm AdS}_{m=0,\hat{\mu}^{2}}\simeq0.4951294$, global AdS$_3$ with $\hat{M}=-1$ is the ground‑state solution of the theory, whereas for $\kappa<\kappa^{\rm AdS}_{m=0,\hat{\mu}^{2}}$ the ground state becomes the $m=0$, $\omega=0$ regular boson star with $\hat{M}(\kappa)<-1$.

The nature of the minimum‑energy solution in the regime where AdS$_3$ is unstable can also be anticipated from the superpotential analysis of Ref.~\cite{Faulkner:2010fh}. Specifically, the lower bound implied by the positivity‑of‑energy theorem~\eqref{GlobalMin}–\eqref{Super:Emin} for $p=1$ (which is relevant for horizonless solutions) is shown as a magenta dot‑dashed curve in Fig.~\ref{fig:m0m1omega0BS}. The superpotential analysis of Ref.~\cite{Faulkner:2010fh}, together with the discussion in Appendix~\ref{secA:superpotentials}, predicts that this curve provides the minimum‑energy bound for a given double‑trace theory (that is, for fixed $\hat{\mu}$, $\kappa$, and $p=1$). Figure~\ref{fig:m0m1omega0BS} shows that this bound lies strictly below (though relatively close to) the AdS$_3$ curve for $\kappa > \kappa^{\rm AdS}_{m=0,\hat{\mu}^{2}}$, and below the $m=0$ zero‑frequency boson star for $\kappa < \kappa^{\rm AdS}_{m=0,\hat{\mu}^{2}}$.

\subsection{Hairy \texorpdfstring{AdS$_3$}{AdS3} black holes with \texorpdfstring{$m = 1$}{m=1}}\label{sec:PhaseDiag-m1:BHs}

A priori, one expects $m=1$ hairy black holes to rotate, \ie to carry non‑vanishing angular momentum $\hat{J}>0$, in contrast with the $m=0$ case where static hairy black hole solutions also exist (see Fig.~\ref{fig:m0_J0-BH}).%
\footnote{Remarkably, as we will show in Section~\ref{sec:PhaseDiag-Total-m1}, this expectation is not always correct: for certain values of the double‑trace parameter $\kappa$ there exist static hairy black holes even for $m=1$ (see the bottom panel of Fig.~\ref{fig:m1TotalPhaseDiag-3k} and Fig.~\ref{fig:m1kappam9_details}). This phenomenon does not occur for generic values of $\kappa$, which is the case studied in the present subsection~\ref{sec:PhaseDiag-m1:BHs}. We therefore postpone the discussion of static $m=1$ black holes to Section~\ref{sec:PhaseDiag-Total-m1}.}

\begin{figure}
    \centering
    \includegraphics[width=0.55\linewidth]{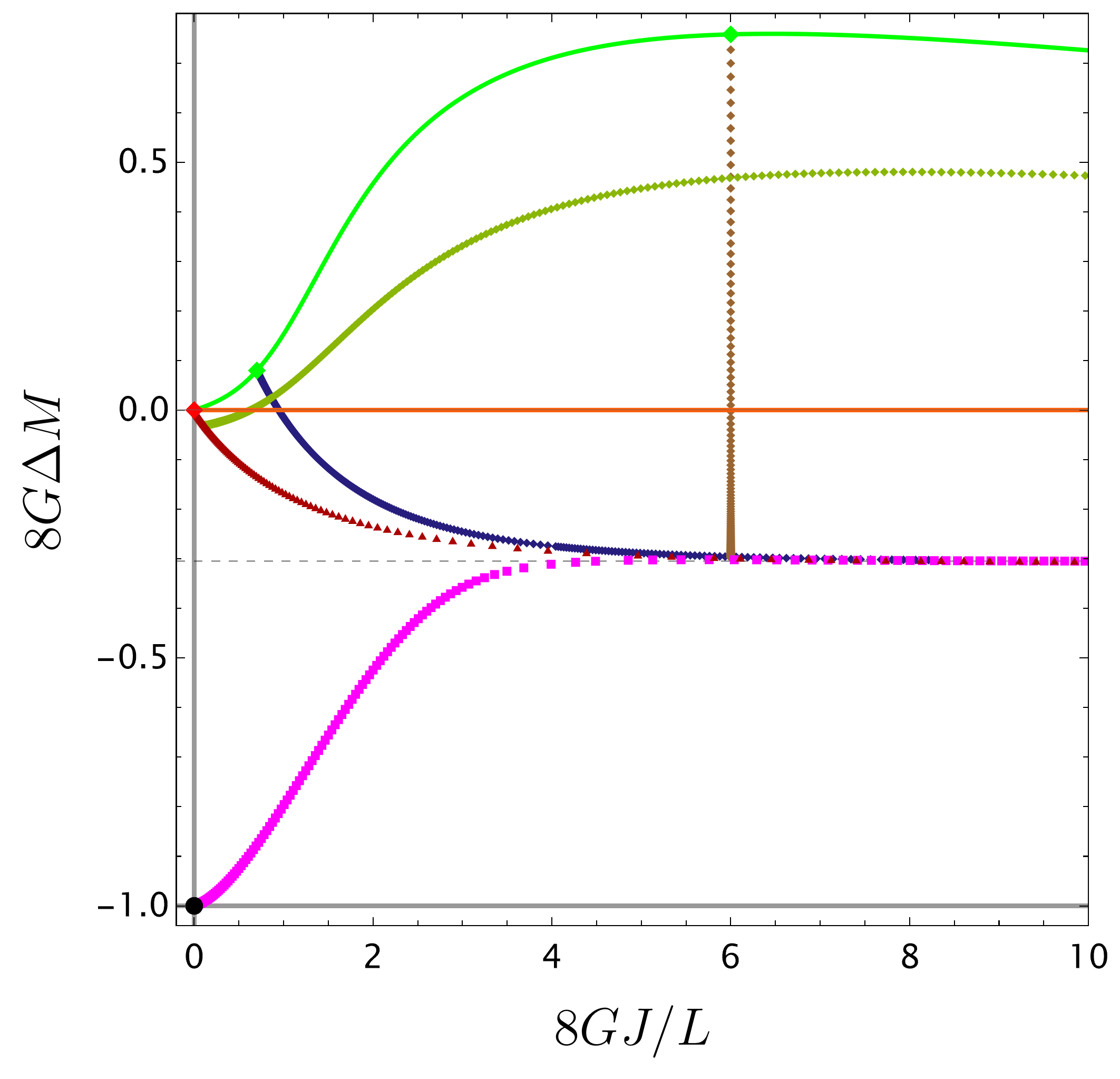}
    \caption{The three 1-parameter sub-families of $\bm{m=1}$ hairy black holes studied in section~\ref{sec:PhaseDiag-m1:BHs}, namely the family with constant: $R_+=3/4$ (navy blue diamonds), $\alpha=0.423475$ (olive green diamonds), and $\hat{J}=6$ (brown diamonds). We also show the 
    regular boson star family (magenta squares) and singular extremal hairy black hole family of section~\ref{sec:NumericalSetup:singBHmJ} (dark-red triangles),  both already displayed in Fig.~\ref{fig:BS-m1}; the dashed black horizontal line has $\Delta \hat{M} \sim -0.305$; see \eqref{mergeBH-BSm1}. This for a theory with $\mu^2L^2 = -15/16$ and $\kappa = -8/10$. We plot $\Delta \hat{M}$ vs  $\hat{J}$ where $\Delta \hat{M} = \hat{M} - \hat{M}^{\hbox{\tiny BTZ}}_{\hbox{\tiny ext}} $ is the mass difference of a given solution with respect to the extremal BTZ mass, $\hat{M}^{\hbox{\tiny BTZ}}_{\hbox{\tiny ext}} =\hat{J}$. The extremal BTZ family is the horizontal orange line with $\Delta\hat{M}=0$ and non-extremal BTZ black holes exist above this line for arbitrarily large $\Delta\hat{M}$. The bright-green curve describes the onset curve $\hat{M}(\hat{J})|^{\hbox{\tiny BTZ}}_{\hbox{\tiny onset (m=1)}}$  of the double-trace $m=1$ instabilty of BTZ black holes as found in \cite{Dias:2025uyk}.  
    Thermodynamic properties of these there sub-families will be presented in Fig.~\ref{fig:m1-hBTZ-Rp075} (family with  $R_+=3/4$), Fig.~\ref{fig:m1-hBTZ-alpha0423475} (family with  $\alpha=0.423475$), and Fig.~\ref{fig:m1-hBTZ-J6} (family with  $\hat{J}=6$) using the same colour code. 
}
    \label{fig:m1:dMJ:3families}
\end{figure}

In this subsection, we carry out a preliminary study of rotating $m=1$ hairy black holes that merge with (or bifurcate from) the spinning BTZ black hole along the one‑parameter curve
\begin{equation}
\hat{M}(\hat{J})\big|^{\text{\tiny BTZ}}_{\text{\tiny onset $(m=1)$}},
\end{equation}
shown as a bright‑green curve in Fig.~\ref{fig:m1:dMJ:3families}. This curve marks the onset of the $m=1$ double‑trace instability of BTZ and was obtained in the linear analysis of Ref.~\cite{Dias:2025uyk}. The corresponding hairy black holes form a two‑parameter family of solutions which, as discussed in Section~\ref{sec:NumericalSetup:BHs}, may be parametrized for instance by the pair $\{R_{+},\alpha\}$ (horizon radius or entropy and scalar condensate amplitude), or equivalently by $\{\hat{M},\hat{J}\}$ (mass and angular momentum).

To explore the structure of this two‑dimensional moduli space, we focus on three representative one‑parameter sub‑families of $m=1$ solutions:  
(i) a family of hairy black holes at fixed horizon radius $R_{+}=3/4$ (equivalently fixed entropy $\hat{S}_{H}=3\pi$, shown as navy‑blue diamonds in our plots);  
(ii) a family at fixed scalar condensate amplitude $\alpha=0.423475$ (olive‑green diamonds); and  
(iii) a family at fixed angular momentum $\hat{J}=6$ (brown diamonds).  
These preliminary investigations allow us to identify the boundaries of the two‑dimensional region in which $m=1$ hairy black holes exist. The particular numerical values chosen here are not special, but rather illustrate generic qualitative features shared by other families with fixed $R_{+}$, $\alpha$, or $\hat{J}$. A complete exploration of the full two‑dimensional parameter space of $m=1$ (and $m=0$) hairy black holes is presented in Section~\ref{sec:PhaseDiag-Total}.

Recall that BTZ black holes form a two‑parameter family with $\hat{M}\geq\hat{J}$, where the line $M=J/L$ corresponds to the one‑parameter family of extremal BTZ black holes of zero temperature. In particular, extremal BTZ black holes saturate the BPS bound $M\geq|J|/L$, unlike extremal Kerr–AdS black holes in higher dimensions. As in the $m=0$ analysis, when discussing the properties of $m=1$ hairy black holes it is convenient to work with the shifted mass
\begin{equation}
\Delta\hat{M}=\hat{M}-\hat{M}^{\text{\tiny BTZ}}_{\text{\tiny ext}},
\qquad
\hat{M}^{\text{\tiny BTZ}}_{\text{\tiny ext}}=\hat{J},
\end{equation}
which measures the deviation from extremality. Similarly, besides presenting the entropy $\hat{S}_{H}=\tfrac{8G}{L}\,S_{H}$, we also make use of the entropy difference
\begin{equation}
\Delta\hat{S}_{H}=\hat{S}_{H}-\hat{S}_{H}^{\text{\tiny BTZ}},
\end{equation}
where $\hat{S}_{H}^{\text{\tiny BTZ}}$ denotes the entropy of the BTZ black hole with the same mass $\hat{M}$ and angular momentum $\hat{J}$ as the hairy black hole under consideration, see \eqref{BTZ:thermoS}, whenever the two solutions coexist.

\begin{figure}
    \centering
   \vskip -1cm
   \includegraphics[width=0.43\linewidth]{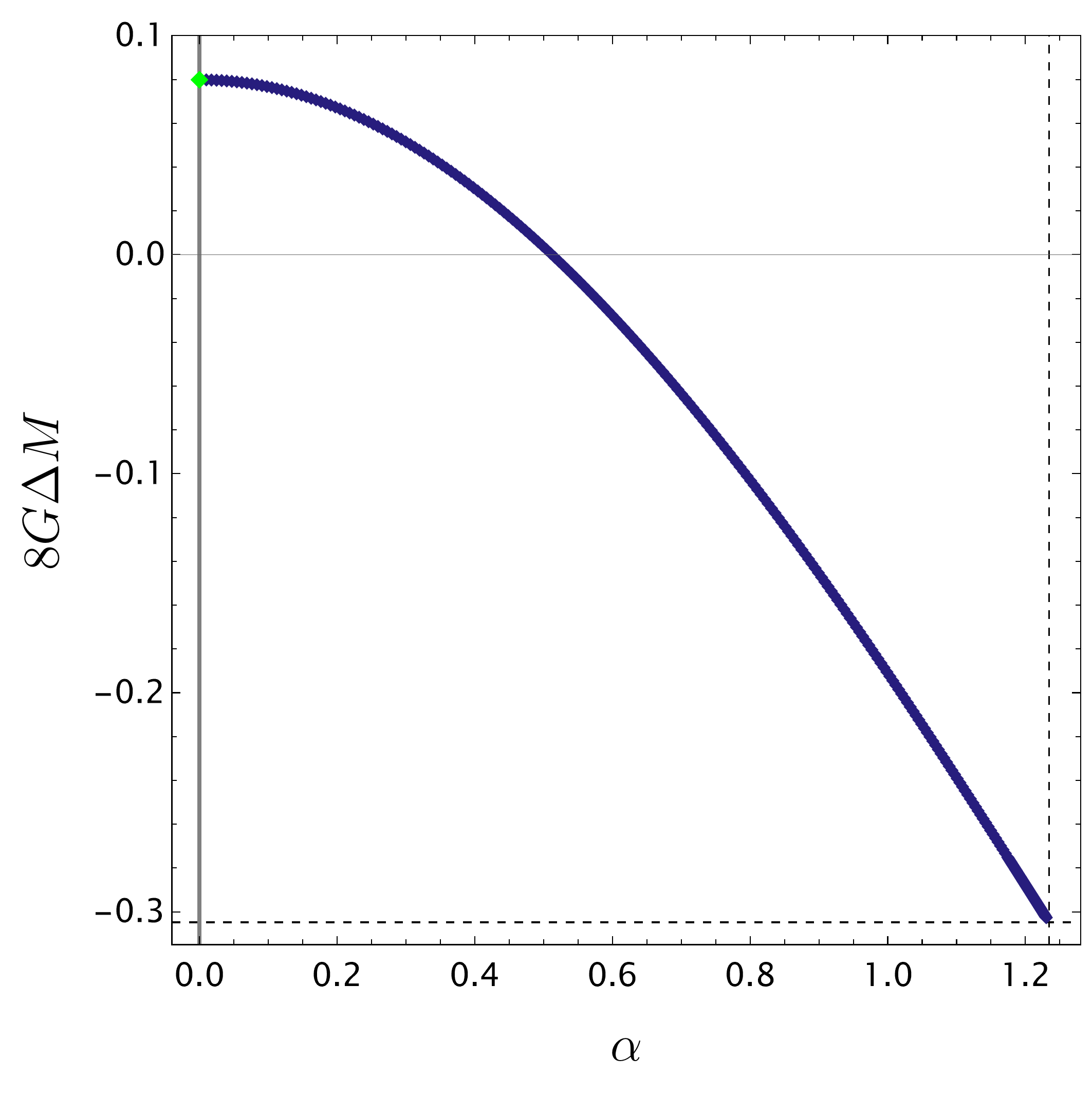}
    \includegraphics[width=0.43\linewidth]{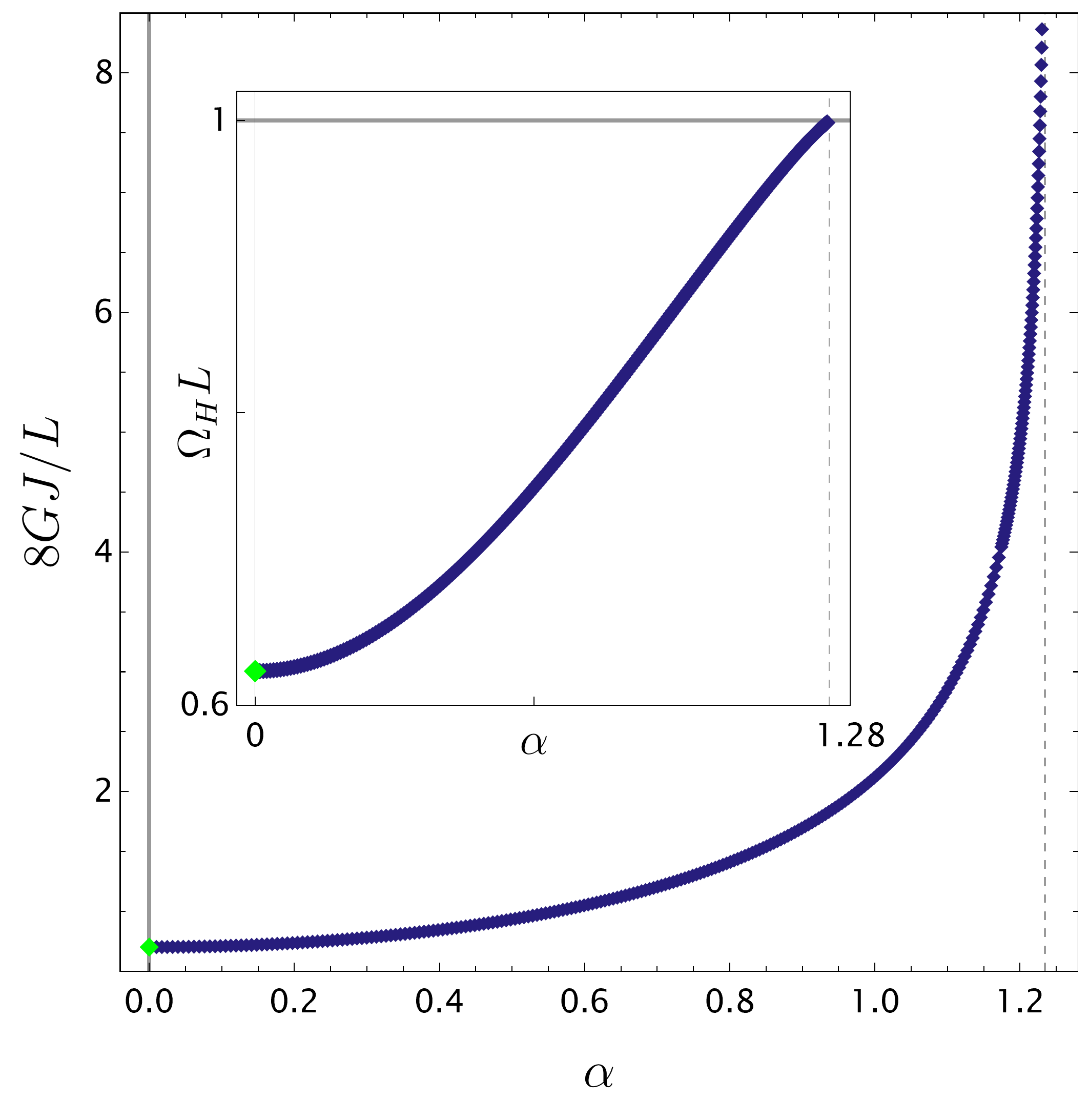}
    \includegraphics[width=0.43\linewidth]{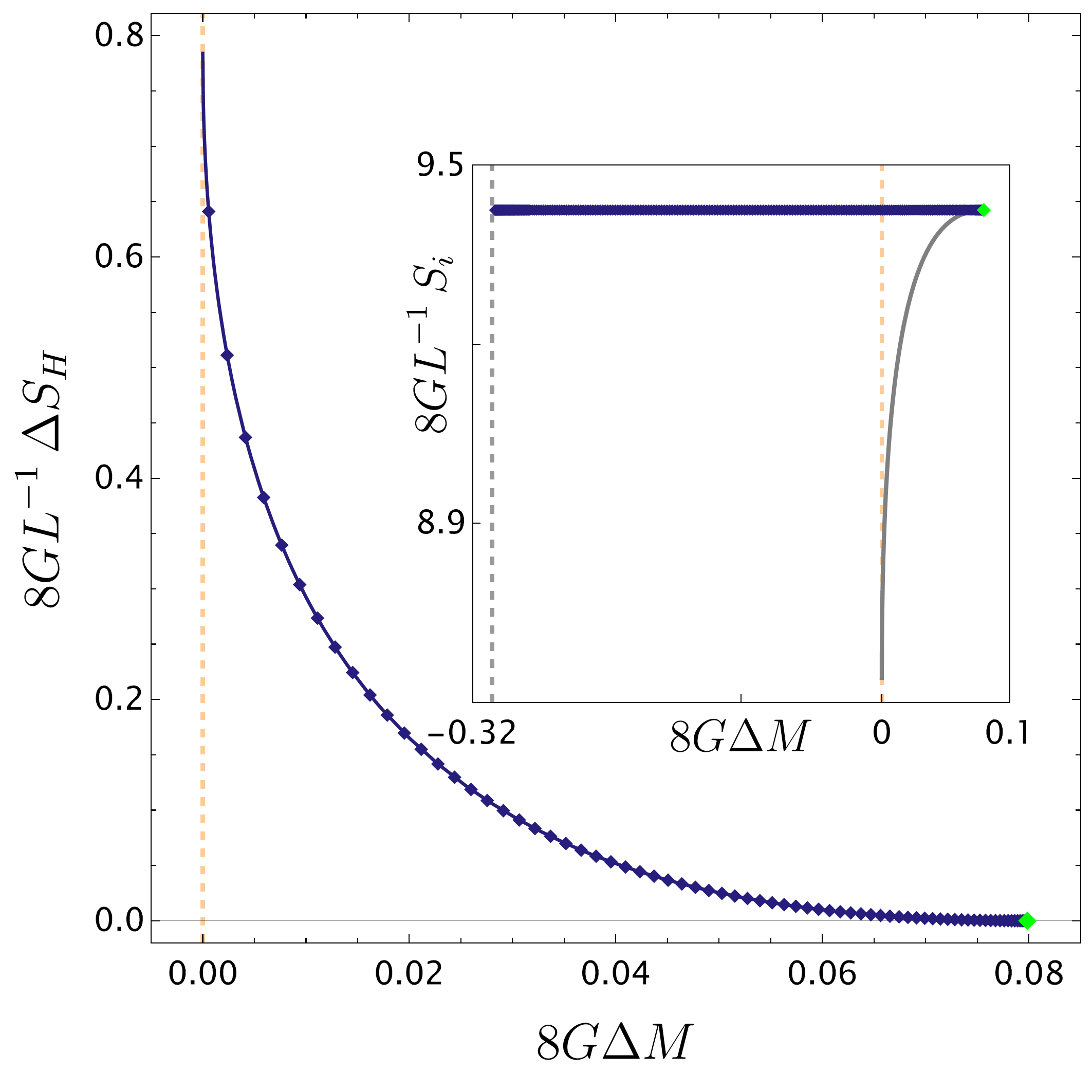}
    \includegraphics[width=0.43\linewidth]{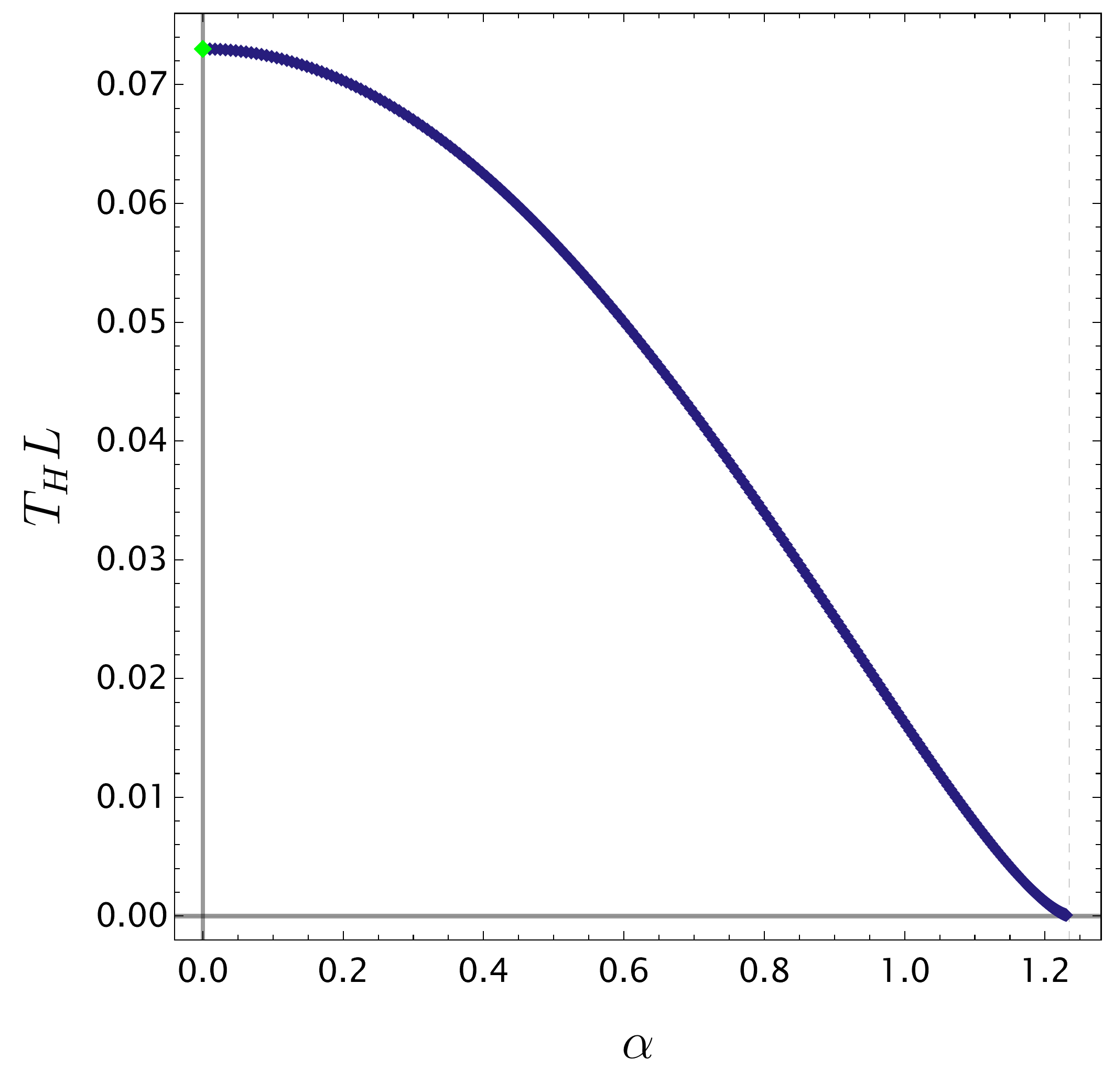}
    \includegraphics[width=0.43\linewidth]{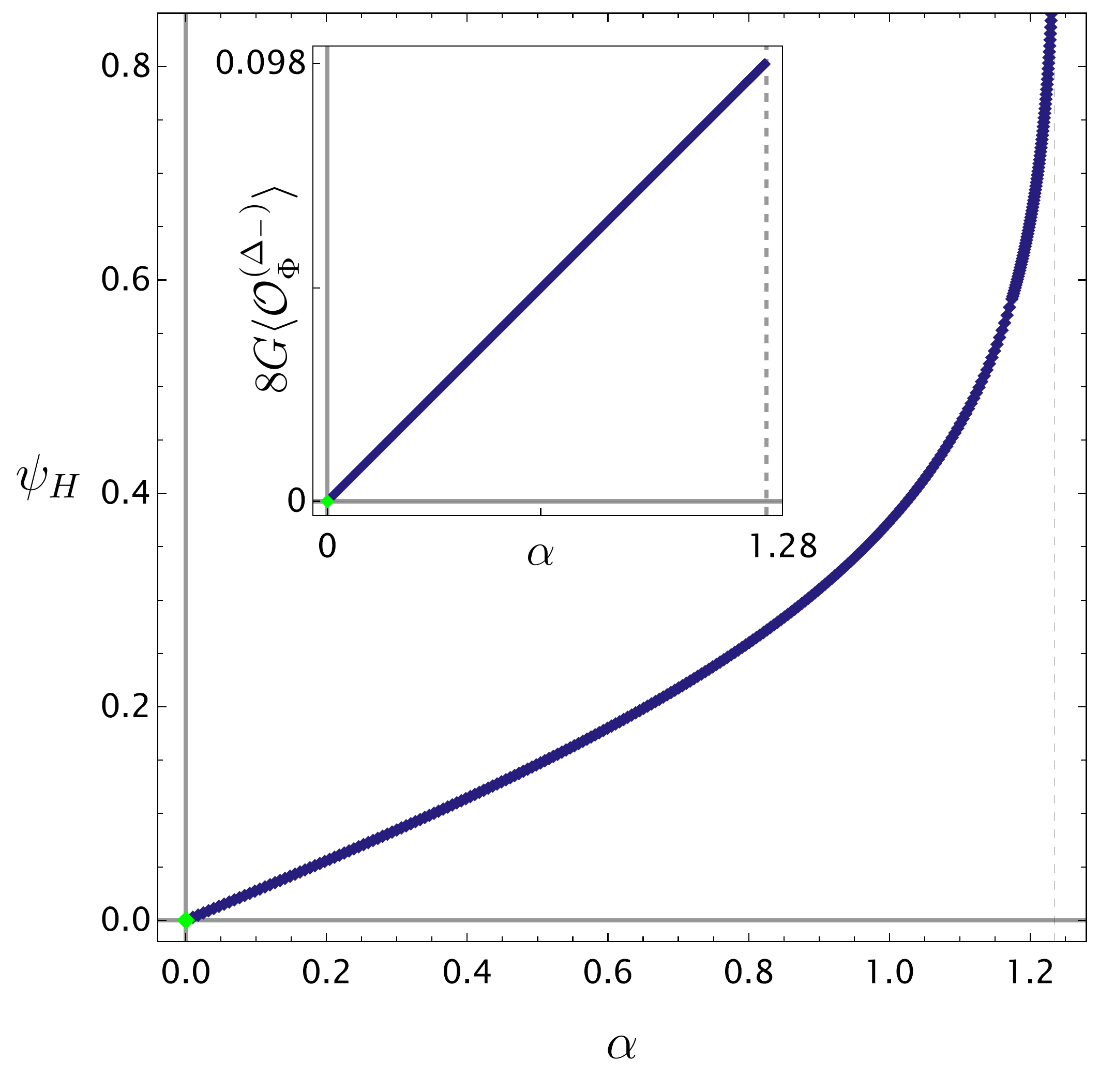}
    \includegraphics[width=0.43\linewidth]{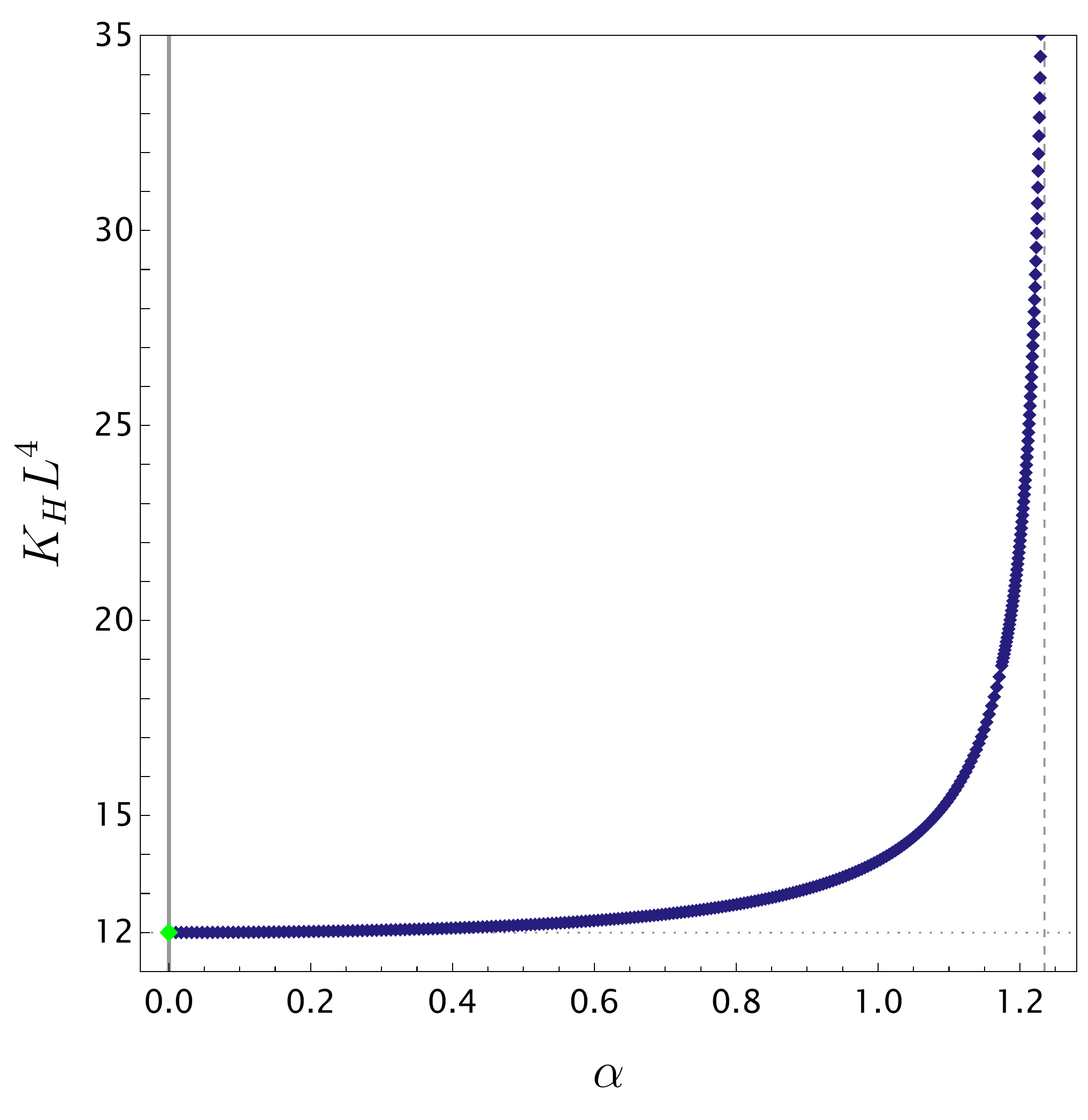}
    \caption{Thermodynamic properties of $\bm{m=1}$ hairy black hole sub-family with $R_+=0.75$ (\ie $\hat{S}_H=3\pi$) for $\kappa = -8/10$, $\mu^2L^2 = -15/16$.  The bright-green diamond is the BTZ instability onset point.  Auxiliary orange dotted lines describe extremal BTZ ($\Delta \hat{M}=0$). $\Delta \hat{M}=\hat{M}- \hat{M}^{\hbox{\tiny BTZ}}_{\hbox{\tiny ext}} =\hat{M} -\hat{J}$ describes how far off the mass of a given hairy black hole is from the extremal BTZ mass.
    $\Delta \hat{S}_H=\hat{S}_H-\hat{S}_H^{\hbox{\tiny BTZ}}$ is the entropy difference between a given hairy black hole and BTZ that has the same  $\hat{M}$ and $\hat{J}$  (when they co-exist). In the inset plot of the middle-left panel, $S_i$ describes either the hairy black hole entropy $S_H=3\pi$ (navy blue diamonds) or the BTZ entropy (grey curve) which meet at the onset bright-green diamond. 
    }
    \label{fig:m1-hBTZ-Rp075}
\end{figure}

\begin{figure}
    \centering
    \vskip -0.5cm
    \includegraphics[width=0.45\linewidth]{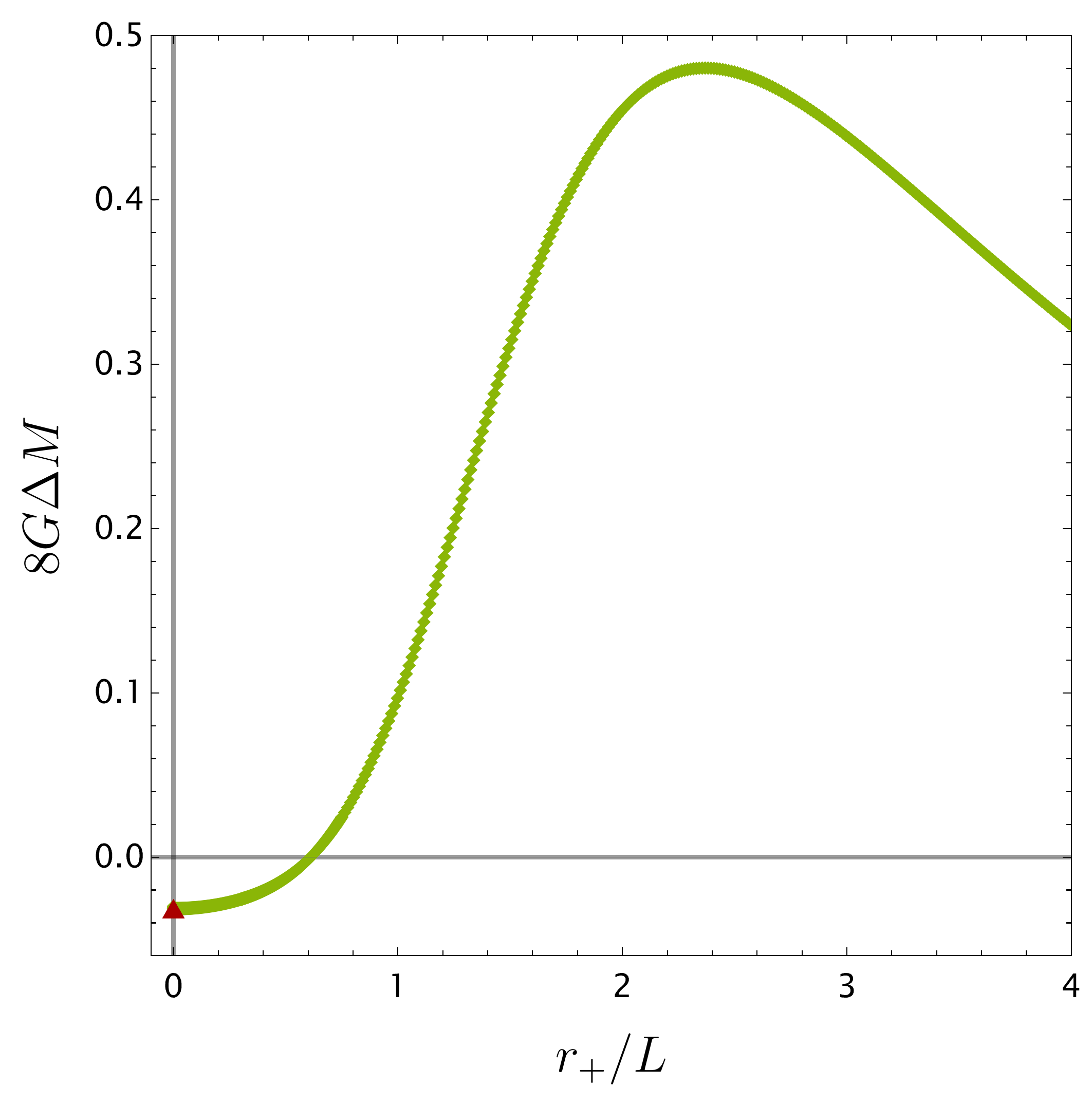}
    \includegraphics[width=0.45\linewidth]{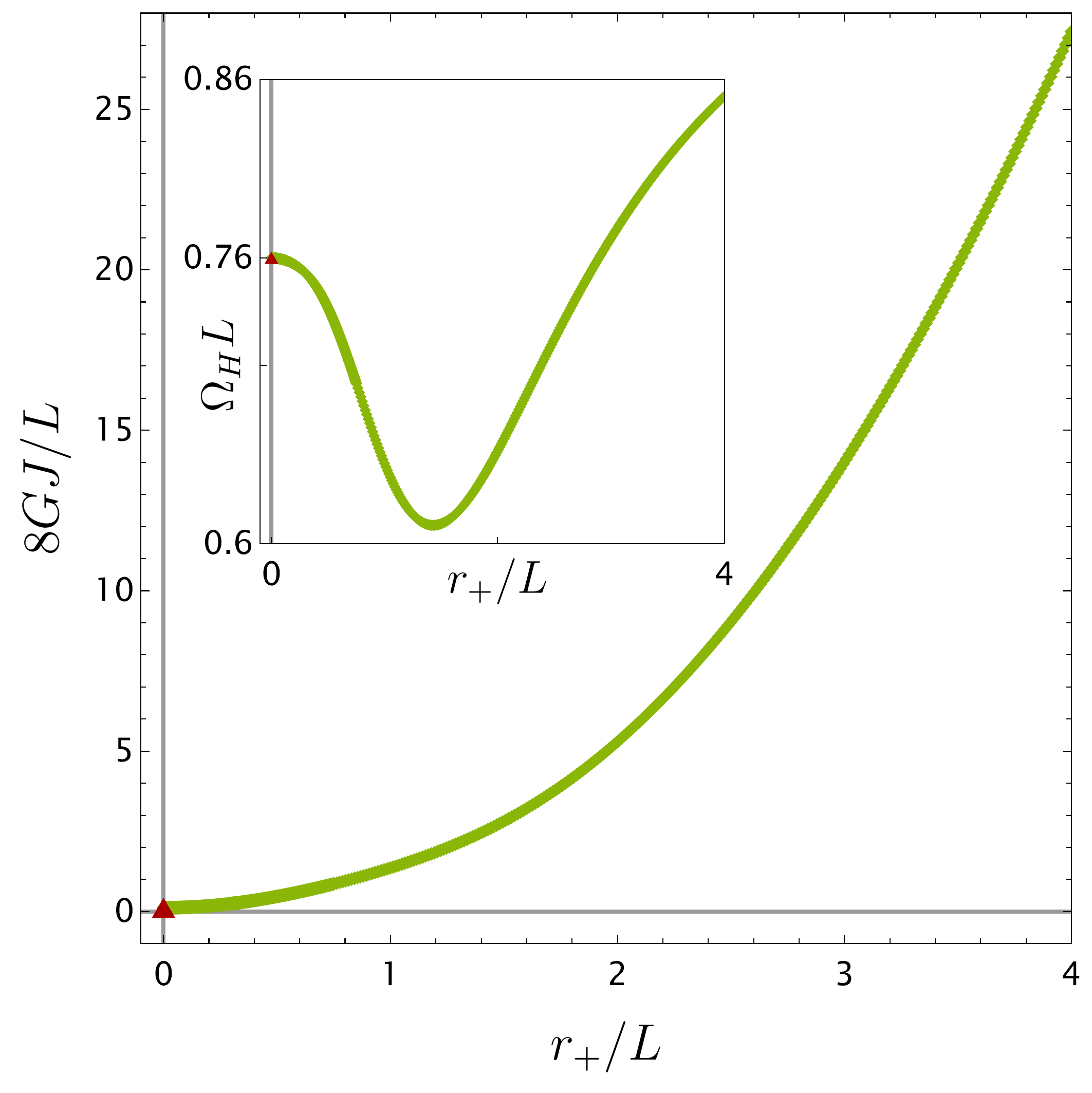}
    \includegraphics[width=0.45\linewidth]{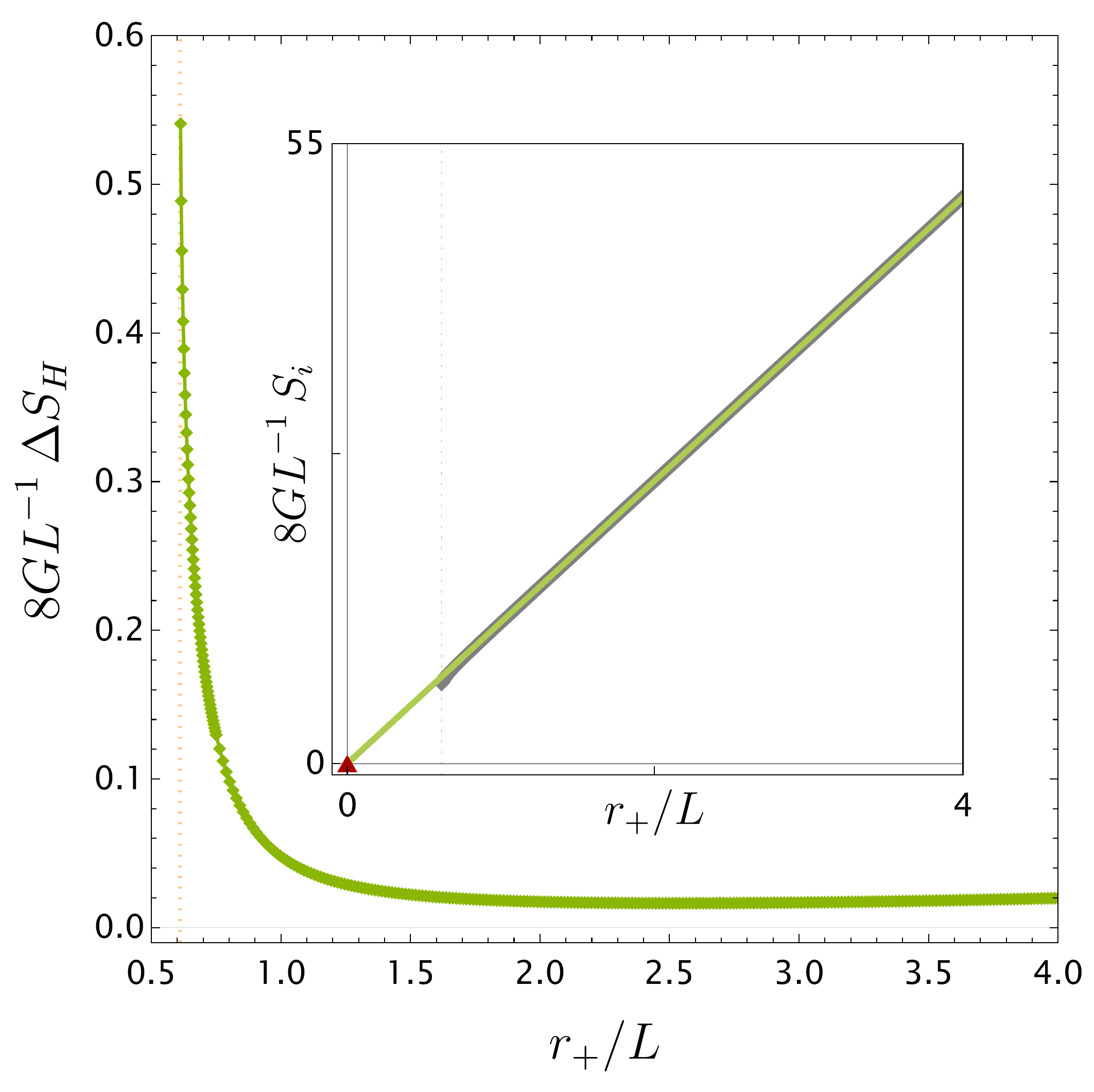}
    \includegraphics[width=0.45\linewidth]{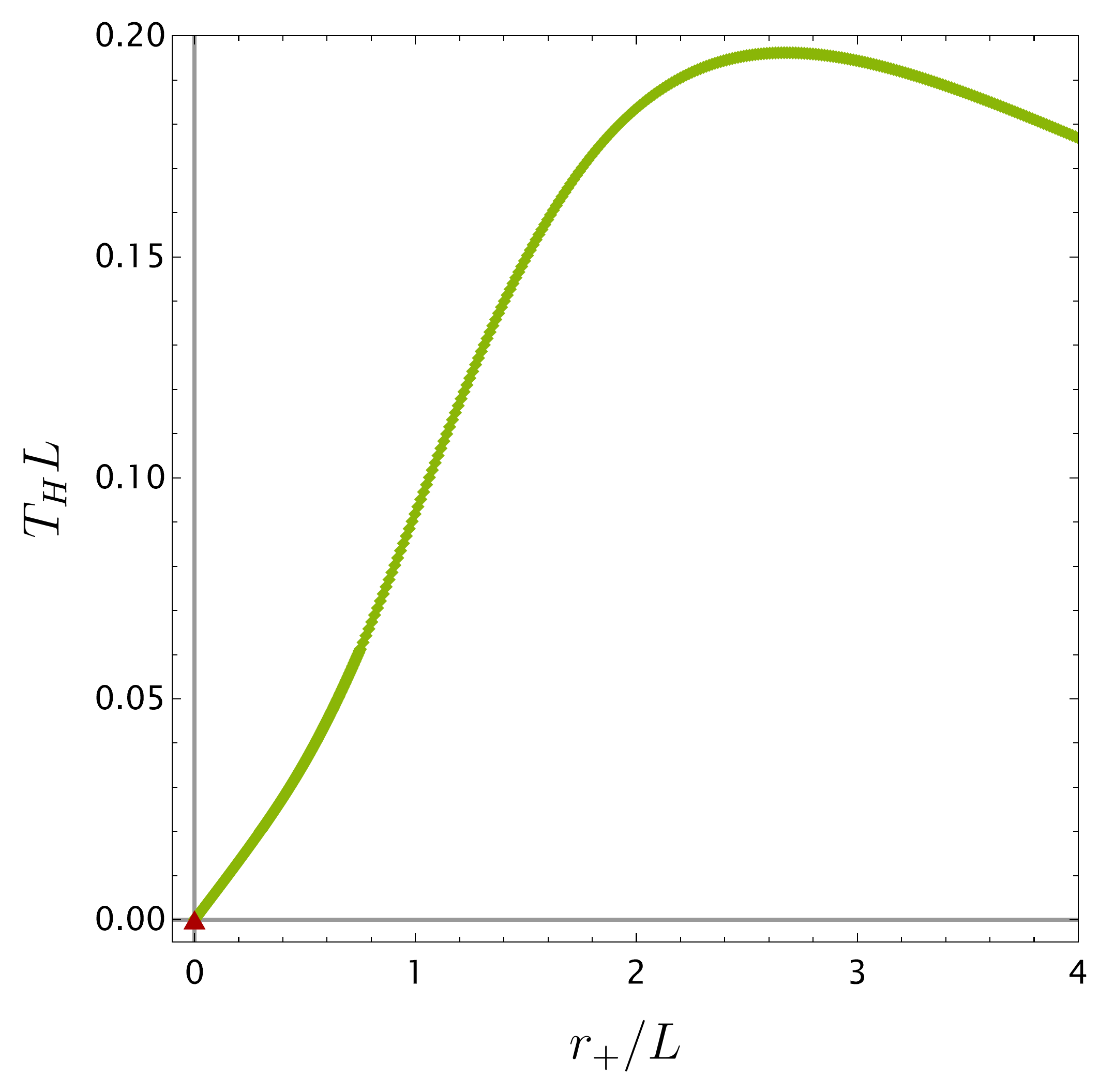}
    \includegraphics[width=0.45\linewidth]{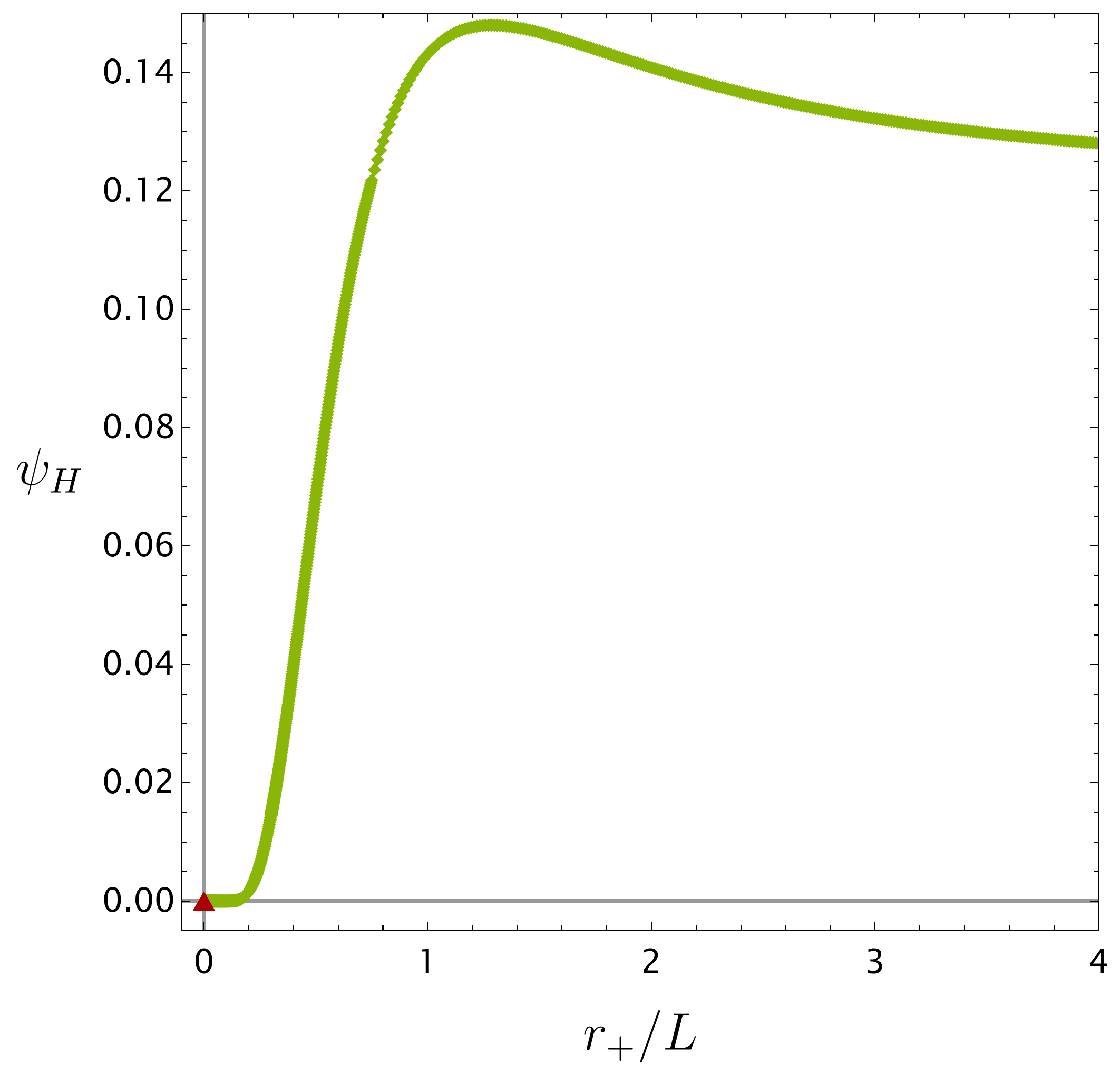}
    \includegraphics[width=0.45\linewidth]{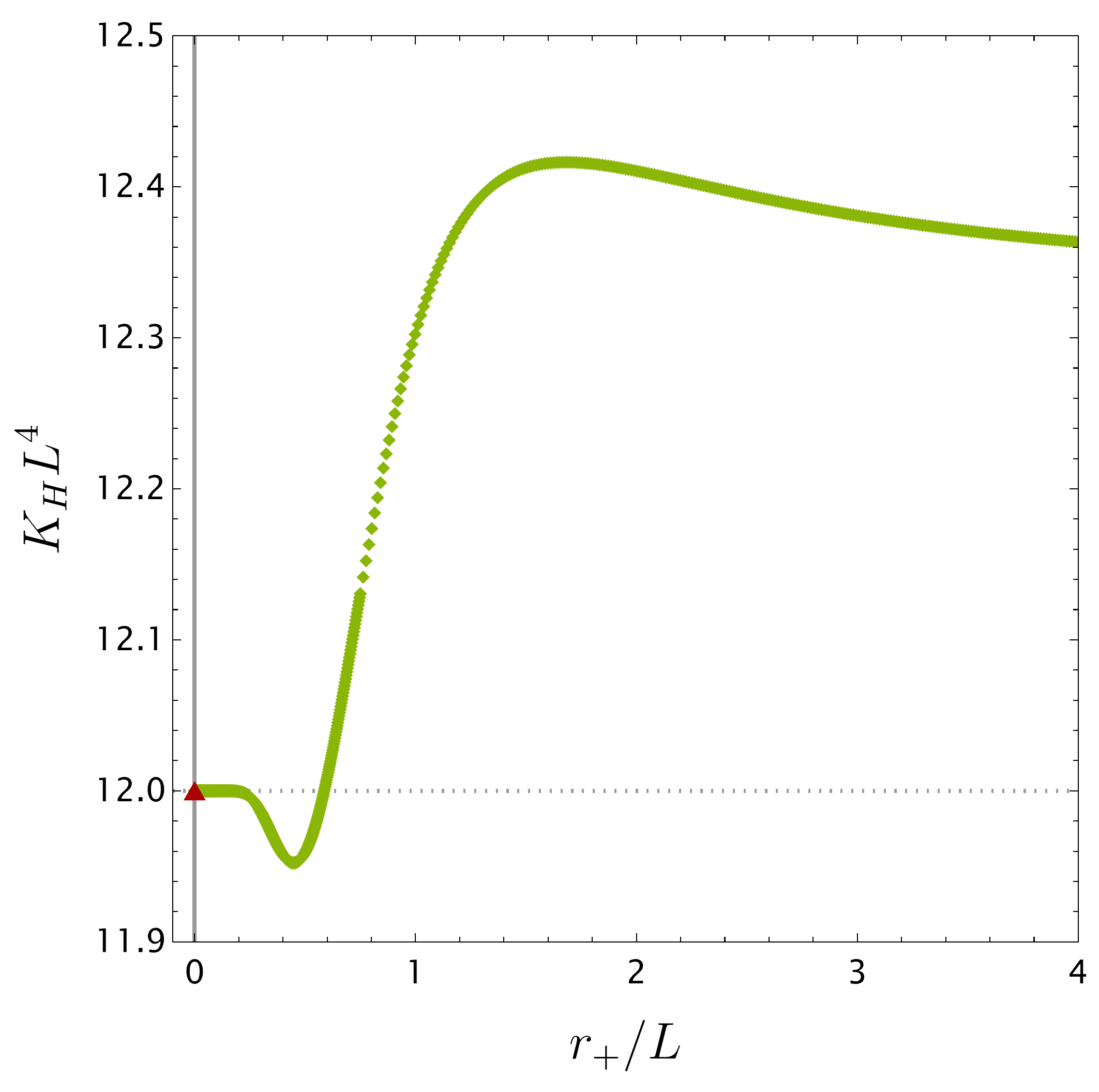}
    \caption{ 
Thermodynamic properties of $\bm{m=1}$ hairy black hole sub-family with $\bm{\alpha =0.423475}$ for $\kappa = -8/10$, $\mu^2L^2 = -15/16$.   $\Delta \hat{M}=\hat{M}- \hat{M}^{\hbox{\tiny BTZ}}_{\hbox{\tiny ext}} =\hat{M} -\hat{J}$ describes how far off the mass of a given hairy black hole is from the extremal BTZ mass. $\Delta \hat{S}_H=\hat{S}_H-\hat{S}_H^{\hbox{\tiny BTZ}}$ is the entropy difference between a given hairy black hole and BTZ that has the same  $\hat{M}$ and $\hat{J}$  (when they co-exist).
Auxiliary orange dotted line describes extremal BTZ ($\Delta \hat{M}=0$).
The dark-red triangle (in the $R_+ \to 0$ limit) represents the singular $m=1$ extremal hairy black hole of section~\ref{sec:NumericalSetup:singBHmJ} with $\Delta\hat{M} \sim -0.0313$, $\hat{J} \sim 0.1197$ and $\hat{\Omega}_H \sim 0.76006$.}
    \label{fig:m1-hBTZ-alpha0423475}
\end{figure}

\begin{figure}
    \centering
    \vskip -0.3cm
    \includegraphics[width=0.45\linewidth]{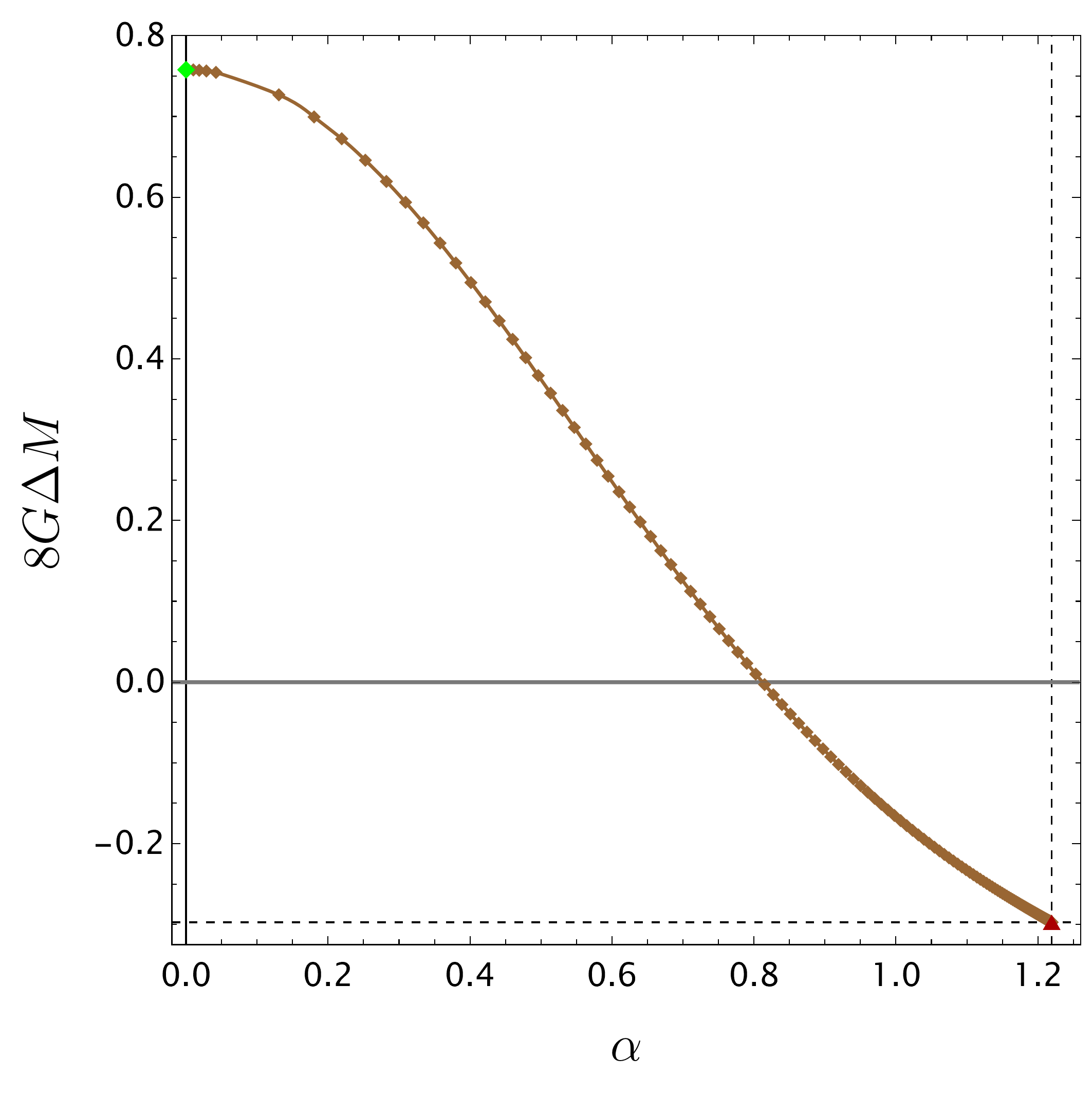}
     \includegraphics[width=0.45\linewidth]{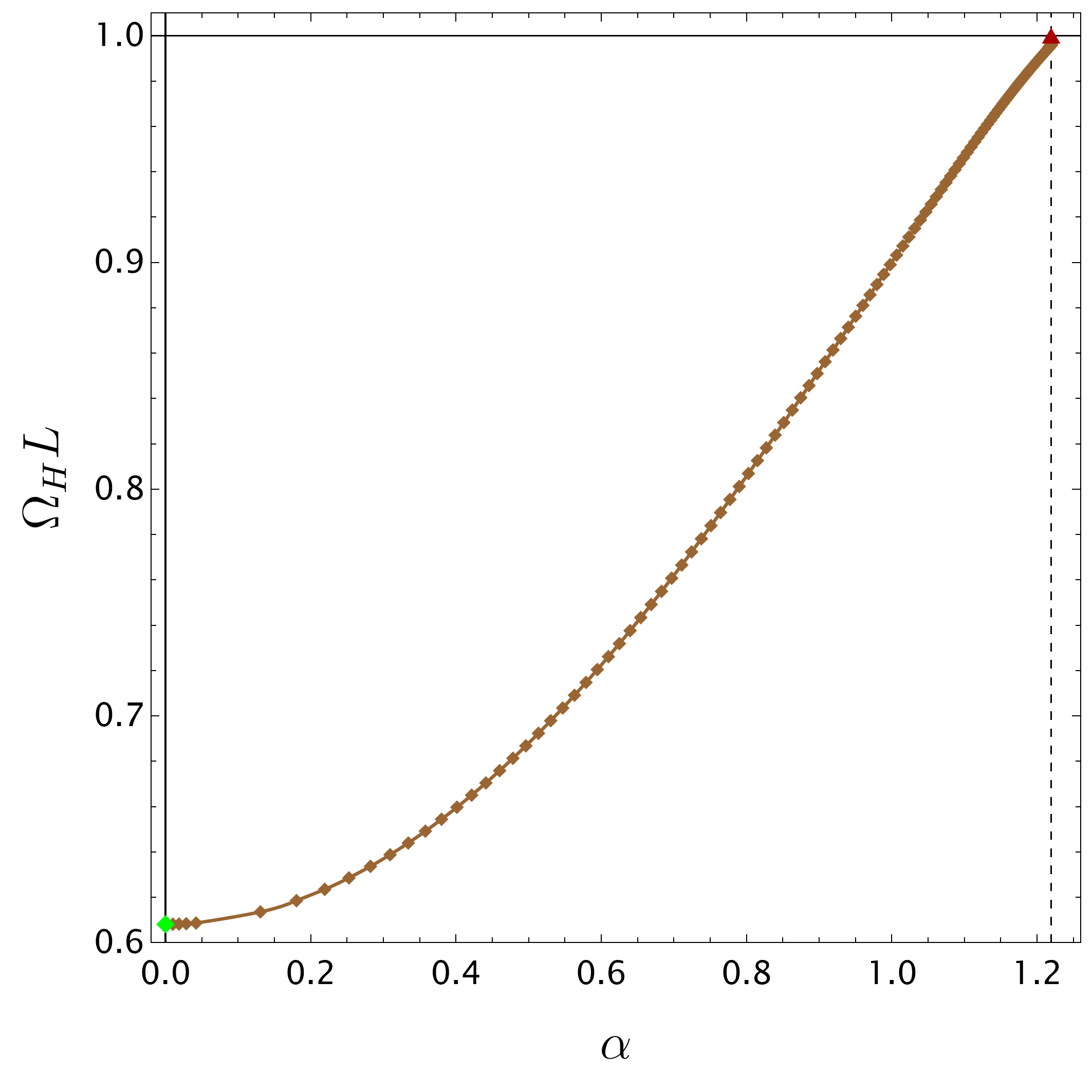}
      \includegraphics[width=0.45\linewidth]{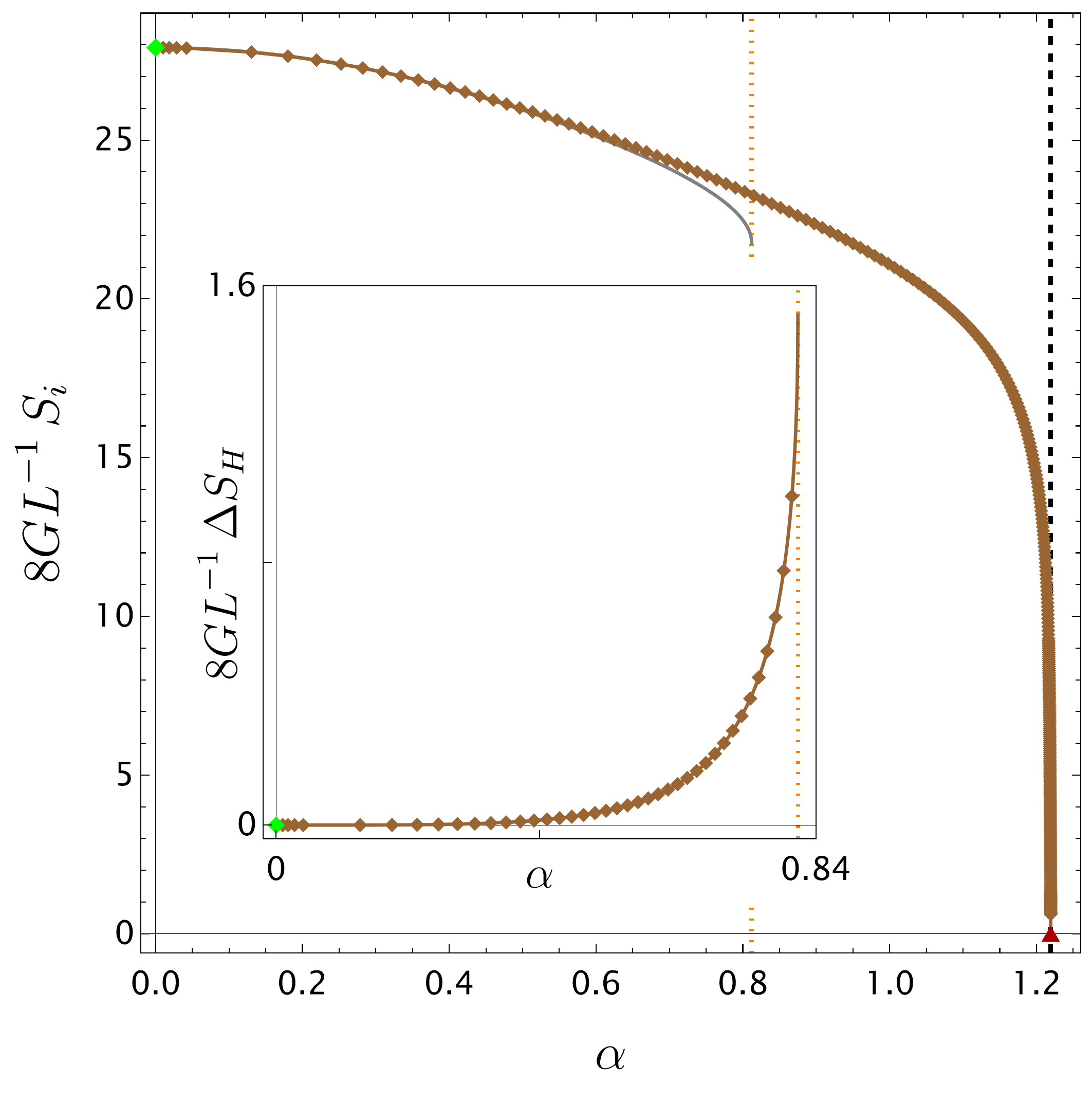}
      \includegraphics[width=0.45\linewidth]{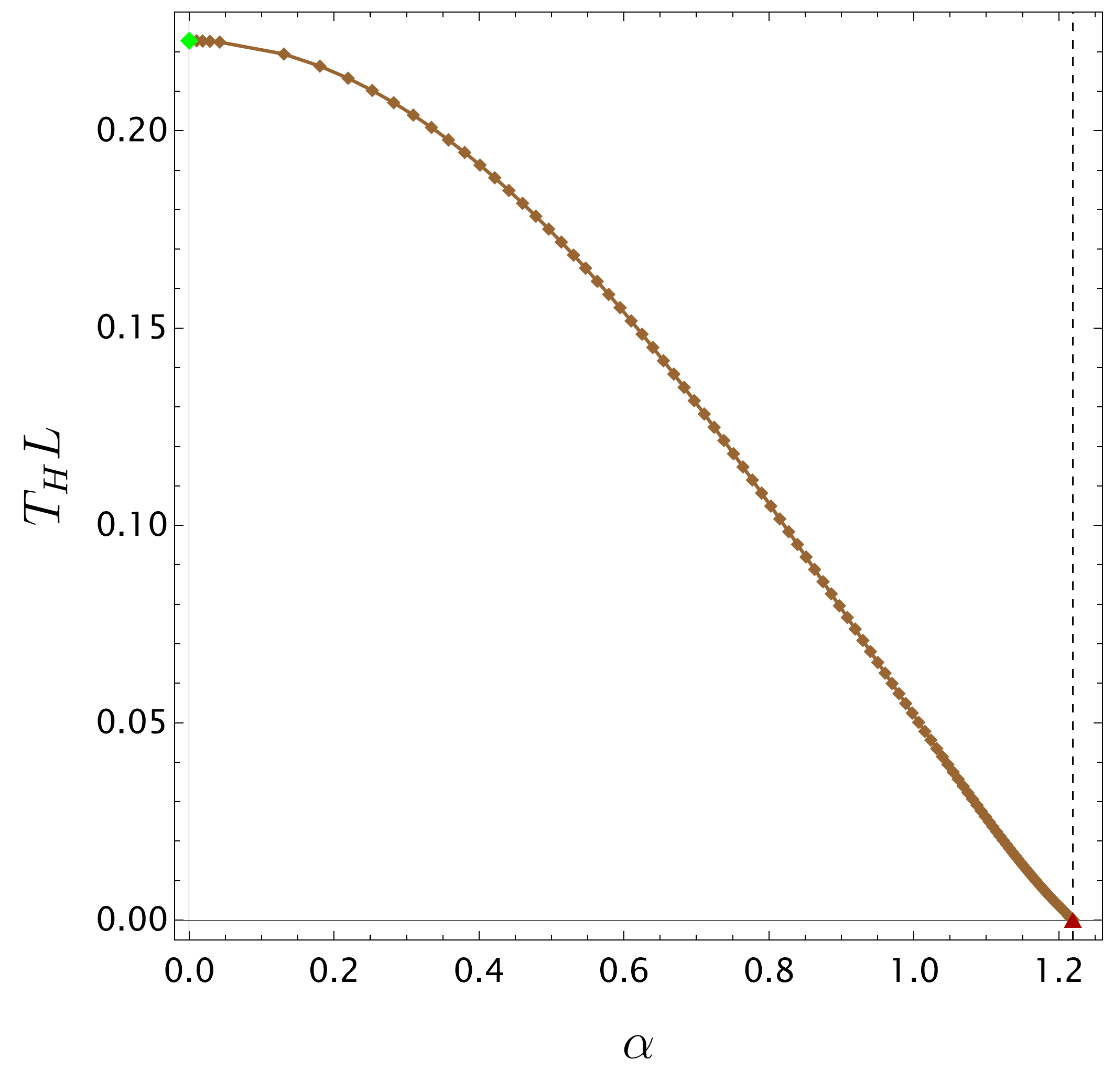}
      \includegraphics[width=0.45\linewidth]{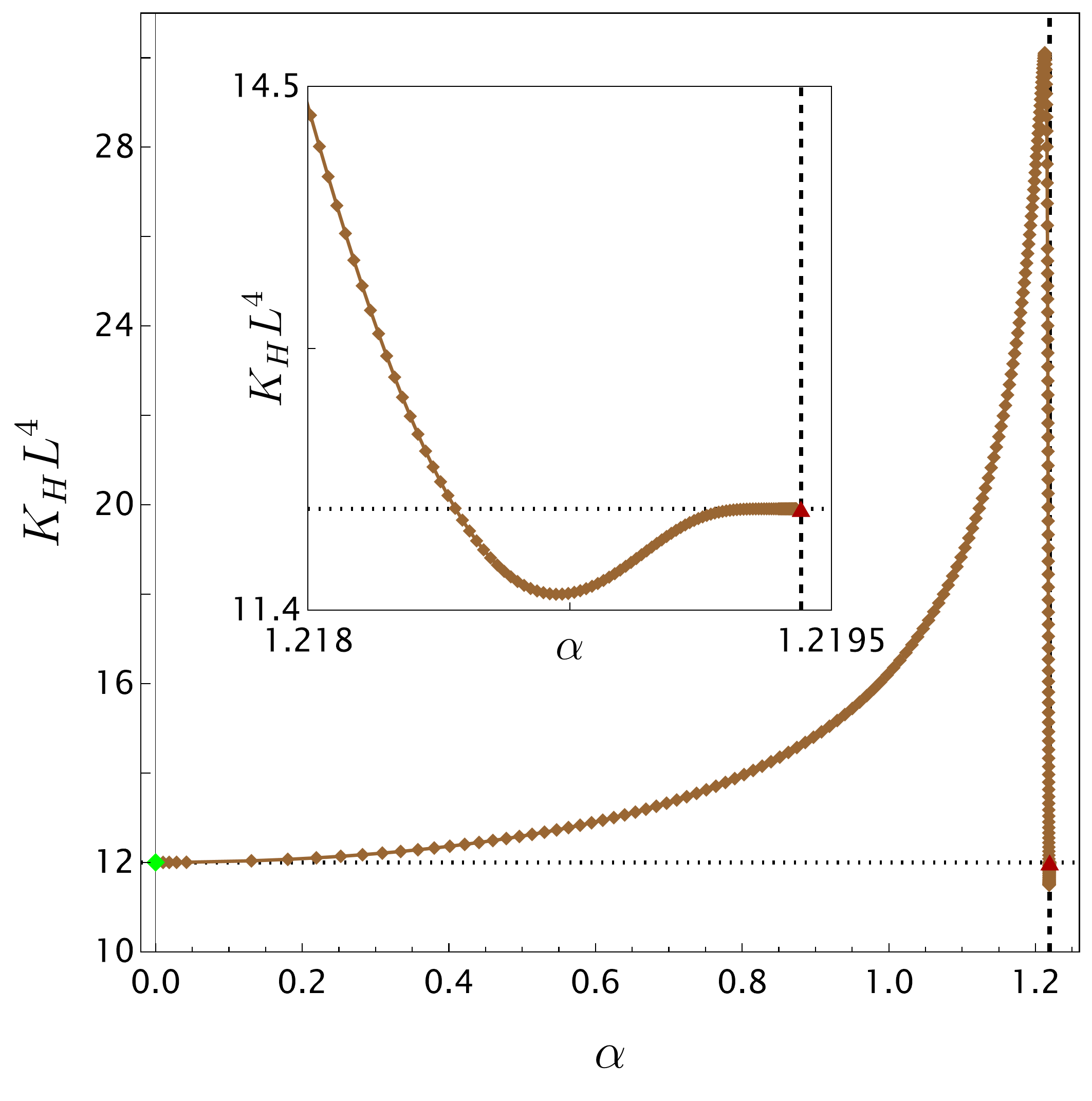}
      \includegraphics[width=0.45\linewidth]{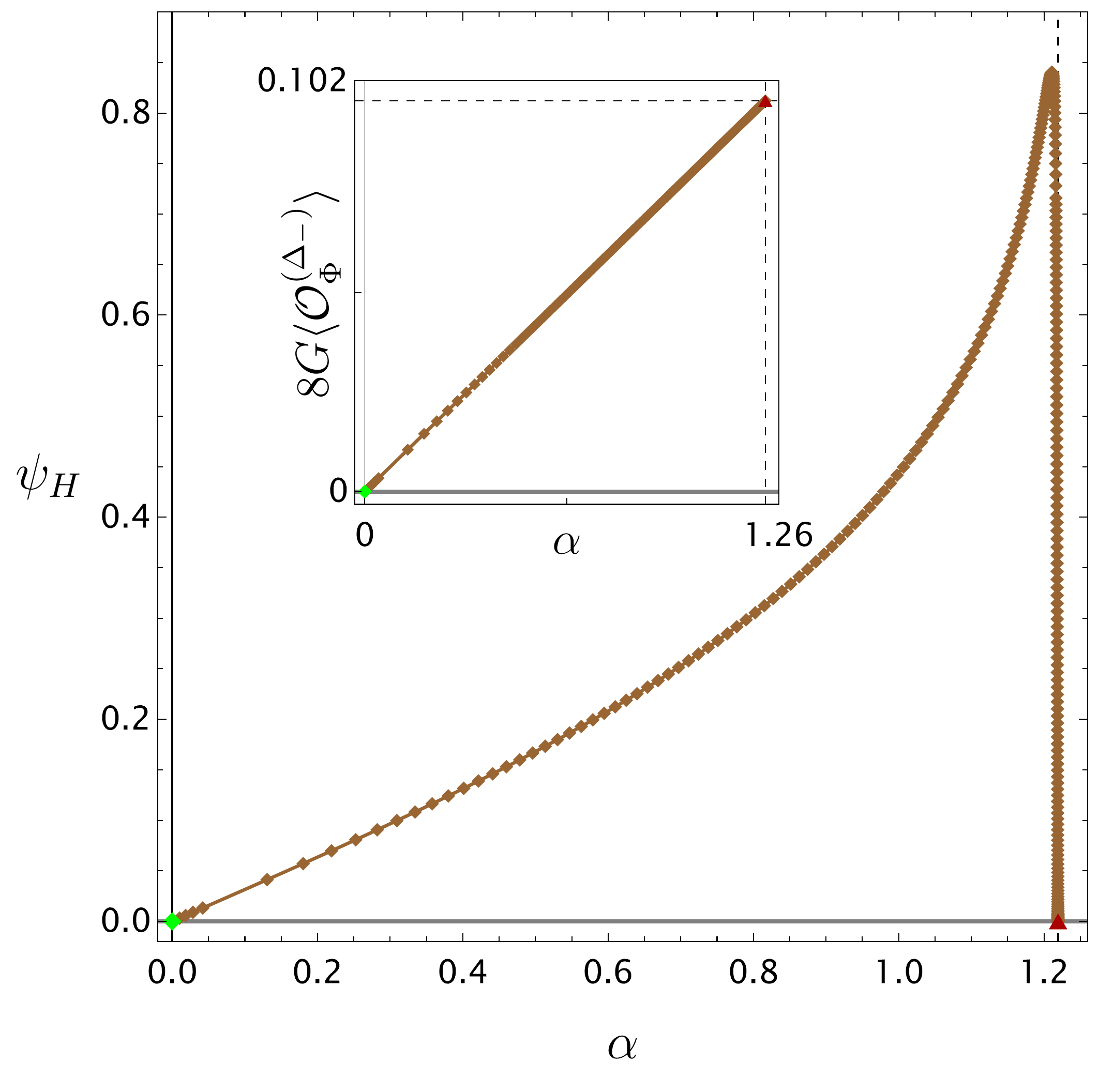}
    \caption{ 
Thermodynamic properties of $\bm{m=1}$ hairy black hole sub-family with $\bm{\hat{J} =6}$ for $\kappa = -8/10$, $\mu^2L^2 = -15/16$.   $\Delta \hat{M}=\hat{M}- \hat{M}^{\hbox{\tiny BTZ}}_{\hbox{\tiny ext}} =\hat{M} -\hat{J}$ describes how far off the mass of a given hairy black hole is from the extremal BTZ mass. $\Delta \hat{S}_H=\hat{S}_H-\hat{S}_H^{\hbox{\tiny BTZ}}$ is the entropy difference between a given hairy black hole and BTZ that has the same  $\hat{M}$ and $\hat{J}$  (when they co-exist).
Auxiliary orange dotted line describes extremal BTZ ($\Delta \hat{M}=0$) and the dark-red triangle represents the singular $m=1$  extremal hairy black hole identified in section~\ref{sec:NumericalSetup:singBHmJ}.
}
\label{fig:m1-hBTZ-J6}
\end{figure}

The three one‑parameter sub‑families of $m=1$ hairy black holes introduced above are displayed in Fig.~\ref{fig:m1:dMJ:3families}. Their thermodynamic properties are analysed in detail in Fig.~\ref{fig:m1-hBTZ-Rp075} for the family with $R_{+}=3/4$, in Fig.~\ref{fig:m1-hBTZ-alpha0423475} for the family with $\alpha=0.423475$, and in Fig.~\ref{fig:m1-hBTZ-J6} for the family with fixed $\hat{J}=6$.

In more detail, Fig.~\ref{fig:m1-hBTZ-Rp075} displays several physical quantities for the sub‑family of $m=1$ hairy black holes with fixed horizon radius $R_{+}=3/4$, plotted as functions of the asymptotic scalar amplitude $\alpha$. Specifically, we show: $\Delta\hat{M}$ (top‑left panel); the angular momentum $\hat{J}$ together with the horizon angular velocity $\hat{\Omega}_{H}$ (top‑right panel, including inset); the entropy difference $\Delta\hat{S}_{H}$ along with the entropy $\hat{S}_{H}$ (middle‑left panel, including inset); the Hawking temperature $\hat{T}_{H}$ (middle‑right panel); the value of the scalar field at the horizon $\psi_{H}$ together with the VEV $\langle\hat{\mathcal{O}}_{\Phi}^{(\Delta_-)}\rangle$ (bottom‑left panel, including inset); and the Kretschmann curvature scalar at the horizon $\hat{K}_{H}$ (bottom‑right panel). We have verified that the first law of black‑hole thermodynamics~\eqref{FirstLawBH} is satisfied with relative errors below $10^{-3}\%$.

The main features of Fig.~\ref{fig:m1-hBTZ-Rp075} are readily apparent, but it is worth emphasizing a few key properties. Families of hairy black holes at fixed entropy $-$ such as the one shown $-$ always bifurcate from the BTZ solution at the onset (bright‑green diamond) of the $m=1$ double‑trace instability, as determined in~\cite{Dias:2025uyk}. This bifurcation corresponds to a second‑order phase transition at which $\Delta\hat{S}_{H}=0$ and the scalar condensate vanishes, \ie
$\alpha=\langle\hat{\mathcal{O}}_{\Phi}^{(\Delta_-)}\rangle=\psi_{H}=0$. The constant‑entropy branch then extends to arbitrarily large angular momentum, much as BTZ itself exists for arbitrarily large $\hat{J}$. In particular, for $\alpha\gtrsim1.2$, a very small increase in $\alpha$ leads to a rapid growth of $\hat{J}$. Along this sub‑family, the scalar condensate increases while the temperature decreases (and the horizon curvature grows), eventually reaching the limiting behaviour $\hat{T}_{H}\to0$ as $\hat{J}\to\infty$, with $\hat{\Omega}_{H}\to1$.

We now turn to Fig.~\ref{fig:m1-hBTZ-alpha0423475}, which shows the sub‑family of $m=1$ hairy black holes at fixed scalar condensate amplitude $\alpha=0.423475$. Several quantities are displayed as functions of the horizon radius $R_{+}$: $\Delta\hat{M}$ (top‑left panel), the angular momentum $\hat{J}$ (top‑right panel), the horizon angular velocity $\hat{\Omega}_{H}$ (middle‑left panel), the Hawking temperature $\hat{T}_{H}$ (middle‑right panel), and the entropy difference $\Delta\hat{S}_{H}$ (bottom panel). We have verified that the first law of black‑hole thermodynamics~\eqref{FirstLawBH} is satisfied, with relative errors below $10^{-3}\%$. Along this family, the VEV $\langle\hat{\mathcal{O}}_{\Phi}^{(\Delta_-)}\rangle$ is fixed by construction, and both the horizon value of the scalar field $\psi_{H}$ and the Kretschmann curvature scalar at the horizon $\hat{K}_{H}$ remain constant.

The main qualitative features of Fig.~\ref{fig:m1-hBTZ-alpha0423475} are again straightforward to identify. In particular, families of $m=1$ hairy black holes at fixed $\alpha$, such as the one shown, always originate at finite angular momentum. Specifically, they begin at the $m=1$ singular boson star branch (dark‑red triangles) discussed in Section~\ref{sec:PhaseDiag-m1:BStar} and shown in Fig.~\ref{fig:BS-m1}. From this starting point, the family extends to arbitrarily large values of $R_{+}$ or $\hat{J}$ (we have verified existence up to $\hat{J}=10$). Crucially, whenever a hairy black hole and a BTZ black hole coexist with the same $(\hat{M},\hat{J})$, the $m=1$ hairy black hole is always entropically favored, \ie
$\Delta\hat{S}_{H}>0$.

The combined analyses of Figs.~\ref{fig:m1-hBTZ-Rp075} and~\ref{fig:m1-hBTZ-alpha0423475} indicate that, in the $\hat{J}$–$\Delta\hat{M}$ plane of Fig.~\ref{fig:m1:dMJ:3families}, the two‑dimensional region populated by $m=1$ hairy black holes is bounded from above by the double‑trace instability onset curve (bright‑green) and from below by the singular $m=1$ rotating extremal hairy black hole curve (dark-red triangles) identified in section~\ref{sec:NumericalSetup:singBHmJ}. To further elucidate how the $m=1$ hairy black hole solutions approach this lower boundary, it is convenient to follow a one‑parameter sub‑family at fixed angular momentum, corresponding to a vertical line in Fig.~\ref{fig:m1:dMJ:3families}.

This is implemented in Fig.~\ref{fig:m1-hBTZ-J6}, where we present a representative sub‑family with $\hat{J}=6$. We plot several quantities as functions of the scalar amplitude $\alpha$: $\Delta\hat{M}$ (top‑left panel), the horizon angular velocity $\hat{\Omega}_{H}$ (top‑right panel), $\Delta\hat{S}_{H}$ together with the entropy $\hat{S}_{H}$ (middle‑left panel, including inset), the Hawking temperature $\hat{T}_{H}$ (middle‑right panel), the Kretschmann curvature scalar at the horizon $\hat{K}_{H}$ (bottom‑left panel), and the scalar‑field value at the horizon $\psi_{H}$ together with the VEV $\langle\hat{\mathcal{O}}_{\Phi}^{(\Delta_-)}\rangle$ (bottom‑right panel, including inset). Again, the first law of thermodynamics~\eqref{FirstLawBH} is satisfied to within relative errors smaller than $10^{-3}\%$.

The behaviour of this constant‑$\hat{J}$ family is clearly illustrated in Fig.~\ref{fig:m1-hBTZ-J6}. As in the previous cases, families of $m=1$ hairy black holes at fixed angular momentum always bifurcate from the BTZ solution at the onset (bright‑green diamond) of the $m=1$ double‑trace instability, as computed in~\cite{Dias:2025uyk}. This bifurcation corresponds to a second‑order phase transition, characterized by $\Delta\hat{S}_{H}=0$ and a vanishing scalar condensate,
\ie $\alpha=\langle\hat{\mathcal{O}}_{\Phi}^{(\Delta_-)}\rangle=\psi_{H}=0$. Away from the onset, the constant‑$\hat{J}$ branch evolves toward lower entropy and temperature, eventually meeting the family of singular $m=1$ rotating extremal hairy black holes (dark-red triangles) $-$ identified in section~\ref{sec:NumericalSetup:singBHmJ} $-$  with $\hat{S}_{H}\to0$, $\hat{T}_{H}\to0$. These features $-$ most transparently visible in Fig.~\ref{fig:m1-hBTZ-J6} $-$ were not apparent from the analyses of Figs.~\ref{fig:m1-hBTZ-Rp075} and~\ref{fig:m1-hBTZ-alpha0423475}. Notably, as the $\alpha$ value is increased, both the Kretschmann curvature and $\psi_H$ increase up to a maximum value, to then decay and meet the values of the singular $m=1$  extremal hairy black holes family: $K_H L^4 = 12$ and $\psi_H = 0$. This is in stark contrast to the $m =0$ case from  Fig.~\ref{fig:m0-hBTZ-J3}, but can be understood by looking at our analysis from subsection \ref{sec:NumericalSetup:SingBS}. Recall that, for $m=1$, we  identify a family of singular extremal hairy black holes with a milder (conical) singularity, that does not cause the Kretschmann or $\psi_H = 0$ to diverge (see section~\ref{sec:NumericalSetup:singBHmJ}). Such family of singular $m=1$ hairy  black holes turns out to be the extremal limit of the $m =1$ non-extremal hairy black holes with $\hat{J} >0$ fixed as we increase $\alpha$.
\section{Full phase diagram of asymptotically \texorpdfstring{AdS$_3$}{AdS3} stationary solutions}\label{sec:PhaseDiag-Total}
In Sections~\ref{sec:PhaseDiag-m0} and~\ref{sec:PhaseDiag-m1} we explored the phase diagram of asymptotically AdS$_3$ hairy solutions with $m=0$ and $m=1$ (the latter illustrating the generic $m\geq1$ case), respectively. In both cases, we studied the boson stars of the system in detail. For $m=0$, this analysis was carried out in Subsections~\ref{sec:PhaseDiag-m0:BStar}–\ref{sec:PhaseDiag-m0:IshWald} and illustrated in Figs.~\ref{fig:BS-m0}–\ref{fig:m0BSevoK} (see also Fig.~\ref{fig:m0m1omega0BS}). For $m=1$ (and more generally $m\geq1$), the corresponding discussion appeared in Subsections~\ref{sec:PhaseDiag-m1:BStar}–\ref{sec:PhaseDiag-m1:IshWald} and Figs.~\ref{fig:BS-m1}–\ref{fig:m1BSevoK} (see also Fig.~\ref{fig:m0m1omega0BS}, which identifies the zero‑frequency boson star lying close to the lower bound implied by the positivity‑of‑energy theorem of Ref.~\cite{Faulkner:2010fh} and Appendix~\ref{secA:superpotentials}).

By contrast, for hairy black holes we have only focused on particular one‑parameter sub‑families of solutions so far, namely those at constant $R_{+}$, constant scalar condensate amplitude $\alpha$, or constant angular momentum $\hat{J}$. We are now ready to perform a full exploration of the two‑dimensional parameter space of hairy black hole solutions and to demonstrate that the specific sub‑families studied previously are not special, but rather illustrate universal qualitative properties of hairy black holes.

When they coexist with BTZ solutions, double‑trace hairy black holes dominate the microcanonical ensemble, although only in special circumstances do they dominate the canonical and grand‑canonical ensembles. We therefore begin by focusing on the microcanonical phase diagram and postpone the discussion of the remaining ensembles to a later stage. Recall that in the microcanonical ensemble the relevant thermodynamic potential is the entropy $\hat{S}_{H}$, and that for fixed mass $\hat{M}$ and angular momentum $\hat{J}$ the dominant configuration is the one with the largest entropy. Accordingly, we represent our solutions in the three‑dimensional microcanonical phase diagram spanned by $(\hat{J},\,\Delta\hat{M},\,\hat{S}_{H})$, where
\begin{equation}
\Delta\hat{M}=\hat{M}-\hat{M}^{\hbox{\tiny BTZ}}_{\hbox{\tiny ext}}
\end{equation}
denotes the mass difference relative to the extremal BTZ black hole with $\hat{M}^{\hbox{\tiny BTZ}}_{\hbox{\tiny ext}}=\hat{J}$.

In Subsection~\ref{sec:PhaseDiag-Total-m0} we analyse the full phase diagram of asymptotically AdS$_3$ stationary solutions with $m=0$, while in Subsection~\ref{sec:PhaseDiag-Total-m1} we carry out the corresponding analysis for $m=1$ (which captures the generic features of the $m\geq1$ sector). In the latter case, we also display the $m=0$ hairy solutions to enable a direct comparison and to determine which families dominate the thermodynamic ensembles.

As throughout this work, we present explicit results for a scalar field mass $\mu^{2}L^{2}=-15/16$. This value is not special, but rather provides a convenient representative of the entire mass range
\begin{equation}
-1<\hat{\mu}^{2}<0,
\end{equation}
for which double‑trace boundary conditions can be imposed while maintaining a normalizable scalar condensate.

Hairy black holes do not exist in the Dirichlet or Neumann limits of the double‑trace boundary condition. This is ultimately because BTZ black holes satisfy $\Omega_{H}L\leq1$ and are therefore not unstable under Dirichlet or Neumann perturbations. By contrast, double‑trace boundary conditions introduce an asymptotic incoming flux that can trigger an instability~\cite{Dias:2025uyk}, allowing for the existence of hairy black holes. Similarly, global AdS$_3$ is stable under Dirichlet and Neumann perturbations~\cite{Ishibashi:2004wx,Dias:2025uyk}.

Nevertheless, regular boson stars in AdS$_3$ do exist for Dirichlet and Neumann boundary conditions and are perturbatively connected to AdS$_3$. In particular, they can be constructed order by order in perturbation theory by back‑reacting a linear normal mode of AdS$_3$, and they also admit fully nonlinear numerical realizations. We study these Dirichlet and Neumann boson stars both analytically and numerically in Appendix~\ref{secA:BStars-DN}. They were also displayed in Figs.~\ref{fig:BS-m0}–\ref{fig:m0BSevoK} for $m=0$ and in Figs.~\ref{fig:BS-m1}–\ref{fig:m1BSevoK} for $m=1$. Dirichlet boson stars are normalizable for any $\hat{\mu}>-1$, whereas Neumann boson stars exist only in the window $-1<\hat{\mu}<0$.

\subsection{Full phase diagram of asymptotically \texorpdfstring{AdS$_3$}{AdS3} stationary solutions with \texorpdfstring{$m=0$}{m=0}}\label{sec:PhaseDiag-Total-m0}

In this subsection, we present the full phase diagram of asymptotically AdS$_3$ stationary solutions with $m=0$ and double‑trace boundary conditions. Recall that $m=0$ hairy black holes form a two‑parameter family of solutions. In the microcanonical ensemble, the appropriate parameters are the conserved mass and angular momentum, $\Delta\hat{M}$ and $\hat{J}$.

\begin{figure}
    \centering
    \includegraphics[width=0.46\linewidth]{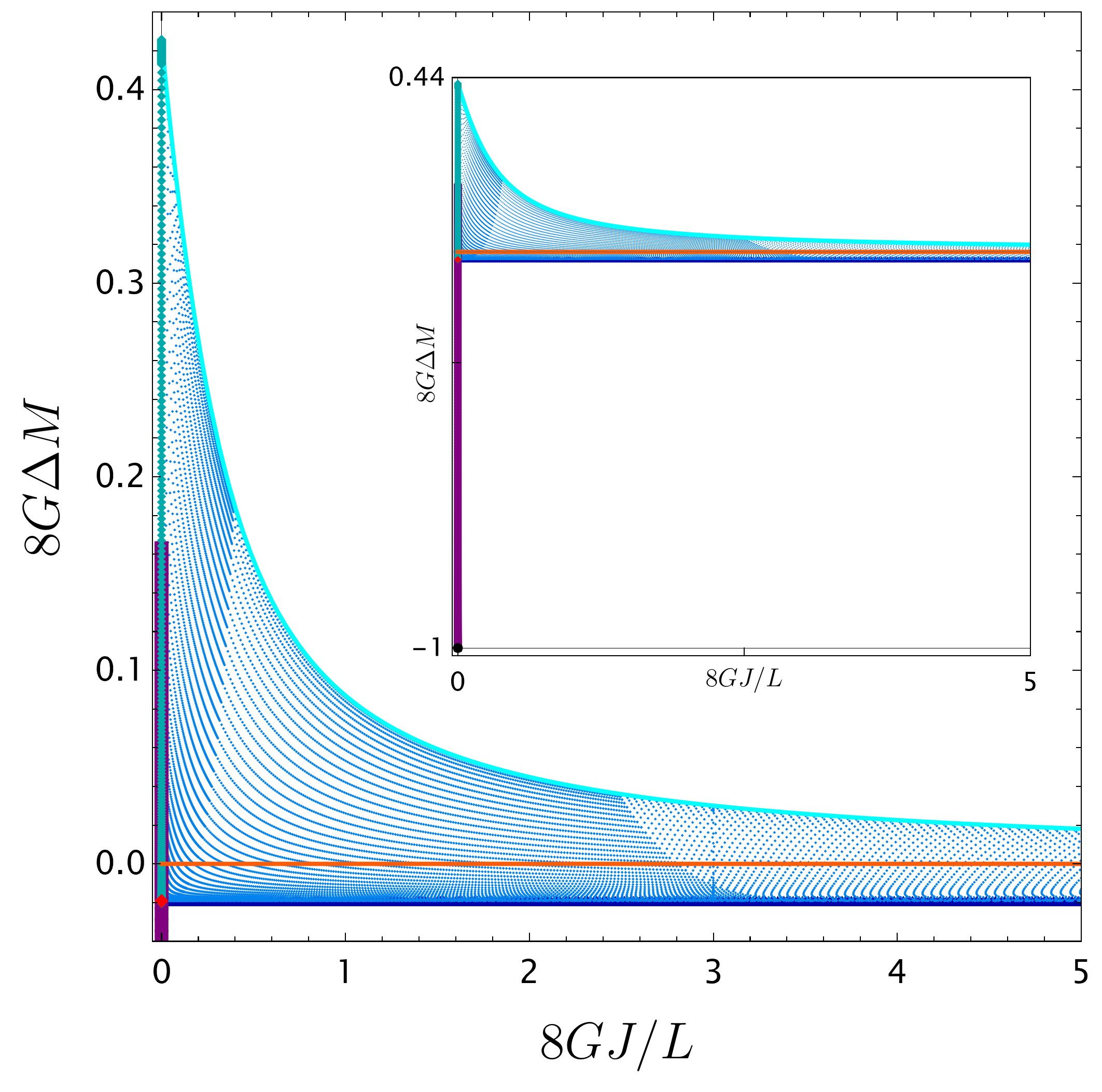}
    \includegraphics[width=0.53\linewidth]{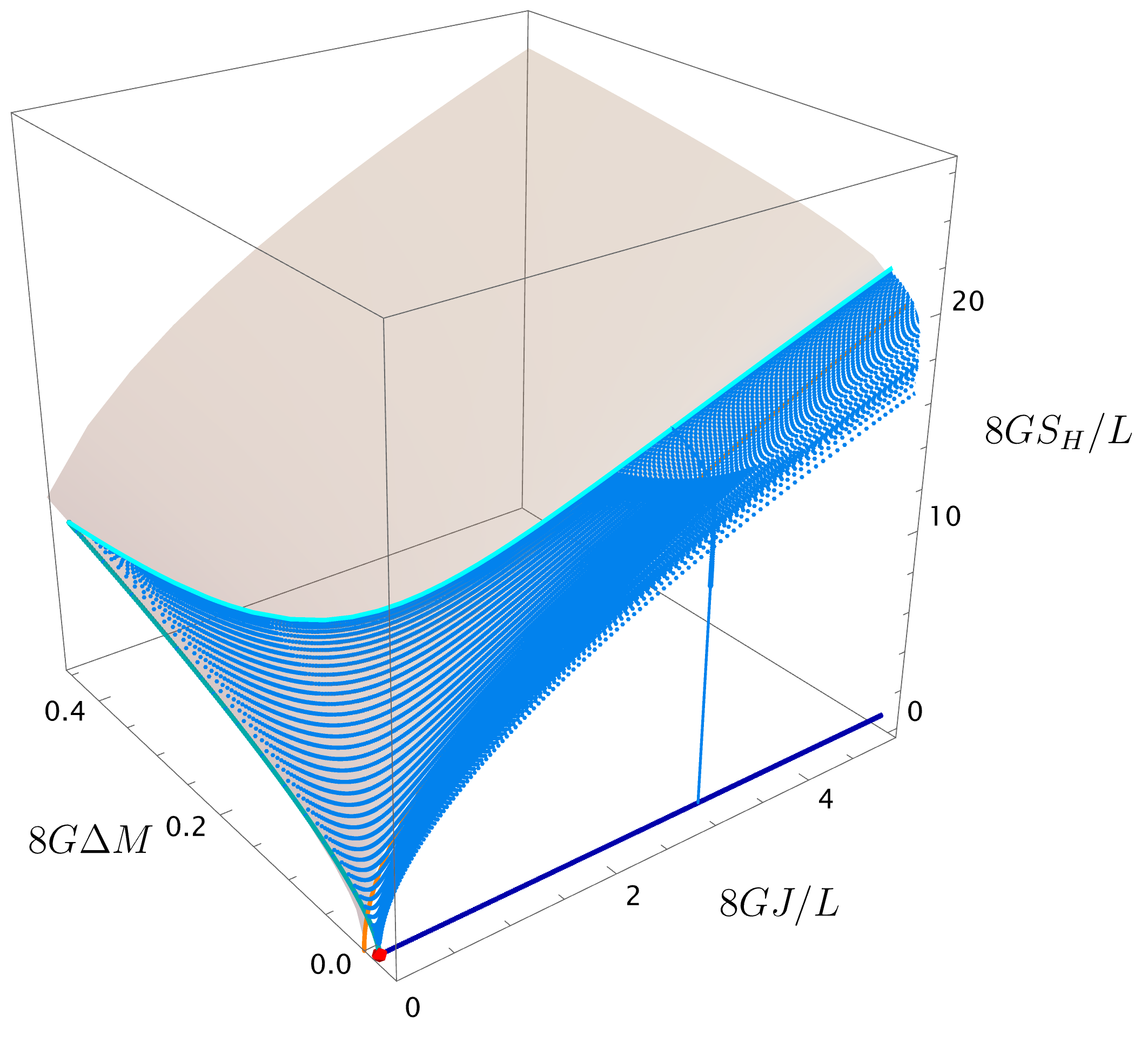}
    \caption{{\bf Microcanonical phase diagram} of solutions with $\bm{m = 0}$, $\mu^2L^2 = -15/16$ and $\bm{\kappa = -4/10>\kappa^{\rm AdS}_{m=0, \hat{\mu}^2=-15/16}\simeq -0.4951294}$ (for which AdS$_3$ is linear mode stable but BTZ can be unstable). When they co-exist, $m=0$ hairy black holes always dominate the microcanonical ensemble (\ie in the right panel, the blue dot hairy surface is always above the grey BTZ surface). The singular $m=0$ extremal hairy black hole (dark blue) has $\Delta \hat{M}(\hat{J})\simeq -0.019$ and it starts at the singular $m=0$ static extremal hairy black hole (red point). This value saturates the minimum energy bound \eqref{GlobalMin} of \cite{Faulkner:2010fh}; see  \eqref{Super:Emin}.
    }
    \label{fig:m0TotalPhaseDiag-k04}
\end{figure}

Figure~\ref{fig:m0TotalPhaseDiag-k04} displays the microcanonical phase diagram for the representative value $\kappa=-0.4>\kappa^{\rm AdS}_{m=0,\hat{\mu}^2}\simeq-0.4951294$, with scalar mass $\hat{\mu}^2=-15/16$. This configuration illustrates the regime $\kappa>\kappa^{\rm AdS}_{m=0,\hat{\mu}^2}$, for which global AdS$_3$ is linearly stable to $m=0$ double‑trace perturbations~\cite{Ishibashi:2004wx,Dias:2025uyk}, while BTZ black holes may nevertheless be unstable~\cite{Dias:2025uyk}. In this figure (and in all analogous phase‑diagram plots), the orange curve at $\Delta\hat{M}=0$ represents the one‑parameter family of extremal BTZ black holes with $\hat{M}=\hat{J}>0$. Regular BTZ black holes exist on and above this curve, \ie for $\Delta\hat{M}\geq0$.

The cyan curve marks the onset of the $m=0$ instability of BTZ, $\Delta\hat{M}(\hat{J})|^{m=0}_{\hbox{\tiny BTZ onset}}$, as computed in~\cite{Dias:2025uyk}. Since static BTZ is unstable to $m=0$ perturbations, this onset curve starts at finite $\hat{M}$ and $\hat{J}=0$ and extends to arbitrarily large $\hat{J}$, asymptotically approaching the extremal BTZ line $\Delta\hat{M}=0$ as $\hat{J}\to\infty$.

The petrol‑green vertical line at $\hat{J}=0$ corresponds to the static $m=0$ hairy black hole branch, whose properties were analysed in detail in subsection~\ref{sec:PhaseDiag-m0:BHsStatic} and illustrated in Figs.~\ref{fig:BS-m0} and~\ref{fig:m0_J0-BH}. These static hairy black holes have vanishing angular momentum but carry a finite $U(1)$ conserved charge $\hat{N}$. The branch originates at the onset of the static BTZ instability (cyan point at $\hat{J}=0$ and $\Delta\hat{M}=\Delta\hat{M}(\hat{J}=0)|^{m=0}_{\hbox{\tiny BTZ onset}}$) and terminates at the  singular $m=0$ static extremal hairy black hole; it is located at $\{\Delta\hat{M},\hat{J}\}\simeq\{-0.019,0\}$ (red diamond) or more precisely at \eqref{sBHm0J0:Thermo}. This singular solution was found using the strategy outlined in section~\ref{sec:NumericalSetup:SingBHm0J0}.

This very same singular $m=0$ static extremal hairy black hole also constitutes one of the endpoints of the regular static $m=0$ boson star branch (purple curve). This family is perturbatively connected to AdS$_3$, as it corresponds to the nonlinear back‑reaction of a linear normal mode of AdS$_3$ (see Fig.~\ref{fig:BS-m0}). Accordingly, the regular static boson star has a second endpoint at global AdS$_3$, represented by the black disk at $\{\Delta\hat{M},\hat{J}\}=\{-1,0\}$. Along this branch, the mass initially increases from $\Delta\hat{M}=-1$ up to a maximum $\Delta\hat{M}\simeq0.018$, after which it decreases and ultimately reaches the singular extremal hairy black hole endpoint at $\Delta\hat{M}\simeq-0.019$, \ie at \eqref{sBHm0J0:Thermo}. While this behaviour is not immediately apparent in the left panel of Fig.~\ref{fig:m0TotalPhaseDiag-k04}, it is clearly visible in Fig.~\ref{fig:BS-m0} of section~\ref{sec:PhaseDiag-m0:BStar}. Recall that these regular static boson stars have $\hat{J}=0$ but non‑vanishing $\hat{N}$, since $\hat{J}=m\,\hat{N}$; see the discussion around~\eqref{FirstLawBStar} and~\eqref{eqn:Omegaeom3}.

Finally, we note that although the detailed properties of the static boson stars and static hairy black holes are more clearly illustrated in Fig.~\ref{fig:BS-m0}, we include them in Fig.~\ref{fig:m0TotalPhaseDiag-k04} in order to present a complete microcanonical phase diagram encompassing all relevant classes of solutions.

We can now summarize the main features of Fig.~\ref{fig:m0TotalPhaseDiag-k04}. The blue dots represent $m=0$ hairy black hole solutions. They populate a two‑dimensional region in the $(\hat{J},\,\Delta\hat{M})$ plane that is bounded from above by the cyan BTZ instability onset curve $\Delta\hat{M}(\hat{J})|^{m=0}_{\hbox{\tiny BTZ onset}}$, from the left by the static line at $\hat{J}=0$ corresponding to the $m=0$ static hairy black holes (petrol‑green curve), and from below by a horizontal line with dark-blue triangles at $\Delta\hat{M}(\hat{J})\simeq-0.019$, more precisely at \eqref{sBHm0J:Thermo}. This lower boundary starts at the  singular $m=0$ static extremal hairy black hole at $\hat{J}=0$ (red diamond; see \eqref{sBHm0J0:Thermo}) and extends to arbitrarily large angular momentum (we analysed solutions up to $\hat{J}=5$) at constant $\Delta\hat{M}(\hat{J})$.

The physical origin of the upper and left boundaries is straightforward. The existence of $m=0$ hairy black holes relies on the fact that BTZ black holes can develop a scalar condensation instability. Beyond the linear regime, once back‑reaction is included, one therefore expects a new branch of hairy black holes bifurcating from the BTZ onset curve (where the scalar condensate vanishes) and extending into the region where BTZ is unstable. The left boundary at $\hat{J}=0$ is equally natural: static BTZ is already unstable to axisymmetric ($m=0$) perturbations, which do not introduce angular momentum. Consequently, $m=0$ hairy black holes exist for all $\hat{J}\geq0$, with a static branch at $\hat{J}=0$ and rotating solutions for $\hat{J}>0$.

The existence of the lower boundary is more subtle. To further understand it, it is useful to examine one‑parameter sub‑families of hairy black holes at fixed angular momentum, corresponding to vertical lines in the left panel of Fig.~\ref{fig:m0TotalPhaseDiag-k04}, and to study how their thermodynamic properties evolve as one moves from the BTZ onset curve downward. This analysis was carried out explicitly for the representative case $\hat{J}=3$ in Fig.~\ref{fig:m0-hBTZ-J3} of section~\ref{sec:PhaseDiag-m0:BHsRotating}. There, we found compelling evidence that as the lower boundary $\Delta\hat{M}(\hat{J})\to-0.019$ is approached, both the temperature and entropy of the hairy black hole tend to zero (with $\hat{\Omega}_{H}\to1$), while the Kretschmann scalar at the horizon diverges. In other words, the lower boundary of Fig.~\ref{fig:m0TotalPhaseDiag-k04} corresponds to a singular, zero‑horizon‑radius and zero-temperature limit of $m=0$ hairy black holes, which explains why the solution terminates there.
In fact, this dark-blue triangle lower boundary describes the singular $m=0$ rotating extremal hairy black hole explicitly constructed in section~\ref{sec:NumericalSetup:singBHm0J} and with thermodynamic observables  given by \eqref{sBHm0J:Thermo}.

Although we are not aware of an analytic argument that enforces an exactly constant mass difference $\Delta \hat M$ along this lower boundary, it is nevertheless noteworthy that \eqref{sBHm0J:Thermo} saturates the positivity-of-energy theorem~\eqref{GlobalMin}, derived in Ref.~\cite{Faulkner:2010fh} and Appendix~\ref{secA:superpotentials}.

As $\hat{J}\to0$, this singular zero‑radius limit reduces precisely to the singular $m=0$ static extremal hairy black hole star (red diamond; see \eqref{sBHm0J0:Thermo} and section~\ref{sec:NumericalSetup:SingBHm0J0}). This static singular solution therefore marks the corner where the left and lower boundaries of the $m=0$ hairy black hole region meet. Thus, the {\it rotating}  extension of the singular $m=0$ static extremal hairy black hole of section~\ref{sec:NumericalSetup:SingBHm0J0} is undoubtedly the singular $m=0$ extremal hairy black hole (dark-blue triangles) of section~\ref{sec:NumericalSetup:singBHm0J}.

The orange curve at $\Delta\hat{M}=0$ corresponds to extremal BTZ black holes, which saturate the BPS bound $\hat{M}\geq|\hat{J}|$, while non‑extremal BTZ solutions exist for $\hat{M}\geq\hat{J}$ (equivalently $\Delta\hat{M}\geq0$). This property is specific to $d=3$; in higher‑dimensional AdS gravity, rotating vacuum black holes reach extremality strictly above the BPS bound. By contrast, Fig.~\ref{fig:m0TotalPhaseDiag-k04} clearly shows that $m=0$ hairy black holes can exist below the BPS line, \ie with negative $\Delta\hat{M}$. This is possible because double‑trace boundary conditions break supersymmetry, allowing violations of the BPS bound $M\geq J/L$ (see also footnote~\ref{footBPS}).

Let us now consider the right panel of Fig.~\ref{fig:m0TotalPhaseDiag-k04}, which displays the entropy as a function of the microcanonical variables $(\Delta\hat{M},\hat{J})$. The BTZ family is represented by the grey surface, which exists for $\Delta\hat{M}\geq0$ and emanates from the orange extremal line. The hairy black hole family is shown by the blue surface, which merges continuously with the BTZ surface along the cyan $m=0$ onset curve via a second‑order phase transition. The primary conclusion is that whenever the two families coexist at the same $(\Delta\hat{M},\hat{J})$, the $m=0$ hairy black holes always have higher entropy than BTZ. Consequently, they dominate the microcanonical ensemble wherever they exist.

In addition, the hairy black hole surface terminates at a singular zero‑entropy limit, corresponding to the bottom dark-blue triangle boundary discussed above. This is the singular $m=0$ extremal hairy black hole line with thermodynamics \eqref{sBHm0J:Thermo}. Although this behaviour is not very sharp visually in the right panel of Fig.~\ref{fig:m0TotalPhaseDiag-k04}, it follows unambiguously from the analysis of constant‑$\hat{J}$ sub‑families such as the $\hat{J}=3$ example shown in Fig.~\ref{fig:m0-hBTZ-J3}. In Fig.~\ref{fig:m0TotalPhaseDiag-k04}, we have in fact included this $\hat{J}=3$ branch, interpolating it down to $\hat{S}_{H}=0$ (\ie the singular extremal hairy black hole \eqref{sBHm0J:Thermo} of section~\ref{sec:NumericalSetup:singBHm0J}), to illustrate how the hairy black hole surface plunges almost vertically toward zero entropy as $\Delta\hat{M}\to-0.019^{+}$. This dramatic drop occurs within an extremely small interval of $\Delta\hat{M}$, corresponding to a very large entropy gradient. As a result, it is technically challenging to generate numerically sufficient constant‑$\hat{J}$ sub‑families to fully resolve this feature, which explains the apparent ``gap’’ visible in the right panel of Fig.~\ref{fig:m0TotalPhaseDiag-k04}.

\begin{figure}
    \centering
     \includegraphics[width=0.47\linewidth]{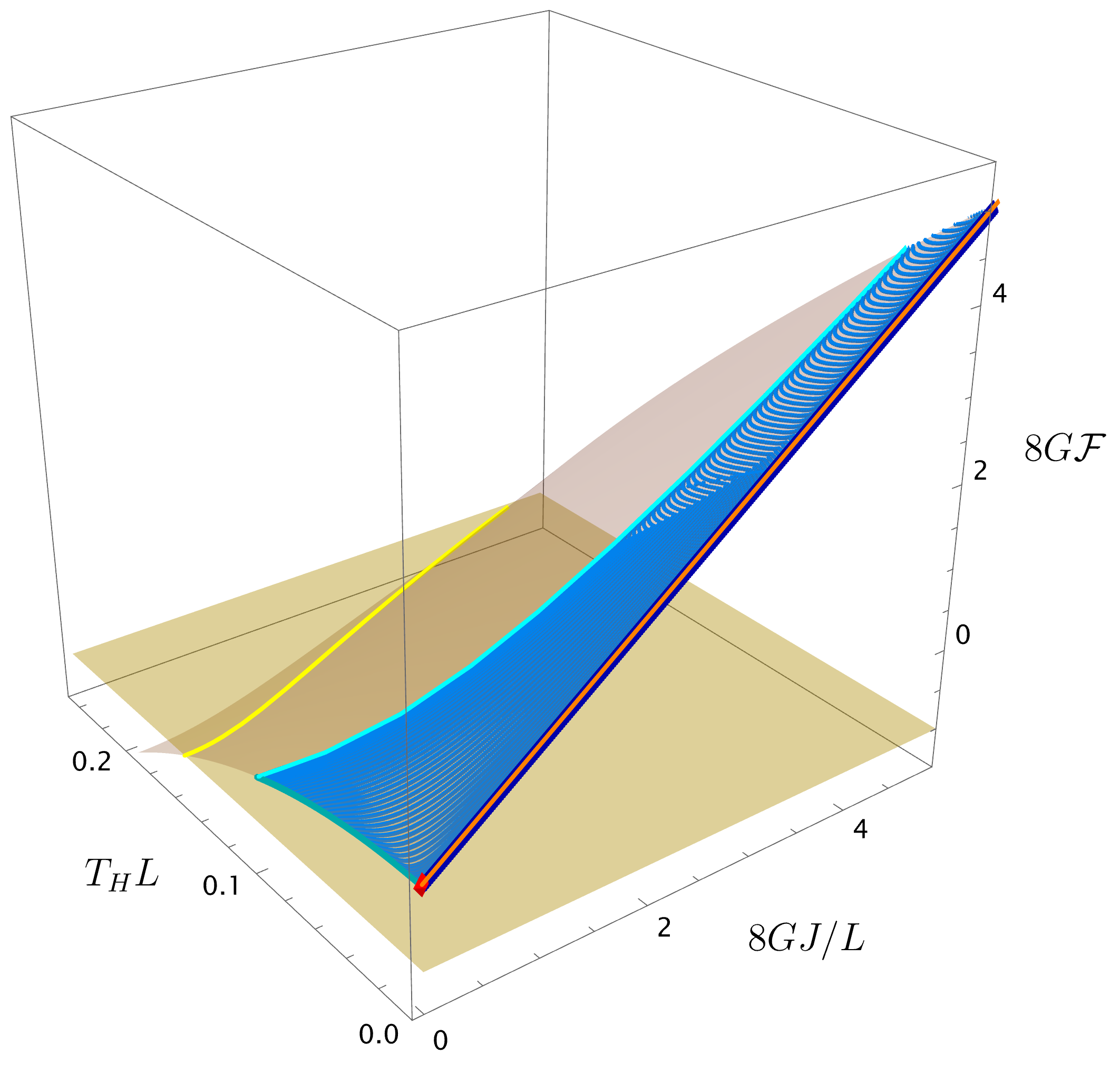} \hspace{0.5cm}
    \includegraphics[width=0.47\linewidth]{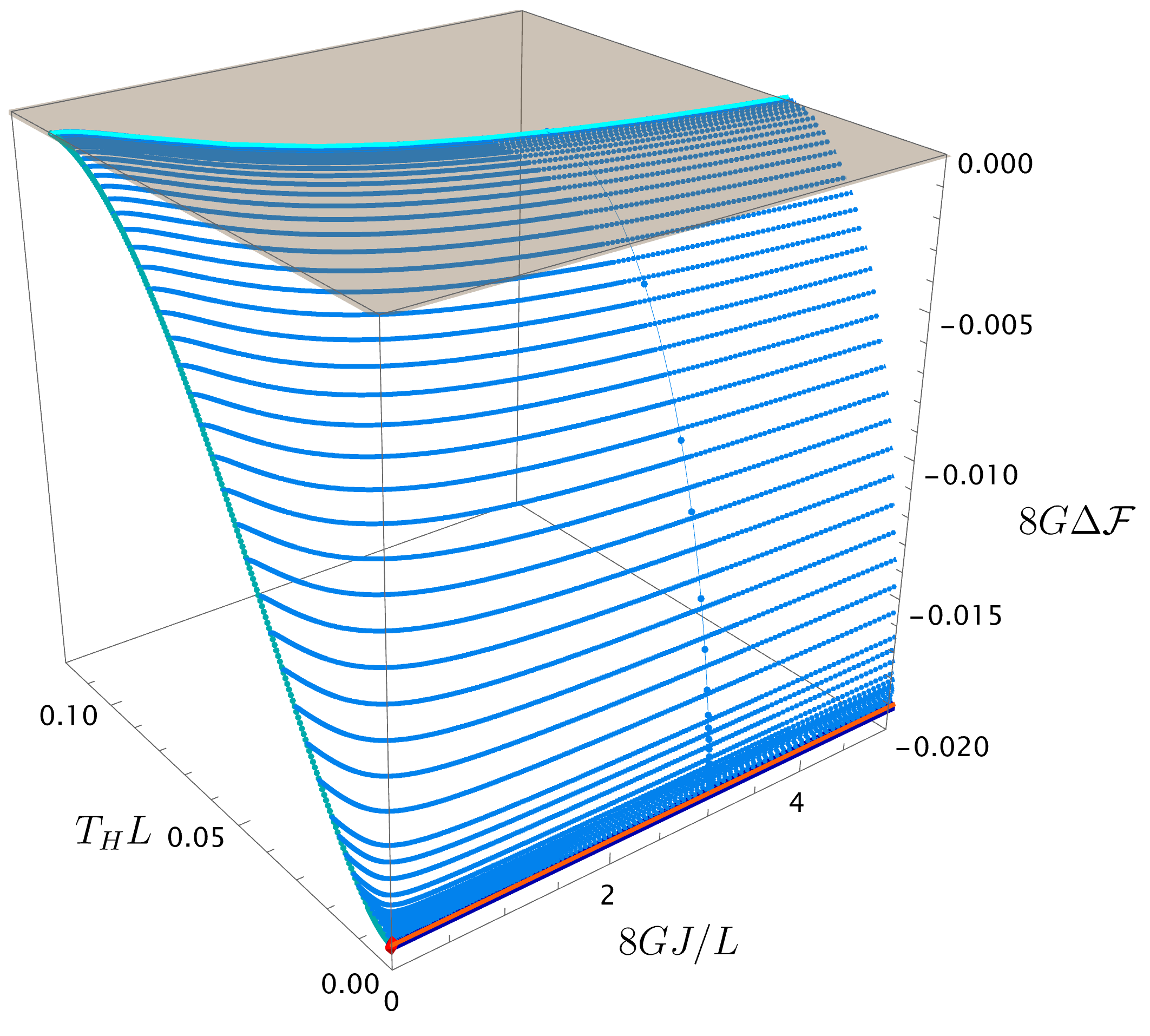}
   \includegraphics[width=0.47\linewidth]{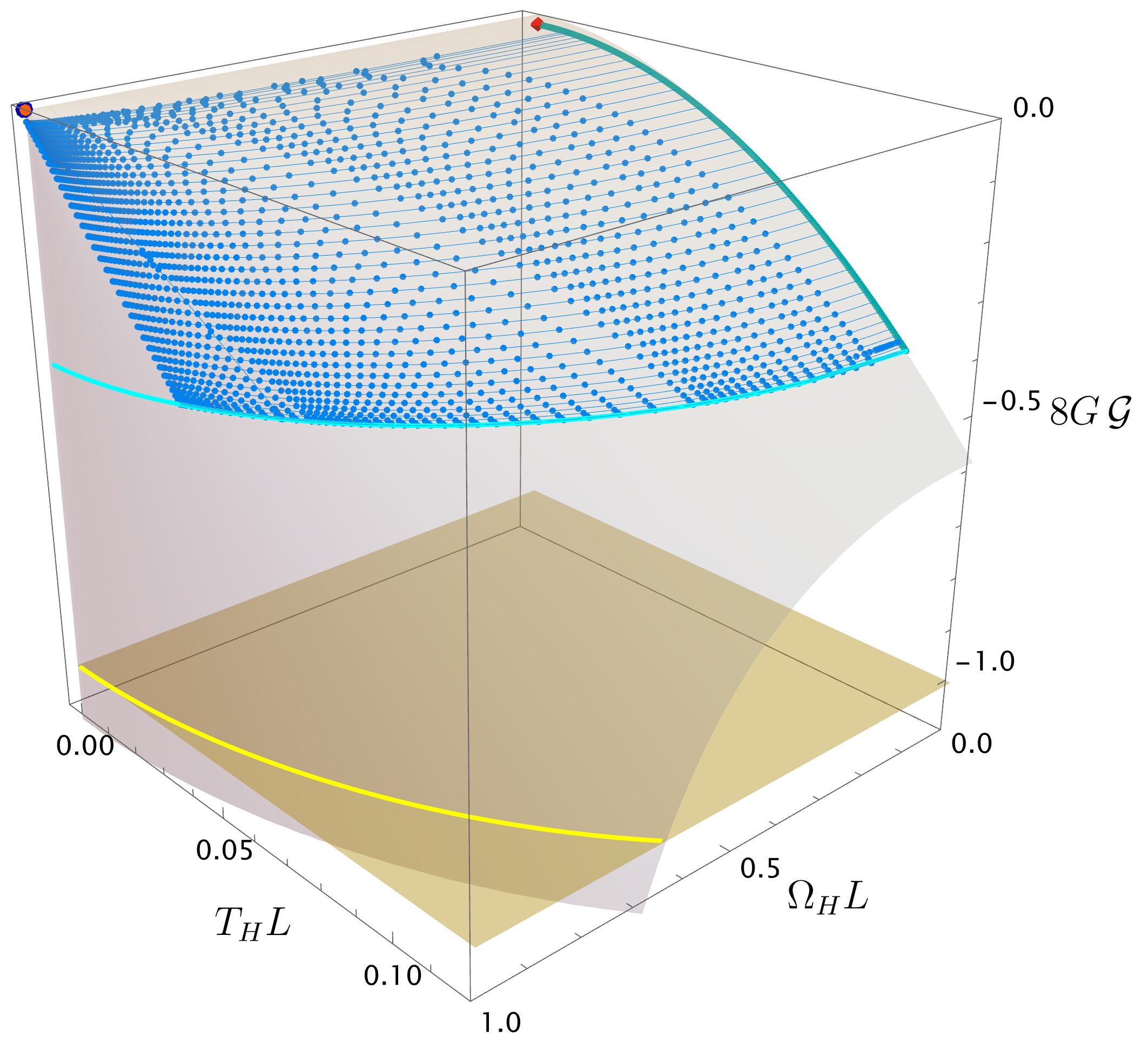} \hspace{0.5cm}
     \includegraphics[width=0.47\linewidth]{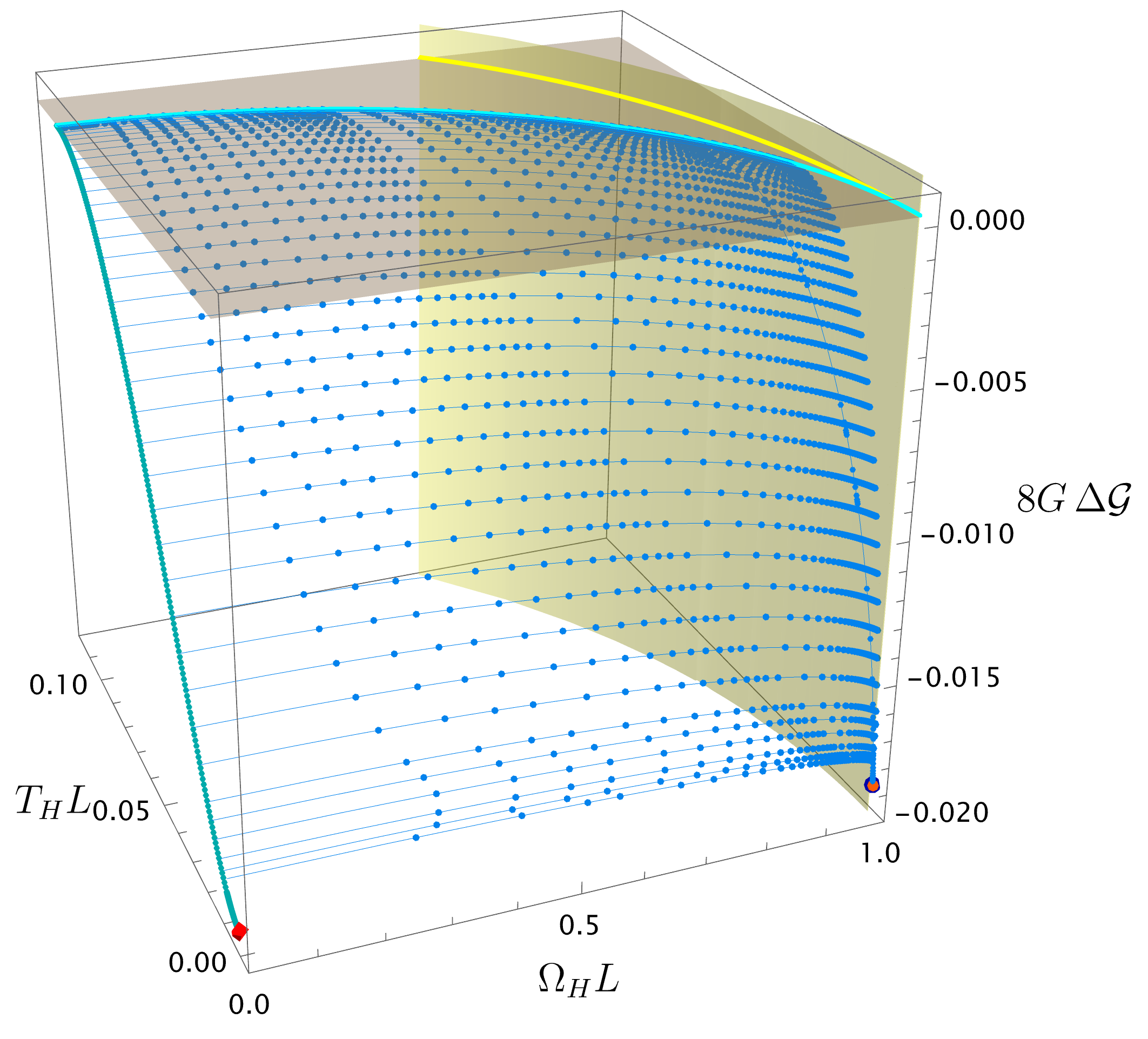}
    \caption{{\bf Canonical (top panels)} and {\bf grand-canonical (bottom panels)} phase diagrams of solutions with $\bm{m = 0}$, $\mu^2L^2 = -15/16$ and $\bm{\kappa = -4/10>\kappa^{\rm AdS}_{m=0, \hat{\mu}^2=-15/16}\simeq -0.4951294}$ (for which AdS$_3$ is linear mode stable but BTZ can be unstable). When they co-exist, $m=0$ hairy black holes (blue dot surface) always have lower Helmholtz and Gibbs free energy than BTZ (grey surface), \ie in the right panels, the blue dot hairy surface is clearly seen to be always below the grey BTZ surface.
    The yellow line, where the dark-yellow AdS$_3$ surface intersects the grey BTZ surface, represents the Hawking-Page  temperature $\hat{T}_{\hbox{\tiny HP}}(\hat{J})$ (top panel), $\hat{T}_{\hbox{\tiny HP}}(\hat{\Omega}_H)$ (bottom panel). For temperatures smaller (larger) than the Hawking-Page one, it is thermal AdS$_3$ (BTZ) that dominates the canonical and grand-canonical ensembles. On the top panels, note that the singular $m = 0$ extremal hairy black hole (dark-blue thick) curve  is very close to the extremal BTZ (orange thin) curve. In the bottom panels, note that all  singular $m = 0$ extremal hairy black holes meet at a point ($\hat{\Omega}_H = 1$, $\hat{T}_H = 0$), which coincides with the orange diamond of extremal BTZ black holes.
    }
\label{fig:m0_kappa_m4o10_DeltaF_DeltaG}
\end{figure}

So far, our discussion has focused on the microcanonical ensemble. We now turn to the canonical and grand‑canonical ensembles, whose phase diagrams are shown in the top and bottom panels, respectively, of Fig.~\ref{fig:m0_kappa_m4o10_DeltaF_DeltaG}. Throughout this discussion we continue to work with $\mu^{2}L^{2}=-15/16$ and $\kappa=-4/10$.

Recall that in the canonical ensemble the temperature $\hat{T}_{H}$ and angular momentum $\hat{J}$ are held fixed. The relevant thermodynamic potential is the Helmholtz free energy,
\begin{equation}
\hat{\mathcal{F}}=\hat{M}-\hat{T}_{H}\hat{S}_{H},
\end{equation}
and the preferred phase is the one that minimizes $\hat{\mathcal{F}}$ at fixed $(\hat{T}_{H},\hat{J})$. Accordingly, the top‑left panel of Fig.~\ref{fig:m0_kappa_m4o10_DeltaF_DeltaG} displays the canonical phase diagram in the $(\hat{T}_{H},\hat{J},\hat{\mathcal{F}})$ space. The dark‑yellow plane at $\hat{\mathcal{F}}=-1$ represents thermal AdS$_3$ (which exists for arbitrary temperature and angular momentum), the grey surface corresponds to BTZ black holes, and the surface formed by blue points describes the $m=0$ hairy black holes.

The latter exist only at sufficiently low temperatures and for arbitrary angular momentum, namely for temperatures below the cyan curve $\hat{T}(\hat{J})|^{m=0}_{\hbox{\tiny BTZ onset}}$, where the $m=0$ hairy black hole branch bifurcates from the BTZ family. Wherever they coexist, hairy $m=0$ black holes have lower Helmholtz free energy than BTZ. This is shown explicitly in the top‑right panel, where we plot the free‑energy difference
\begin{equation}
\Delta\hat{\mathcal{F}}=\hat{\mathcal{F}}-\hat{\mathcal{F}}^{\hbox{\tiny BTZ}},
\end{equation}
computed at the same $(\hat{T}_{H},\hat{J})$ (thermal AdS$_3$ is not displayed in this plot).

Nevertheless, whenever $m=0$ hairy black holes exist, it is thermal AdS$_3$ that dominates the canonical ensemble, since it has the lowest free energy among all available solutions. At higher temperatures, where the hairy black holes cease to exist, the system undergoes a Hawking–Page transition at the temperature $\hat{T}_{\hbox{\tiny HP}}(\hat{J})$, shown as the yellow curve in the top‑left panel. For $\hat{T}(\hat{J})>\hat{T}_{\hbox{\tiny HP}}(\hat{J})$, the BTZ black hole has lower Helmholtz free energy than thermal AdS$_3$ and therefore dominates the canonical ensemble.

We now turn to the grand‑canonical ensemble, shown in the bottom panels of Fig.~\ref{fig:m0_kappa_m4o10_DeltaF_DeltaG}, where the system is studied at fixed temperature and fixed angular velocity. The relevant thermodynamic potential is the Gibbs free energy,
\begin{equation}
\hat{\mathcal{G}}
   = \hat{M}
     - \hat{T}_{H}\hat{S}_{H}
     - \hat{\Omega}_{H}\hat{J},
\end{equation}
and the dominant thermal phase is the one that minimizes $\hat{\mathcal{G}}$ at fixed $(\hat{T}_{H},\hat{\Omega}_{H})$. As in the canonical ensemble, the dark‑yellow and grey surfaces describe thermal AdS$_3$ and BTZ black holes, respectively. The yellow curve $\hat{T}_{\hbox{\tiny HP}}(\hat{\Omega}_{H})$ denotes the Hawking–Page transition in the grand‑canonical ensemble: for $\hat{T}(\hat{\Omega}_{H})<\hat{T}_{\hbox{\tiny HP}}(\hat{\Omega}_{H})$ thermal AdS$_3$ dominates, whereas for higher temperatures BTZ black holes are thermodynamically preferred.

\begin{figure}
\vskip -0.8cm
    \centering
    \includegraphics[width=0.4\linewidth]{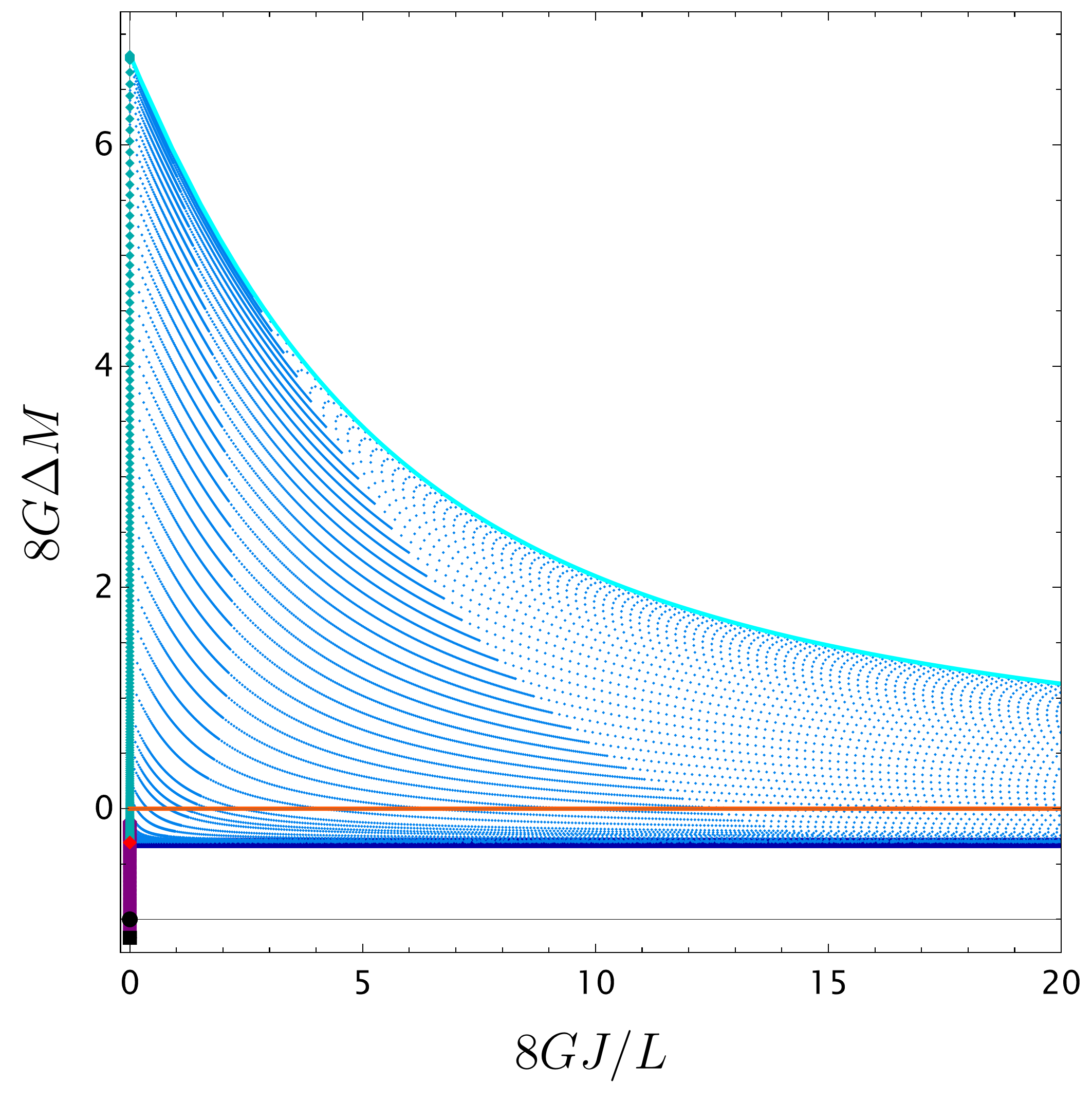}
    \includegraphics[width=0.5\linewidth]{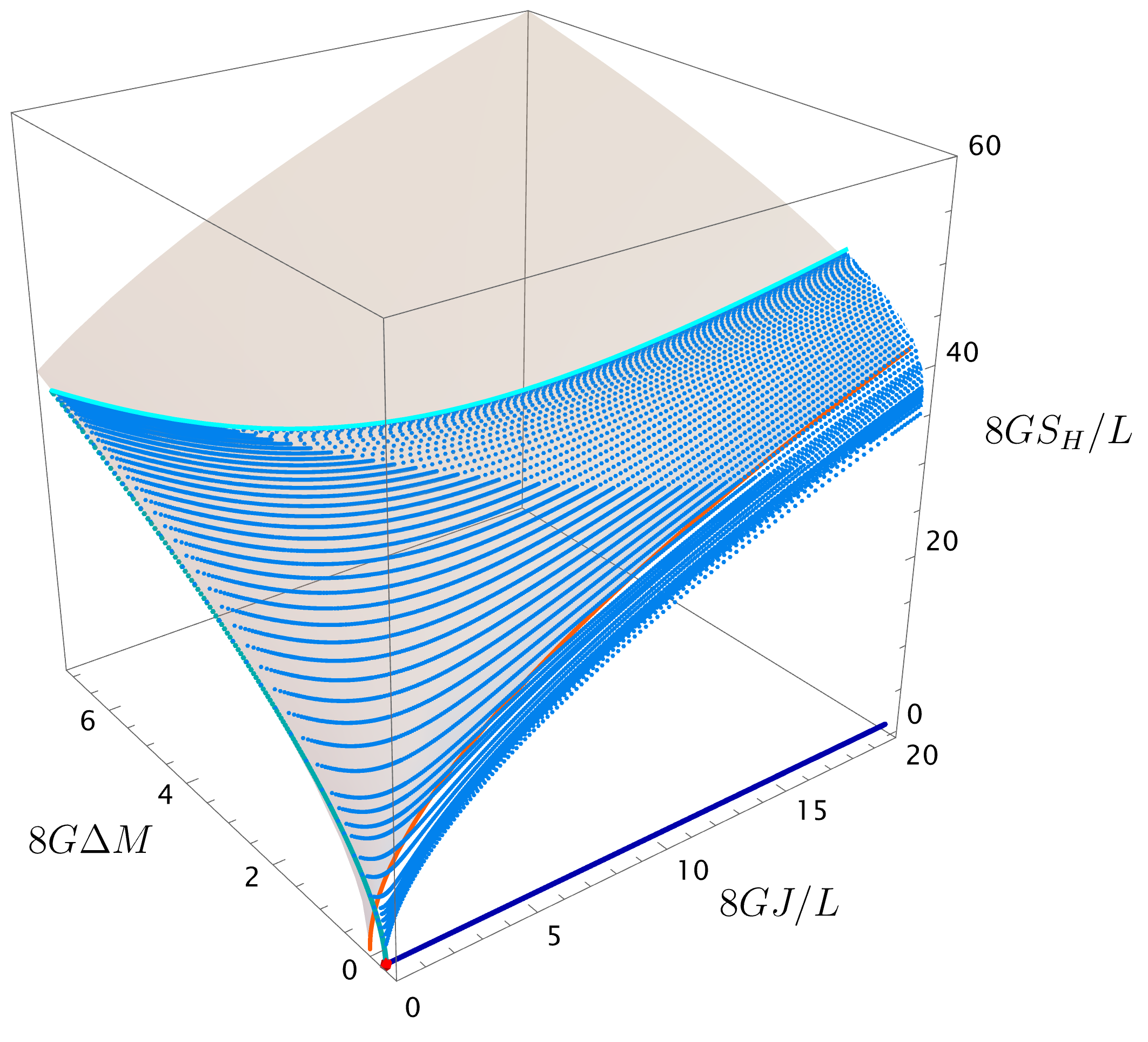}
    \includegraphics[width=0.4\linewidth]{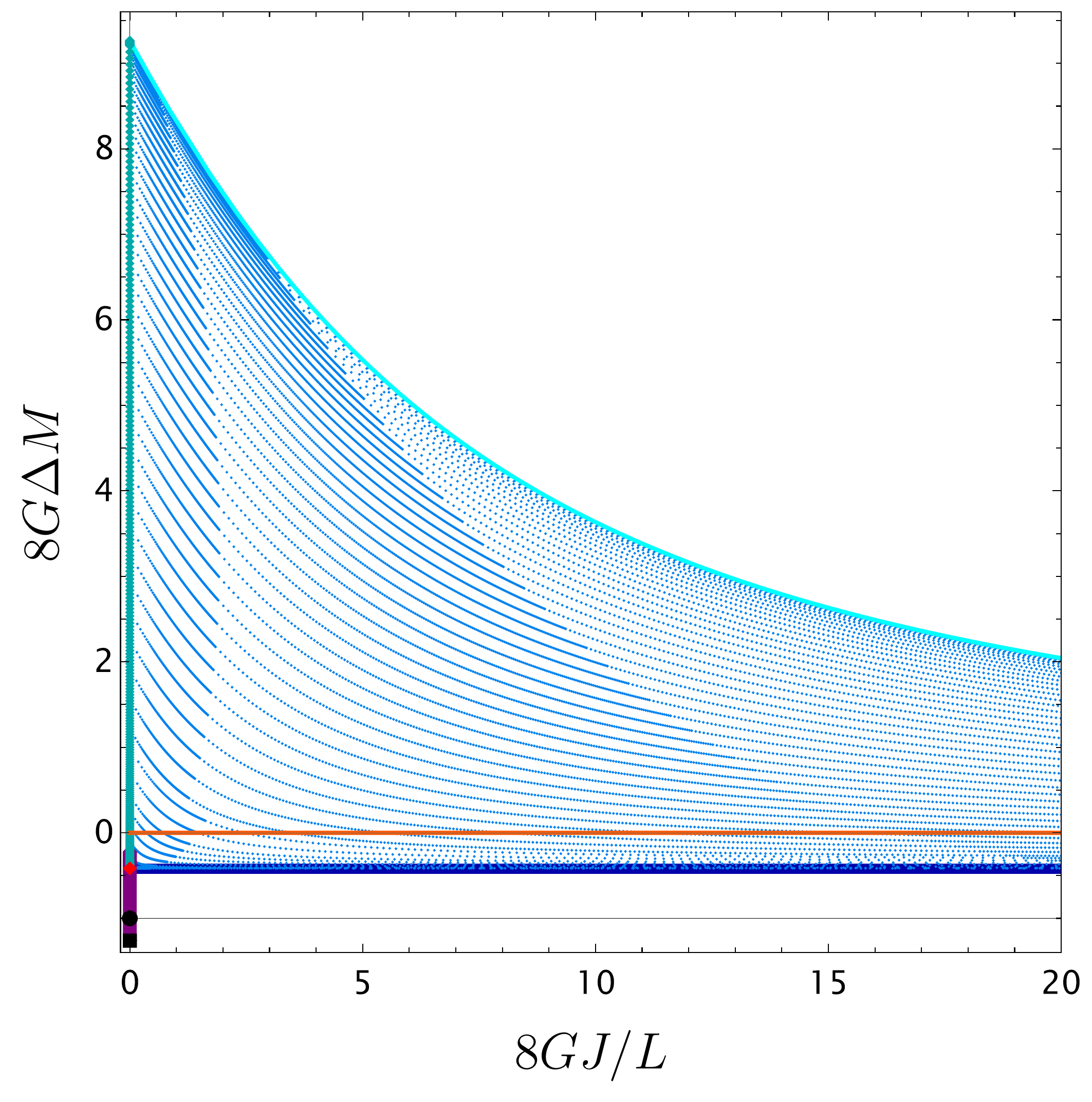}
    \includegraphics[width=0.5\linewidth]{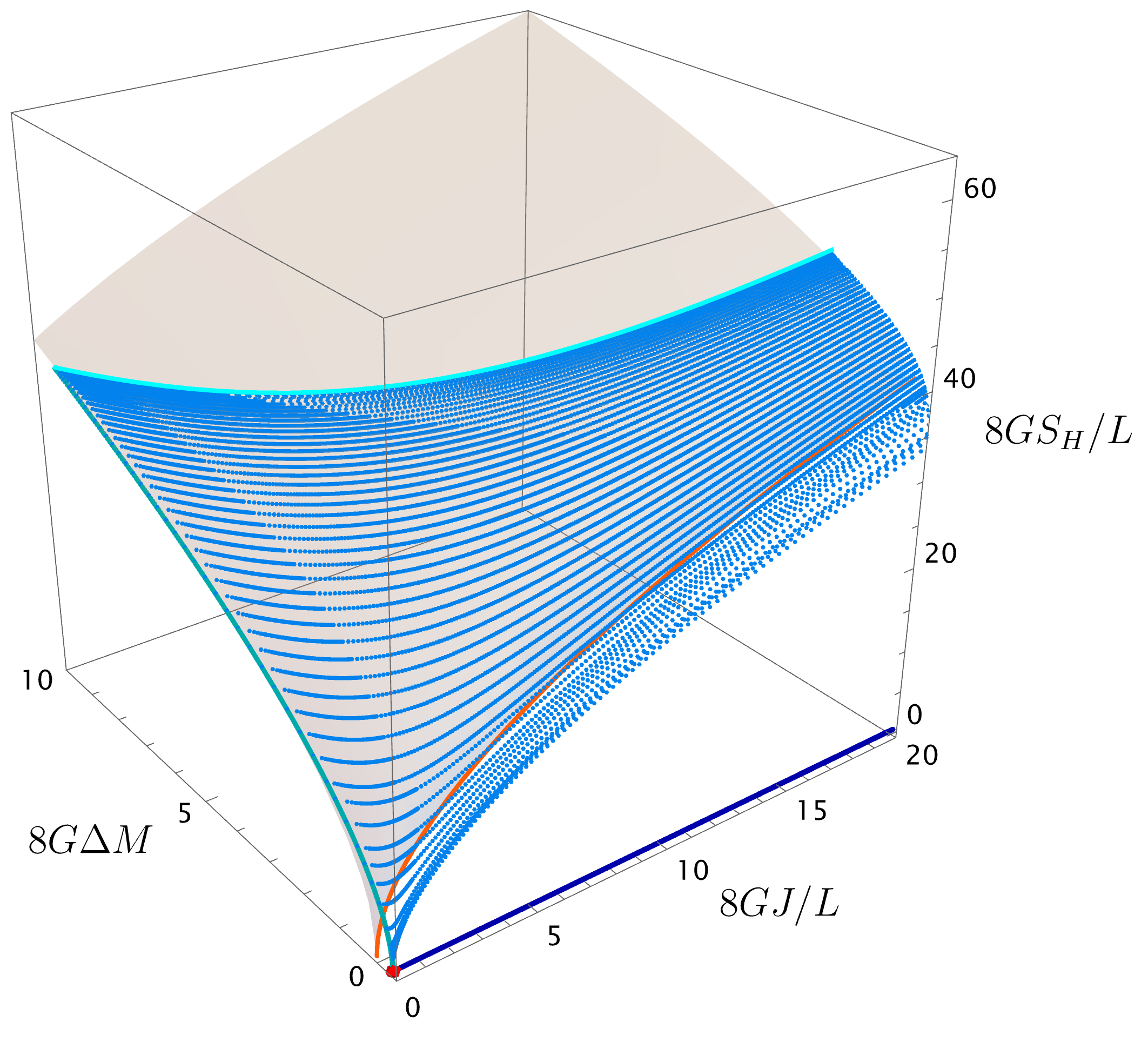}
    \includegraphics[width=0.4\linewidth]{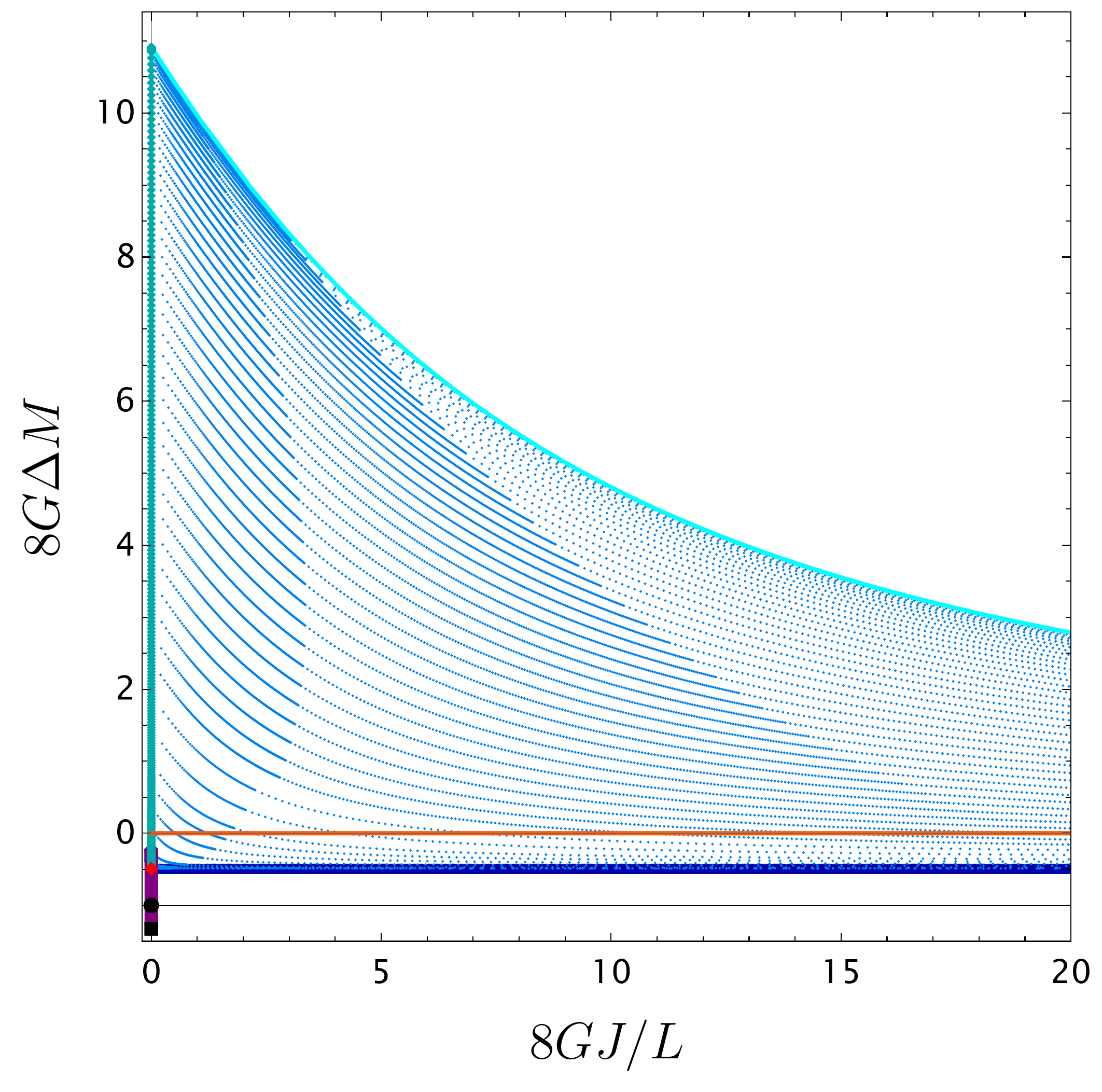}
    \includegraphics[width=0.5\linewidth]{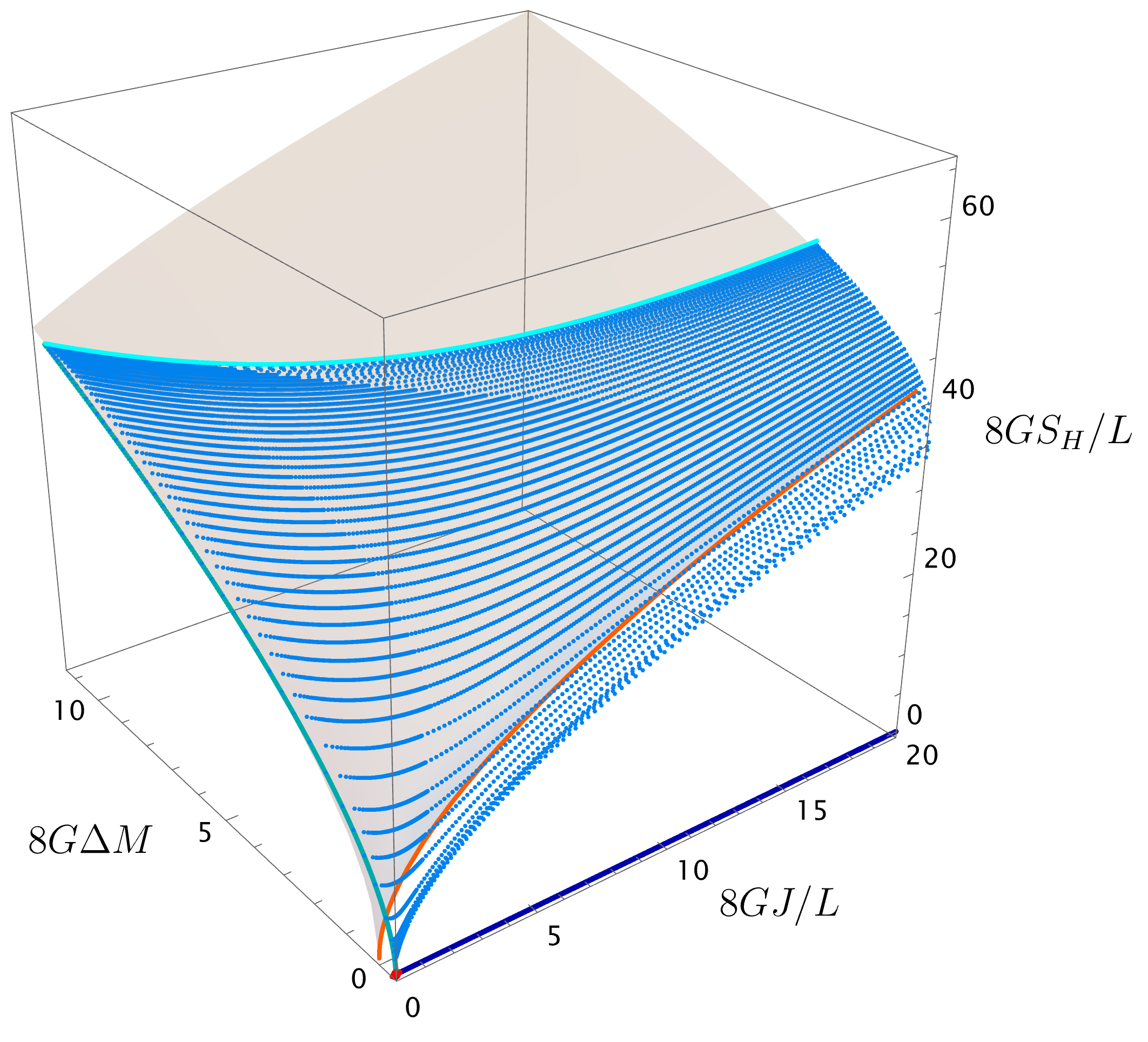}
    \caption{\footnotesize{\bf Microcanonical phase diagram} of solutions with $\bm{m = 0}$, $\mu^2L^2 = -15/16$ and $\bm{\kappa = -8/10}$ {\bf (top panels)}, $\bm{\kappa = -0.8639}$ {\bf (middle panels)} and $\bm{\kappa = -9/10}$ {\bf (bottom panels)}.  In the three cases we have $\bm{\kappa<\kappa^{\rm AdS}_{m=0, \hat{\mu}^2}\simeq -0.4951294}$ (thus, AdS$_3$ is unstable and BTZ can be unstable or linear mode stable). When they co-exist, $m=0$ hairy black holes always dominate the microcanonical ensemble (\ie in the right panel, the blue dot hairy surface is always above the grey BTZ surface). The singular $m=0$ extremal hairy black hole (dark blue) has $\Delta \hat{M}(\hat{J})$ given by  $-0.305934$ (top), $-0.416027$ (middle) and $-0.490047$ (bottom) and it reduces to the singular $m=0$ static solution when $J\to 0$ (red point).
    These values saturate the positivity-of-energy minimum energy bound \eqref{GlobalMin}-\eqref{Super:Emin} of Appendix~\ref{secA:superpotentials} \cite{Faulkner:2010fh}.
 }
    \label{fig:m0TotalPhaseDiag-3k}
\end{figure}

As is most clearly seen in the bottom‑right panel of Fig.~\ref{fig:m0_kappa_m4o10_DeltaF_DeltaG}, where we plot the Gibbs free‑energy difference between hairy and BTZ black holes,
\begin{equation}
\Delta\hat{\mathcal{G}}
   = \hat{\mathcal{G}}
     - \hat{\mathcal{G}}^{\hbox{\tiny BTZ}},
\end{equation}
evaluated at the same $(\hat{T}_{H},\hat{\Omega}_{H})$, the $m=0$ hairy black holes always have lower $\hat{\mathcal{G}}$ than BTZ whenever the two coexist. Nevertheless, in this region of the phase diagram thermal AdS$_3$ still has the smallest Gibbs free energy and therefore dominates the grand‑canonical ensemble.

Up to this point (Figs.~\ref{fig:m0TotalPhaseDiag-k04} and~\ref{fig:m0_kappa_m4o10_DeltaF_DeltaG}), we have focused on a theory with $\kappa=-0.4$ and $\hat{\mu}^{2}=-15/16$. For any other value of the double‑trace parameter satisfying $\kappa>\kappa^{\rm AdS}_{m=0,\hat{\mu}^{2}=-15/16}\simeq-0.4951294$, for which AdS$_3$ is linearly stable, the microcanonical, canonical, and grand‑canonical phase diagrams are qualitatively similar to those shown in these figures.

One may then ask whether these phase diagrams change qualitatively when the double‑trace coupling is such that $\kappa<\kappa^{\rm AdS}_{m=0,\hat{\mu}^{2}}$, where AdS$_3$ becomes unstable to the $m=0$ Ishibashi–Wald instability~\cite{Ishibashi:2004wx,Dias:2025uyk} (while BTZ may be either unstable or linearly stable, depending on parameters~\cite{Dias:2025uyk}). The answer is negative for the microcanonical ensemble, but affirmative for the canonical and grand‑canonical ensembles.

Indeed, the qualitative structure of the microcanonical phase diagram remains similar to that shown in Fig.~\ref{fig:m0TotalPhaseDiag-k04}, independently of the stability properties of AdS$_3$. This is illustrated in Fig.~\ref{fig:m0TotalPhaseDiag-3k} for $\kappa=-0.8$ (top panels), $\kappa=-0.8639$ (middle panels), and $\kappa=-0.9$ (bottom panels), all of which satisfy $\kappa<\kappa^{\rm AdS}_{m=0,\hat{\mu}^{2}=-15/16}\simeq-0.4951294$. We present these three values explicitly because they will play a role in a later section. These phase diagrams are qualitatively similar to Fig.~\ref{fig:m0TotalPhaseDiag-k04} and require no further discussion.

It is nevertheless important to recall that, because AdS$_3$ at $\{\Delta\hat{M},\hat{J}\}=\{-1,0\}$ is now unstable to the $m=0$ Ishibashi–Wald instability~\cite{Ishibashi:2004wx,Dias:2025uyk}, the theory possesses a new ground state, \ie a new minimum‑energy solution. This ground state is a regular $m=0$, $\omega=0$ boson star with $\Delta\hat{M}(\kappa)<-1$, which was analysed in detail in subsection~\ref{sec:PhaseDiag-m0:IshWald} and Fig.~\ref{fig:m0BSevoK}, and is also anticipated by the superpotential analysis of Ref.~\cite{Faulkner:2010fh} (see Appendix~\ref{secA:superpotentials}). We do not revisit this analysis here and instead refer the reader to the aforementioned sections.

The fact that the $m=0$ zero‑frequency boson star is now the true ground state of the theory has profound implications for the canonical and grand‑canonical phase diagrams. A detailed discussion of these consequences is postponed to the next subsection~\ref{sec:PhaseDiag-Total-m1}, where we analyse Fig.~\ref{m0_m1_kappa_m8o10_DeltaF_DeltaG}.

\newpage
\subsection{Full phase diagram of asymptotically \texorpdfstring{AdS$_3$}{AdS3} stationary solutions with \texorpdfstring{$m=1$}{m=1}}\label{sec:PhaseDiag-Total-m1}

\begin{figure}
    \centering
    \vskip -1.3cm
    \includegraphics[width=0.41\linewidth]{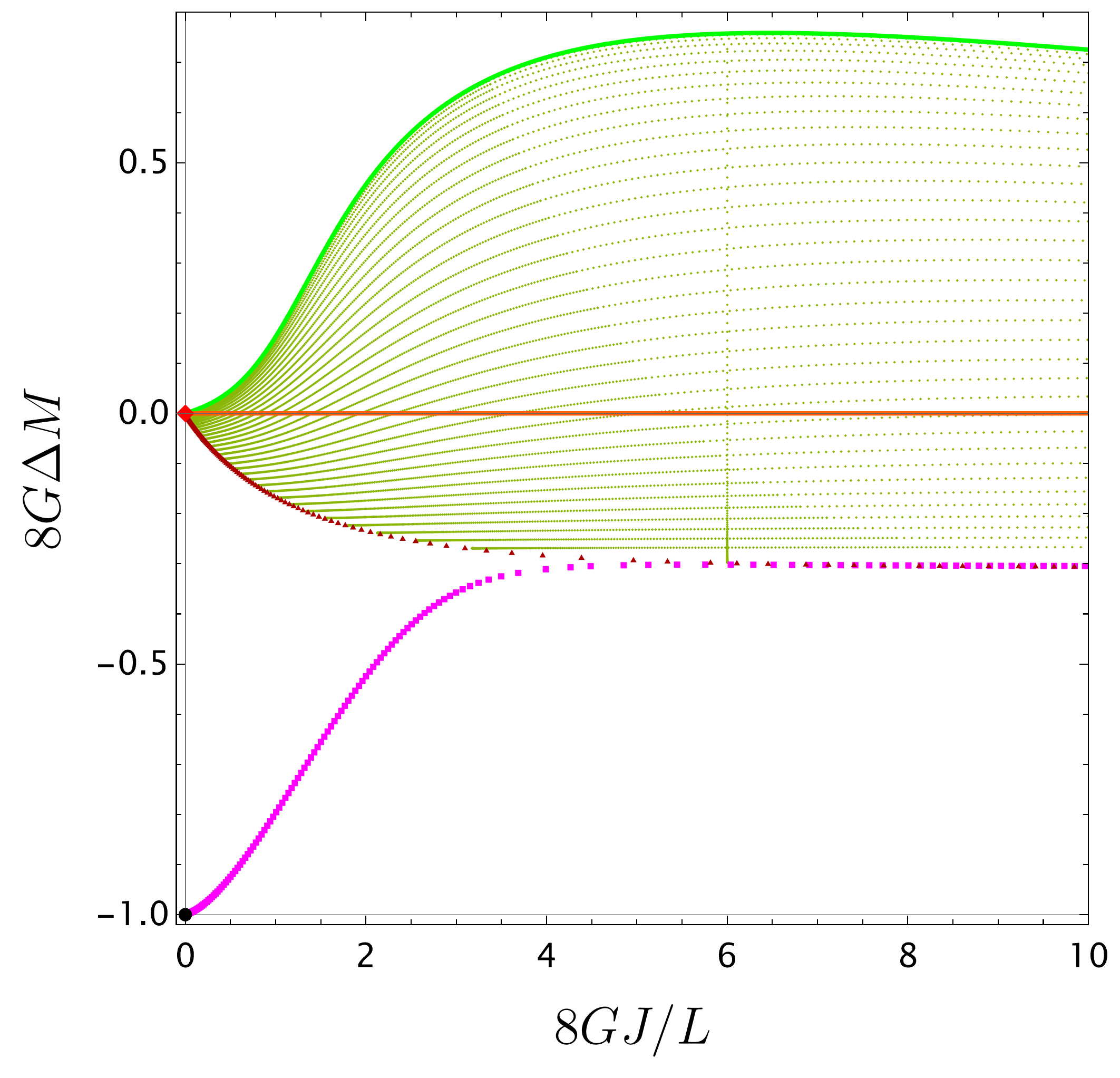}
    \includegraphics[width=0.5\linewidth]{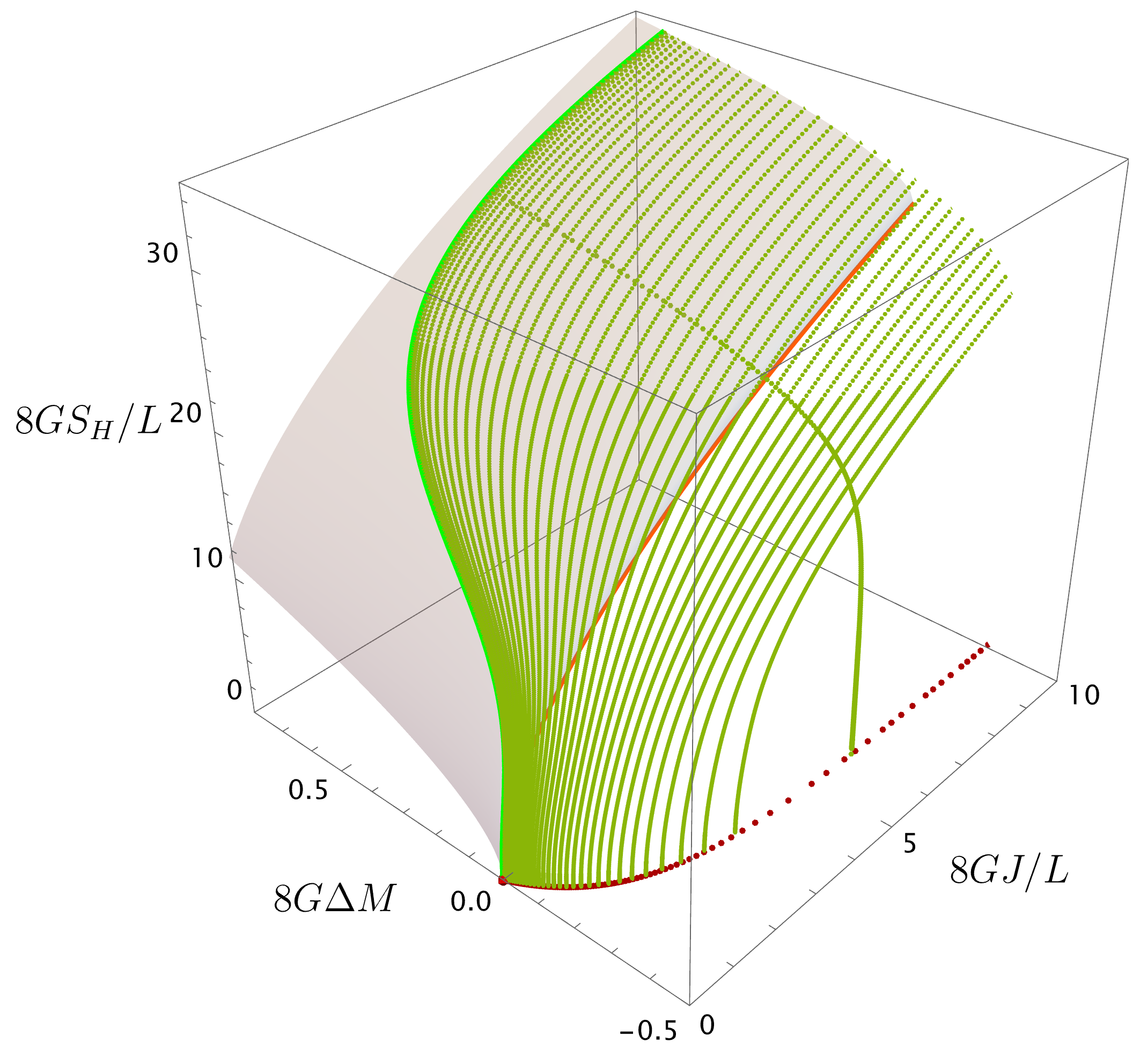}
     \includegraphics[width=0.41\linewidth]{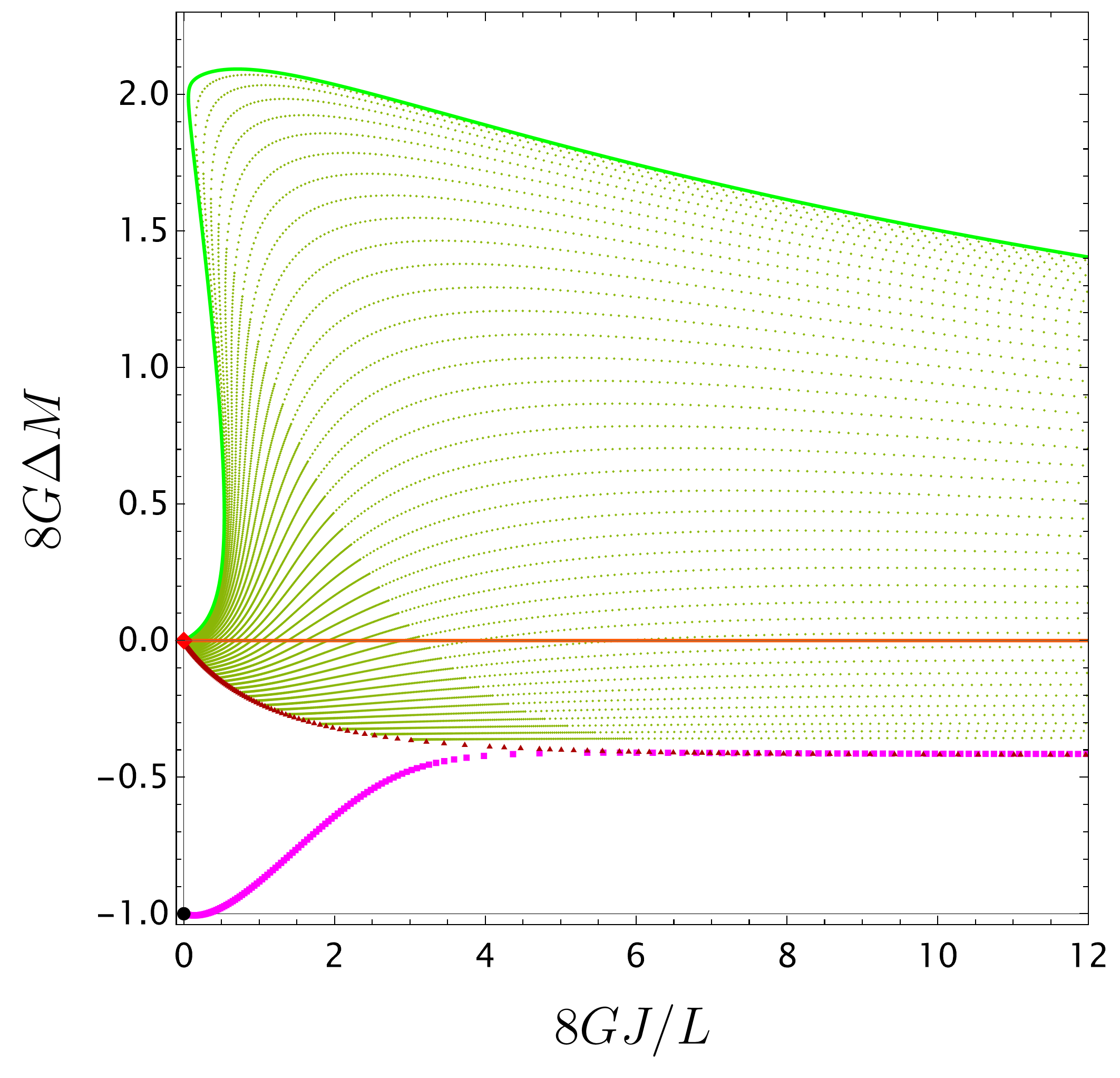}
     \includegraphics[width=0.5\linewidth]{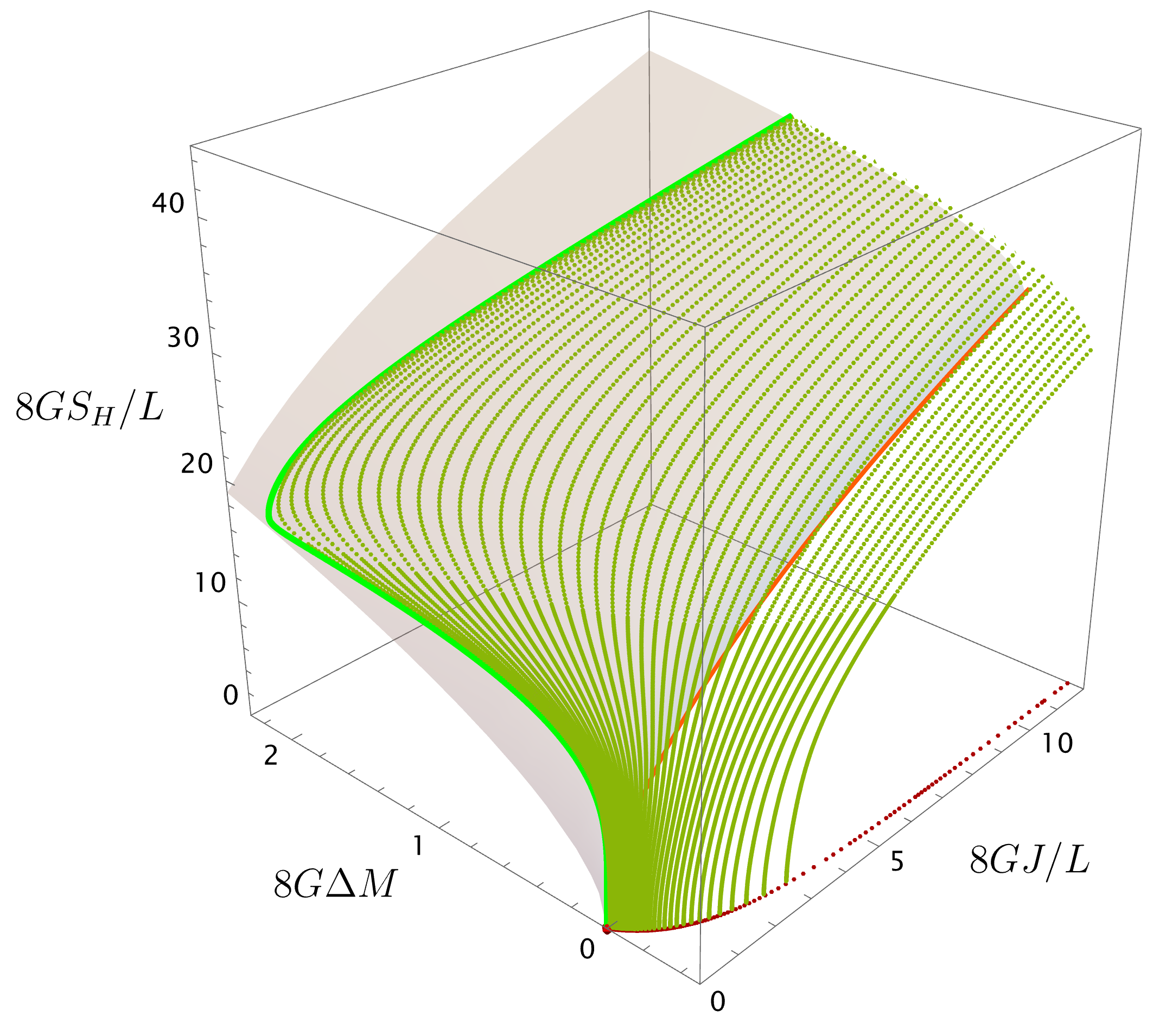}
     \includegraphics[width=0.41\linewidth]{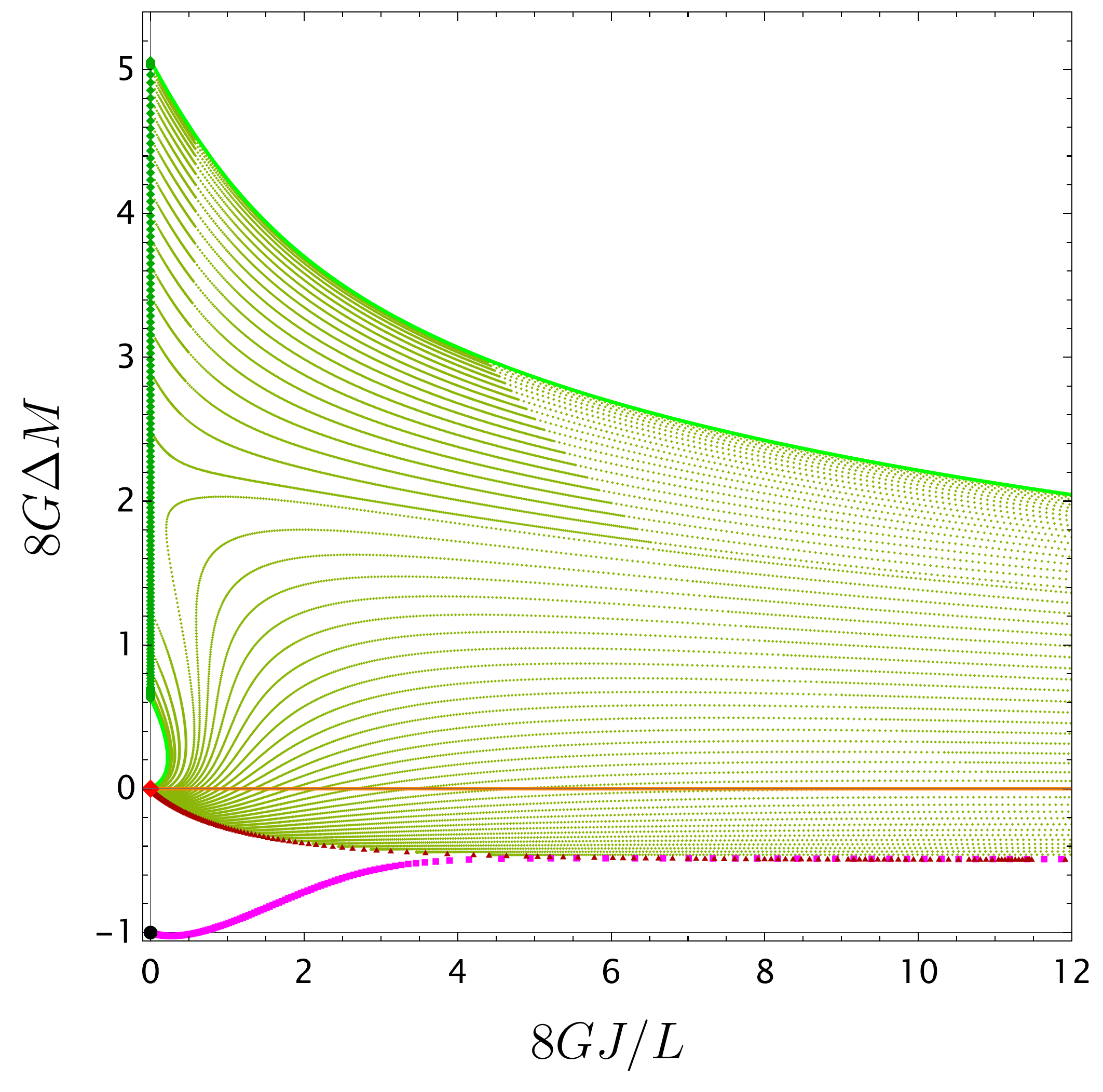}
     \includegraphics[width=0.5\linewidth]{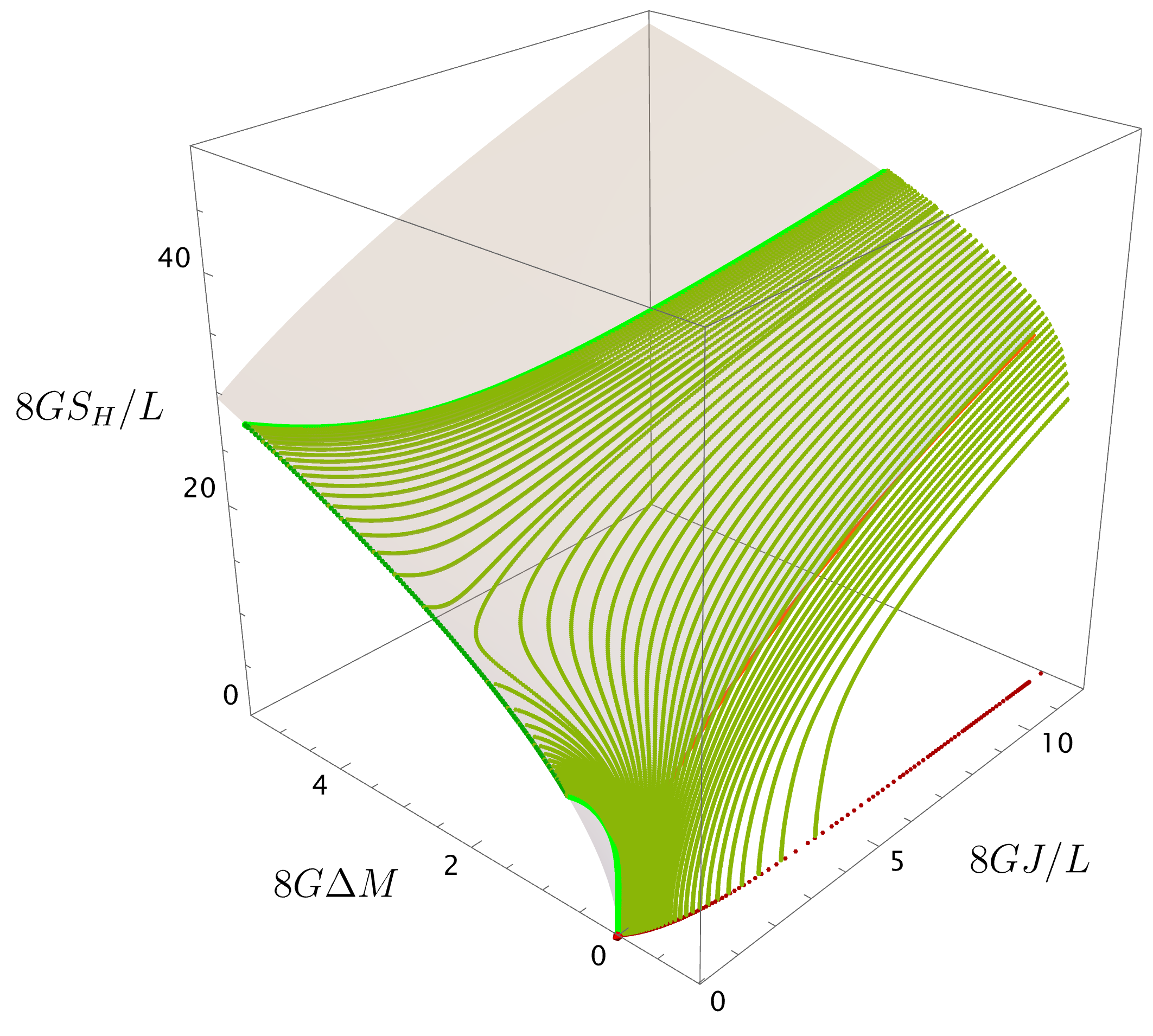}
    \caption{\footnotesize {\bf Microcanonical phase diagram} of solutions with $\bm{m = 1}$, $\mu^2L^2 = -15/16$ and $\bm{\kappa = -8/10}$ {\bf (top panels)}, $\bm{\kappa = -0.8639}$ {\bf (middle panels)} and $\bm{\kappa = -9/10}$ {\bf (bottom panels)}.  In the three cases we have $\bm{\kappa<\kappa^{\rm AdS}_{m=1, \hat{\mu}^2}\simeq -1.009837}$ (thus, AdS$_3$ is linear mode stable to $m=1$ and BTZ can be unstable or linear mode stable). When they co-exist, $m=1$ hairy black holes always dominate the microcanonical ensemble over BTZ (\ie in the right panel, the green dot hairy surface is always above the grey BTZ surface). We also show: extremal BTZ (orange), the BTZ $m=1$ onset (bright-green), the singular $m =1$ extremal hairy black holes (dark red triangles/circles), which correspond to the $R_+ \to 0$ limit of $m=1$ hairy black holes, $m=1$ regular boson star (magenta squares) and AdS$_3$ (black circle). Finally, in the bottom panels the dark green diamonds represent the static $m=1$ hairy black holes.}
    \label{fig:m1TotalPhaseDiag-3k}
\end{figure}

BTZ black holes can be unstable not only to $m=0$ double‑trace perturbations, but also to non‑axisymmetric perturbations with $m\geq1$. However, the onset of the $m=1$ instability occurs at values of the double‑trace parameter $\kappa$ that are smaller (more negative) than those associated with the $m=0$ instability, and this ordering persists for higher values of $m$~\cite{Dias:2025uyk}. 

For example, in the reference case with $\mu^{2}L^{2}=-15/16$, if $\kappa=-0.4$ (the value considered in the previous subsection~\ref{sec:PhaseDiag-Total-m0}), BTZ black holes can be unstable to $m=0$ modes but remain stable to $m\geq1$ perturbations. By contrast, for $\kappa=-0.8$ $-$ a value we will use in the present subsection to illustrate our results $-$ BTZ solutions can be unstable to both $m=0$ and $m=1$ modes. In such cases, one expects the phase diagram of the theory to include BTZ black holes as well as both $m=0$ and $m=1$ hairy black hole branches. The latter two should bifurcate from BTZ along the respective onset curves
\begin{equation}
\Delta\hat{M}=\Delta\hat{M}(\mu^{2},\kappa;\hat{J})\big|^{m=0}_{\hbox{\tiny BTZ onset}},
\qquad
\Delta\hat{M}=\Delta\hat{M}(\mu^{2},\kappa;\hat{J})\big|^{m=1}_{\hbox{\tiny BTZ onset}},
\end{equation}
which were determined with high precision in Ref.~\cite{Dias:2025uyk}. Our fully nonlinear numerical solutions agree with these onset curves in the limit where the scalar condensate vanishes.

In this subsection we confirm that these expectations are indeed realized and then analyse the competition among the various thermodynamic phases to determine which configurations dominate in a given ensemble and region of parameter space. For a fixed value of $\kappa$, the theory also admits regular boson stars and singular extremal hairy black holes for each value of $m$. These solutions were already discussed in detail in subsections~\ref{sec:PhaseDiag-m0:BStar}–\ref{sec:PhaseDiag-m0:IshWald} for $m=0$ and subsections~\ref{sec:PhaseDiag-m1:BStar}–\ref{sec:PhaseDiag-m1:IshWald} for $m=1$. Accordingly, in the present subsection we only refer to boson stars when they play a direct role in shaping the phase diagram of hairy black holes.

Once the $m\geq1$ instabilities are present, three qualitatively different types of phase diagrams can occur, depending on the value of $\kappa$ (for fixed $m\geq1$). This distinction traces back to the fact that, already at the linear level, the BTZ instability onset curve for $m\geq1$ can exhibit three qualitatively different shapes as $\kappa$ is varied. These possibilities were analysed in detail in Ref.~\cite{Dias:2025uyk} (see in particular Figs.~9 and~11 therein). For this reason, we illustrate our microcanonical results for $\hat{\mu}^{2}=-15/16$ and three representative values of the double‑trace coupling,
\begin{equation}
\kappa=\{-0.8,\,-0.8639,\,-0.9\},
\end{equation}
which correspond to three distinct theories. As a first indication of the qualitative differences between these cases, note that for the latter value of $\kappa$ the theory admits static $m=1$ hairy black holes (see below), whereas for the former two values all $m=1$ hairy black holes necessarily rotate.

Figure~\ref{fig:m1TotalPhaseDiag-3k} shows the microcanonical phase diagram for $\kappa=-0.8$ (top panels), $\kappa=-0.8639$ (middle panels), and $\kappa=-0.9$ (bottom panels). All three values lie in the interval
\begin{equation}
\kappa^{\rm AdS}_{m=1,\hat{\mu}^{2}}
   \simeq -1.009837
   < \kappa
   < \kappa^{\rm AdS}_{m=0,\hat{\mu}^{2}}
   \simeq -0.4951294,
\end{equation}
with $\hat{\mu}^{2}=-15/16$. They therefore illustrate the regime in which AdS$_3$ is linearly stable against $m=1$ double‑trace perturbations (but unstable to $m=0$ ones)~\cite{Ishibashi:2004wx,Dias:2025uyk}, while BTZ can nevertheless be unstable to $m=1$ modes. Consistent with the results of Ref.~\cite{Dias:2025uyk}, whenever BTZ is unstable to $m=1$ perturbations it is necessarily also unstable to $m=0$ modes.

In Fig.~\ref{fig:m1TotalPhaseDiag-3k}, the orange curve at $\Delta\hat{M}(\hat{J})=0$ represents the one‑parameter family of extremal BTZ black holes with $\hat{M}=\hat{J}$. Regular BTZ solutions exist at and above this curve, \ie for $\Delta\hat{M}(\hat{J})\geq0$. The bright‑green curve denotes the onset of the $m=1$ instability of BTZ,
$\Delta\hat{M}(\kappa;\hat{J})\big|^{m=1}_{\hbox{\tiny BTZ onset}}$,
as computed in Ref.~\cite{Dias:2025uyk}. A comparison of the top, middle, and bottom panels shows that this onset curve can take three qualitatively different forms. For $\kappa=-0.8$ (top panels), it is single‑valued, with a unique value of $\Delta\hat{M}$ for each $\hat{J}$. For $\kappa=-0.8639$ (middle panels), it develops an $S$‑shape with two turning points, implying that for a finite range of $\hat{J}$ there are three distinct values of $\Delta\hat{M}$ on the onset curve. Finally, for $\kappa=-0.9$ (bottom panels), the onset curve splits into two disconnected branches, one of which has a distinctive semicircular shape.

The magenta curve in Fig.~\ref{fig:m1TotalPhaseDiag-3k} represents the regular $m=1$ boson star. Since all three values of $\kappa$ satisfy $\kappa>\kappa^{\rm AdS}_{m=1,\hat{\mu}^{2}}$, this branch is perturbatively connected to AdS$_3$, shown as the black disk at $\{\Delta\hat{M},\hat{J}\}=\{-1,0\}$. As discussed in subsections~\ref{sec:PhaseDiag-m1:BStar}–\ref{sec:PhaseDiag-m1:IshWald} and illustrated in Figs.~\ref{fig:BS-m1}–\ref{fig:m1BSevoK}, this regular boson star does not directly affect the phase structure of $m=1$ hairy black holes and is therefore not shown in the right panels of Fig.~\ref{fig:m1TotalPhaseDiag-3k}.

By contrast, the dark‑red triangle curve corresponds to the singular $m=1$ rotating extremal hairy black hole family, constructed using the numerical strategy of section~\ref{sec:NumericalSetup:singBHmJ}, and whose properties play a crucial role in the $m=1$ hairy black hole phase diagram. This curve, discussed in detail in subsections~\ref{sec:PhaseDiag-m1:BStar}–\ref{sec:PhaseDiag-m1:IshWald}, starts at the singular vacuum BTZ geometry (red diamond at $\{\Delta\hat{M},\hat{J}\}=\{0,0\}$), which is also the point where the bright‑green BTZ onset curve originates.

We can now describe the main features of Fig.~\ref{fig:m1TotalPhaseDiag-3k}. For concreteness, we focus first on the top panels with $\kappa=-0.8$, noting that the same qualitative considerations apply $-$ upon minor modifications $-$ to the middle and bottom panels. The green dots denote $m=1$ hairy black hole solutions, which occupy a region bounded from above by the bright‑green BTZ onset curve and from below by the dark‑red singular $m=1$ extremal hairy black hole curve. These two boundaries meet at $\{\Delta\hat{M},\hat{J}\}=\{0,0\}$ and extend to arbitrarily large angular momentum (we verified this up to $\hat{J}\simeq10$–$12$). The upper boundary reflects the fact that $m=1$ hairy black holes owe their existence to the instability of BTZ to scalar condensation: once back‑reaction is included, a branch of hairy solutions necessarily bifurcates from BTZ at the onset curve and extends into the unstable region. The lower boundary arises because the zero‑horizon‑radius (zero‑entropy and zero-temperature) limit of the $m=1$ hairy black holes is reached precisely at the singular $m=1$ extremal hairy black hole of section~\ref{sec:NumericalSetup:singBHmJ}. This behaviour is most transparently seen by examining constant‑$\hat{J}$ sub‑families of $m=1$ hairy black holes $-$ vertical lines in left panels of Fig.~\ref{fig:m1TotalPhaseDiag-3k} $-$ as carried out explicitly for $\hat{J}=6$ in Fig.~\ref{fig:m1-hBTZ-J6} of subsection~\ref{sec:PhaseDiag-m1:BHs}. There, we found clear evidence that, as the lower boundary is approached, both the temperature and entropy tend to zero (with $\hat{\Omega}_{H}\to1$).

A comparison between the $m=1$ phase diagram in Fig.~\ref{fig:m1TotalPhaseDiag-3k} and the $m=0$ phase diagram shown in Fig.~\ref{fig:m0TotalPhaseDiag-k04} reveals substantial qualitative differences, underscoring the fact that the $m=0$ and $m\geq1$ sectors of the theory are physically distinct.

Finally, the discussion above for the top panels with $\kappa=-0.8$ also applies, with suitable adjustments, to the middle panels corresponding to $\kappa=-0.8639$, where the bright‑green onset curve exhibits an $S$‑shape. For instance, following a family of $m=1$ hairy black holes with fixed $\hat{J}=0.5$ from the upper onset point downwards, one encounters first a window of $\Delta\hat{M}$ where hairy black holes exist, then an intermediate range of $\Delta\hat{M}$ for which no such solutions are present, and finally a second branch of hairy black holes emerging at lower masses. This lower branch crosses $\Delta\hat{M}=0$ and extends down to the singular $m=1$ extremal hairy black hole at $\hat{J}=0.5$ and negative $\Delta\hat{M}$.

As $\kappa$ is decreased further, the left turning point of the onset curve in the middle panels of Fig.~\ref{fig:m1TotalPhaseDiag-3k} eventually reaches the $\hat{J}=0$ axis (a case not shown). For smaller values of $\kappa$, the bright‑green onset curve
$\Delta\hat{M}(\kappa;\hat{J})|^{m=1}_{\hbox{\tiny BTZ onset}}$
then splits into two disconnected branches, one of which forms a lower branch with a characteristic semicircular shape. This situation is illustrated for $\kappa=-0.9$ in the bottom panels of Fig.~\ref{fig:m1TotalPhaseDiag-3k}. This qualitative change has noteworthy consequences.

In particular, the upper branch of the onset curve now resembles closely the $m=0$ onset curve displayed in Fig.~\ref{fig:m0TotalPhaseDiag-k04}. Most notably, \emph{static} BTZ black holes are now also unstable to $m=1$ perturbations. As a result, there exists a finite window of $\Delta\hat{M}$ at $\hat{J}=0$ (indicated by the dark‑green segment in the bottom panels of Fig.~\ref{fig:m1TotalPhaseDiag-3k}) where \emph{static} $m=1$ hairy black holes are present. In other words, although the scalar condensate carries azimuthal dependence with $m=1$ and thus orbits the horizon, the resulting hairy black hole has vanishing total angular momentum. Such a configuration is highly non‑trivial and would not have been anticipated a priori.

\begin{figure}
    \centering
    \includegraphics[width=0.47\linewidth]{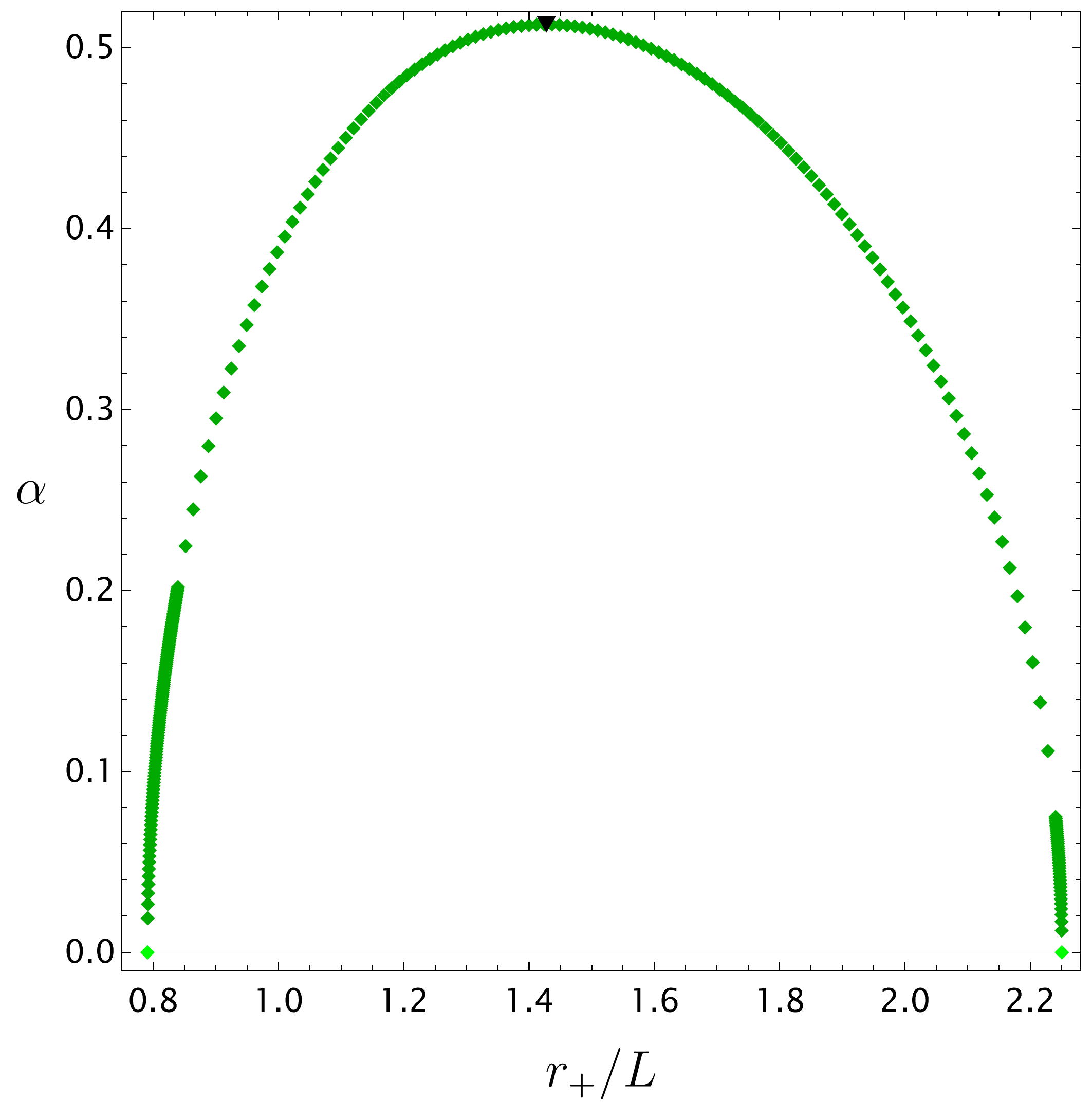}
    \includegraphics[width=0.49\linewidth]{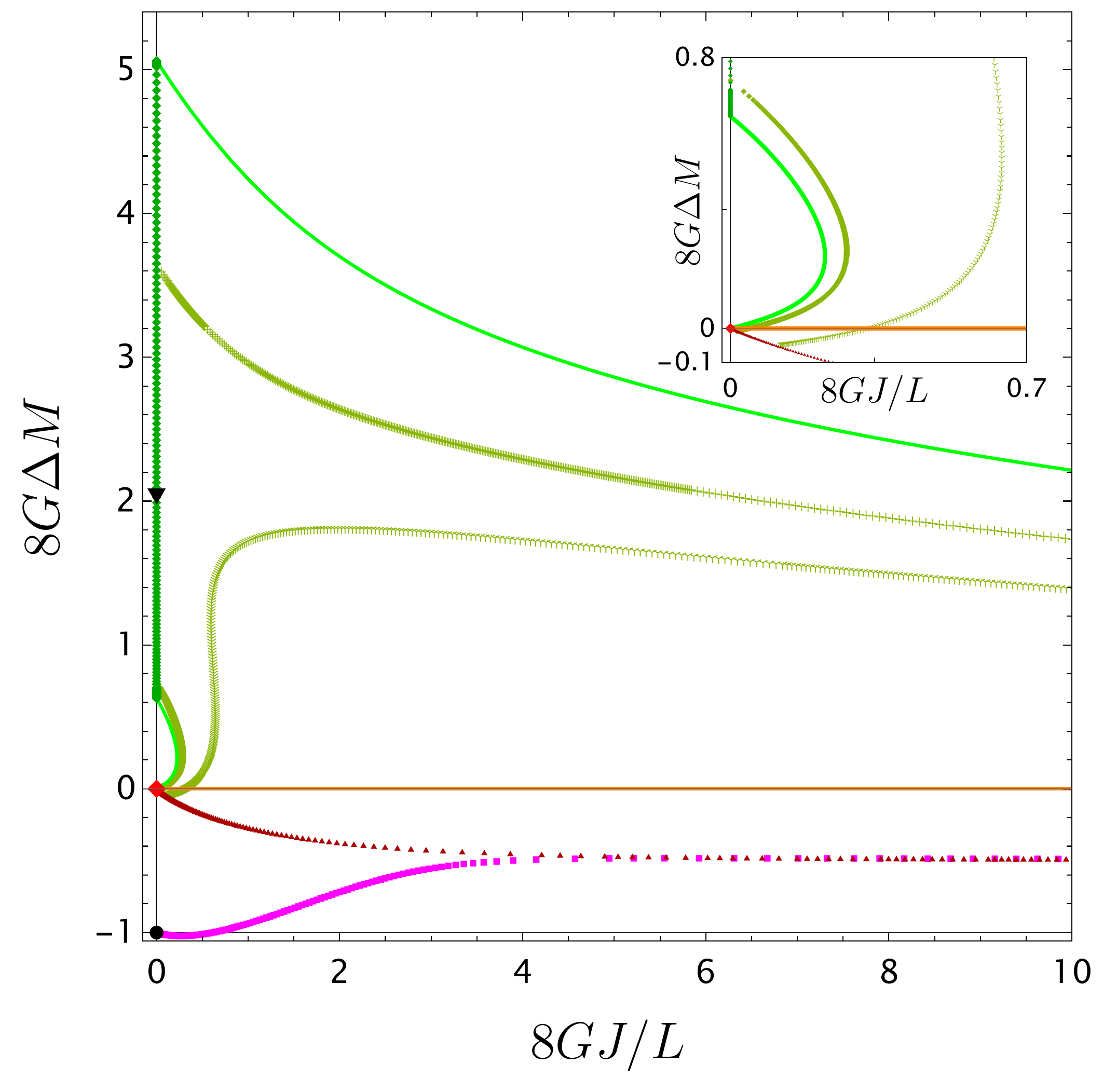}
    \caption{Features of the static $m=1$ hairy black holes of the bottom row in Fig.~\ref{fig:m1TotalPhaseDiag-3k}. \textbf{Left plot}: evolution of $\alpha$ with $R_+$ along the $J=0$ and $\omega=0$ family of hairy black holes (dark green). For $J=0$ and $\omega=0$, the maximum $\alpha$ allowed is $\alpha_{\rm max} \sim 0.513$ (black inverted triangle) and occurs at $R_+ \sim 1.43$. \textbf{Right plot}: selected constant $\alpha$  families of hairy black holes, with constant $\alpha \sim 0.23$ (diamonds), $\alpha \sim 0.55$ (y-cross) and  $\alpha \sim 0.40$ (+-cross). The value $\alpha_{\rm max}$ for the $J=0$ family of solutions is attained for $\Delta \hat{M} \sim 2.03$ (black inverted triangle).
    }
    \label{fig:m1kappam9_details}
\end{figure}

These static $m=1$ hairy black holes are examined in more detail in Fig.~\ref{fig:m1kappam9_details}. In the left panel we plot the asymptotic scalar amplitude $\alpha$ as a function of the horizon radius $R_{+}$, following the branch from the lower onset point (bright‑green diamond at $R_{+}\simeq0.79$) to the upper onset point (bright‑green diamond at $R_{+}\simeq2.25$). As expected, the scalar amplitude vanishes at both onset points, while in between it reaches a maximum at the black inverted triangle, located at approximately $\{R_{+},\alpha\}\simeq\{1.43,0.513\}$. This same configuration is highlighted in the right panel of Fig.~\ref{fig:m1kappam9_details}.

This right panel also displays three representative one‑parameter sub‑families of $m=1$ hairy black holes at fixed values of $\alpha$ (shown in green). The $\alpha=0.23$ sub‑family follows closely the semicircular lower branch of the onset curve and describes solutions that begin at a static $m=1$ hairy black hole (with $R_{+}\simeq0.86$, and more generally for $R_{+}\lesssim1.43$) and terminate at a singular boson star as $R_{+}\to0$. The green $S$‑shaped curve corresponds to the $\alpha=0.55$ sub‑family, which starts at the singular boson star and extends to arbitrarily large angular momentum. Finally, the $\alpha=0.4$ sub‑family consists of solutions lying below the upper bright‑green branch of the onset curve; these start from a static $m=1$ hairy black hole (with $R_{+}\simeq1.91$, and more generally for $R_{+}\gtrsim1.43$) and display a monotonically decreasing $\Delta\hat{M}$ as $\hat{J}$ increases.

An interesting by‑product of this analysis is the appearance of solution non‑uniqueness not only for rotating configurations, but also for static ones. For certain values of $\kappa$ (for example $\kappa=-0.9$), black holes with identical mass can include static BTZ solutions as well as both static $m=0$ and static $m=1$ hairy black holes.

The right panels of Fig.~\ref{fig:m1TotalPhaseDiag-3k} display the entropy of the black hole solutions as a function of the microcanonical variables $\Delta\hat{M}$ and $\hat{J}$. The BTZ family appears as a grey surface, defined for $\Delta\hat{M}\geq0$ and originating from the orange extremal line. The $m=1$ hairy black hole family is represented by a surface of green points that merges continuously with the BTZ surface along the bright‑green $m=1$ onset curve
$\Delta\hat{M}(\kappa;\hat{J})|^{m=1}_{\hbox{\tiny BTZ onset}}$
via a second‑order phase transition. A robust conclusion, valid for all values of $\kappa$ shown, is that whenever $m=1$ hairy black holes coexist with BTZ at fixed $(\Delta\hat{M},\hat{J})$, they always have higher entropy than BTZ. In addition, these solutions possess a zero‑horizon‑radius limit $-$ the singular $m=1$ extremal hairy black hole limit of subsection~\ref{sec:NumericalSetup:singBHmJ} $-$ where both the entropy and temperature vanish.

This behaviour is clearly visible in the right panels of Fig.~\ref{fig:m1TotalPhaseDiag-3k} for relatively small angular momentum, say $\hat{J}\lesssim3$. For larger $\hat{J}$, although not explicitly resolved in the plots, the green surface still descends almost vertically $-$ \ie with a very large entropy gradient for a small change in $\Delta\hat{M}$ $-$ toward zero entropy, where it terminates at the dark‑red singular $m=1$ extremal hairy black hole curve. To illustrate this behaviour, we include in the top‑right panel of Fig.~\ref{fig:m1TotalPhaseDiag-3k} the constant‑$\hat{J}=6$ sub‑family (for $\kappa=-0.8$) already discussed in Fig.~\ref{fig:m1-hBTZ-J6}. Completing the figure with additional constant‑$\hat{J}$ sub‑families would be computationally expensive and would not add significant qualitative insight; the apparent gap between the lowest displayed green points and the dark‑red curve is therefore a numerical artifact.

For values of $\kappa$ smaller than $-0.9$ (with $\hat{\mu}^2=-15/16$ fixed), the phase diagram remains qualitatively similar to that shown in the bottom panels of Fig.~\ref{fig:m1TotalPhaseDiag-3k}. As $\kappa$ decreases, the upper branch of the bright‑green $m=1$ onset curve moves progressively farther from the orange extremal BTZ line, while the radius of the semicircular lower branch shrinks. This trend is evident, for example, in the right panel of Fig.~6 and the left panel of Fig.~9 in Ref.~\cite{Dias:2025uyk}. For $\kappa\lesssim-0.97$ the lower semicircular branch disappears entirely, leaving only the upper onset branch. In this regime the phase diagram becomes qualitatively similar to that of the $m=0$ case (cf.~Fig.~\ref{fig:m0TotalPhaseDiag-3k}), with the $m=1$ onset curve lying increasingly far above the extremal BTZ line as $\kappa$ decreases further. This qualitative behaviour persists even for $\kappa<\kappa^{\rm AdS}_{m=1,\hat{\mu}^2=-15/16}\simeq-1.009837$, where AdS$_3$ itself is unstable to $m=1$ double‑trace perturbations.

Figure~\ref{fig:m1TotalPhaseDiag-3k} pertains to a theory with $\hat{\mu}^2=-15/16$ and $m=1$. We expect the qualitative features of this phase diagram to persist for any scalar mass in the range $-1<\hat{\mu}^2<0$, for which double‑trace boundary conditions yield normalizable solutions. Moreover, the phase diagram for any $m\geq1$ is expected to be qualitatively similar to the one shown in Fig.~\ref{fig:m1TotalPhaseDiag-3k}. This expectation is strongly supported by the linear analysis presented in our companion paper~\cite{Dias:2025uyk}, which demonstrated that the structure of instabilities and their onset curves is qualitatively universal throughout the range $-1<\hat{\mu}^2<0$ and for all $m\geq1$.

\begin{figure}
    \centering
    \includegraphics[width=0.47\linewidth]{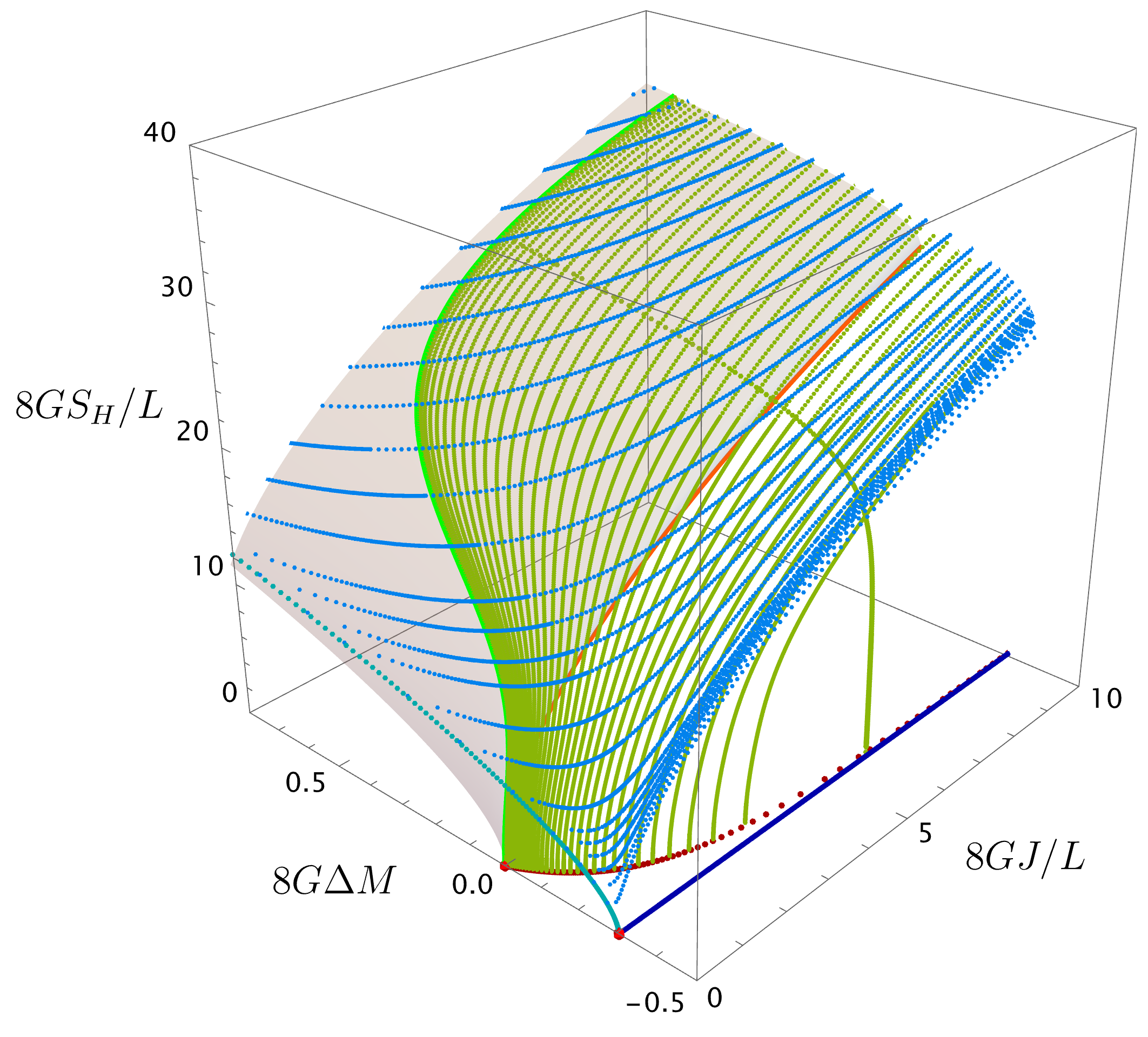}
    \includegraphics[width=0.47\linewidth]{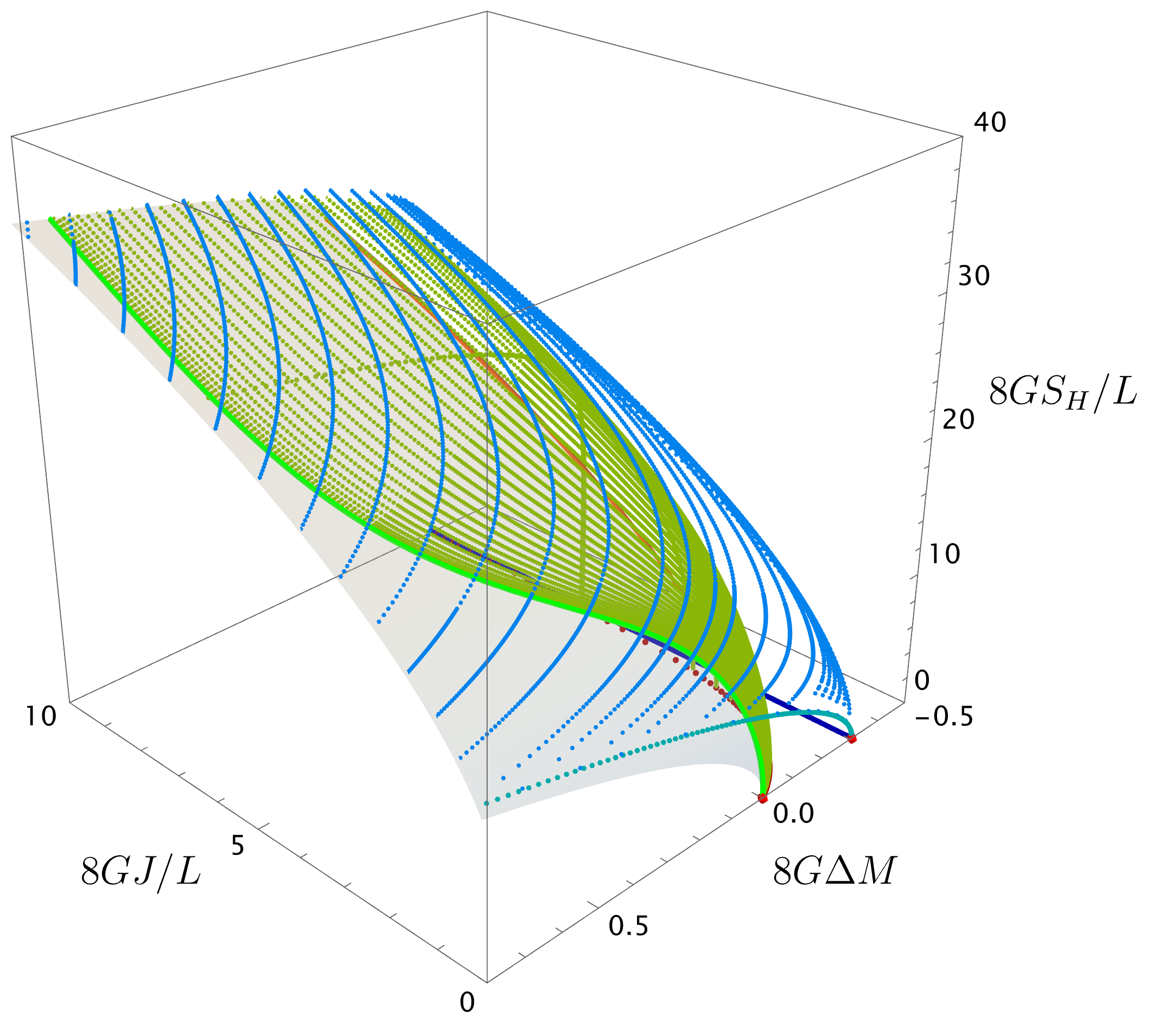}
      \includegraphics[width=0.47\linewidth]{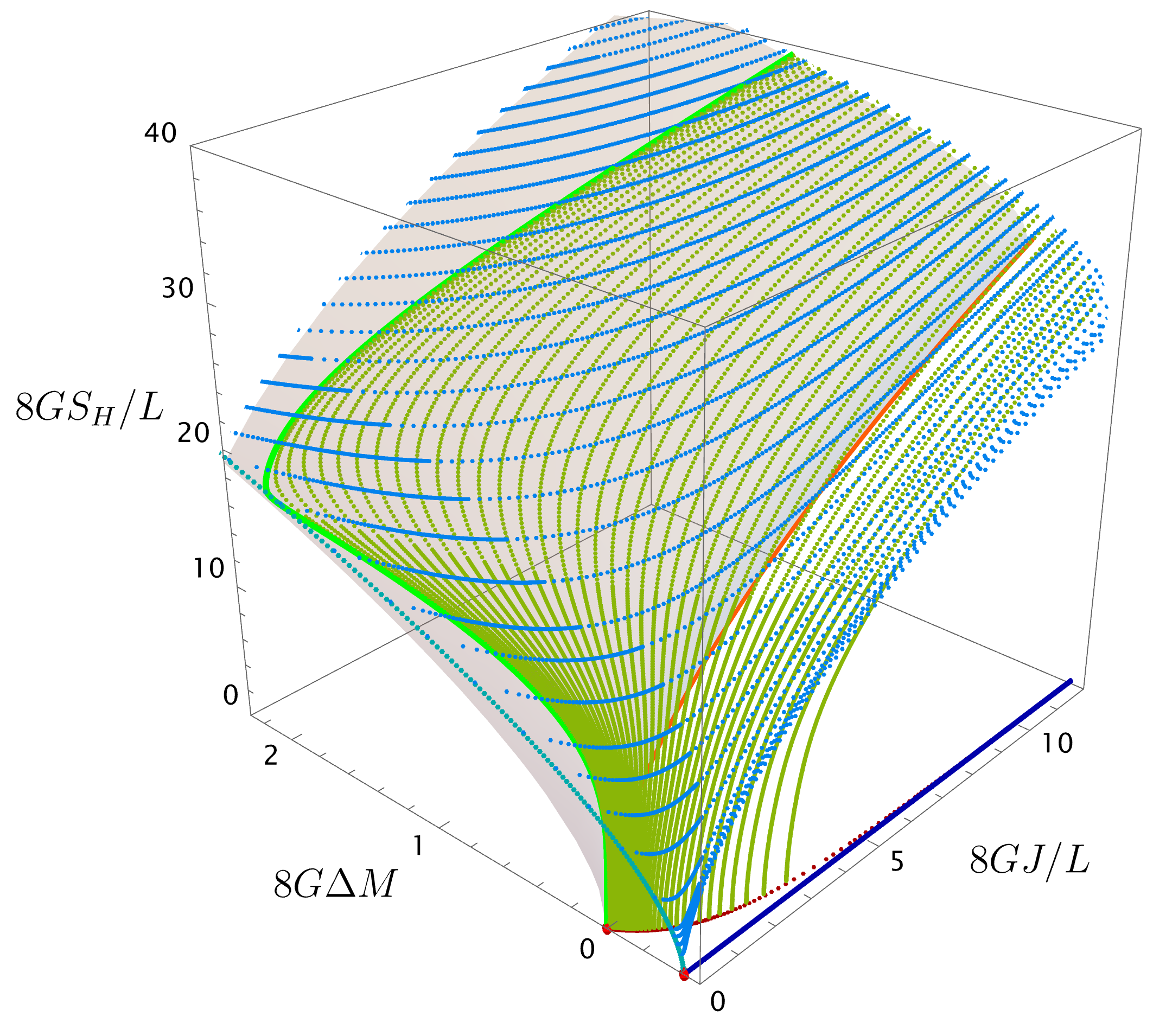}
    \includegraphics[width=0.47\linewidth]{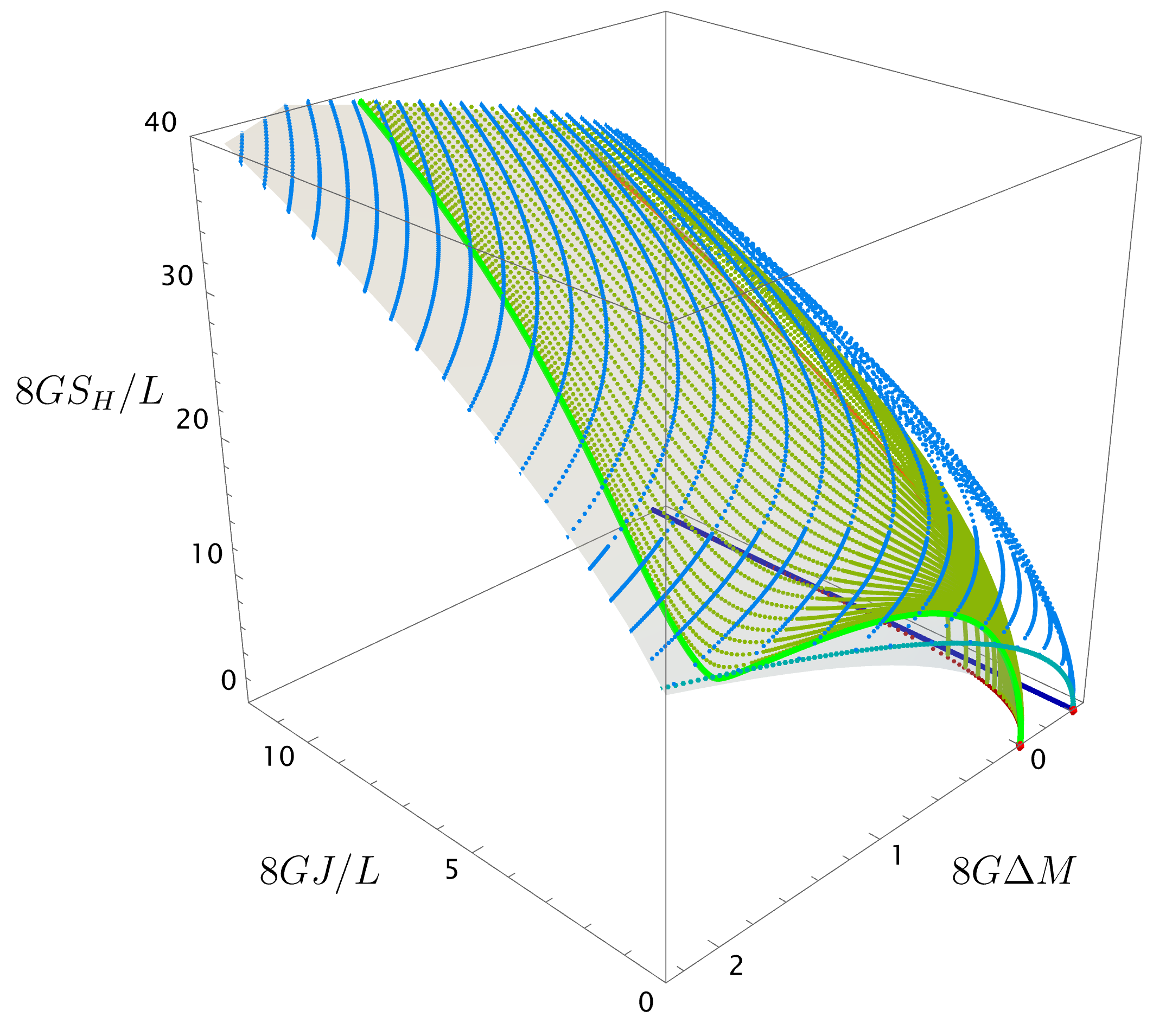}
      \includegraphics[width=0.47\linewidth]{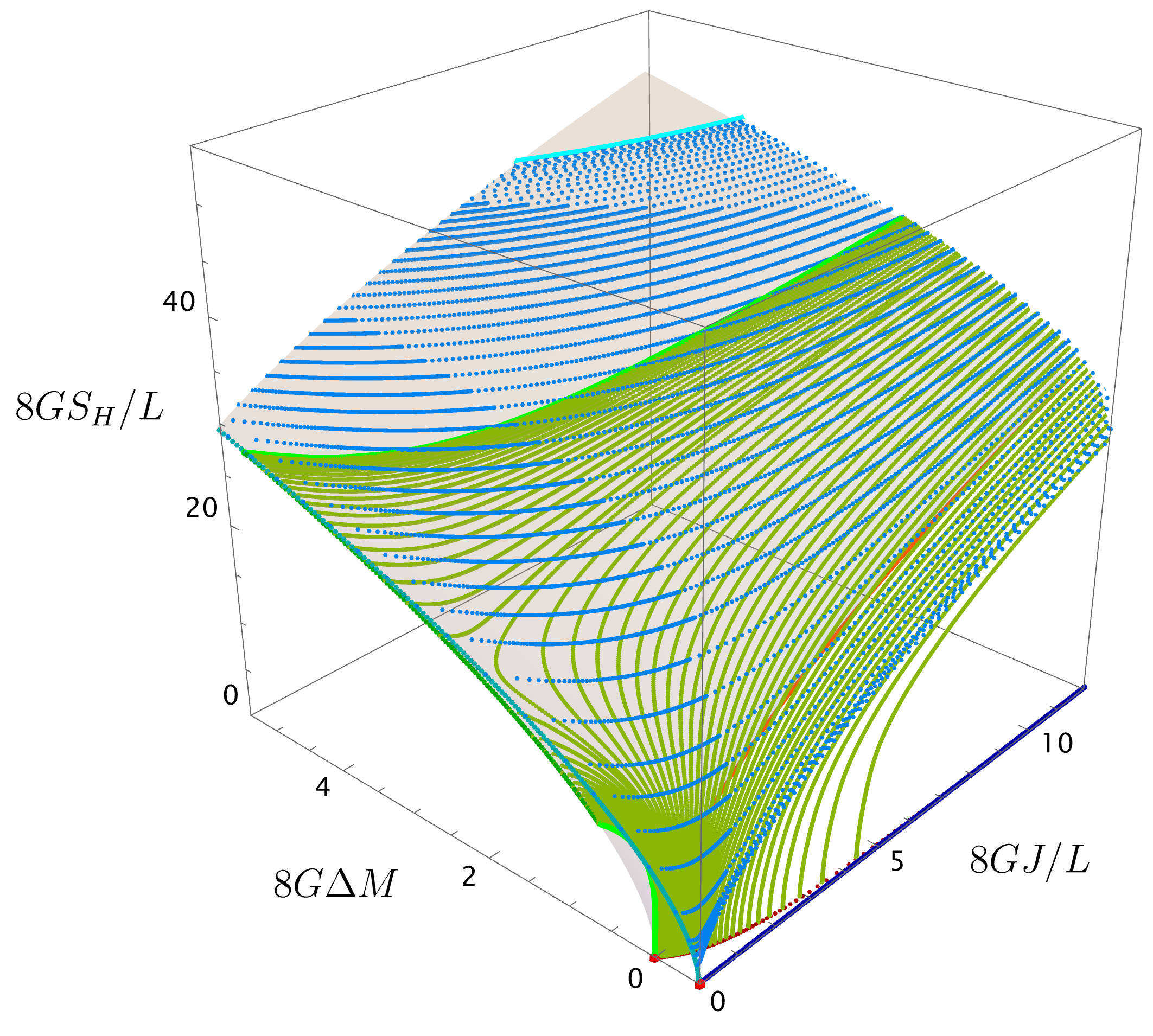}
    \includegraphics[width=0.47\linewidth]{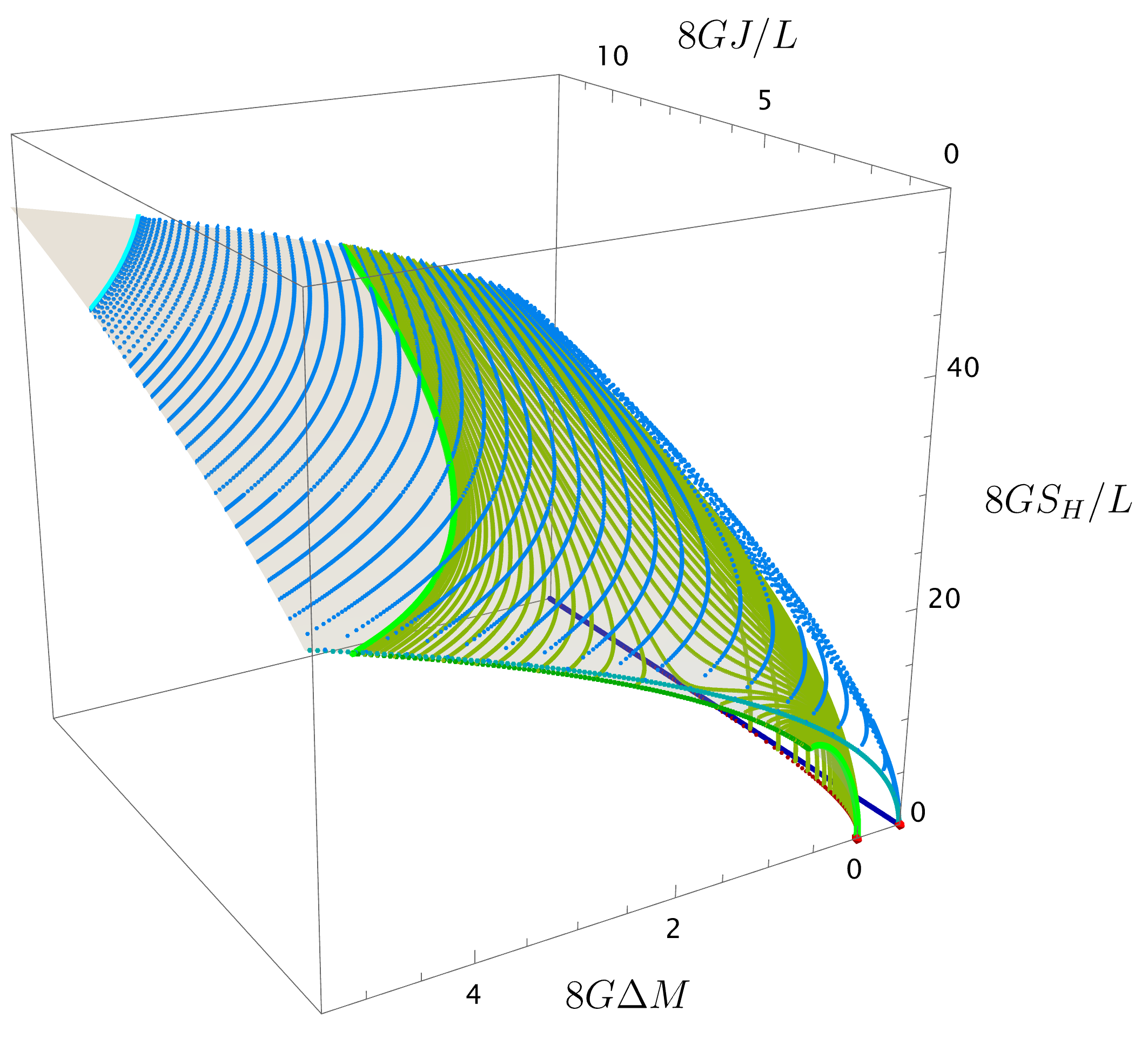}
    \caption{{\bf Microcanonical phase diagram} of solutions of AdS$_3$-Einstein gravity with a massive scalar field ($\mu^2L^2 = -15/16$) with double-trace parameter $\bm{\kappa = -8/10}$ {\bf (top panels)}, $\bm{\kappa = -0.8639}$ {\bf (middle panels)} and $\bm{\kappa = -9/10}$ {\bf (bottom panels)}.  This figure merges Figs.~\ref{fig:m0TotalPhaseDiag-3k}~and~\ref{fig:m1TotalPhaseDiag-3k} to show the thermal competition between BTZ (grey surface) and  $m=0$ (blue dots) and $m=1$ (green dots) hairy black holes. When they co-exist (we just plot this range), the  $m=0$ hairy black holes have higher entropy that the $m=1$ hairy black holes with the same mass and angular momentum, and both have higher entropy than BTZ. (The left and right plots are the same but shown from different viewpoints).
    }
    \label{fig:m0m1FinalPhaseDiag-3k}
\end{figure}

A further fundamental aspect must still be addressed. Whenever the $m=1$ BTZ instability is present, the BTZ black hole is necessarily also unstable to the $m=0$ sector of double‑trace perturbations. This is precisely the case for the values $\kappa=\{-0.8,-0.8639,-0.9\}$ shown in Fig.~\ref{fig:m1TotalPhaseDiag-3k}. Consequently, in addition to the $m=1$ hairy black holes displayed in that figure, the theory must also admit $m=0$ hairy black hole solutions. Anticipating this situation, we already presented the corresponding $m=0$ phase diagrams for $\kappa=\{-0.8,-0.8639,-0.9\}$ in Fig.~\ref{fig:m0TotalPhaseDiag-3k}.

The natural question is therefore which of the three thermodynamic phases $-$ BTZ, $m=0$ hairy black holes, or $m=1$ hairy black holes $-$ dominates the microcanonical ensemble for a given pair $\{\Delta\hat{M},\hat{J}\}$. To answer this question, it is necessary to display these three phases simultaneously. This is achieved in Fig.~\ref{fig:m0m1FinalPhaseDiag-3k}, where we superpose the $m=0$ and $m=1$ hairy black hole families with BTZ for $\kappa=-0.8$ (top panels), $\kappa=-0.8639$ (middle panels), and $\kappa=-0.9$ (bottom panels). The left and right panels correspond to the same three‑dimensional plot of $\{\Delta\hat{M},\hat{J}\}$ versus entropy $\hat{S}_H$, viewed from different perspectives to improve visibility. For clarity, we display only the region where $m=1$ hairy black holes exist (green surface), together with the overlapping portion of the $m=0$ hairy black hole family (blue surface). We do not show the full extent of the $m=0$ solutions nor the entire region $\Delta\hat{M}\geq0$ occupied by BTZ, since our goal is to compare entropies only in regions where all three phases coexist.

Inspection of Fig.~\ref{fig:m0m1FinalPhaseDiag-3k} reveals several robust features. First, for any given $\hat{J}$, the $m=0$ BTZ instability onset curve always lies farther from extremality than the $m=1$ onset curve, \ie
\begin{equation}
\Delta\hat{M}(\kappa;\hat{J})\big|^{m=0}_{\hbox{\tiny BTZ onset}}
>
\Delta\hat{M}(\kappa;\hat{J})\big|^{m=1}_{\hbox{\tiny BTZ onset}},
\end{equation}
a pattern that persists for higher $m$ as well,
\begin{equation}
\Delta\hat{M}\big|^{m}_{\hbox{\tiny BTZ onset}}>
\Delta\hat{M}\big|^{m+1}_{\hbox{\tiny BTZ onset}}
\end{equation}
for all $m\geq0$~\cite{Dias:2025uyk}. This hierarchy explains why any BTZ solution that is unstable to $m=1$ (or higher‑$m$) perturbations is automatically unstable to $m=0$ perturbations, while the converse need not hold.

A second important feature is that, for fixed $\hat{J}$, the $m=0$ hairy black hole solutions extend to more negative values of $\Delta\hat{M}$ than the $m=1$ solutions (recall that regular BTZ black holes do not exist for $\Delta\hat{M}<0$). Specifically, the $m=1$ hairy black holes terminate at the singular $m=1$  extremal hairy black hole (sExtHBH) of section~\ref{sec:NumericalSetup:singBHmJ} with
\begin{equation}
\Delta\hat{M}=\Delta\hat{M}(\kappa;\hat{J})|^{m=1}_{\hbox{\tiny sExtHBH}},
\end{equation}
whereas the $m=0$ hairy black holes extend further, down to the singular $m=0$ extremal hairy black hole of section~\ref{sec:NumericalSetup:singBHm0J}  with
\begin{equation}
\Delta\hat{M}=\Delta\hat{M}(\kappa)|^{m=0}_{\hbox{\tiny sExtHBH}},
\end{equation}
and one always finds
\begin{equation}
\Delta\hat{M}|^{m=0}_{\hbox{\tiny sExtHBH}}
<
\Delta\hat{M}|^{m=1}_{\hbox{\tiny sExtHBH}}.
\end{equation}
Notice, however, that as $\hat{J}$ increases, these values seem to approach each other asymptotically, as it can be seen in the plots.

The most important conclusion emerging from Fig.~\ref{fig:m0m1FinalPhaseDiag-3k} is, however, thermodynamic. For any point $\{\Delta\hat{M},\hat{J}\}$ at which BTZ, $m=0$ hairy black holes, and $m=1$ hairy black holes coexist, the BTZ entropy is always the smallest. More significantly, the entropy of the $m=0$ hairy black hole is always strictly larger than that of the $m=1$ hairy black hole (and likewise larger than that of any $m>1$ hairy black hole, not shown). In other words, in entropy ordering one has
\begin{equation}
\hat{S}_{H}^{(m=0)} > \hat{S}_{H}^{(m=1)} > \hat{S}_{H}^{\rm BTZ}.
\end{equation}

It follows that, for any $\{\Delta\hat{M},\hat{J}\}$ lying between the $m=0$ BTZ instability onset curve and the extremal BTZ line $\Delta\hat{M}=0$, as well as for all negative $\Delta\hat{M}$ where they exist, the $m=0$ hairy black holes dominate the microcanonical ensemble. By contrast, for
\begin{equation}
\Delta\hat{M}(\hat{J})>\Delta\hat{M}(\kappa;\hat{J})|^{m=0}_{\hbox{\tiny BTZ onset}},
\end{equation}
BTZ is the only available phase and thus trivially dominates. Together with the explicit construction of the hairy black hole solutions, this constitutes one of the main results of this work.

It is worth emphasizing the physical distinction between the two classes of solutions. The $m=0$ static and rotating hairy black holes described in Section~\ref{sec:PhaseDiag-Total-m0} are time‑independent and axisymmetric, \ie they are stationary black holes in the strict and original sense of the term. In contrast, the $m\geq1$ hairy black holes $-$ whether rotating or static (for values of $\kappa$ where static solutions exist) $-$ are neither time‑independent nor axisymmetric. They are instead quasi‑periodic solutions in the sense that the Killing horizon generator
\begin{equation}
K=\partial_{T}+\hat{\Omega}_{H}\partial_{\phi}
\end{equation}
is a Killing vector of the full spacetime–scalar configuration. In this sense, the $m\geq1$ hairy black holes provide explicit examples of {\it black resonators}.

Black resonators were originally discovered in the context of superradiant instabilities of rotating AdS$_{d\geq4}$ black holes with Dirichlet boundary conditions, both in AdS‑Einstein–scalar systems~\cite{Dias:2011at,Stotyn:2011ns,Ishii:2018oms,Ishii:2021xmn} and in pure AdS‑Einstein gravity~\cite{Dias:2015rxy}. In those cases, black resonators owe their existence to superradiant amplification. In the present AdS$_3$ setting, by contrast, the $m\geq1$ black resonators arise because double‑trace boundary conditions inject both energy and angular momentum into the spacetime from the asymptotic boundary (see Section~5.2 of~\cite{Dias:2025uyk}). We also note that the first explicit example of a double‑trace $m=1$ hairy black hole $-$ essentially identical to the solutions studied here $-$ was obtained in Ref.~\cite{Iizuka:2015vsa}, where it was already interpreted as a black resonator in AdS$_3$.

Although we have not performed a linear stability analysis of the hairy black holes themselves, we conjecture that $m=0$ hairy black holes are linearly stable to all $m\geq0$ double‑trace perturbations, while any $m\geq1$ hairy black hole is unstable to $(m-1)$ modes and hence ultimately to $m=0$ perturbations. The evidence for this conjecture is twofold. First, any BTZ solution that is unstable to $m$ modes is necessarily unstable to all lower‑$m$ modes, a property that one expects to persist for the corresponding hairy solutions, except that an $m$ hairy black hole is constructed to be marginally stable to the $m$‑mode that generates it. Second, whenever $m\geq1$ hairy black holes exist at fixed $\{\hat{M},\hat{J}\}$, $m=0$ hairy black holes also exist and are always entropically favored. In any scenario where a linear instability can develop, it is therefore natural to expect that $m\geq1$ hairy black holes are metastable and will evolve $-$ possibly through a cascade of decreasing $m$ $-$ toward a stable $m=0$ hairy black hole.

\begin{figure}
    \centering
    \includegraphics[width=0.47\linewidth]{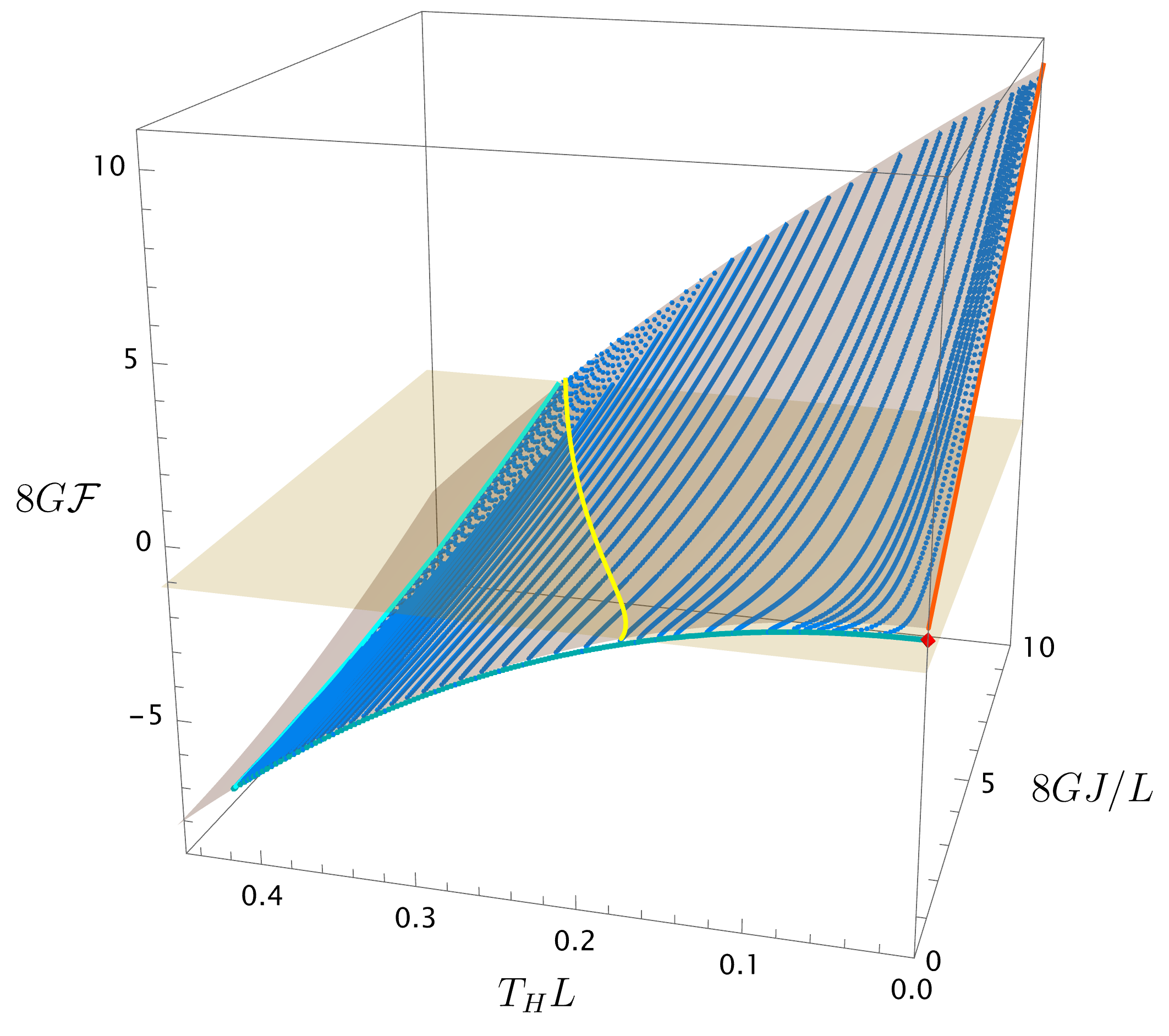}
    \includegraphics[width=0.47\linewidth]{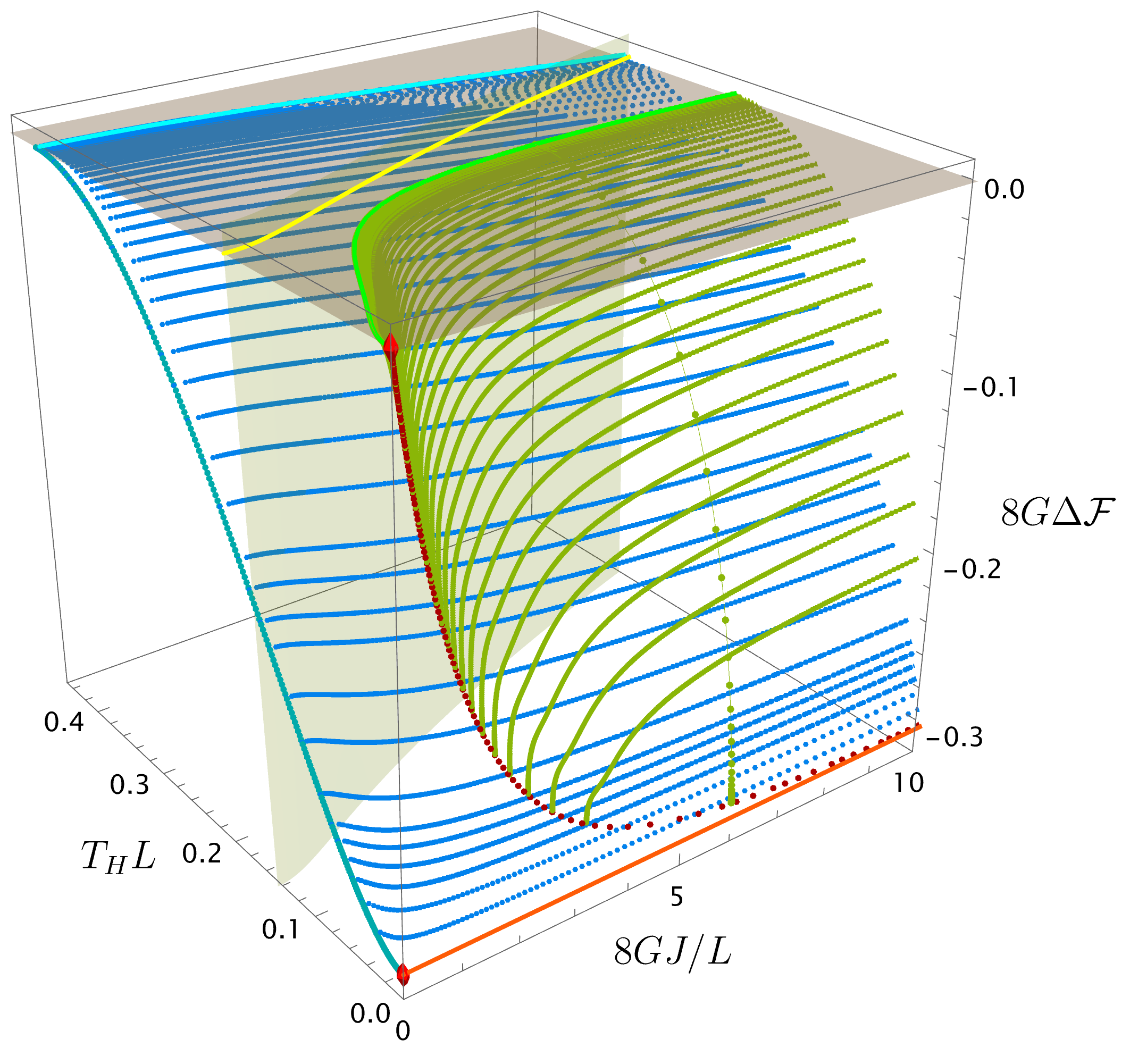}
    \includegraphics[width=0.47\linewidth]{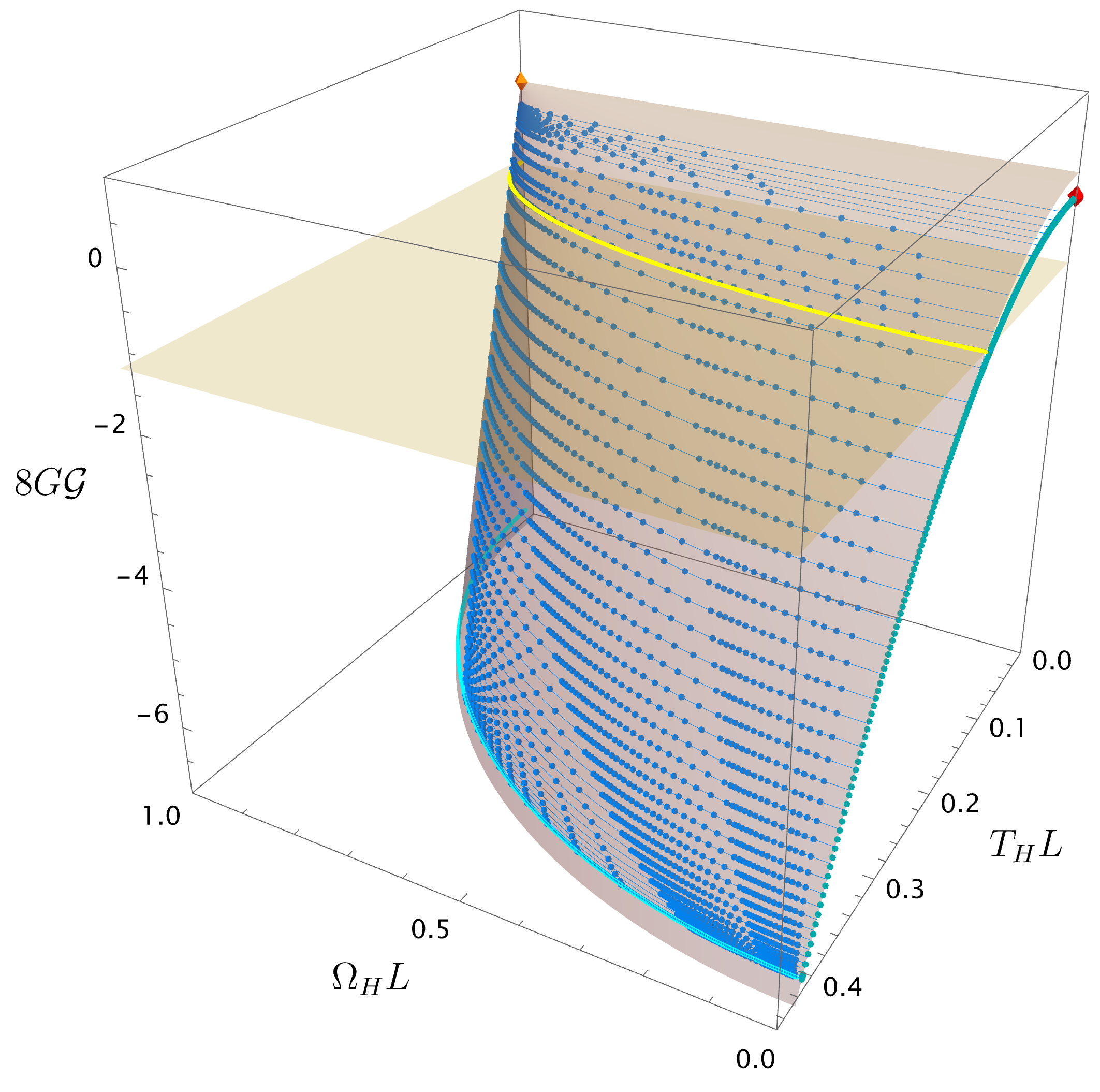}
    \includegraphics[width=0.47\linewidth]{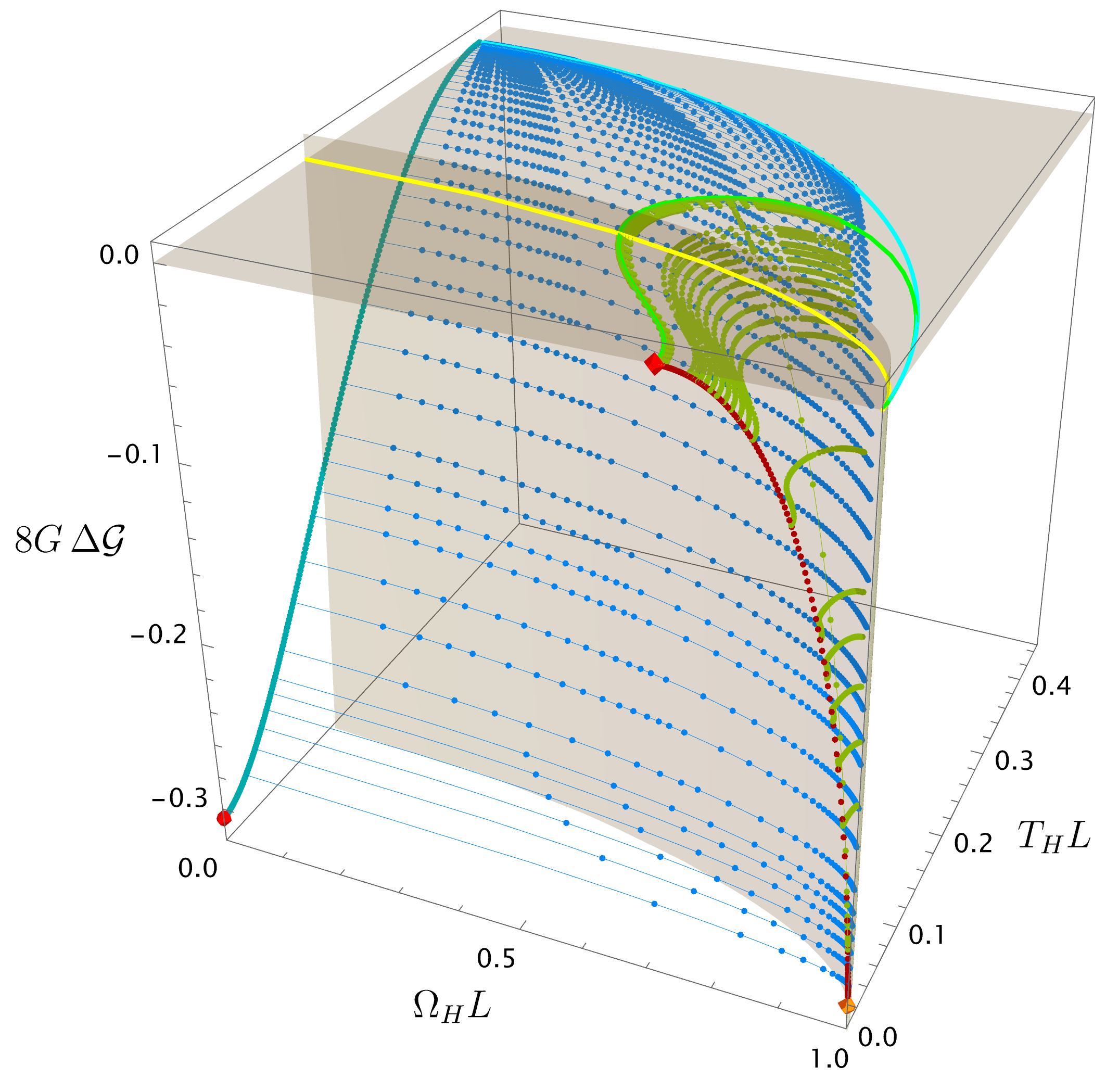}
    \caption{{\bf Canonical (top panels)} and {\bf grand-canonical (bottom panels)} phase diagrams of solutions with $\mu^2L^2 = -15/16$ and $\bm{\kappa = -8/10}$ (\ie $\kappa^{\rm AdS}_{m=1, \hat{\mu}^2}<\kappa <\kappa^{\rm AdS}_{m=0, \hat{\mu}^2}$) for which AdS$_3$ is unstable to $m=0$ (but not $m=1$) modes and BTZ can be stable/unstable both to $m=0$ and $m=1$ modes. When they co-exist, $\bm{m = 0}$ hairy black holes (blue dot surface) always have lower Helmholtz $(\hat{\mathcal{F}})$ and Gibbs $(\hat{\mathcal{G}})$ free energy than BTZ (dark-yellow or grey surface) and $\bm{m = 1}$ hairy black holes (green dot surface). That is, in the right panels (where we plot $\Delta \hat{\mathcal{F}}=\hat{\mathcal{F}}-\hat{\mathcal{F}}^{\hbox{\tiny BTZ}}$ or $\Delta \hat{\mathcal{G}}=\hat{\mathcal{G}}-\hat{\mathcal{G}}^{\hbox{\tiny BTZ}}$), the blue dot surface is clearly seen to be always below the grey BTZ and green dot surfaces).
    The yellow line represents the Hawking-Page critical temperature $\hat{T}_{\hbox{\tiny HP}}(\hat{J})$ (top panels) or $\hat{T}_{\hbox{\tiny HP}}(\hat{\Omega}_H)$ (bottom panels). For temperatures smaller than the Hawking-Page, it is the thermal $m=0$ zero-frequency boson star (faded-yellow surface with $\hat{\mathcal{F}}=\hat{\mathcal{G}}\simeq -1.16$) that dominates the canonical and grand-canonical ensembles. For intermediate temperatures between the yellow and cyan curves, it is $m=0$ hairy black hole family that dominates the canonical and grand-canonical ensembles. For temperatures higher than the cyan curve (\ie the merger curve of BTZ with $m=0$ hairy black holes), it is BTZ that has lower $\hat{\mathcal{F}}$ and lower $\hat{\mathcal{G}}$.
    Note that, for clarity, we do not display the singular $m=0$ extremal hairy black hole in these plots since this curve (point) is very close (on top) of extremal BTZ in the top (bottom) panels (see Fig.~\ref{fig:m0_kappa_m4o10_DeltaF_DeltaG} for $\kappa=-0.4$ case).
    }
\label{m0_m1_kappa_m8o10_DeltaF_DeltaG}
\end{figure}

Thus far our discussion has focused on the microcanonical ensemble. To conclude this subsection, we briefly address the canonical and grand‑canonical ensembles for theories with
\begin{equation}
\kappa^{\rm AdS}_{m=1,\hat{\mu}^{2}}\simeq-1.009837
<
\kappa
<
\kappa^{\rm AdS}_{m=0,\hat{\mu}^{2}}\simeq-0.4951294
\end{equation}
(with $\hat{\mu}^{2}=-15/16$). As a representative example we consider $\kappa=-0.8$, whose canonical and grand‑canonical phase diagrams are shown in the top and bottom panels of Fig.~\ref{m0_m1_kappa_m8o10_DeltaF_DeltaG}. This analysis is necessary because the competition among thermodynamic phases depends sensitively on $\kappa$, and the results for $\kappa=-0.4$ shown in Fig.~\ref{fig:m0_kappa_m4o10_DeltaF_DeltaG} are not universal.

In Fig.~\ref{m0_m1_kappa_m8o10_DeltaF_DeltaG}, the faded‑green surface represents the $m=0$ zero‑frequency boson star, which is the minimum‑energy solution of the $\kappa=-0.8$ theory and has $\hat{\mathcal{F}}=\hat{\mathcal{G}}\simeq-1.16$ (see Fig.~\ref{fig:m0m1omega0BS}), well below the $\hat{\mathcal{F}}=\hat{\mathcal{G}}=-1$ of AdS$_3$. The grey surface corresponds to BTZ, while the blue and green surfaces represent the $m=0$ and $m=1$ hairy black holes, respectively. The $m=0$ and $m=1$ hairy black hole branches bifurcate from BTZ at the cyan and bright‑green curves, respectively. The yellow curve denotes the Hawking–Page transition between the $m=0$ hairy black holes and the $m=0$ zero‑frequency boson star, with $\hat{T}_{\hbox{\tiny HP}}(\hat{J})$ shown in the top panels and $\hat{T}_{\hbox{\tiny HP}}(\hat{\Omega}_{H})$ in the bottom panels.

The conclusions are clear. Whenever $m=0$ hairy black holes exist, neither BTZ nor $m=1$ hairy black holes dominate the canonical or grand‑canonical ensembles. In the canonical ensemble, for temperatures $\hat{T}(\hat{J})<\hat{T}_{\hbox{\tiny HP}}(\hat{J})$ the $m=0$ zero‑frequency boson star (not AdS$_3$) dominates. For intermediate temperatures between the yellow and cyan curves, the $m=0$ hairy black hole is dominant. At sufficiently high temperatures (above the cyan curve), BTZ dominates. An entirely analogous pattern holds in the grand‑canonical ensemble.

Finally, lowering $\kappa$ below $-0.8$ modifies this structure further. As $\kappa$ decreases, both the yellow Hawking–Page curve and the cyan hairy BTZ onset curve shift to higher temperatures, but the Hawking–Page curve does so more rapidly because the energy of the zero‑frequency boson star decreases significantly (see Fig.~\ref{fig:m0m1omega0BS}). Eventually, one reaches values of $\kappa$ for which the Hawking–Page transition always occurs at higher temperature than the hairy black hole onset. In this regime, the $m=0$ zero‑frequency boson star dominates at low temperatures and BTZ dominates at high temperatures, with $m=0$ hairy black holes never dominating the canonical or grand‑canonical ensembles - unlike in the intermediate $\kappa=-0.8$ case.

\section{Conclusion and final discussions \label{sec:Conc}}
In this paper we have studied AdS$_3$ Einstein gravity coupled to a massive scalar field with mass in the window between the Breitenlohner–Freedman (BF) and unitarity bounds,
\begin{equation}
-1<\mu^{2}L^{2}<0,
\end{equation}
where both asymptotic modes $\alpha\,z^{\Delta_-}$ and $\beta\,z^{\Delta_+}$ are normalizable. We imposed double‑trace boundary conditions $\beta=\kappa\,\alpha$, corresponding to double‑trace deformations in the dual CFT$_2$. Both global AdS$_3$ and BTZ black holes can become unstable under such boundary conditions when the double‑trace coupling $\kappa$ is sufficiently negative.

The double‑trace instability of BTZ black holes (and of AdS$_3$) exhibits features that are markedly different from those encountered in more familiar black‑hole instabilities. Perhaps the most striking one is that whenever BTZ (or AdS$_3$) is unstable to non‑axisymmetric modes with $m\geq1$, it is already unstable to all lower modes, including the axisymmetric $m=0$ sector. However, there exist regions of parameter space where BTZ (or AdS$_3$) is unstable only to $m=0$ perturbations. In this sense, the double‑trace instability provides an example of an unconventional ``finite‑$m$ instability’’, whose first known realization arose in the context of superradiant instabilities of rotating black strings~\cite{Dias:2022mde}. This behaviour contrasts sharply with standard black‑hole instabilities, where instability under a low‑$m$ mode typically implies instability under all higher‑$m$ modes. For instance, Kerr–AdS$_4$ black holes with angular velocity just above the Hawking–Reall bound $\Omega_{+}L=1$ are unstable to arbitrarily large $m$, but remain stable to $m=1$ or $m=2$ perturbations; lower‑$m$ instabilities only appear for significantly larger $\Omega_{+}L$~\cite{Dias:2013sdc,Cardoso:2013pza,Ishii:2020muv,Dias:2022mde}.

As in many other black‑hole instabilities, the onset of the double‑trace instability signals a bifurcation, for each $m$, to new families of hairy solutions. The main results of this paper can be summarized as follows:
\begin{itemize}

\item For values of $\kappa$ for which AdS$_3$ is linearly stable, we constructed regular boson stars with $m\geq0$ that are perturbatively connected to AdS$_3$, in the sense that they arise as the nonlinear back‑reaction of an AdS$_3$ normal mode (see Section~\ref{sec:PhaseDiag-m0:BStar} and Fig.~\ref{fig:BS-m0} for $m=0$, and Section~\ref{sec:PhaseDiag-m1:BStar} and Fig.~\ref{fig:BS-m1} for $m\geq1$). Such regular boson stars also exist in the Dirichlet and Neumann limits of the boundary conditions (see Appendix~\ref{secA:BStars-DN}).

\item For values of $\kappa$ where AdS$_3$ develops the Ishibashi–Wald instability~\cite{Ishibashi:2004wx,Dias:2025uyk}, we found a new class of regular boson stars (with $m\geq0$) that are \emph{not} perturbatively connected to AdS$_3$ (see Sections~\ref{sec:PhaseDiag-m0:IshWald} and~\ref{sec:PhaseDiag-m1:IshWald}). In this regime, unstable AdS$_3$ can naturally decay into these boson stars, which may have energy lower than that of AdS$_3$ itself (see Figs.~\ref{fig:m0BSevoK} and~\ref{fig:m1BSevoK}). More precisely, it is the zero‑frequency limit of these boson stars $-$ not AdS$_3$ $-$ that provides the true ground state of the theory for $\kappa<\kappa^{\rm AdS}_{m=0,\hat{\mu}^{2}}$. This ground state is anticipated by the superpotential analysis of Ref.~\cite{Faulkner:2010fh}, revisited in Appendix~\ref{secA:superpotentials}, and is shown explicitly in Fig.~\ref{fig:m0m1omega0BS}. 

\item The $m=0$ zero‑frequency boson stars lie above, but close to, the minimum‑energy bound implied by the positivity‑of‑energy theorem~\eqref{GlobalMin} of Ref.~\cite{Faulkner:2010fh} for the theory with \textit{anti‑periodic} boundary conditions (see Appendix~\ref{secA:superpotentials} and Fig.~\ref{fig:m0m1omega0BS} for details). That is, the energy of the theory is bounded from below, and for sufficiently small values of the double‑trace parameter, the $m=0$ zero‑frequency boson star realises the ground state. 

\item We explicitly constructed singular $m=0$ static/rotating extremal hairy black holes (see Sections~\ref{sec:NumericalSetup:SingBHm0J0}~and~\ref{sec:NumericalSetup:singBHm0J} and Figs.~\ref{fig:dMJ:3families},~\ref{fig:m0TotalPhaseDiag-k04}~and~\ref{fig:m0TotalPhaseDiag-3k}) and singular $m\geq 1$ extremal hairy black holes (see Section~\ref{sec:NumericalSetup:singBHmJ} and Figs.~\ref{fig:BS-m1},~\ref{fig:m1:dMJ:3families}~and~\ref{fig:m1TotalPhaseDiag-3k}), which arise as limiting solutions of $m=0$ or $m\geq 1$ non-extremal hairy black holes, respectively. These solutions imply non‑uniqueness not only for rotating configurations, but also for static ones. 

\item The singular $m=0$ extremal hairy black holes saturate—sharply—the minimum‑energy bound implied by the positivity‑of‑energy theorem~\eqref{GlobalMin} of Ref.~\cite{Faulkner:2010fh} for the theory with \textit{periodic} boundary conditions (see Appendix~\ref{secA:superpotentials} for details). That is, the energy of the theory is bounded from below, and the singular $m=0$ extremal hairy black hole family realises the ground state.  

\item We found static and rotating hairy black holes for all $m\geq0$, bifurcating from the BTZ family at the onset of the double‑trace instability. For $m=0$, see Sections~\ref{sec:PhaseDiag-m0:BHs} and~\ref{sec:PhaseDiag-Total-m0}; for $m\geq1$, see Sections~\ref{sec:PhaseDiag-m1:BHs} and~\ref{sec:PhaseDiag-Total-m1}. Static $m=0$ hairy black holes were anticipated by the linear analysis of~\cite{Dias:2025uyk} and are illustrated in Figs.~\ref{fig:BS-m0} and~\ref{fig:m0_J0-BH}. Interestingly, for certain ranges of $\kappa$ we also find static hairy black holes even for $m>0$ (see the bottom panel of Fig.~\ref{fig:m1TotalPhaseDiag-3k}).

\item Whenever they coexist with BTZ and $m\geq1$ hairy black holes, the static and rotating $m=0$ hairy black holes always dominate the microcanonical ensemble (see Section~\ref{sec:PhaseDiag-Total-m1}, Fig.~\ref{fig:m0TotalPhaseDiag-k04}, and Fig.~\ref{fig:m0m1FinalPhaseDiag-3k}).

\item Hairy black holes never dominate the canonical or grand‑canonical ensembles when $\kappa>\kappa^{\rm AdS}_{m=0,\hat{\mu}^{2}}$. In that regime, thermal AdS$_3$ dominates at low temperatures and BTZ dominates at high temperatures (see Fig.~\ref{fig:m0_kappa_m4o10_DeltaF_DeltaG}). However, for $\kappa<\kappa^{\rm AdS}_{m=0,\hat{\mu}^{2}}$ there exists a window of $\kappa$ where $m=0$ hairy black holes dominate the canonical and grand‑canonical ensembles at intermediate temperatures (see Fig.~\ref{m0_m1_kappa_m8o10_DeltaF_DeltaG}). In these theories, the thermal phase dominance sequence is: zero‑frequency boson star at low temperature, $m=0$ hairy black hole at intermediate temperature, and BTZ at high temperature.
\end{itemize}

A correct determination of the dominant thermodynamic phase requires computing the \emph{physical}, conserved mass and angular momentum, as well as the corresponding entropy. These quantities can be computed either via the covariant Noether‑charge (covariant phase‑space) formalism~\cite{Lee:1990nz,Wald:1993nt,Iyer:1994ys,Wald:1999wa} or by holographic renormalization~\cite{Balasubramanian:1999re,deHaro:2000vlm,Skenderis:2002wp,Papadimitriou:2005ii}. In the presence of mixed (Robin) boundary conditions, it is essential to include not only divergent counterterms but also appropriate finite counterterms to ensure a well‑posed variational principle. Failure to do so leads to incorrect definitions of energy and angular momentum and to spurious scalar contributions in the first law. To avoid such ambiguities, in Appendix~\ref{secA:HoloRen} we provide a detailed holographic‑renormalization computation of all conserved charges, and in Appendix~\ref{secA:NoetherFirstLaw} we reproduce the same results using the Noether‑charge formalism. Both methods agree and yield the expressions \eqref{Mass:Def}–\eqref{FirstLaw} employed throughout this work.

Although we did not explicitly study linear stability, we conjecture that $m=0$ hairy black holes are linearly stable under all double‑trace perturbations, whereas any $m\geq1$ hairy black hole is unstable to perturbations with $\tilde m<m$, and in particular to $\tilde m=0$ modes. This conjecture is supported by two independent observations. First, any BTZ black hole unstable to an $m$‑mode is necessarily unstable to all lower‑$m$ modes, a property we expect to persist for the associated hairy solutions (with the caveat that an $m$‑hairy black hole is marginally stable to the mode that generates it). Second, whenever $m\geq1$ hairy black holes exist at fixed $\{\hat{M},\hat{J}\}$, $m=0$ hairy black holes also exist and always possess higher entropy.

If this picture is correct, the time evolution of a system perturbed by double‑trace interactions should proceed through a cascade of metastable configurations with decreasing $m$, ultimately settling into a stable $m=0$ hairy black hole. This behaviour contrasts sharply with the evolution driven by conventional superradiant instabilities, such as those of Kerr–AdS$_4$, where numerical studies~\cite{Chesler:2018txn,Chesler:2021ehz} indicate a cascade toward increasingly higher‑$m$ black resonators~\cite{Dias:2011at,Stotyn:2011ns,Dias:2015rxy,Ishii:2018oms,Ishii:2021xmn}, potentially culminating in a ``grey galaxy’’ configuration~\cite{Kim:2023sig}.

Finally, the AdS$_3$/BTZ double‑trace instability and its associated hairy boson stars and hairy black holes provide a valuable prototype for analogous phenomena in higher‑dimensional AdS spacetimes. In particular, static charged AdS$_4$ black holes are known to exhibit double‑trace instabilities for both $m=0$ and $m\geq1$ modes~\cite{Faulkner:2010gj,Dias:2013bwa,Katagiri:2020mvm,Harada:2023cfl,Kinoshita:2023iad}. Although further work is required, our results suggest that similar cascades of double‑trace instabilities and novel hairy solutions may be generic in higher dimensions (work is in progress and we will report it somewhere else).

While our analysis focused on double‑trace deformations, general multi‑trace deformations in AdS$_3$/CFT$_2$ are also of considerable interest~\cite{Klebanov:1999tb,Witten:2001ua,Amsel:2007im,Sever:2002fk,Berkooz:2002ug,Hertog:2004dr,Martinez:2004nb,Hertog:2004ns,Amsel:2006uf,Faulkner:2010fh,Faulkner:2010gj,Witten:2003ya,Ishibashi:2004wx,Marolf:2006nd,Compere:2008us}. Our results may provide useful guidance for future studies of multi‑trace perturbations and the corresponding families of hairy solutions.

Moreover, in dimensional reductions of eleven‑dimensional or type‑II supergravities, scalar fields with multi‑trace boundary conditions often arise in the lower‑dimensional effective theories. Our analysis indicates that the bald black holes of such systems should generically be unstable to the formation of hairy black holes of the type constructed in this paper.

\begin{acknowledgments}
The authors would like to acknowledge Paolo Arnaudo, Javier Carballo, Kostas Skenderis and Ben Withers for insightful discussions.
O.J.C.D. acknowledges financial support from the  STFC consolidated grant~ST/X000583/1. D.S.G acknowledges financial support from a STFC Ph.D scholarship.  
J.~E.~S. has been partially supported by STFC consolidated grant ST/X000664/1 and by Hughes Hall College. The authors also acknowledge the use of the IRIDIS High Performance Computing Facility, and associated support services at the University of Southampton, in the completion of this work.
\end{acknowledgments}

\appendix
\section{Arbitrarily negative \texorpdfstring{$\kappa < 0$}{kappa<0} theories admit ground states}\label{secA:superpotentials}

We impose double‑trace boundary conditions of the form
\begin{equation}
  \beta = \kappa\,\alpha ,
\end{equation}
with $\kappa<0$. In the dual CFT$_2$, this corresponds to deforming the action by a term
\begin{equation}
  -\kappa \int \mathcal{O}^{\dagger}\mathcal{O}.
\end{equation}
At first sight, allowing $\kappa$ to become arbitrarily negative appears dangerous, as it suggests the possibility that the energy of the theory becomes unbounded from below, thus precluding the existence of a ground state. However, it was shown in \cite{Faulkner:2010fh} (see also \cite{Hertog:2004ns,Cheng:2005wk,Amsel:2006uf}) that this conclusion is not generic: provided certain conditions are satisfied, the energy can remain bounded even for arbitrarily negative $\kappa$. In this appendix, we demonstrate that the theories studied in this paper satisfy precisely those conditions.

Numerically, we find that all solutions with $\omega \neq 0$ and $m \neq 0$ have higher energy than those with $\omega = m = 0$. Henceforth, we restrict attention to the case $m = \omega = 0$. Note that this includes both static and rotating ground states.

For our setup we have $d=3$ and choose
\begin{equation}
  \hat{\mu}^2 = -\frac{15}{16},
\end{equation}
so that
\begin{equation}
  \Delta_- = \frac{3}{4},
  \qquad
  \Delta_+ = \frac{5}{4},
\end{equation}
and the scalar potential reads\footnote{To match the normalization conventions used in \cite{Faulkner:2010fh}, we rescale the scalar field as $\Phi\to\phi/\sqrt{2}$. This overall normalization does not affect the equations of motion, and all equations imported from \cite{Faulkner:2010fh} have been adjusted accordingly.}
\begin{equation}
  V(\Phi) = -1 - \frac{15}{16}\,\Phi^2.
\end{equation}
The generalised minimum energy theorem of \cite{Faulkner:2010fh} states that the total energy obeys the bound 
\begin{equation}
\label{eqn:E_min_condition}
  8\pi G\,\left(E-p\,E_{\hbox{\tiny AdS$_3$}}-\frac{|J|}{L}\right) \geq \oint I\,{\rm d}S\quad\text{with}\quad I\equiv \frac{W(\alpha)}{2}+ \frac{3\,s_c}{16}\,|\alpha|^{8/3}
\end{equation}
where the integral is taken over a constant‑time slice of the AdS$_3$ conformal boundary, \ie over the unit circle $S^1$ at large radius, and $E_{\rm AdS_3}$ is the Casimir energy of AdS$_3$. With periodic boundary conditions for the Witten-Nester spinor, we have $p=0$, while for anti-periodic boundary conditions we have $p=1$. Therefore, to establish boundedness of the energy, it suffices to show that the integrand in~\eqref{eqn:E_min_condition} is bounded from below. This requires determining two quantities: the function $W(\alpha)$ and the constant $s_c$.

By definition, $W(\alpha)$ encodes the double‑trace boundary condition via
\begin{equation}
\label{eqn:W_definition}
  \beta = \frac{{\rm d}W}{{\rm d}\alpha}.
\end{equation}
For $\beta=\kappa\alpha$, this integrates immediately to
\begin{equation}
  W(\alpha)=\frac{\kappa}{2}\,\alpha^2 .
\end{equation}

The coefficient $s_c$ is defined as the \emph{largest} value of $s$ for which an auxiliary superpotential $P_s(\Phi)$ exists globally. Following \cite{Faulkner:2010fh}, we restrict to $\Phi>0$ and demand that $P_s(\Phi)$ be an even function, which allows extension to $\Phi<0$.

The superpotential obeys the nonlinear differential equation
\begin{equation}
\label{eqn:Ps_equation}
  \frac{1}{2}\,P_s'(\Phi)^2 - 2\,P_s(\Phi)^2
  = -1 - \frac{15}{16}\,\Phi^2 ,
\end{equation}
and near $\Phi=0$ must have the asymptotic expansion
\begin{equation}
\label{eqn:Ps_expansion}
  P_s(\Phi)
  = \frac{1}{\sqrt{2}}
    + \frac{3\,\Phi^2}{4\sqrt{2}}
    - s\,\frac{3\,\Phi^{8/3}}{2^{19/6}}
    + \cdots .
\end{equation}
Selecting the positive branch appropriate to~\eqref{eqn:Ps_expansion}, one finds
\begin{equation}
\label{eqn:Ps_positive_root_eqn}
  P_s'(\Phi)
   = \sqrt{2}\,
     \sqrt{2 P_s(\Phi)^2
           - \frac{15}{16}\,\Phi^2
           - 1 } .
\end{equation}

For numerical convenience, we define
\begin{equation}
  y_s(\Phi) \equiv P_s'(\Phi).
\end{equation}
Differentiating~\eqref{eqn:Ps_positive_root_eqn} and eliminating $P_s(\Phi)$ yields a first‑order equation for $y_s(\Phi)$:
\begin{equation}
\label{eqn:ys_eqn}
  y_s'(\Phi)
   = \sqrt{8
             + \frac{15}{2}\,\Phi^2
             + 4\,y_s(\Phi)^2 } -\frac{15\,\Phi}{8\,y_s(\Phi)}.
\end{equation}
From~\eqref{eqn:Ps_expansion}, the boundary condition near $\Phi=0$ is
\begin{equation}
  y_s(\Phi)
  = \frac{3 \Phi }{2 \sqrt{2}}-\frac{s}{2^{1/6}} \Phi ^{5/3}-\frac{5}{3} 2^{1/6} s^2 \Phi ^{7/3}+\frac{27-80 s^3}{2^{1/2}12}\Phi ^3
    + \mathcal{O}(\Phi^{11/3}) .
\end{equation}

Global existence of the superpotential requires
\begin{equation}
  y_s(\Phi) > 0 \qquad \text{for all }\Phi>0,
\end{equation}
since~\eqref{eqn:ys_eqn} becomes singular when $y_s(\Phi)=0$. As shown in \cite{Faulkner:2010fh}, solutions with different values of $s$ do not cross except at singular points. Consequently, once a value $s_1$ is found for which $y_{s_1}(\Phi)>0$ for all $\Phi$, all smaller values $s<s_1$ also admit globally defined solutions.

\begin{figure}[ht]
    \centering
    \includegraphics[width=0.5\linewidth]{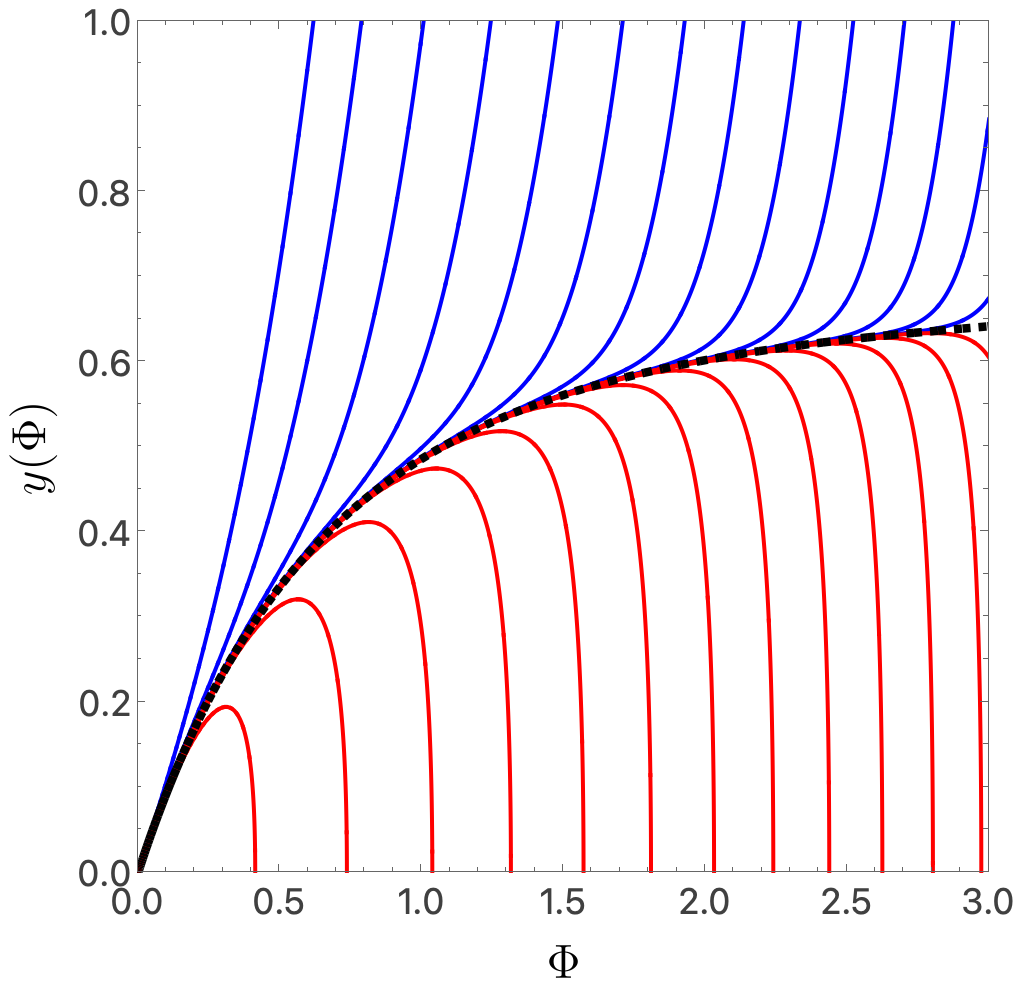}
    \caption{Solutions to \eqref{eqn:ys_eqn}. Blue curves are some selected generic solutions. The orange black dashed line corresponds to $y_s(\Phi)$ with $s_c = 0.551079483443393$, the solution with the largest value of $s_c$ which exists for all $\Phi$.}
    \label{fig:mum15o16_yVSPsiplot}
\end{figure}

Proceeding numerically using this criterion, we find
\begin{equation}
0.551079483443393<s_c<0.551079483443394 \,.
\end{equation}
Representative solutions are shown in Fig.~\ref{fig:mum15o16_yVSPsiplot}.
The precise numerical value beyond the quoted precision depends on integration accuracy; what is crucial is that
\begin{equation}
  s_c > 0\,.
\end{equation}

With all ingredients determined, the integrand in~\eqref{eqn:E_min_condition} becomes
\begin{equation}
  I= \frac{\kappa}{4}\,\alpha^2+ \frac{3\,s_c}{16}\,|\alpha|^{8/3},\qquad s_c>0 .
\end{equation}

If $\kappa\geq0$, the integrand is manifestly non‑negative and admits a unique global minimum at $\alpha=0$, independently of the value of $s_c$. In the full gravitational problem, on‑shell solutions relate $\alpha$ and $\kappa$ through the equations of motion, \ie one has $\alpha=\alpha(\kappa)$. However, when the goal is to establish a lower bound on the energy functional~\eqref{eqn:E_min_condition}, it suffices to treat $\alpha$ as an unconstrained real variable and minimize the integrand with respect to $\alpha$ at fixed $\kappa$. The resulting bound is therefore obtained off‑shell. On‑shell configurations need not saturate this bound, since the dynamically determined relation $\alpha(\kappa)$ does not in general extremise the integrand. Nevertheless, the bound guarantees that all on‑shell solutions have energies greater than or equal to the derived minimum, which is precisely the desired result.

Let us now consider $\kappa<0$. Since the function is even, it suffices to restrict to $\alpha\geq0$. In this case, we have
\begin{equation}
  I'(\alpha)
   = -\frac{|\kappa|}{2}\,\alpha
     + \frac{s_c}{2}\,\alpha^{5/3},
  \qquad
  I''(\alpha)
   = -\frac{|\kappa|}{2}
     + \frac{5\,s_c}{6}\,\alpha^{2/3}.
\end{equation}
The stationary points satisfy
\begin{equation}
  I'(\alpha)=0
  \quad\Longrightarrow\quad
  \alpha\in
  \left\{0,
         \left(\frac{|\kappa|}{s_c}\right)^{3/2}
  \right\}.
\end{equation}
Since $I''(0)=-|\kappa|/2<0$, $\alpha=0$ is a local maximum. By contrast,
\begin{equation}
  I''\!\left(
        \left(\frac{|\kappa|}{s_c}\right)^{3/2}
      \right)
   = \frac{|\kappa|}{3} > 0,
\end{equation}
so the second critical point corresponds to a minimum. By symmetry, a second minimum occurs at
\begin{equation}
  \alpha = -\left(\frac{|\kappa|}{s_c}\right)^{3/2}.
\end{equation}

We conclude that the integrand in~\eqref{eqn:E_min_condition} is bounded from below for all $\kappa\in\mathbb{R}$. In particular, for $\kappa<0$ the energy attains a new, lower global minimum at
\begin{equation}\label{GlobalMin}
  8 \pi G \, \left(E-p\,E_{\hbox{\tiny AdS$_3$}}-\frac{|J|}{L}\right)\geq -\frac{\pi  \kappa ^4}{8 s_c^3} \iff E-p\,E_{\hbox{\tiny AdS$_3$}} -\frac{|J|}{L}\geq -\frac{1}{8G}\frac{\kappa ^4}{8 s_c^3},
\end{equation}
corresponding physically to the appearance of a new ground state. Thus, even arbitrarily negative values of the double‑trace coupling $\kappa$ lead to a consistent theory with a well‑defined ground state.

This analytical bound agrees with our numerical findings. Indeed, the minimum energy $ (\hat{E}_{\rm min}-p\,\hat{E}_{\hbox{\tiny AdS$_3$}}-|\hat{J}|) = -\kappa^4/8s_c^3$ (recall that $\hat{E}=8G E$ and $\hat{J}=8G J/L$) evaluates to
\begin{align}\label{Super:Emin}
   \hat{E}_{\rm min}(\kappa)-p\,E_{\hbox{\tiny AdS$_3$}}-|\hat{J}|= \begin{cases}
        -0.019121 & \kappa =-4/10, \\
        -0.305934 & \kappa = -8/10, \\
        -0.416027 & \kappa = -8639/10000, \\
        -0.490047 & \kappa = -9/10\,.
    \end{cases}
\end{align}

For $p=0$, the above bound matches sharply the numerical masses of the singular $m=0$ static and rotating extremal hairy black holes with the same value of $\kappa$; see \eqref{sBHm0J0:Thermo} and \eqref{sBHm0J:Thermo}, and the red diamond in Figs.~\ref{fig:BS-m0},~\ref{fig:m0_J0-BH},~\ref{fig:dMJ:3families}, and Fig.~\ref{fig:m0TotalPhaseDiag-k04} for $\kappa=-0.4$, as well as Fig.~\ref{fig:m0TotalPhaseDiag-3k} (including its caption) for the remaining three values of $\kappa$ appearing in \eqref{Super:Emin}.

On the other hand, for $p=1$, the bound~\eqref{GlobalMin} is \emph{not} sharp, although it is satisfied by all smooth boson stars. Recall that the minimum‑energy boson star we find for any value of $\kappa$ is the regular $m=0$ static zero‑frequency boson star, whose energy as a function of $\kappa$ is shown as the black curve in Fig.~\ref{fig:m0m1omega0BS} (and also in Fig.~\ref{fig:m0BSevoK}). In Fig.~\ref{fig:m0m1omega0BS}, we also display the lower bound implied by~\eqref{GlobalMin} for $p=1$ as a magenta dotted line. Consistent with the discussion above, this bound lies below the black zero‑frequency boson star for $\kappa < \kappa^{\hbox{\tiny AdS}}_{m=0,\hat{\mu}}$ (where the latter exists), and below the Casimir energy of AdS$_3$ for $0 > \kappa > \kappa^{\hbox{\tiny AdS}}_{m=0,\hat{\mu}}$.\footnote{For the values of $\kappa$ shown, the zero‑frequency boson stars of Fig.~\ref{fig:m0m1omega0BS} have energies
\begin{align}
\hat{E} = \begin{cases}
-1.167142 & \kappa = -8/10, \\
-1.261255 & \kappa = -8639/10000, \\
-1.326726 & \kappa = -9/10 \,,
\end{cases}
\end{align}
all of which lie above the corresponding values of \eqref{Super:Emin} for $p=1$ and $\hat{J}=0$.}

\section{\texorpdfstring{AdS$_3$}{AdS3} boson stars with Dirichlet and Neumann boundary conditions \label{secA:BStars-DN}}

In this section, we perturbatively construct the ground‑state hairy boson stars with $m=0$ and $m=1$, imposing either Dirichlet ($\alpha=0$) or Neumann ($\beta=0$) boundary conditions on the scalar field. The coefficients $\alpha$ and $\beta$ were introduced in~\eqref{FGscalar}. By \emph{ground‑state} boson stars we mean those configurations with the lowest energy for a given scalar mass $\hat{\mu}$ and azimuthal number $m$, \ie those arising as the nonlinear back‑reaction of the global AdS$_3$ normal mode with the lowest radial overtone $n=0$ (which, equivalently, corresponds to the mode with no radial nodes in the wavefunction).

We perform this construction for normal modes of global AdS$_3$ subject to either Dirichlet ($\alpha=0$) or Neumann ($\beta=0$) boundary conditions. In both cases, the corresponding normal‑mode frequencies are known analytically and are given by~\cite{Burgess:1984ti,Dias:2025uyk}
\begin{subequations}\label{AdS3-wG}
\begin{align}
   \hat{\omega}_{\pm}^{\hbox{\tiny AdS}}\big|_{\hbox{\tiny Dir}}
      &= \pm\left(m+2n+\Delta_{+}\right),\\
   \hat{\omega}_{\pm}^{\hbox{\tiny AdS}}\big|_{\hbox{\tiny Neu}}
      &= \pm\left(m+2n+\Delta_{-}\right),
\end{align}
\end{subequations}
where $n$ labels the radial overtone, and $n=0$ corresponds to the ground state. For definiteness, and in order to facilitate comparison with the rest of the paper, we focus on the scalar mass $\hat{\mu}^2=-15/16$. Recall, however, that Dirichlet boson stars exist for any scalar mass above the Breitenlohner–Freedman (BF) bound, $\hat{\mu}^2>-1$, while Neumann boson stars exist only in the window between the BF and unitarity bounds, $-1<\hat{\mu}^2<0$. For $\hat{\mu}^2=-15/16$ and $n=0$, \eqref{AdS3-wG} reduces to
\begin{subequations}\label{AdS3-w}
\begin{align}
   \hat{\omega}_{\pm}^{\hbox{\tiny AdS}}\big|_{\hbox{\tiny Dir}}
      &= \pm\left(m+\tfrac{5}{4}\right), \label{AdS3-wDir}\\
   \hat{\omega}_{\pm}^{\hbox{\tiny AdS}}\big|_{\hbox{\tiny Neu}}
      &= \pm\left(m+\tfrac{3}{4}\right). \label{AdS3-wNeu}
\end{align}
\end{subequations}

In what follows, we construct boson stars with $\omega\geq0$, $m=0$ or $m=1$, and either Dirichlet or Neumann boundary conditions. This yields four distinct perturbative solutions, which we later compare against the fully nonlinear numerical solutions.

To obtain these boson stars perturbatively, we solve the coupled system of ordinary differential equations for the fields $\{f,g,\Omega,\psi\}$ order by order in a small‑amplitude expansion using methods similar to those detailed \eg in \cite{Dias:2011at,Dias:2011tj,Dias:2016pma}. Regularity at the origin requires the boundary conditions~\eqref{BCs:OriginBSm0} for $m=0$ or~\eqref{BCs:OriginBSm} for $m=1$, while the asymptotic boundary conditions are $\alpha=0$ (Dirichlet) or $\beta=0$ (Neumann), as specified in~\eqref{UVexpansion:m15o16}. For each field equation, the two integration constants are completely fixed by these boundary conditions.

We carry out perturbation theory around global AdS$_3$, using the scalar‑field amplitude $\epsilon$ at infinity as the expansion parameter. Explicitly, we take
\begin{equation}
\epsilon \equiv \beta
\quad\text{for Dirichlet boundary conditions},\qquad
\epsilon \equiv \alpha
\quad\text{for Neumann boundary conditions}.
\end{equation}
Since the energy‑momentum tensor depends quadratically on the scalar field, the first non‑trivial back‑reaction on the metric fields arises at $\mathcal{O}(\epsilon^{2})$. The structure of the equations of motion~\eqref{4ODEs} then implies that only even powers of $\epsilon$ enter the metric functions, while only odd powers contribute to the scalar field. We also determine the perturbative corrections to the frequency $\omega$, which at linear order is given by the normal‑mode frequencies~\eqref{AdS3-w} and receives higher‑order corrections at each step in the expansion.

Before proceeding, it is worth explaining why we do \emph{not} perform an analogous perturbative construction for boson stars obeying double‑trace boundary conditions, which were instead obtained numerically in Sections~\ref{sec:PhaseDiag-m0:BStar} and~\ref{sec:PhaseDiag-m1:BStar}. In principle, such a perturbative construction could be attempted. However, already at leading order one encounters a fundamental obstacle: the normal‑mode frequencies associated with double‑trace boundary conditions cannot be obtained analytically. Instead, for given values of $\hat{\mu}$, $\kappa$, and $m$, the frequency must satisfy a transcendental quantization condition, namely~(3.15) of our companion paper~\cite{Dias:2025uyk}. While this condition reproduces~\eqref{AdS3-wG} in the Dirichlet and Neumann limits, it has no closed‑form solution for generic double‑trace boundary conditions and must be solved numerically. For this reason, we restrict the perturbative analysis in this section to the analytically tractable Dirichlet and Neumann cases.

\subsection{Static Dirichlet and Neumann boson stars with \texorpdfstring{$m =0$}{m=0}}\label{secA:BStars-DNm0}
For $m=0$, the residual gauge freedom~\eqref{ResidualGaugeFreedom:m0}, together with the equations of motion~\eqref{4ODEs}, forces the angular velocity to vanish identically, $\Omega(R)=0$. Consequently, $m=0$ boson stars are necessarily static and carry zero angular momentum, $\hat{J}=0$.

\begin{figure}[t]
    \centering
    \includegraphics[width=0.4\linewidth]{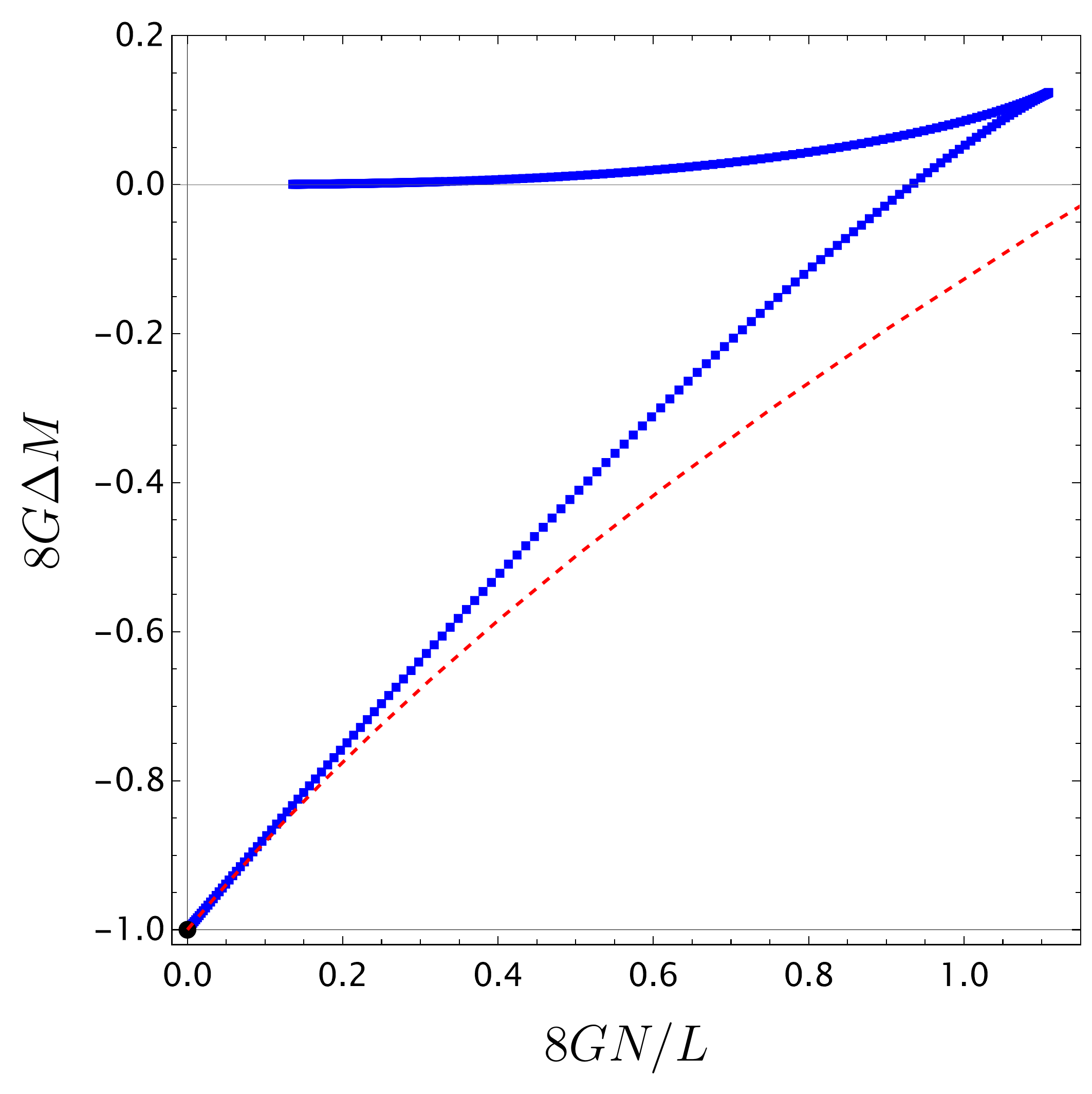}
    \includegraphics[width=0.4\linewidth]{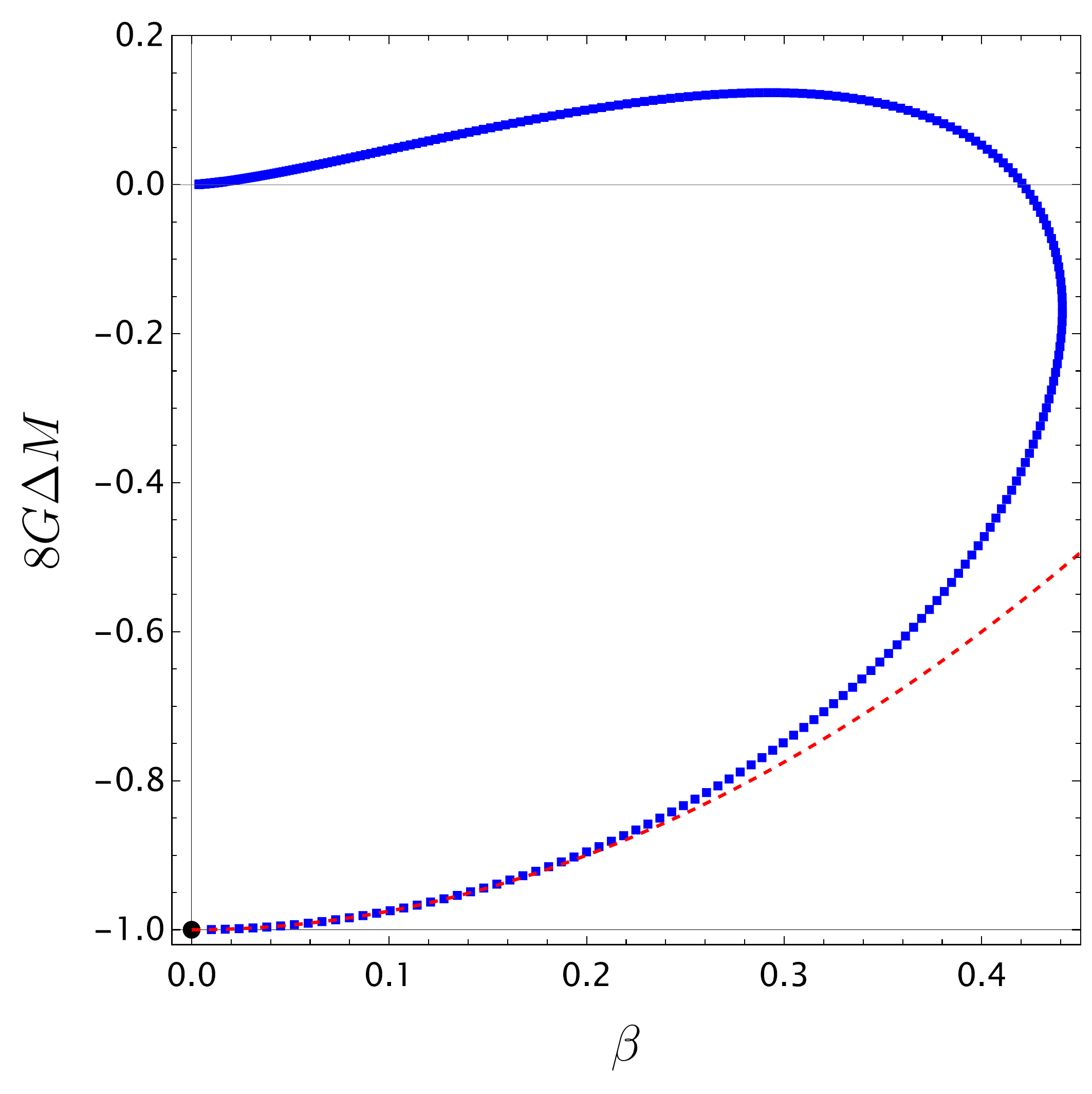}
    \includegraphics[width=0.4\linewidth]{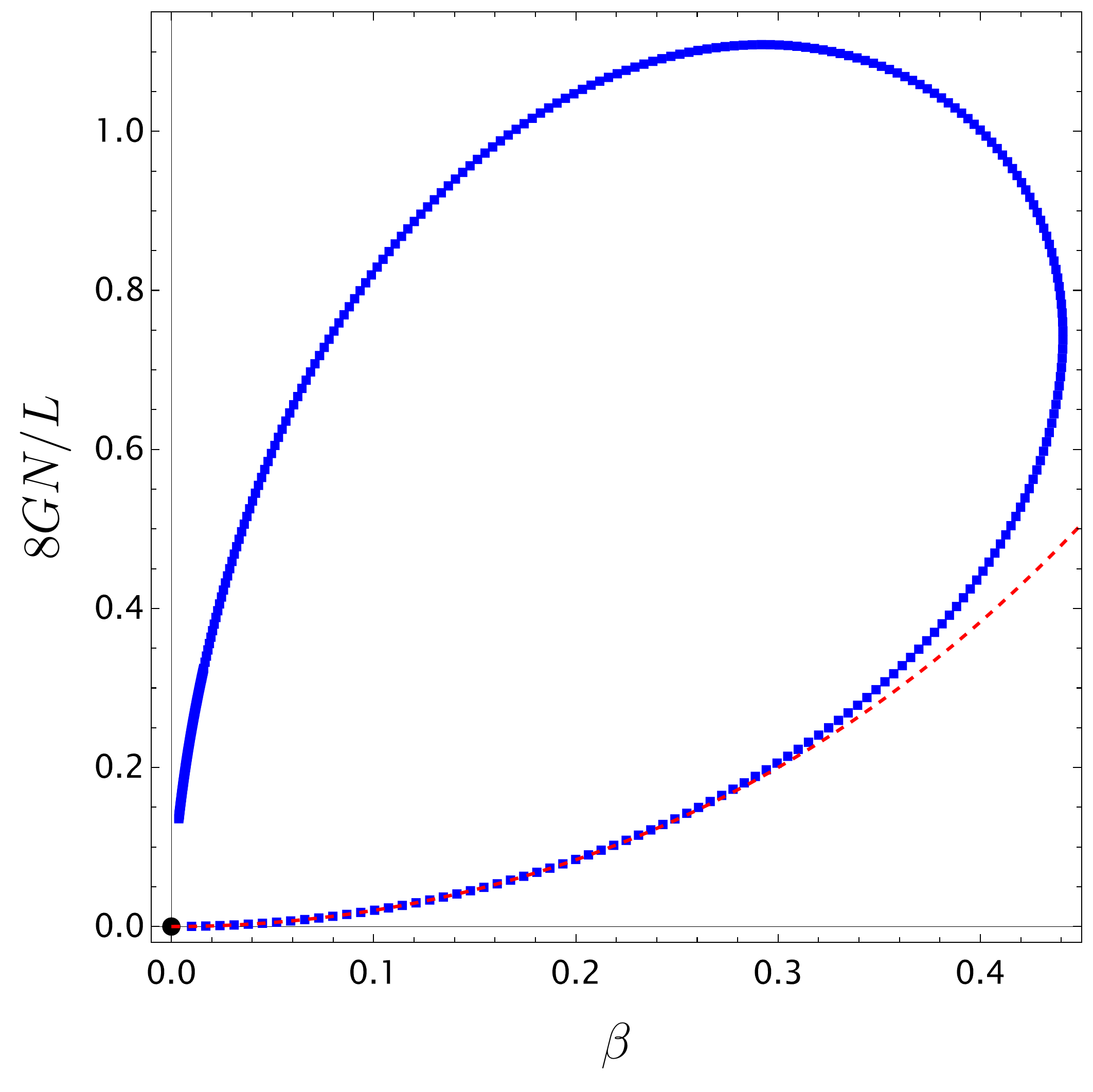}
    \includegraphics[width=0.4\linewidth]{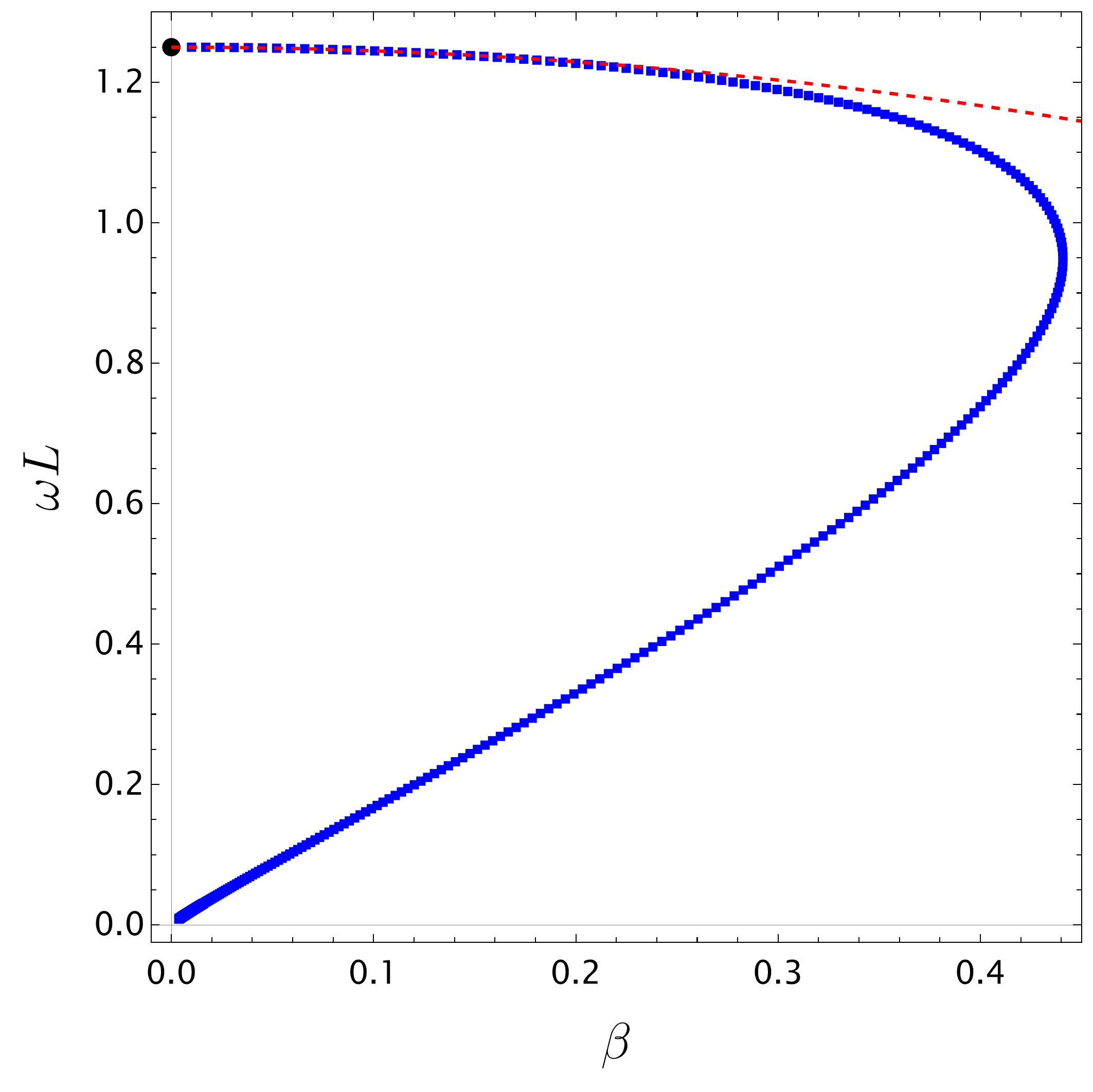}
    \caption{Physical properties (mass $\hat{M}$, $U(1)$ conserved charge $\hat{N}$, frequency $\hat{\omega}$ and asymptotic scalar field amplitude $\beta$) of {\bf Dirichlet} $\bm{m = 0}$ boson stars with $\mu^2L^2 = -15/16$. The red dashed line describes the perturbative result \eqref{eqn:Dir_m0_BS_metricexp_M}, \eqref{eqn:Dir_m0_BS_metricexp_N} and\eqref{eqn:Dir_m0_BS_metricexp_omega}. The blue square curve describes the exact numerical result found using the methods described in section~\ref{sec:NumericalSetup:RegBS}.}
\label{fig:m0_DirBS_numerics_VS_perturbation}
\end{figure}
\paragraph{Dirichlet boundary conditions ($\alpha=0$).}
Setting $m=0$ and imposing Dirichlet boundary conditions $(\alpha=0)$, we obtain the following perturbative expansions in the small parameter $\epsilon\equiv\beta$:
\begin{subequations}
\begin{align}
\label{eqn:Dir_m0_BS_metricexp_omega}
\hat{\omega}
  &= \frac{5}{4}
     - \frac{25}{48}\,\beta^{2}
     + \mathcal{O}(\beta^{4}), \\[1mm]
f(R)
  &= 1+R^{2}
     + \frac{5}{2}
       \left(\frac{1}{(1+R^{2})^{1/4}}-1\right)\beta^{2}
     + \mathcal{O}(\beta^{4}), \\[1mm]
g(R)
  &= 1
     - \frac{5\,\beta^{2}}{2(1+R^{2})^{5/4}}
     + \mathcal{O}(\beta^{4}), \\[1mm]
\psi(R)
  &= \frac{\beta}{(1+R^{2})^{5/8}}
     + \psi_{3}(R)\,\beta^{3}
     + \mathcal{O}(\beta^{5}),
\end{align}
\end{subequations}
where
\begin{multline}
\psi_{3}(R)
 = -\frac{5}{8(1+R^{2})^{15/8}}
   \Bigg\{
     26 + 20R^{2}
     - 5 (1+R^{2})^{1/4}\!\left[3+\pi(1+R^{2})\right]
\\
     + 10(1+R^{2})^{5/4}
       \left[
         \arctan\!\left((1+R^{2})^{1/4}\right)
         - \log\!\left(
           \frac{(1+(1+R^{2})^{1/4})\sqrt{1+\sqrt{1+R^{2}}}}
                {\sqrt{1+R^{2}}}
         \right)
       \right]
   \Bigg\}.
\end{multline}

From these expansions, we compute the conserved quantities defined in
Eqs.~\eqref{N:Def} and~\eqref{Mass:Def}, obtaining
\begin{subequations}
\begin{align}
\label{eqn:Dir_m0_BS_metricexp_M}
\hat{M}
  &= -1 + \frac{5}{2}\,\beta^{2}
     + \mathcal{O}(\beta^{4}), \\
\label{eqn:Dir_m0_BS_metricexp_N}
\hat{N}
  &= 2\,\beta^{2}
     + \frac{5}{36}\!\left(15\pi-71+60\log 2\right)\beta^{4}
     + \mathcal{O}(\beta^{6}).
\end{align}
\end{subequations}
As a non‑trivial consistency check, we verify that these quantities satisfy the first law of boson star thermodynamics~\eqref{FirstLawBStar} up to $\mathcal{O}(\beta^{3})$ (since $\hat{\omega}$, $\hat{M}$ and $\hat{N}$ are known only up to $\mathcal{O}(\beta^{4})$). The thermodynamic properties of these Dirichlet $m=0$ boson stars are shown in Fig.~\ref{fig:m0_DirBS_numerics_VS_perturbation}. For sufficiently small $\beta$, the perturbative predictions (red dashed curves) are in excellent agreement with the fully nonlinear numerical solutions (blue squares).

\paragraph{Neumann boundary conditions ($\beta=0$).} 
We now turn to the case $m=0$ with Neumann boundary conditions $(\beta=0)$. The perturbative expansions are written in terms of the small parameter $\epsilon\equiv\alpha$ and take the form
\begin{subequations}
\begin{align}
\label{eqn:Neu_m0_BS_metricexp_omega}
\hat{\omega}
  &= \frac{3}{4}
     + \frac{9}{16}\,\alpha^{2}
     + \mathcal{O}(\alpha^{4}), \\[1mm]
f(R)
  &= 1+R^{2}
     + \frac{3}{2}\!\left((1+R^{2})^{1/4}-1\right)\alpha^{2}
     + \mathcal{O}(\alpha^{4}), \\[1mm]
g(R)
  &= 1
     - \frac{3\,\alpha^{2}}{2(1+R^{2})^{3/4}}
     + \mathcal{O}(\alpha^{4}), \\[1mm]
\psi(R)
  &= \frac{\alpha}{(1+R^{2})^{3/8}}
     + \psi_{3}(R)\,\alpha^{3}
     + \mathcal{O}(\alpha^{5}),
\end{align}
\end{subequations}
with
\begin{multline}
\psi_{3}(R)
 = \frac{3}{16(1+R^{2})^{11/8}}
   \Bigg\{
     3
     - 2(1+R^{2})^{1/4}
     - 3\pi(1+R^{2})
\\
     + 3(1+R^{2})
       \left[
         2\arctan\!\left((1+R^{2})^{1/4}\right)
         + \log\!\left(
           \frac{(1+(1+R^{2})^{1/4})^{2}(1+\sqrt{1+R^{2}})}
                {1+R^{2}}
         \right)
       \right]
   \Bigg\}.
\end{multline}

\begin{figure}[t]
    \centering
    \includegraphics[width=0.4\linewidth]{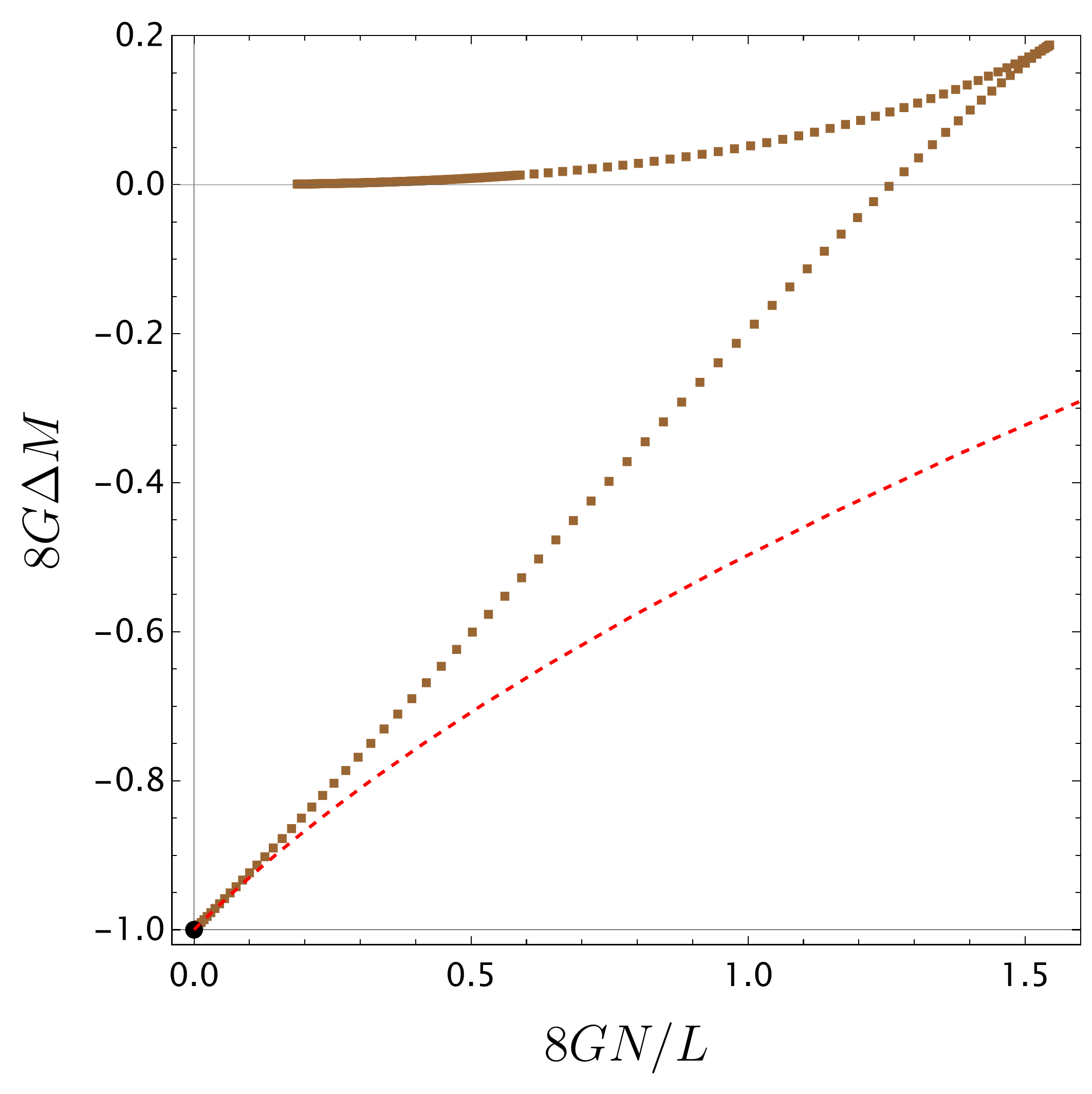}
    \includegraphics[width=0.4\linewidth]{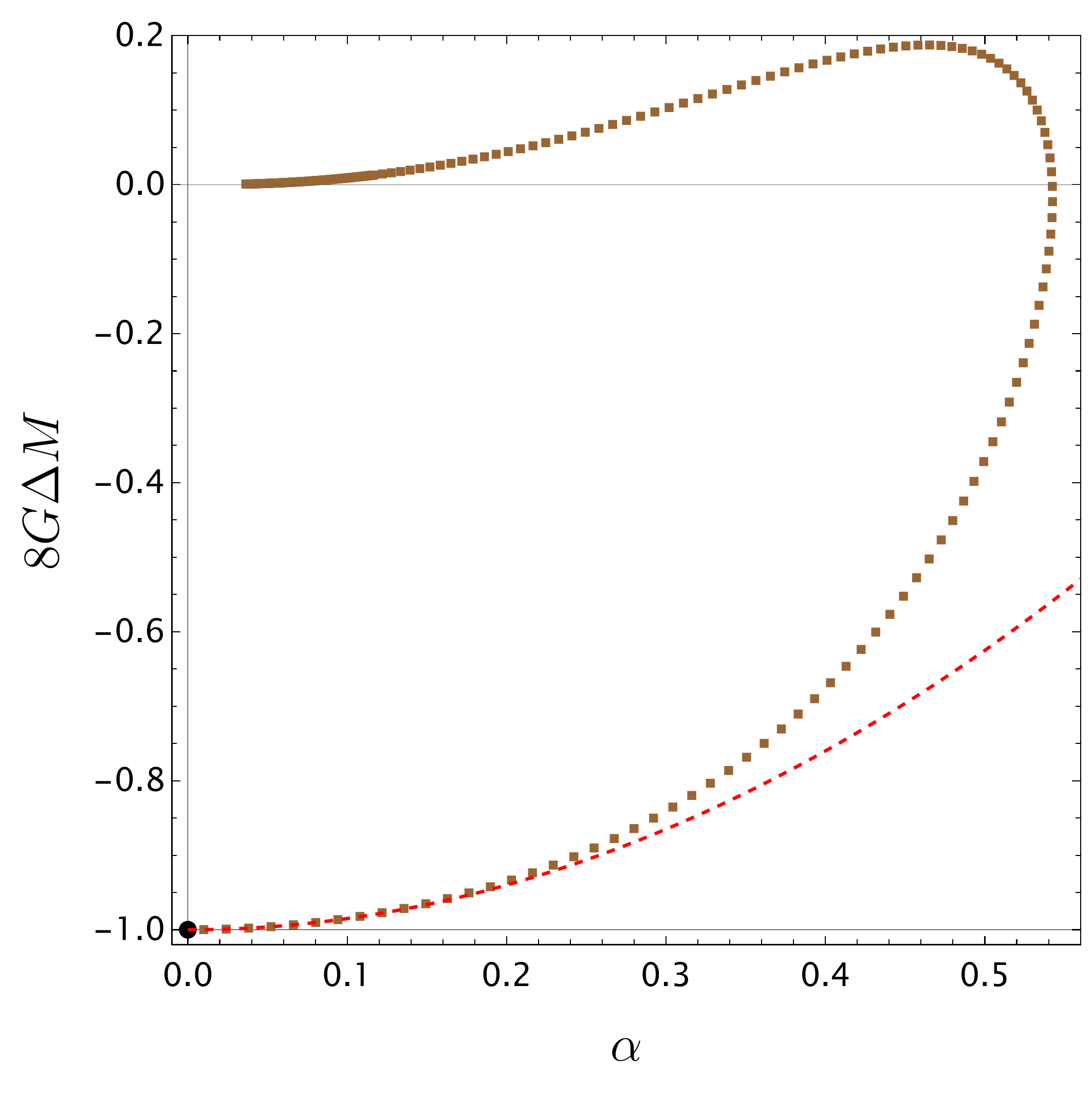}
    \includegraphics[width=0.4\linewidth]{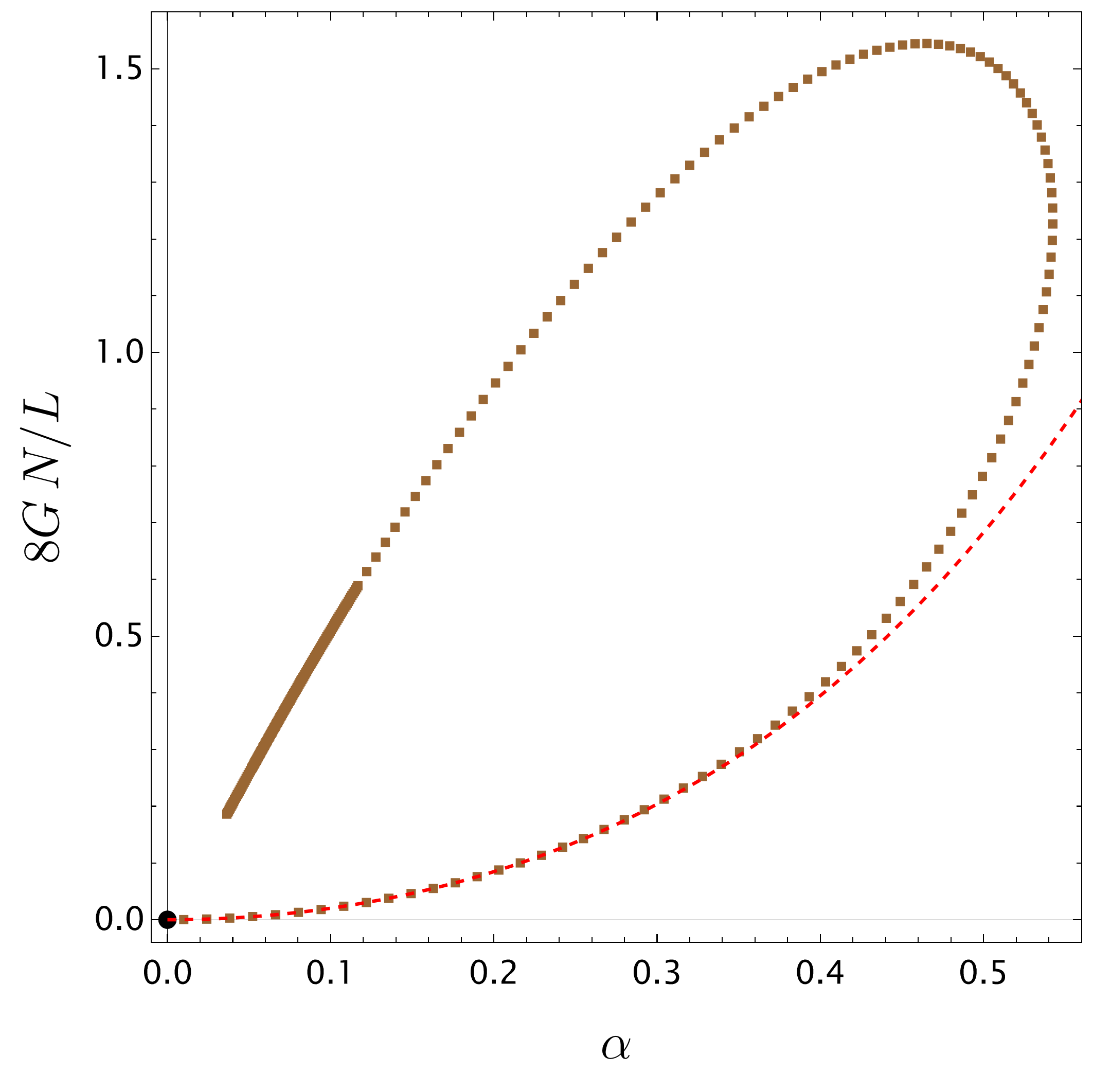}
    \includegraphics[width=0.4\linewidth]{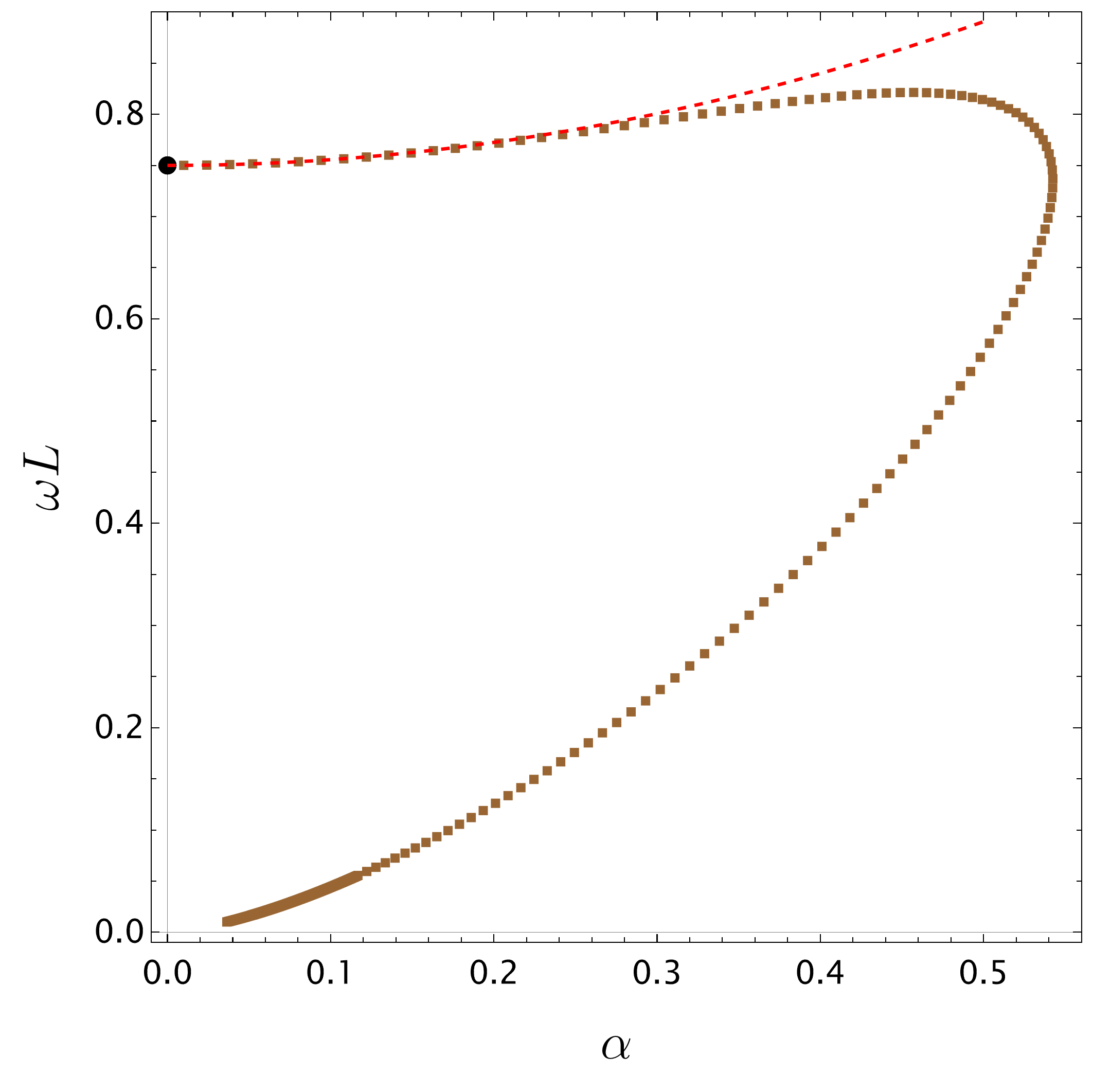}
        \caption{Physical properties (mass $\hat{M}$, $U(1)$ conserved charge $\hat{N}$, frequency $\hat{\omega}$ and asymptotic scalar field amplitude $\alpha$) of {\bf Neumann} $\bm{m = 0}$ boson stars with $\mu^2L^2 = -15/16$. The red dashed line describes the perturbative result \eqref{eqn:Neu_m0_BS_metricexp_M}, \eqref{eqn:Neu_m0_BS_metricexp_N} and \eqref{eqn:Neu_m0_BS_metricexp_omega}. The brown square curve describes the exact numerical result found using the methods described in section~\ref{sec:NumericalSetup:RegBS}.}    \label{fig:m0_NeuBS_numerics_VS_perturbation}
\end{figure}

Using these expansions, the conserved quantities \eqref{N:Def} and~\eqref{Mass:Def} are
\begin{subequations}
\begin{align}
\label{eqn:Neu_m0_BS_metricexp_M}
\hat{M}
  &= -1 + \frac{3}{2}\,\alpha^{2}
     + \mathcal{O}(\alpha^{4}), \\
\label{eqn:Neu_m0_BS_metricexp_N}
\hat{N}
  &= 2\,\alpha^{2}
     + \frac{3}{4}\!\left(5-3\pi+12\log 2\right)\alpha^{4}
     + \mathcal{O}(\alpha^{6}).
\end{align}
\end{subequations}
These quantities satisfy the first law of thermodynamics~\eqref{FirstLawBStar} up to $\mathcal{O}(\alpha^{3})$ (given that $\hat{\omega}$, $\hat{M}$, and $\hat{N}$ are computed up to $\mathcal{O}(\alpha^{4})$). The thermodynamic properties of the Neumann $m=0$ boson stars are displayed in Fig.~\ref{fig:m0_NeuBS_numerics_VS_perturbation}. Once again, for sufficiently small amplitudes, the perturbative results (red dashed curves) are in very good agreement with the exact numerical solutions (brown squares).

\subsection{Rotating Dirichlet and Neumann boson stars with \texorpdfstring{$m=1$}{m=1}}
\label{secA:BStars-DNm1}

For $m=1$, we exploit the residual gauge freedom~\eqref{ResidualGaugeFreedom:m} to fix $\Omega(\infty)=0$ in our perturbative construction.

\paragraph{Dirichlet boundary conditions ($\alpha=0$).}
Setting $m=1$ and imposing Dirichlet boundary conditions, the perturbative expansions in the small parameter $\epsilon\equiv\beta$ take the form
\begin{subequations}
\begin{align}
\label{eqn:Dir_m1_BS_metricexp_omega}
\hat{\omega}
  &= \frac{9}{4}
     - \frac{20}{21}\,\beta^{2}
     + \mathcal{O}(\beta^{4}), \\[1mm]
f(R)
  &= 1+R^{2}
     + \frac{1}{10}
       \left(
         \frac{36+25R^{2}}{(1+R^{2})^{5/4}}-36
       \right)\beta^{2}
     + \mathcal{O}(\beta^{4}), \\[1mm]
g(R)
  &= 1
     - \frac{4+5R^{2}}{2(1+R^{2})^{9/4}}\,\beta^{2}
     + \mathcal{O}(\beta^{4}), \\[1mm]
\hat{\Omega}(R)
  &= \frac{4}{5R^{2}}
     \left(1-\frac{1}{(1+R^{2})^{5/4}}\right)\beta^{2}
     + \mathcal{O}(\beta^{4}), \\[1mm]
\psi(R)
  &= \frac{R}{(1+R^{2})^{9/8}}\,\beta
     + \psi_{3}(R)\,\beta^{3}
     + \mathcal{O}(\beta^{5}),
\end{align}
\end{subequations}
where
\begin{multline}
\psi_{3}(R)
  = \frac{1}{4200\,R(1+R^{2})^{27/8}}
    \Bigg\{
      500
      - 42498R^{2}
      - 72903R^{4}
      - 32740R^{6}
\\
      + 5(1+R^{2})^{5/4}(3302R^{2}-100)
      + 51637R^{3/2}(1+R^{2})^{9/4}
\\
      \times
      \left[
        4\,{}_2F_1\!\left(\frac14,\frac14,\frac54,-\frac{1}{R^{2}}\right)
        + \sqrt{R}\bigl(2\log R-\log(1+R^{2})\bigr)
      \right]
    \Bigg\}.
\end{multline}

Using these expressions, the conserved quantities defined in
Section~\ref{sect:thermodynamics} are found to be
\begin{subequations}
\begin{align}
\label{eqn:Dir_m1_BS_metricexp_M}
\hat{M}
  &= -1 + \frac{18}{5}\,\beta^{2}
     + \mathcal{O}(\beta^{4}), \\[1mm]
\label{eqn:Dir_m1_BS_metricexp_N}
\hat{N}
  = \hat{J}
  &= \frac{8}{5}\,\beta^{2}
     + \gamma_{D}\,\beta^{4}
     + \mathcal{O}(\beta^{6}),
\end{align}
\end{subequations}
where $\gamma_{D}\simeq1.84633345$ arises from a linear combination of Meijer-G functions. These quantities satisfy the first law of boson star thermodynamics~\eqref{FirstLawBStar} up to $\mathcal{O}(\beta^{3})$, consistent with the order to which $\hat{\omega}$, $\hat{M}$, and $\hat{N}$ are known.

\begin{figure}[t]
    \centering
    \includegraphics[width=0.4\linewidth]{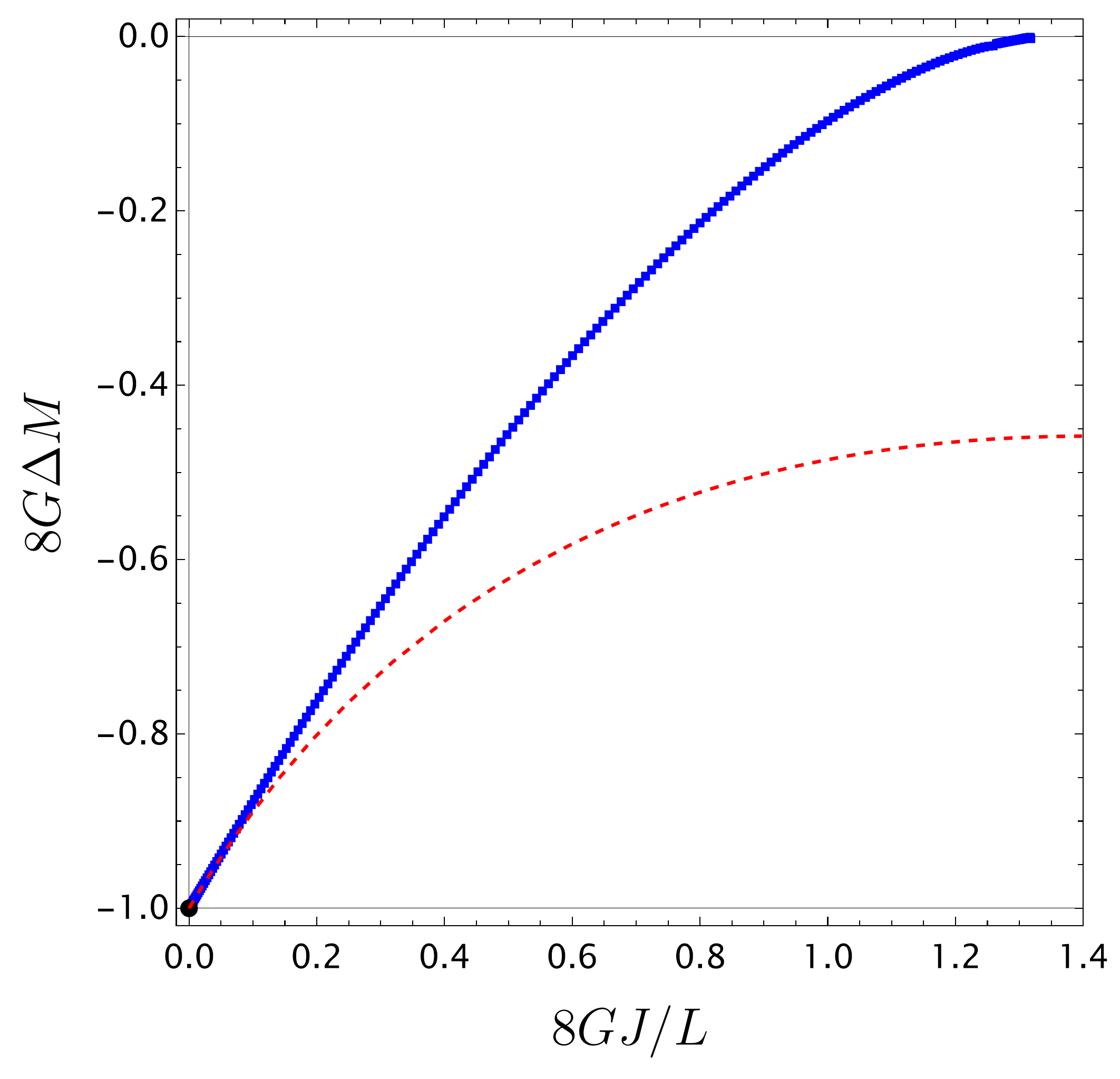}
    \includegraphics[width=0.4\linewidth]{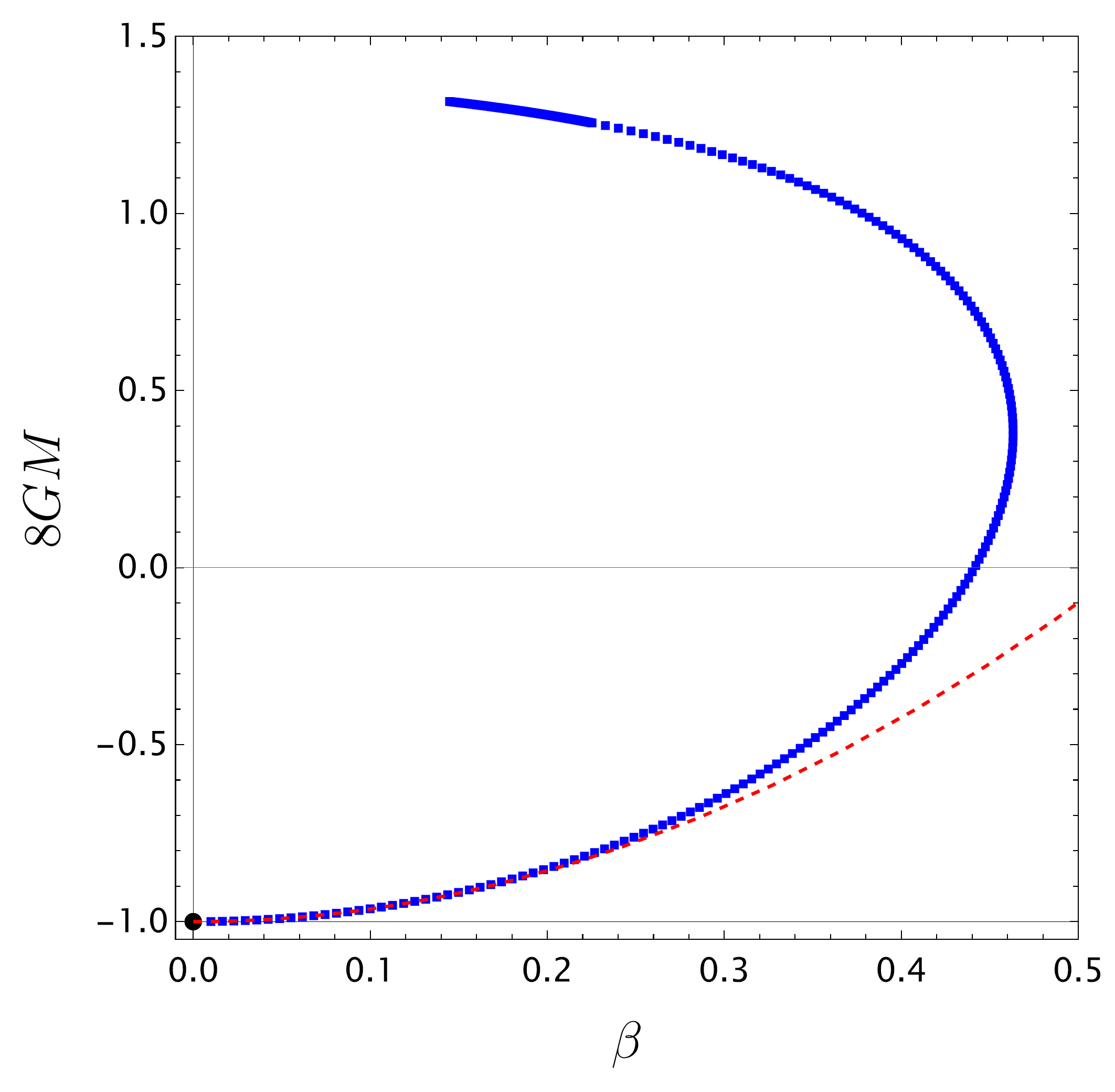}
    \includegraphics[width=0.4\linewidth]{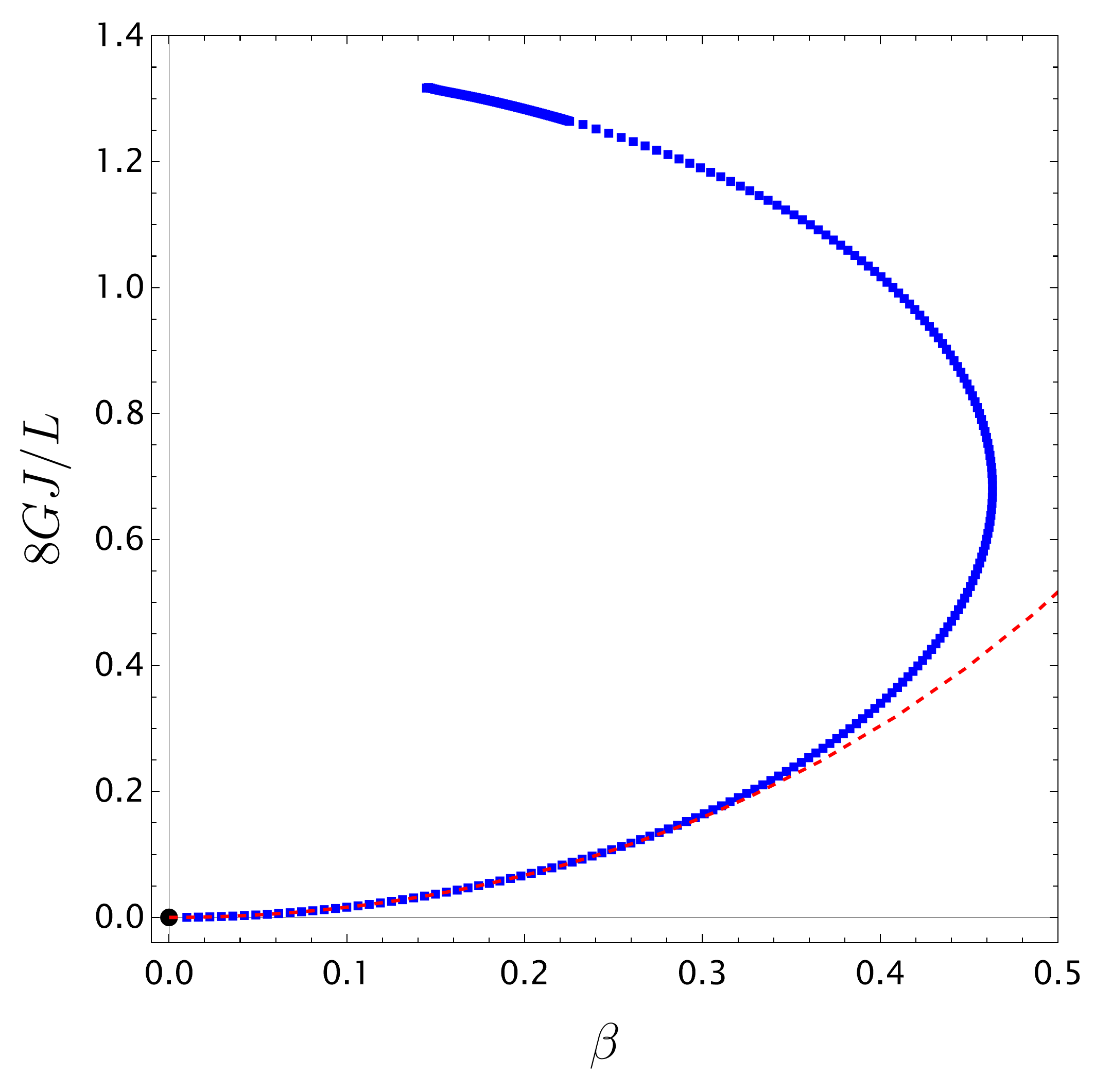}
    \includegraphics[width=0.4\linewidth]{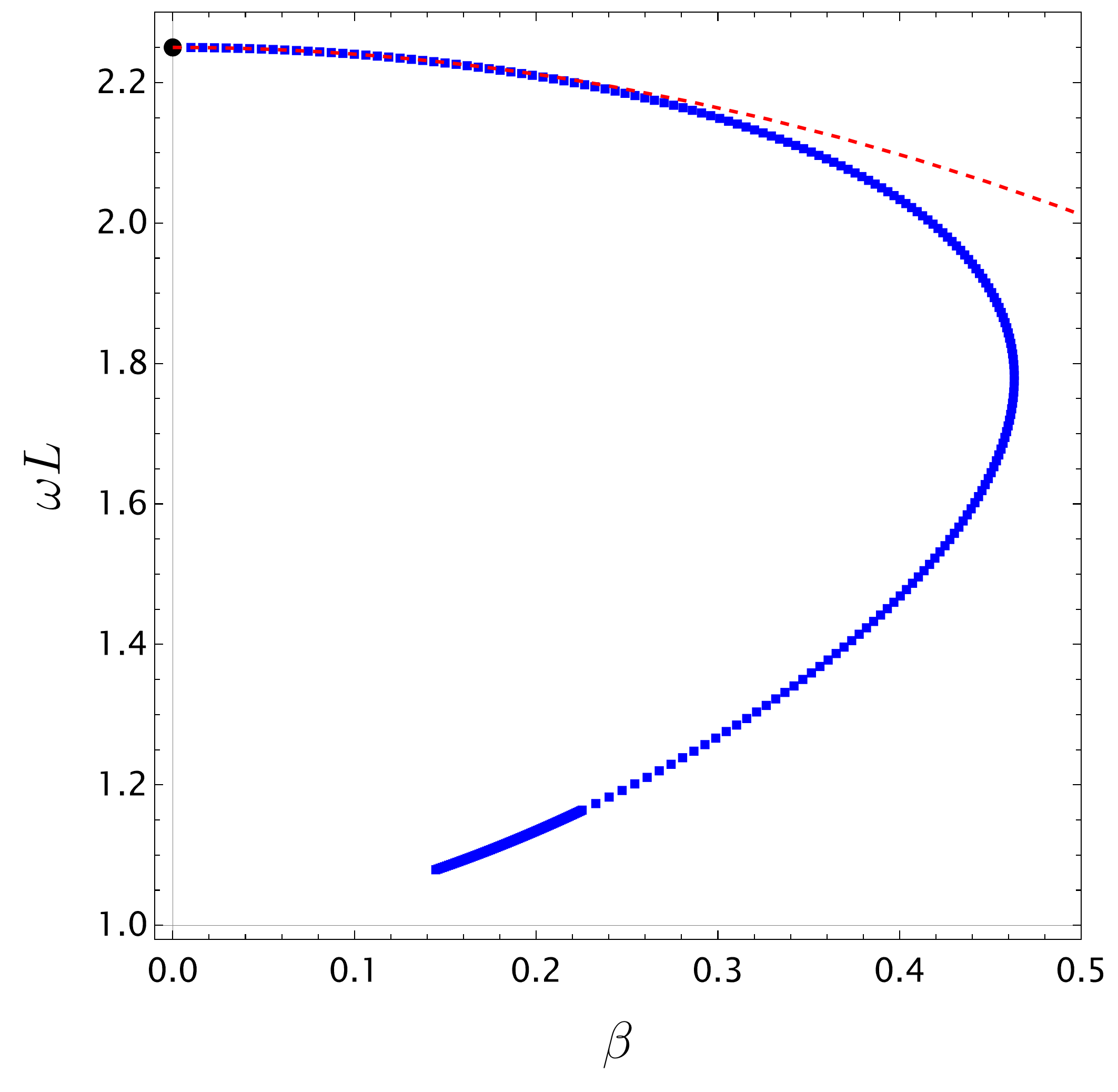}
    \caption{Physical properties (mass $\hat{M}$, angular momentum $\hat{J}$, frequency $\hat{\omega}$ and asymptotic scalar field amplitude $\beta$) of {\bf Dirichlet} $\bm{m = 1}$ boson stars with $\mu^2L^2 = -15/16$. The red dashed line describes the perturbative result \eqref{eqn:Dir_m1_BS_metricexp_M}, \eqref{eqn:Dir_m1_BS_metricexp_N} and\eqref{eqn:Dir_m1_BS_metricexp_omega}. The blue square curve describes the exact numerical result found using the methods described in section~\ref{sec:NumericalSetup:RegBS}.}    \label{fig:m1_DirBS_numerics_VS_perturbation}
\end{figure}

The thermodynamic properties of these Dirichlet $m=1$ boson stars are displayed in Fig.~\ref{fig:m1_DirBS_numerics_VS_perturbation}. For sufficiently small $\beta$, the perturbative results (red dashed curves) are in very good agreement with the fully nonlinear numerical solutions (brown squares).

\vskip 0.3cm

\paragraph{Neumann boundary conditions ($\beta=0$).}
For $m=1$ with Neumann boundary conditions, the perturbative expansions in $\epsilon\equiv\alpha$ read
\begin{subequations}
\begin{align}
\label{eqn:Neu_m1_BS_metricexp_omega}
\hat{\omega}
  &= \frac{7}{4}
     - \frac{3}{10}\,\alpha^{2}
     + \mathcal{O}(\alpha^{4}), \\[1mm]
f(R)
  &= 1+R^{2}
     + \frac{1}{6}
       \left(
         \frac{28+9R^{2}}{(1+R^{2})^{3/4}}-28
       \right)\alpha^{2}
     + \mathcal{O}(\alpha^{4}), \\[1mm]
g(R)
  &= 1
     - \frac{4+3R^{2}}{2(1+R^{2})^{7/4}}\,\alpha^{2}
     + \mathcal{O}(\alpha^{4}), \\[1mm]
\hat{\Omega}(R)
  &= \frac{4}{3R^{2}}
     \left(1-\frac{1}{(1+R^{2})^{3/4}}\right)\alpha^{2}
     + \mathcal{O}(\alpha^{4}), \\[1mm]
\psi(R)
  &= \frac{R}{(1+R^{2})^{7/8}}\,\alpha
     + \psi_{3}(R)\,\alpha^{3}
     + \mathcal{O}(\alpha^{5}),
\end{align}
\end{subequations}
where
\begin{multline}
\psi_{3}(R)
  = \frac{1}{360\,R(1+R^{2})^{21/8}}
    \Bigg\{
      -1093R^{4}
      -108\bigl((1+R^{2})^{3/4}-1\bigr)
\\
      +2R^{2}\!\left[681(1+R^{2})^{3/4}-755\right]
      +307\sqrt{R}(1+R^{2})^{7/4}
\\
      \times
      \left[
        4\,{}_2F_1\!\left(\frac34,\frac34,\frac74,-\frac{1}{R^{2}}\right)
        + 3R^{3/2}\bigl(2\log R-\log(1+R^{2})\bigr)
      \right]
    \Bigg\}.
\end{multline}

The corresponding thermodynamic quantities are
\begin{subequations}
\begin{align}
\label{eqn:Neu_m1_BS_metricexp_M}
\hat{M}
  &= -1 + \frac{14}{3}\,\alpha^{2}
     + \mathcal{O}(\alpha^{4}), \\[1mm]
\label{eqn:Neu_m1_BS_metricexp_N}
\hat{N}
  = \hat{J}
  &= \frac{8}{3}\,\alpha^{2}
     + \gamma_{N}\,\alpha^{4}
     + \mathcal{O}(\alpha^{6}),
\end{align}
\end{subequations}
with $\gamma_{N}\simeq2.44514636$, again arising from a linear combination of Meijer-G functions. As in the Dirichlet case, these quantities satisfy the first law~\eqref{FirstLawBStar} up to $\mathcal{O}(\alpha^{3})$.

\begin{figure}[ht]
    \centering
    \includegraphics[width=0.4\linewidth]{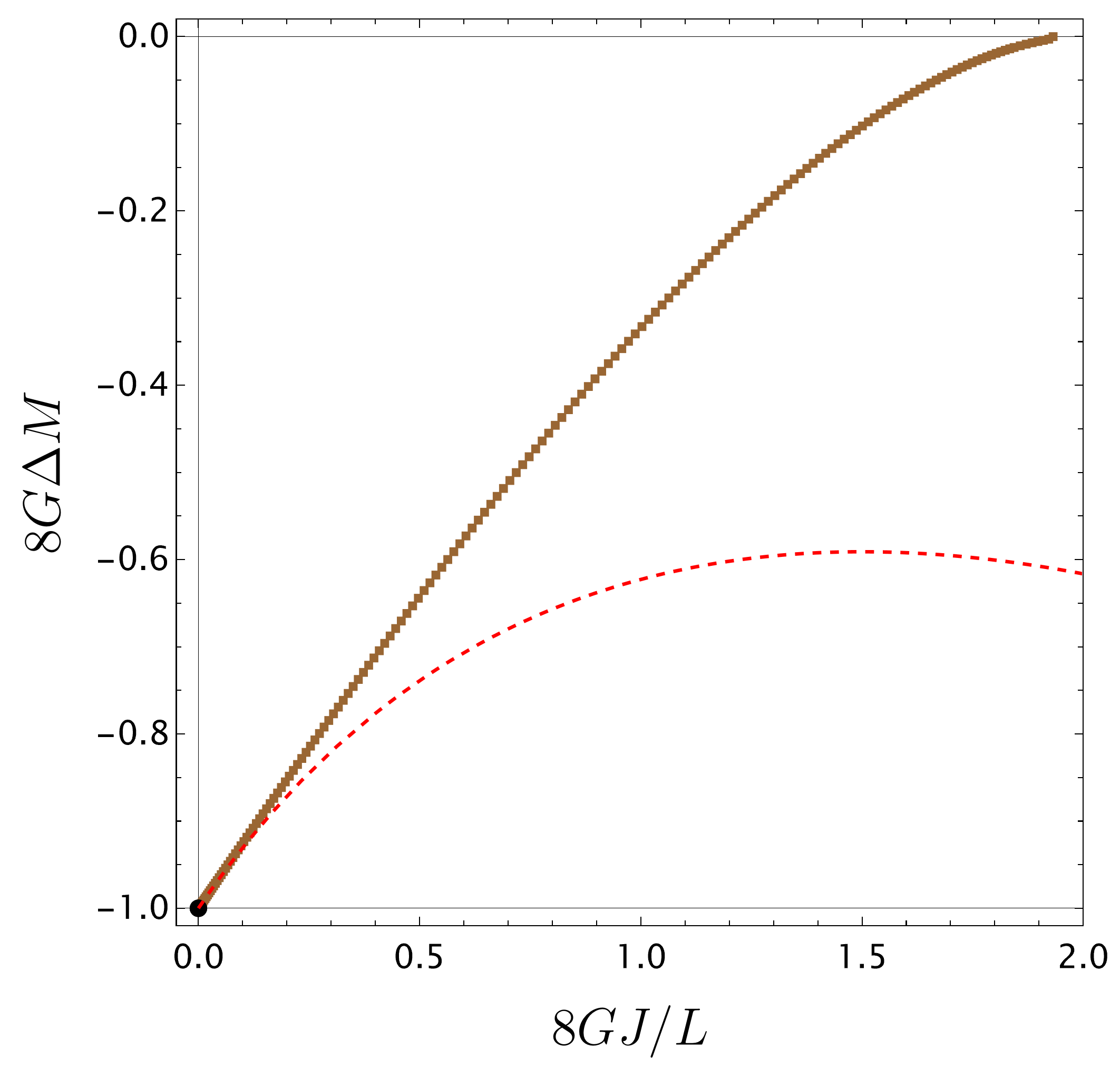}
    \includegraphics[width=0.4\linewidth]{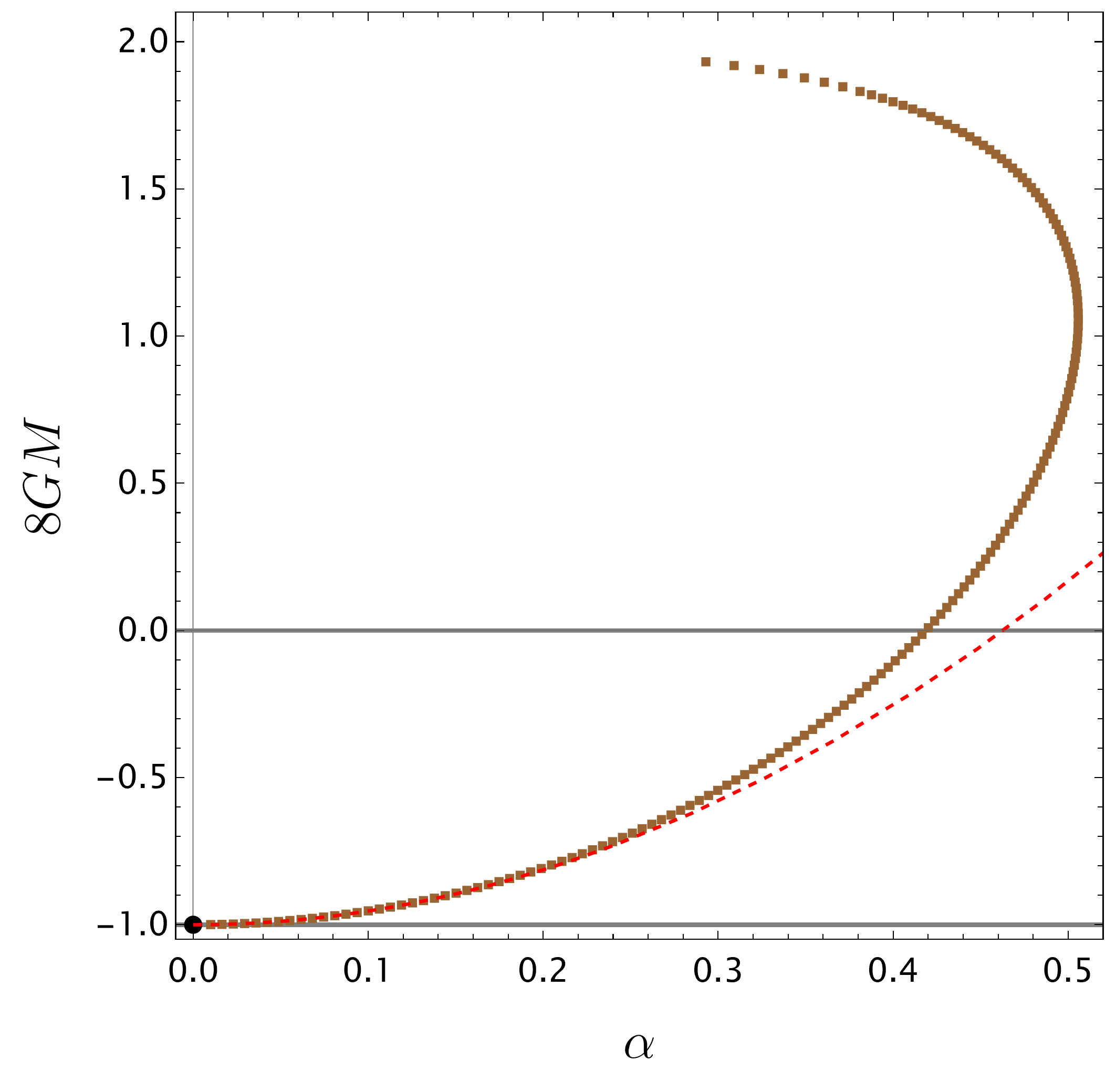}
    \includegraphics[width=0.4\linewidth]{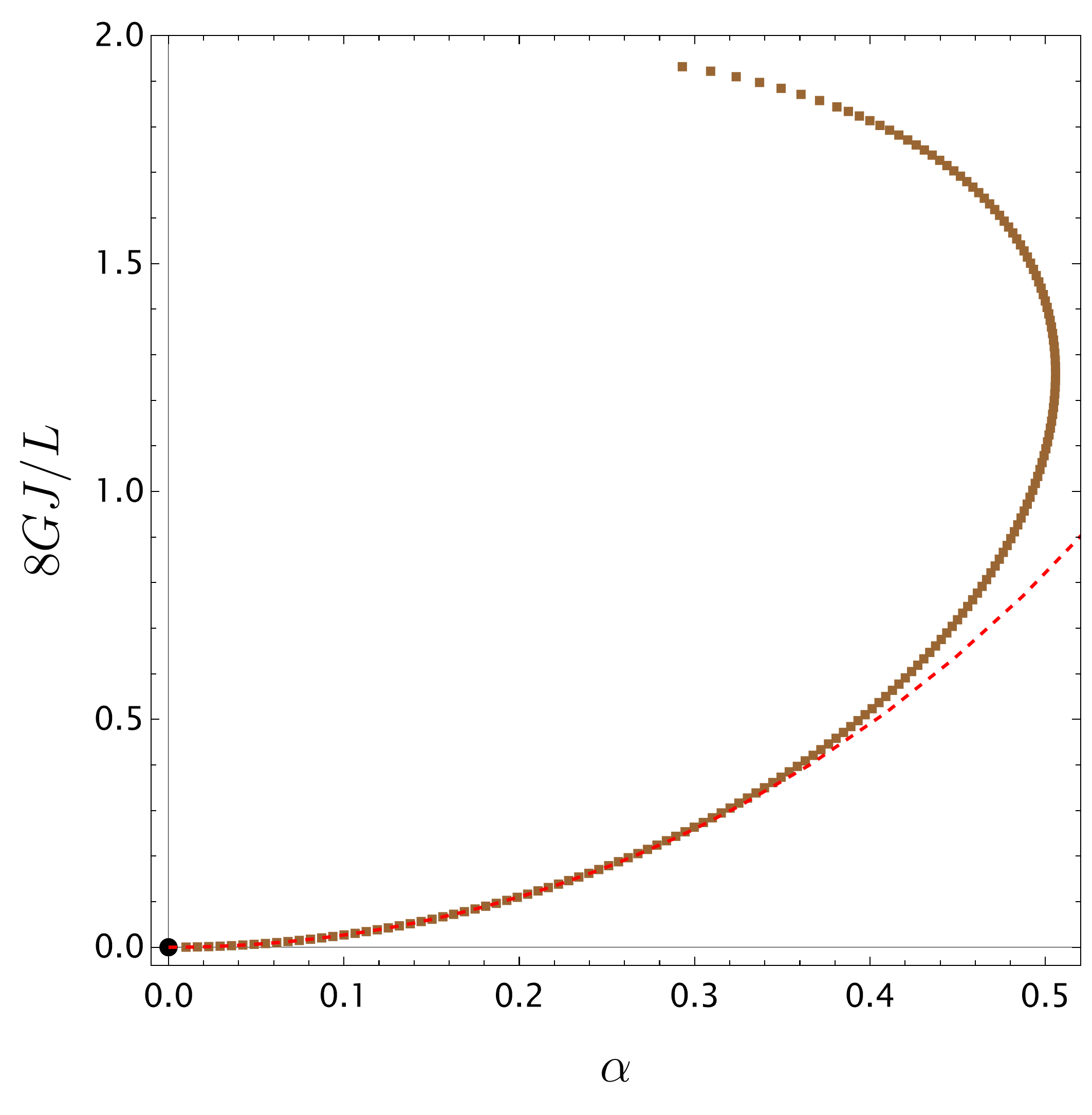}
    \includegraphics[width=0.4\linewidth]{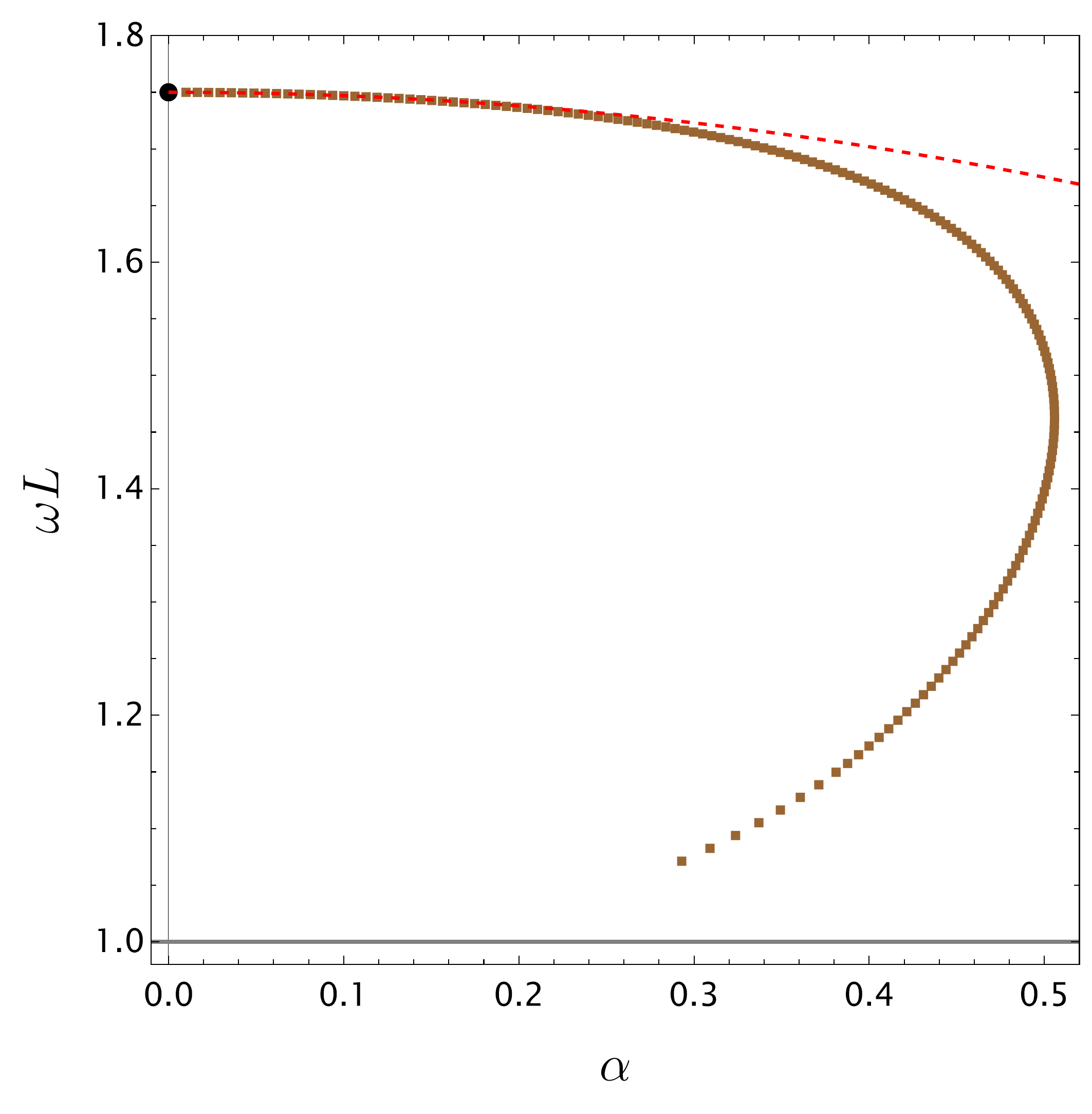}
    \caption{Physical properties (mass $\hat{M}$, angular momentum  $\hat{J}$, frequency $\hat{\omega}$ and asymptotic scalar field amplitude $\alpha$) of {\bf Neumann} $\bm{m = 1}$ boson stars with $\mu^2L^2 = -15/16$. The red dashed line describes the perturbative result \eqref{eqn:Neu_m1_BS_metricexp_M}, \eqref{eqn:Neu_m1_BS_metricexp_N} and \eqref{eqn:Neu_m1_BS_metricexp_omega}. The brown square curve describes the exact numerical result found using the methods described in section~\ref{sec:NumericalSetup:RegBS}.}  
\label{fig:m1_NeuBS_numerics_VS_perturbation}
\end{figure}

The thermodynamic properties of these Neumann $m=1$ boson stars are shown in Fig.~\ref{fig:m1_NeuBS_numerics_VS_perturbation}. For sufficiently small amplitudes, the perturbative predictions (red dashed curves) are in excellent agreement with the exact numerical solutions (brown squares).

The boson stars of Figs.~\ref{fig:m1_DirBS_numerics_VS_perturbation} and \ref{fig:m1_NeuBS_numerics_VS_perturbation} are expected to extend further back to $\beta=0$ but this time with $\hat{M}\neq -1$. We have not used our computational resources to complete this task since our main purpose for this appendix was to check the validity of the analytical approximations, which is accomplished with the given data.

\section{Holographic renormalization and holographic conserved charges\label{secA:HoloRen}}

We consider asymptotically AdS$_{d+1}$ spacetimes $\mathcal{M}$ with $d=2$. In a neighbourhood of the conformal boundary $\partial\mathcal{M}$, any such spacetime can be expressed in Fefferman–Graham (FG) coordinates $\{\tau,\phi,z\}$ as
\begin{align}\label{HoloRen:FG}
\mathrm{d}s^2 =\frac{L^2}{z^2}\left[\mathrm{d}z^2-A(\tau,\phi ,z){\rm d}\tau^2+B(\tau,\phi ,z) \Big({\rm d}\phi-\Omega(\tau,\phi ,z){\rm d}\tau \Big)^2\right]
\end{align}
where $z$ is the radial Fefferman–Graham coordinate and the conformal boundary $\partial\mathcal{M}$ is located at $z=0$. Note that requiring the spacetime to be asymptotically AdS$_3$ fixes only the boundary conformal structure. Accordingly, the functions $A$, $B$, and $\Omega$ admit the asymptotic expansions
 \begin{align}\label{HoloRen:FGexp}
& A(\tau,\phi ,z)=A_0(\tau,\phi) + \cdots + A_2(\tau,\phi) z^2 + a_2(\tau,\phi) z^2 \ln z + \cdots\,, \nonumber \\ 
& B(\tau,\phi ,z)=B_0(\tau,\phi) + \cdots + B_2(\tau,\phi) z^2 + b_2(\tau,\phi) z^2 \ln z + \cdots\,,   \qquad  \\
& \Omega{(\tau,\phi ,z)}=\Omega_0(\tau,\phi) + \cdots + \Omega_2(\tau,\phi) z^2 + \omega_2(\tau,\phi) z^2 \ln z + \cdots\, \nonumber
\end{align}
where $A_{0}$, $B_{0}$, and $\Omega_{0}$ are not fixed by the equations of motion and encode the boundary metric data, while the subleading coefficients indicated by the ellipses are determined recursively by solving the equations of motion order by order in a $z$‑expansion. The coefficient $B_{2}$ is also fixed locally by the equations of motion, whereas $A_{2}$ and $\Omega_{2}$ are not. Instead, $A_{2}$ and $\Omega_{2}$ are determined only once the full nonlinear equations of motion are solved subject to appropriate inner boundary conditions - namely regularity at the origin in the soliton case, or regularity at the black‑hole horizon.

Towards the end of this appendix, we will impose Dirichlet boundary conditions on the metric so that the induced boundary geometry is exactly the Einstein static universe. This amounts to setting
\begin{equation}
A_{0}=1,\ B_{0}=1,\ \Omega_{0}=0,
\end{equation}
corresponding to a reference frame that does not rotate at infinity. Until then, however, we keep the boundary metric completely general.

In even boundary dimensions - such as the $d=2$ case considered here - the Fefferman–Graham expansion of the metric generically includes logarithmic terms of the form $z^{d}\ln z$. The coefficients of these terms are determined by the metric variation of the gravitational conformal (Weyl) anomaly $\mathcal{A}$. Pure AdS gravity in $d=2$ is special in that the gravitational conformal anomaly vanishes identically. However, when matter fields are present, a non‑vanishing conformal anomaly may still arise even in $d=2$~\cite{Petkou:1999fv,deHaro:2000vlm,Kanitscheider:2006zf,Grumiller:2008qz,Skenderis:2009nt,Bzowski:2015pba}.

Our system also includes a neutral complex scalar field of mass $\mu$, which is related to the conformal dimension $\Delta$ of the dual operator according to
\begin{equation}
\mu^{2}L^{2}=\Delta(\Delta-2).
\end{equation}
The asymptotic Fefferman–Graham expansion of the scalar field $\Phi$ (and similarly for its complex conjugate $\Phi^{\dagger}$) takes the form
\begin{align}\label{HoloRen:FGexpScalar}
\Phi(\tau,\phi,z)
 ={}& \alpha(\tau,\phi)\,z^{\Delta_-}
      + \cdots
      + \beta(\tau,\phi)\,z^{\Delta_+}
      + \varphi(\tau,\phi)\,z^{\Delta_+}\ln z
      + \cdots , \\
&\text{where}\qquad
\Delta_{\pm}=1\pm\sqrt{1+\mu^{2}L^{2}}. \nonumber
\end{align}
Here, $\alpha(\tau,\phi)$ and $\beta(\tau,\phi)$ are free coefficients, while the omitted terms and the logarithmic coefficient $\varphi(\tau,\phi)$ are determined as functions of the free data
\begin{equation}
\{\alpha,\beta,A_{0},B_{0},\Omega_{0},A_{2},\Omega_{2}\}
\end{equation}
and their derivatives by solving the equations of motion order by order in a $z$‑expansion. The logarithmic term proportional to $\varphi(\tau,\phi)$ appears only for special discrete values of $\Delta_{\pm}$, and is often associated with the presence of a matter conformal anomaly $\mathcal{A}_{\Phi}$~\cite{Petkou:1999fv,deHaro:2000vlm,Kanitscheider:2006zf,Grumiller:2008qz,Skenderis:2009nt,Bzowski:2015pba}.

Before proceeding further, we discuss the logarithmic terms appearing in the metric expansion~\eqref{HoloRen:FGexp} and in the scalar expansion~\eqref{HoloRen:FGexpScalar}. Such terms are generically present whenever the system exhibits gravitational and/or scalar conformal anomalies; a comprehensive discussion can be found in~\cite{Bzowski:2015pba} and references therein. The values of the conformal dimension (equivalently, of the scalar mass) for which logarithmic terms occur form a discrete subset. More details on the conditions for the absence of logarithmic terms in the asymptotic expansions~\eqref{HoloRen:FGexp} and~\eqref{HoloRen:FGexpScalar} can be found in~\cite{Bzowski:2015pba}.

In what follows, we will assume that no logarithms are present. That is, we restrict attention to scalar field masses for which
\begin{equation}
a_{2}=b_{2}=\omega_{2}\equiv0
\quad\text{in~\eqref{HoloRen:FGexp}},\qquad
\varphi\equiv0
\quad\text{in~\eqref{HoloRen:FGexpScalar}}.
\end{equation}
From a technical standpoint, this assumption is extremely convenient: our numerical construction relies on pseudospectral methods, which exhibit exponential convergence (as the number of grid points is increased) if and only if the underlying solution admits no logarithmic terms. From a physical perspective, this restriction is not severe. Within the double‑trace mass range~\eqref{2xTrace:rangeMass}, logarithms only appear for a discrete set of scalar masses, \eg $\hat{\mu}^2= 3/4$. Moreover, since the qualitative properties of our hairy solutions are continuous and do not depend sensitively on the precise value of $\hat{\mu}$ within this range, it is natural to expect that solutions at these exceptional values do not exhibit qualitatively different behaviour from those studied here.

To ensure stability of the AdS$_{d+1}$ vacuum under scalar perturbations, the scalar field mass must satisfy the Breitenlohner–Freedman (BF) bound,
\begin{equation}
\mu^{2}\geq \mu^{2}_{\rm BF},
\qquad
\mu^{2}_{\rm BF}L^{2}=-\frac{d^{2}}{4}=-1
\quad (d=2),
\end{equation}
as established in~\cite{Breitenlohner:1982jf,Mezincescu:1984ev}. Whenever this bound is obeyed, one can always impose Dirichlet boundary conditions (BCs) on the scalar field, fixing the coefficient $\alpha(\tau,\phi)$ in the asymptotic expansion. The corresponding mode is normalizable, and solving the bulk equations of motion subject to appropriate inner boundary conditions allows one to read off the vacuum expectation value (VEV), proportional to $\beta(\tau,\phi)$, of the dual CFT operator $\mathcal{O}_{\Phi}^{(\Delta_+)}$ with conformal dimension $\Delta_+$.

If the Dirichlet BC is homogeneous, $\alpha(\tau,\phi)=0$, the system is unsourced; nevertheless, one may also prescribe a non‑vanishing source $\alpha(\tau,\phi)$ as a Dirichlet BC, which then induces a non‑trivial VEV. 

A qualitatively new situation arises when the scalar mass lies in the window
\begin{align}\label{HoloRen:rangeMass}
\mu^{2}_{\rm BF}L^{2}
\;<\;
\mu^{2}L^{2}
\;<\;
\mu^{2}_{\rm BF}L^{2}+1,
\qquad
\mu^{2}_{\rm BF}L^{2}=-1.
\end{align}
Only in this mass range are both asymptotic modes $z^{\Delta_{\pm}}$ normalizable. The upper bound is equivalent to the condition that $\Delta_{-}$ lies above the unitarity bound, $\Delta_{-}\geq(d-2)/2$. Consequently, for scalar masses in the range~\eqref{HoloRen:rangeMass}, one may still impose Dirichlet BCs fixing $\alpha(\tau,\phi)$ (the so‑called \emph{standard quantization}), but one is also allowed to impose Neumann BCs, $\beta(\tau,\phi)=0$ (the \emph{alternative quantization}), or more generally mixed boundary conditions that relate $\beta(\tau,\phi)$ to $\alpha(\tau,\phi)$. In these latter cases, the dual operator $\mathcal{O}_{\Phi}^{(\Delta_-)}$ has conformal dimension $\Delta_{-}$.

Of particular importance is the mixed boundary condition corresponding to a double‑trace deformation
\begin{equation}\label{HoloRen:2xTrBC}
\beta(\tau,\phi)=\kappa\,\alpha(\tau,\phi).
\end{equation}
This boundary condition is dual to deforming the CFT action according to
\begin{equation}
S_{\rm CFT}\;\longrightarrow\;
S_{\rm CFT}
 - \int \mathrm{d}^{2}x\,\kappa\,
   \mathcal{O}_{\Phi}^{(\Delta_-)}
   \mathcal{O}_{\Phi^{\dagger}}^{(\Delta_-)} ,
\end{equation}
where the expectation value
$\big\langle \mathcal{O}_{\Phi}^{(\Delta_-)}\big\rangle$
is proportional to $\alpha$; see Eq.~\eqref{HoloRen:ScalarVEVs} and
Refs.~\cite{Witten:2001ua,Berkooz:2002ug,Mueck:2002gm,Sever:2002fk,Papadimitriou:2007sj,Faulkner:2010fh,Faulkner:2010gj}.
\footnote{The double‑trace boundary condition is a special case of a more general multi‑trace BC of the form $\beta(\tau,\phi)=\kappa\,\alpha(\tau,\phi)^{\,n-1}$, with integer $n$ satisfying $2\leq n\leq d/\Delta_{-}$ ($d=2$ in the present case). Such boundary conditions correspond to multi‑trace deformations of the dual CFT, discussed in detail in \cite{Witten:2001ua, Sever:2002fk}.}

In what follows, we work exclusively in the scalar‑mass range~\eqref{HoloRen:rangeMass} and impose double‑trace boundary conditions, since these are precisely the conditions satisfied by the hairy solutions studied in the main text. Note that standard quantization, $\alpha=0$, corresponds formally to the limit $\kappa\to\infty$, while alternative quantization, $\beta=0$, corresponds to the opposite limit $\kappa=0$.

Ultimately, we are interested on computing the conserved charges of asymptotically AdS$_3$ gravitational hairy solutions of the Einstein scalar field theory described by the bulk action \eqref{eqn:action} and subject to double-trace BCs.
For that, one needs to take the bulk action  \eqref{eqn:action} of the system and add the required local and covariant boundary counterterms that render the final action finite and yield a well-posed variational problem. This is systematically done following a procedure known as \textit{holographic renormalization} \cite{Balasubramanian:1999re, deHaro:2000vlm,Skenderis:2002wp,Papadimitriou:2005ii}. The required counterterms that do the job depend also on the boundary conditions imposed on the bulk fields. As stated above, we restrict our attention to scalar masses $\mu^2 L^2$ in the range \eqref{HoloRen:rangeMass} such that the expansions \eqref{HoloRen:FGexp} and  \eqref{HoloRen:FGexpScalar}. have no logarithmic terms ($a_2=b_2=\omega_2\equiv 0$  and $\varphi=0$ ). 
We will be mainly interested on the case where we impose double-trace BCs on the scalar fields: holographic renormalization for similar systems was studied in \cite{Gegenberg:2003jr,Marolf:2006nd,Papadimitriou:2007sj,Faulkner:2010gj,Marolf:2012vvz,Caldarelli:2016nni,Harada:2023cfl}, often for particular values of $d$ and/or $\mu$.

In these conditions, holographic renormalization instructs us to consider a radial infrared cutoff $z=\epsilon$ and first define the regularized action that is the sum of the the bulk action and the boundary Gibbons-Hawking term: $S_{\hbox{\tiny reg}}=\left(S_{\hbox{\tiny bulk}}+S_{\hbox{\tiny GH}}\right)|_{z=\epsilon}$  \cite{Balasubramanian:1999re, deHaro:2000vlm,Skenderis:2002wp}. The Gibbons-Hawking boundary term $S_{\hbox{\tiny GH}}$ is necessary to get an action which only depends on first derivatives of the metric, and it guarantees that the variational problem with Dirichlet boundary conditions for the metric is well-defined \cite{Gibbons:1976ue,Brown:1992br}.  It is well known that $S_{\hbox{\tiny reg}}$ diverges when the cutoff is simply removed by taking the limit $z=\epsilon\to 0$. The divergence of  $S_{\hbox{\tiny reg}}$ is cancelled by adding judiciously chosen counterterms $S_{\hbox{\tiny \!{\it ct}}}$ such that the so-called renormalizable action $S_{\hbox{\tiny ren}}=\underset{\epsilon \to 0}{\lim} \left(S_{\hbox{\tiny reg}}+S_{\hbox{\tiny \!{\it ct}}}\right)$ is finite. The counterterm action might also include a {\it finite} contribution since $S_{\hbox{\tiny \!{\it ct}}}$ needs to be such that not only it eliminates the divergences of $S_{\hbox{\tiny reg}}$, but it must also render a well-posed variational principle, \ie one must also have $\delta S_{\hbox{\tiny ren}}=0$ when we impose the desired boundary conditions on the scalar field (in addition to Dirichlet boundary conditions on the metric).

For the system of interest, namely a locally asymptotically AdS$_3$ spacetime with a scalar field with double-trace boundary conditions, the renormalizable Lorentzian action has at most five counterterms:\footnote{For reference, for generic $d\geq 2$ one has $S_{\hbox{\tiny \!{\it ct},1}}= \frac{1}{2 \kappa_{\hbox{\tiny {\it N}}}^2} \int _{\partial \mathcal{M}} \mathrm{d}^{d}x\sqrt{-h}\:\frac{2 (1-d)}{L}$ and $\delta S_{\hbox{\tiny \!{\it ct},1}}= -\frac{1}{2 \kappa_{\hbox{\tiny {\it N}}}^2}\int _{\partial \mathcal{M}} \mathrm{d}^{d}x\sqrt{-h}\, \frac{2 (1-d)}{L} h_{ab}\delta g^{ab}$ and further counterterms are required to renormalize the gravitational action.} 
\begin{align}\label{HoloRen:Sren}
S_{\hbox{\tiny ren}}=&\: \underset{\epsilon \to 0}{\lim} \Big( S_{\hbox{\tiny bulk}}+S_{\hbox{\tiny GH}}+ S_{\hbox{\tiny \!{\it ct},1}} +S_{\hbox{\tiny \!{\it ct},2}}+ S_{\hbox{\tiny \!{\it ct},3}}+ S_{\hbox{\tiny \!{\it ct},$\mathcal{A}$}} + S_{\hbox{\tiny \!{\it ct},$\mathcal{A}_{\Phi}$}}  \Big) \nonumber \\
   =&\: \underset{\epsilon \to 0}{\lim} \, \frac{1}{2 \kappa_{\hbox{\tiny {\it N}}}^2} \bigg\{\int_{\mathcal{M}} \mathrm{d}^{3}x\, \sqrt{-g}\left[R - 2\Lambda - 2 \nabla_\alpha \Phi \nabla^\alpha \Phi^{\dagger} - 2\mu^2\Phi\Phi^{\dagger} \right]+ \int _{\partial \mathcal{M}} \mathrm{d}^{2}x\sqrt{-h}\,2K \nonumber \\
   &\:  \hspace{0.5cm} -\lambda_1 \int _{\partial \mathcal{M}} \mathrm{d}^{2}x\sqrt{-h}\:\frac{2}{L} 
    - \lambda_2 \int _{\partial \mathcal{M}} \mathrm{d}^{2}x\sqrt{-h}\: \frac{\Delta _-}{L}\, \Phi  \Phi^{\dagger}
 \\
     &\: \hspace{0.5cm}  - \lambda_3 \int _{\partial \mathcal{M}} \mathrm{d}^{2}x\sqrt{-h}\, \frac{L} {\Delta _-}(n^a \nabla_a\Phi) (n^b \nabla_b\Phi^{\dagger}) + \int _{\partial \mathcal{M}} \mathrm{d}^{2}x\sqrt{-h} (\cdots) \ln z \bigg\} \bigg|_{z=\epsilon} , \nonumber
\end{align}
where $\kappa_{\hbox{\tiny {\it N}}}^2=8\pi G$ ($G$ being Newton's constant), $S_{\hbox{\tiny bulk}}$ is the bulk action $S$ given in \eqref{eqn:action}, $\Lambda=-\frac{d(d-1)}{2L^2}=-\frac{1}{L^2}$ ($d=2$) and $V(\Phi\Phi^{\dagger})$ is the scalar potential (in our case we take it to simply describe the mass $\mu$ of the complex scalar fields \ie $V(\Phi\Phi^{\dagger})=\mu^2 \Phi\Phi^{\dagger}$). The induced metric on the boundary $\partial \mathcal{M}$ with outward unit normal $n^a$ ($n_a n^a=1$) is $h^{ab}=g^{ab}-n^a n^b$  and $K=g^{ab}K_{ab}$ is the trace of the extrinsic curvature $K_{ab}=+h_a{}^{c} \nabla_c n_b $ (second fundamental form). The Gibbons-Hawking boundary term $S_{\hbox{\tiny GH}}$ is required for reasons already described above \cite{Gibbons:1976ue,Brown:1992br}. 
The (cosmological) counterterm $S_{\hbox{\tiny \!{\it ct},1}} $ is necessary to kill an infrared divergence as $z=\epsilon\to 0$ that is present already in pure AdS gravity. Similarly, the counterterm $S_{\hbox{\tiny \!{\it ct},$\mathcal{A}$}} $ is necessary to kill a logarithmic divergence $-$ related to the gravitational conformal anomaly $\mathcal{A}$ $-$ that already appears in pure AdS gravity. In $d=2$ pure gravity,  $\mathcal{A}$ is given by a topological invariant $\int R_{\hbox{\tiny $(\partial)$}}$ and therefore its variation w.r.t. the boundary metric vanishes ($\delta  S_{\hbox{\tiny \!{\it ct},$\mathcal{A}$}} =0$) but it can be present when gravity couples to matters fields. On the other hand, the counterterm $S_{\hbox{\tiny \!{\it ct},$\mathcal{A}_{\Phi}$}} $ is required to kill logarithmic divergences associated to the matter conformal anomaly $\mathcal{A}_{\Phi}$. As stated above, in our analysis we  assume that we work with conformal dimensions for which the gravitational and matter conformal anomaly terms vanish, $S_{\hbox{\tiny \!{\it ct},$\mathcal{A}$}} =0$ and $S_{\hbox{\tiny \!{\it ct},$\mathcal{A}_{\Phi}$}} =0$.
Consequently, these two terms will not contribute to the holographic stress tensor (and thus to the conserved charges) in our setup and therefore it is not enlightening to explicitly display the $(\cdots)$ expression in the last line of \eqref{HoloRen:Sren}.
The counterterm $S_{\hbox{\tiny \!{\it ct},2}}$ is required to kill an UV divergence that appears when the scalar field is present. On the other hand, $S_{\hbox{\tiny \!{\it ct},3}}$ is a counterterm whose finite contribution is required to guarantee that we have a well-posed variational problem whereby $\delta S_{\hbox{\tiny ren}}=0$ when we impose double-trace boundary conditions in the scalar field. The coefficients $\lambda_1, \lambda_2$ and $\lambda_3$ in \eqref{HoloRen:Sren} are simply equal to unity for double-trace BCs but we keep them for a while to help presenting the results and better identify the need of each counterterm. Moreover, their presence will also allow to include the analysis independent of the scalar field BCs for most of the analysis, namely till \eqref{HoloRen:variationSfinal0}: double-trace, Dirichlet and Neumann BCs will simply correspond to particular choices of the coefficients $\lambda_{1,2,3}$ that will be specified later.

We also need to compute  $\delta S_{\hbox{\tiny ren}}$. Taking the variation of \eqref{HoloRen:Sren} one gets
\begin{align}\label{HoloRen:variationS}
\delta S_{\hbox{\tiny ren}}=&\, \delta S_{\hbox{\tiny bulk}}+\delta S_{\hbox{\tiny GH}}+ \delta S_{\hbox{\tiny \!{\it ct},1}} + \delta S_{\hbox{\tiny \!{\it ct},3}} + \delta S_{\hbox{\tiny \!{\it ct},3}} + \delta  S_{\hbox{\tiny \!{\it ct},$\mathcal{A}$}} + \delta  S_{\hbox{\tiny \!{\it ct},$\mathcal{A}_{\Phi}$}} \nonumber \\
   =&\, \frac{1}{2 \kappa_{\hbox{\tiny {\it N}}}^2}\int _{\mathcal{M}}d^{3}x \left(\delta  \sqrt{-g}\right)\Big[R-2 \Lambda -2\nabla_c\Phi \nabla^c\Phi^{\dagger}-2 V \Big] \nonumber \\
&+ \frac{1}{2 \kappa_{\hbox{\tiny {\it N}}}^2}\int _{\mathcal{M}}d^{3}x\sqrt{-g}\left[ g^{ab} \delta R_{ab} + R_{ab} \delta g^{ab} -2\,\delta \Big(\nabla_c \Phi \nabla^c \Phi^{\dagger}+ V \Big)\right] \nonumber \\
&+ \frac{1}{\kappa_{\hbox{\tiny {\it N}}}^2}\int _{\mathcal{M}}d^{3}x\sqrt{-g} \left[\nabla^2\Phi^{\dagger}-\frac{\partial V}{\partial \Phi }\right]\delta \Phi
+ \frac{1}{\kappa_{\hbox{\tiny {\it N}}}^2}\int _{\mathcal{M}}d^{3}x\sqrt{-g} \left[\nabla^2\Phi -\frac{\partial V}{\partial \Phi^{\dagger}}\right]\delta \Phi^{\dagger} \nonumber \\
& +\frac{1}{2 \kappa_{\hbox{\tiny {\it N}}}^2}\int _{\partial \mathcal{M}} \mathrm{d}^{2}x\sqrt{-h}\, n_c\left[-2(\nabla^c \Phi^{\dagger})\delta \Phi -2(\nabla^c\Phi) \delta \Phi^{\dagger}\right] \nonumber \\
& +\frac{1}{\kappa_{\hbox{\tiny {\it N}}}^2}\int _{\partial \mathcal{M}} \mathrm{d}^{2}x \left(\delta  \sqrt{-h}\right)K+\frac{1}{\kappa_{\hbox{\tiny {\it N}}}^2}\int _{\partial \mathcal{M}} \mathrm{d}^{2}x\sqrt{-h}\,\delta K
\nonumber \\
& - \lambda_1 \, \frac{1}{2 \kappa_{\hbox{\tiny {\it N}}}^2} \delta \Big[  \int _{\partial \mathcal{M}} \mathrm{d}^{2}x\sqrt{-h}\:\frac{2}{L} \Big]
    - \lambda_2 \, \frac{1}{2 \kappa_{\hbox{\tiny {\it N}}}^2} \delta \Big[  \int _{\partial \mathcal{M}} \mathrm{d}^{2}x\sqrt{-h}\: \frac{\Delta _-}{L}\, \Phi  \Phi^{\dagger}\Big]
\nonumber \\
     &\, - \lambda_3\,\frac{1}{2\kappa_{\hbox{\tiny {\it N}}}^2} \delta \Big[ \int _{\partial \mathcal{M}} \mathrm{d}^{2}x\sqrt{-h}\, \frac{L} {\Delta _-} (n^a \nabla_a\Phi) (n^b \nabla_b\Phi^{\dagger}) \Big],
\end{align}
where we have used integration by parts to write the variation of the scalar field bulk action as the third and fourth lines and, as discussed above,  one simply has $\delta  S_{\hbox{\tiny \!{\it ct},$\mathcal{A}$}}=0$ and $\delta  S_{\hbox{\tiny \!{\it ct},$\mathcal{A}_{\Phi}$}}=0$ in our analysis.

We can simplify \eqref{HoloRen:variationS} considerably.  For that use use the following textbook identities: 
\small{
\begin{align}\label{HoloRen:variationSaux}
& \bullet \delta\sqrt{-g}=-\frac{1}{2}\sqrt{-g}\, g_{ab}\delta g^{ab}\,,
\\  
& \bullet \delta\sqrt{-h}=-\frac{1}{2}\sqrt{-h}\, h_{ab}\delta h^{ab}=-\frac{1}{2}\sqrt{-h}\, h_{ab}\delta g^{ab} \,,
\\  
& \bullet \delta R_{ab}=\nabla_c\delta \Gamma ^c{}_{ab}-\nabla_b\delta \Gamma ^c{}_{ac} \:\:\hbox{(Palatini identity)} \quad \hbox{with}\quad 
\delta \Gamma^a{}_{bc}=\frac{1}{2} g^{ad}\big(\nabla_b\delta g_{dc}-\nabla_d\delta g_{bc}+\nabla_c\delta g_{db}\big) \nonumber\\
& \qquad \Rightarrow 
\frac{1}{2 \kappa_{\hbox{\tiny {\it N}}}^2}\int _{\mathcal{M}}d^{3}x\sqrt{-g}\, g^{ab}\delta R_{ab}
=\frac{1}{2 \kappa_{\hbox{\tiny {\it N}}}^2}\int _{\mathcal{M}}d^{3}x\sqrt{-g}\, \nabla_c\left[g^{ab} \delta \Gamma ^c{}_{ab}-g^{ca} \delta \Gamma ^d{}_{ad}\right] \nonumber \\
& \hspace{5.5cm} 
=-\frac{1}{2 \kappa_{\hbox{\tiny {\it N}}}^2}\int _{\partial \mathcal{M}} \mathrm{d}^{2}x\sqrt{-h}\, n_a\left(\nabla_b\delta g^{ab}-g_{cb}\nabla^a\delta g^{bc}\right),
\end{align}
\begin{align}\label{HoloRen:variationSaux2}
& \bullet\delta K=\delta  \left(g^{ab} K_{ab}\right)=g^{ab} \delta K_{ab} + K_{ab} \delta g^{ab}
=-\frac{1}{2} K^{ab} \delta g_{ab} -\frac{1}{2}n^a\left(\nabla^b\delta g_{ab}-g^{cb} \nabla_a\delta g_{bc}\right)+ D_a \mathcal{C}^a \nonumber \\
& \hspace{0.5cm} \hbox{where} \quad 
\mathcal{C}_a\equiv \delta n_a-\frac{1}{2}n^b \delta g_{ab} =-\frac{1}{2} n^b h^c{}_a  \delta g_{cb}  \quad \hbox{is a vector normal to $n^a$ and its covariant} \nonumber \\
&
 \hspace{0.5cm}  \hbox{derivative w.r.t. to $h$  is}
\quad D_a\mathcal{C}^a = h^c{}_a \nabla_c\left(h^a{}_d \mathcal{C}^d\right)=\frac{1}{2} K n^a  n^b \delta g_{ab} -\frac{1}{2} n^b h^{ca} \nabla_c\delta g_{ab}-\frac{1}{2}K^{ba}\delta g_{ab}  \nonumber \\
& \quad \Rightarrow  
 \delta S_{\hbox{\tiny GH}}=\frac{1}{\kappa_{\hbox{\tiny {\it N}}}^2}\int _{\partial \mathcal{M}} \mathrm{d}^{2}x \left(\delta  \sqrt{-h}\right)K+\frac{1}{\kappa_{\hbox{\tiny {\it N}}}^2}\int _{\partial \mathcal{M}} \mathrm{d}^{2}x\sqrt{-h}\, \delta K \nonumber \\
& \hspace{1cm} =-\frac{1}{2 \kappa_{\hbox{\tiny {\it N}}}^2}\int _{\partial \mathcal{M}} \mathrm{d}^{2}x\sqrt{-h}\, K h_{ab}\delta g^{ab}
+\frac{1}{2 \kappa_{\hbox{\tiny {\it N}}}^2}\int _{\partial \mathcal{M}} \mathrm{d}^{2}x\sqrt{-h}\, \left[ K_{ab} \delta g^{ab} +n_a\left(\nabla_b\delta g^{ab}-g_{cb}\nabla^a\delta g^{bc}\right)\right]  \nonumber \\
& \hspace{1.45cm}+\overbrace{\frac{1}{\kappa_{\hbox{\tiny {\it N}}}^2}\int _{\partial \mathcal{M}} \mathrm{d}^{2}x\sqrt{-h}\, D_a\mathcal{C}^a}^{=0 \: \hbox{\tiny (total derivative)}} \,,
\\
& \bullet \delta \left(-2\nabla_a\Phi \nabla^b\Phi^{\dagger}-2 V \right)=-\big(\nabla_a\Phi^{\dagger} \nabla_b\Phi +\nabla_a\Phi  \nabla_b\Phi^{\dagger}\big)\delta g^{\alpha \beta } \,,
\\   
& \bullet \delta S_{\hbox{\tiny \!{\it ct},1}}=-\lambda_1 \frac{1}{2 \kappa_{\hbox{\tiny {\it N}}}^2} \int _{\partial \mathcal{M}} \mathrm{d}^{2}x \left(\delta  \sqrt{-h}\right) 
\frac{2}{L}
= \lambda_1 \frac{1}{2 \kappa_{\hbox{\tiny {\it N}}}^2}\int _{\partial \mathcal{M}} \mathrm{d}^{2}x\sqrt{-h}\, \frac{2}{L} h_{ab}\delta g^{ab} \,,\nonumber\\
& \bullet \delta S_{\hbox{\tiny \!{\it ct},2}}=-\lambda_2 \frac{1}{2 \kappa_{\hbox{\tiny {\it N}}}^2} \delta \int _{\partial \mathcal{M}} \mathrm{d}^{2}x\sqrt{-h}\, \frac{\Delta _-}{L}\, \Phi  \Phi^{\dagger}
\nonumber \\   
&\hspace{1.3cm} = \lambda_2 \frac{1}{2\kappa_{\hbox{\tiny {\it N}}}^2} \int _{\partial \mathcal{M}} \mathrm{d}^{2}x\sqrt{-h}\, \frac{1}{2} \frac{\Delta _- }{ L} \Phi  \Phi^{\dagger} 
h_{ab}\delta g^{ab} 
+ \lambda_2 \frac{1}{2\kappa_{\hbox{\tiny {\it N}}}^2} \int _{\partial \mathcal{M}} \mathrm{d}^{2}x\sqrt{-h}\, \frac{\Delta _- }{ L} \left(\Phi^{\dagger} \delta \Phi + \Phi  \delta \Phi^{\dagger} \right)\,,
\\   
& \bullet \delta S_{\hbox{\tiny \!{\it ct},3}}=
-\lambda_3\, \frac{1}{2\kappa_{\hbox{\tiny {\it N}}}^2} \int _{\partial \mathcal{M}} \mathrm{d}^{2}x\sqrt{-h}\left(\frac{1}{2} \frac{L} {\Delta _-} (n^c \nabla_c\Phi^{\dagger}) (n^d \nabla_d\Phi)   h_{ab}
\right)\delta g^{ab} \nonumber\\
& 
\hspace{1.7cm}  +\lambda_3\,\frac{1}{2\kappa_{\hbox{\tiny {\it N}}}^2} \int _{\partial \mathcal{M}} \mathrm{d}^{2}x\sqrt{-h}\, \frac{L} {\Delta _-}  (n^c \nabla_c\Phi^{\dagger}) (n^d \nabla_d\Phi)  n_a n_a \delta g^{ab} \nonumber\\
& 
\hspace{1.7cm} -\lambda_3\,\frac{1}{2\kappa_{\hbox{\tiny {\it N}}}^2} \int _{\partial \mathcal{M}} \mathrm{d}^{2}x\sqrt{-h}\,\frac{L} {\Delta _-}\left[(n^a \nabla_a\Phi^{\dagger}) n^b \nabla_b\delta \Phi +(n^a \nabla_a\delta \Phi^{\dagger})(n^b \nabla_b\Phi)  \right], 
\end{align}
where we used the divergence theorem $\int _{\mathcal{M}}\mathrm{d}^{d+1}x\,\sqrt{-g}\,\nabla_{\mu }X^{\mu }=\int _{\partial \mathcal{M}} \mathrm{d}^{d}x\,\sqrt{-h}\,n_{\mu } X^{\mu }$ and further observed that  $\int _{\partial \mathcal{M}} \mathrm{d}^{2}x \, n_a n_a \delta g^{ab}=0$ because the variation of the metric at the boundary is orthogonal to the boundary's normal. Additionally, note that the contribution $\int _{\mathcal{M}} g^{ab}\delta R_{ab}$ (after using divergence's theorem) is precisely canceled by a term in $\delta S_{\hbox{\tiny GH}}$ and this justifies the choice of Gibbons-Hawking boundary term.

It follows that  \eqref{HoloRen:variationS} reads:
\begin{align}\label{HoloRen:variationS2}
\delta S_{\hbox{\tiny ren}}= &\,\delta S_{\hbox{\tiny bulk}} +\delta S_{\hbox{\tiny GH}}  + \delta S_{\hbox{\tiny \!{\it ct},1}} + \delta S_{\hbox{\tiny \!{\it ct},3}} + \delta S_{\hbox{\tiny \!{\it ct},3}}
\nonumber \\
= &\, \frac{1}{2 \kappa_{\hbox{\tiny {\it N}}}^2}\int _{\partial \mathcal{M}} \mathrm{d}^{2}x\sqrt{-h} \bigg[ 
 K_{ab}-K h_{ab}
 +\lambda_1\, \frac{2}{L}h_{ab}
+\lambda_2 \, \frac{1}{2} \frac{\Delta _-}{L} \Phi \Phi ^{\star}  h_{ab} \nonumber \\
&  \hspace{3.3cm}
-\lambda_3\,\frac{1}{2} \frac{L} {\Delta _-} h_{ab} n^d(\nabla_d\Phi ) n^c\left(\nabla_c\Phi^{\dagger}\right)
 \bigg] \delta g^{ab}
\\
&
 +\frac{1}{2 \kappa_{\hbox{\tiny {\it N}}}^2}\int _{\partial \mathcal{M}} \mathrm{d}^{2}x\sqrt{-h}\biggl[
 \lambda_1\, n^a\left(-2 \delta \Phi^{\dagger} \nabla_a\Phi -2 \delta \Phi  \nabla_a\Phi^{\dagger}\right)
 -\lambda_2\, \frac{\Delta _-}{L}\left(\Phi  \delta \Phi^{\dagger}+\delta \Phi  \Phi^{\dagger}\right) \nonumber \\
&  \hspace{3.7cm}
 -\lambda_3\,\frac{L} {\Delta _-}\left(n^b(\nabla_b\Phi ) n^a\left(\nabla_a\delta \Phi^{\dagger}\right)+n^b(\nabla_b\delta \Phi ) n^a\left(\nabla_a\Phi^{\dagger}\right)\right)
 \biggr]  \nonumber
\end{align}

We are now ready to introduce the asymptotic expansions \eqref{HoloRen:FGexp}-\eqref{HoloRen:FGexpScalar} into \eqref{HoloRen:Sren} and \eqref{HoloRen:variationS2} to get the final expressions for $S_{\hbox{\tiny ren}}$ and $\delta S_{\hbox{\tiny ren}}$.
Introducing the asymptotic expansions \eqref{HoloRen:FGexp}-\eqref{HoloRen:FGexpScalar} and after doing the radial integration encoded in the bulk integral, $\int_{\mathcal{M}} \mathrm{d}^{3}x \sqrt{-g}\, \mathcal{L}_{bulk}= \int _{\partial \mathcal{M}} \mathrm{d}^{2}x\sqrt{-h} \int_{z=\epsilon} \mathrm{d}z \frac{L}{z} \mathcal{L}_{bulk}$,  \eqref{HoloRen:Sren} reads:
\begin{align}\label{HoloRen:SrenFinal}
& S_{\hbox{\tiny ren}}=  \frac{L}{\kappa_{\hbox{\tiny {\it N}}}^2}\int _{\partial \mathcal{M}} \mathrm{d}^{2}x \bigg[\sqrt{A_0(\tau,\phi)}\sqrt{B_0(\tau,\phi)}
\\
& \hspace{1.3cm} \times \left(\frac{\lambda_1-1}{z^2} 
+ \frac{(\Delta_+-{\Delta_-})-2\lambda_1- \Delta_-(\lambda_2+\lambda_3)}{2\,z^{\Delta_+-\Delta_- }}\,\alpha(\tau,\phi)\alpha^{\dagger}(\tau,\phi)\right) + \mathcal{O}(z^0) \bigg]\bigg|_{z=\epsilon}\,. \nonumber
\end{align}
On the other hand, after doing integration by parts (to eliminate first and second derivative terms of variations) and using the equations of motion,  \eqref{HoloRen:variationS2} reduces to
{\small
\begin{align}\label{HoloRen:variationSfinal0}
 & \hspace{-0.3cm} \delta S_{\hbox{\tiny ren}}= \frac{L}{2 \kappa_{\hbox{\tiny {\it N}}}^2}  \int _{\partial \mathcal{M}} \mathrm{d}^{2}x \sqrt{A_0(\tau,\phi)}\sqrt{B_0(\tau,\phi)}\bigg\{  
(\cdots)\delta A_0(\tau,\phi)+(\cdots)\delta B_0(\tau,\phi)+(\cdots)\delta \Omega_0(\tau,\phi)
\nonumber\\
&  \hspace{0.8cm} 
+\frac{(\Delta_+-\Delta_-)-2\lambda_1+\Delta_-(\lambda_2+\lambda_3)}{z^{\Delta_+-\Delta _-}} 
 \Big[ \alpha^{\dagger}(\tau,\phi) \, \delta\alpha(\tau,\phi) +\alpha(\tau,\phi) \, \delta\alpha^{\dagger}(\tau,\phi) \Big]
 \nonumber\\
&  \hspace{0.8cm} 
- \Big( (\Delta_+-\Delta_-)^2+4\Delta_-\Delta_+ \lambda_1 -2\, \Delta_- \lambda_2 - 2 \Delta_+ \lambda_3\Big)
 \Big[ \beta^{\dagger}(\tau,\phi) \, \delta\alpha(\tau,\phi) +\beta(\tau,\phi) \, \delta\alpha^{\dagger}(\tau,\phi) \Big]
\nonumber\\
&  \hspace{0.8cm}  +\Big(( \Delta_-(\Delta_+-\Delta_-)-2\Delta_-\Delta_+ \lambda_1+\Delta_- \lambda_2+\Delta_+ \lambda_3  \Big)\nonumber\\
&  \hspace{1.1cm} \times
\Big[  \Big( \alpha^{\dagger}(\tau,\phi)\delta\beta(\tau,\phi) - \beta^{\dagger}(\tau,\phi) \,\delta \alpha(\tau,\phi) \Big)
  + \Big( \alpha(\tau,\phi)\, \delta\beta^{\dagger}(\tau,\phi) -  \beta(\tau,\phi)  \,\delta \alpha^{\dagger}(\tau,\phi) \Big) \Big]
 \nonumber  \\
&  \hspace{0.8cm} 
+ \mathcal{O}\left(z^{\Delta_+-\Delta_- }\right) 
\bigg\}\bigg|_{z=\epsilon}  
\end{align}
}

Here, the $(\cdots)$ terms vanish when we impose Dirichlet boundary condition on the metric which require that the variations $\delta A_0(\tau,\phi) =\delta B_0(\tau,\phi) =\delta \Omega_0(\tau,\phi)$ vanish. On the other hand, we still have to impose the boundary conditions on the scalar field. In the main text, we are interested on double-trace  BCs and associated double-trace variations, namely
\begin{align}\label{HoloRen:2xTrBCs}
 \bigg\{ & 
\begin{array}{c}
 \beta(\tau,\phi)=\kappa \,\alpha(\tau,\phi)\,,   \\
 \beta^{\dagger}(\tau,\phi)=\kappa \,\alpha^{\dagger}(\tau,\phi)\,   \\
\end{array}   
\qquad \Rightarrow \quad
\bigg\{ 
\begin{array}{c}
\delta \beta(\tau,\phi)=\kappa \,\delta\alpha(\tau,\phi)\,,   \\
\delta \beta^{\dagger}(\tau,\phi)=\kappa \,\delta\alpha^{\dagger}(\tau,\phi)\,.   \\
\end{array}   
\end{align}
which we impose onwards (unless otherwise stated).
Then, the second to last line in \eqref{HoloRen:variationSfinal0} becomes  $\big[\alpha^{\dagger}(\delta\beta-\kappa \,\delta\alpha)+\alpha(\delta\beta^{\dagger}-\kappa \,\delta\alpha^{\dagger})\big]$ for  double-trace BCs and thus the associated  term vanishes for double-trace variations \eqref{HoloRen:2xTrBCs}. In \eqref{HoloRen:variationSfinal0} we are thus left with two contributions, namely one proportional to $ (\alpha^{\dagger} \,\delta\alpha +\alpha\,\delta\alpha^{\dagger})$ and the other proportional to $ \left[ \beta^{\dagger} \, \delta\alpha +\beta \, \delta\alpha^{\dagger} \right]=\kappa \left(  \alpha^{\dagger} \,\delta\alpha +\alpha\,\delta\alpha^{\dagger} \right)$ that do not vanish after imposing all the BCs and their variations.
To have a well variational problem for the double-trace boundary conditions,  $\delta S_{\hbox{\tiny ren}}=0$, one must then choose the counterterm coefficients $\lambda_{1,2,3}$ to be such that these two contributions also vanish. These coefficients $\lambda_{1,2,3}$ must also be such that the divergences in \eqref{HoloRen:SrenFinal} are eliminated, \ie that the renormalizable action is finite. Altogether, $\underset{\epsilon \to 0}{\lim} \,S_{\hbox{\tiny ren}}=\hbox{finite}$ and $ \delta S_{\hbox{\tiny ren}}=0$ requires that we fix $\lambda_1=\lambda_2=\lambda_3\equiv 1$. Note that, as stated previously, the metric variation of the gravitational conformal anomaly vanishes in $d=2$ and this is why it does not appear in \eqref{HoloRen:variationSfinal0}. As stated before, this further implies that the FG asymptotic expansion does not contain a logarithmic coefficient $h_{(2)} z^2 \ln z$.\footnote{For completeness, note that our analysis is universal in the sense that it also covers the cases where we impose Dirichlet or Neumann boundary conditions. Indeed, for homogeneous Neumann BCs, $\beta=\beta^{\dagger}=0$ and $\delta\beta=\delta\beta^{\dagger}=0$,  our analysis applies straightforwardly by simply setting $\kappa=0$ in our discussion. In particular, we can effectively still set $\lambda_1=\lambda_2=\lambda_3\equiv 1$ to have a finite renormalizable action and a well posed variational principle with Neumann BCs for the scalar field. Note however that in this case we also have the simpler option to set $\lambda_3\equiv 0$ as long as we set $\lambda_2\equiv 2$ (and still $\lambda_1\equiv 1$). This is because the third line in \eqref{HoloRen:variationSfinal0} vanishes for Neumann BC: $ \left[ \beta^{\dagger} \, \delta\alpha +\beta \, \delta\alpha^{\dagger} \right]=0$. On the other hand, for homogeneous Dirichlet BC one has $\alpha=\alpha^{\dagger}=0$ and $\delta\alpha=\delta\alpha^{\dagger}=0$. Thus, all contributions of \eqref{HoloRen:variationSfinal0} vanish straightforwardly, \ie we have $\delta S_{\hbox{\tiny ren}}=0$ for any value of $\lambda_{1,2,3}$. On the other hand, the second term of \eqref{HoloRen:SrenFinal} vanishes and finiteness of $S_{\hbox{\tiny ren}}$ simply requires that we set $\lambda_1\equiv 1$. Summarizing, for Dirichlet BCs all the analysis till \eqref{HoloRen:variationSfinal0} does apply as long as we set $\lambda_1\equiv 1$ and $\lambda_2=\lambda_3 \equiv 0$.} For the above reasons, onwards we set  $\lambda_1=\lambda_2=\lambda_3\equiv 1$ in all previous expressions.

We are now ready to compute the expectation values of the dual operators. With that aim, note that the boundary metric at $\partial \mathcal{M}$ ($z=0$) is given by
\begin{align}\label{HoloRen:bdryMetric}
\mathrm{d}s^2_{\hbox{\tiny $(\partial)$}} = \gamma_{ij}^{\hbox{\tiny $(\partial)$}}dx^i dx^j =-A_0(\tau,\phi)d\tau^2+B_0(\tau,\phi) \Big(d\phi-\Omega_0(\tau,\phi)d\tau \Big)^2
\end{align}
where $x^i=\{\tau,\phi\}$ are the boundary coordinates and we can write $\mathcal{F}_0\equiv \{A_0,B_0,\Omega_0\}$ as a function of the contravariant metric components $\gamma^{ij}_{\hbox{\tiny $(\partial)$}}$ as 
\begin{align}\label{HoloRen:bdryMetricFasG}
A_0(\tau,\phi)=-\frac{1}{ \gamma^{\tau \tau}_{\hbox{\tiny $(\partial)$}}}, \qquad 
B_0(\tau,\phi)=-\frac{\gamma^{\tau \tau}_{\hbox{\tiny $(\partial)$}} }{ \gamma^{\tau \tau}_{\hbox{\tiny $(\partial)$}} \gamma^{\phi\phi}_{\hbox{\tiny $(\partial)$}} -\big( \gamma^{\tau \phi}_{\hbox{\tiny $(\partial)$}} \big)^2 }, \qquad
\Omega_0(\tau,\phi)=\frac{\gamma^{\tau \phi}_{\hbox{\tiny $(\partial)$}}}{ \gamma^{\tau \tau}_{\hbox{\tiny $(\partial)$}} },
\end{align} 
It follows that, using the chain rule, the variation of  $\mathcal{F}_0\equiv \{A_0,B_0,\Omega_0\}$ can be written in terms of the variation of the contravariant metric variation $\delta \gamma^{ij}_{\hbox{\tiny $(\partial)$}}$ as 
\begin{align}\label{HoloRen:bdryMetricdeltaFasG}
\delta \mathcal{F}_0=\frac{\partial \mathcal{F}_0}{\partial \gamma^{\tau \tau}_{\hbox{\tiny $(\partial)$}} } \, \delta \gamma^{\tau \tau}_{\hbox{\tiny $(\partial)$}}
+\frac{\partial \mathcal{F}_0}{\partial \gamma^{\tau \phi}_{\hbox{\tiny $(\partial)$}} } \, \delta \gamma^{\tau \phi}_{\hbox{\tiny $(\partial)$}}
+\frac{\partial \mathcal{F}_0}{\partial \gamma^{\phi\phi}_{\hbox{\tiny $(\partial)$}} } \, \delta \gamma^{\phi\phi}_{\hbox{\tiny $(\partial)$}}
\end{align} 
We can now replace these relations into \eqref{HoloRen:variationSfinal0} to get
\begin{align}\label{HoloRen:variationSfinal}
 & \hspace{-0.8cm}  \delta S_{\hbox{\tiny ren}}=   \int _{\partial \mathcal{M}} \mathrm{d}^{2}x \sqrt{-\gamma_{\hbox{\tiny $(\partial)$}}}\bigg[  
\frac{1}{2}\big\langle  \mathcal{T}_{ij} \big\rangle \delta \gamma^{ij}_{\hbox{\tiny $(\partial)$}} 
+\big\langle  \mathcal{O}_{\Phi}^{\hbox{\tiny $(\Delta_-)$}} \big\rangle \Big( \delta\beta(\tau,\phi)- \kappa\, \delta\alpha(\tau,\phi) \Big) \nonumber \\
& \hspace{3.5cm} +\big\langle \mathcal{O}_{\Phi^{\dagger}}^{\hbox{\tiny $(\Delta_-)$}}\big\rangle \Big( \delta\beta^{\dagger}(\tau,\phi)- \kappa\, \delta\alpha^{\dagger}(\tau,\phi) \Big)
\bigg]
\end{align}
Here, $\big\langle\mathcal{T}_{ij} \big\rangle = \frac{2}{\sqrt{-\gamma_{\hbox{\tiny $(\partial)$}}}} \frac{\partial \delta S_{\hbox{\tiny ren}}}{\delta \gamma^{ij}_{\hbox{\tiny $(\partial)$}} }$  is  the holographic stress tensor (gravitational VEV) with components:
{\small
\begin{align}\label{HoloRen:GravVEVs}
     \big\langle\mathcal{T}_{\tau \tau} \big\rangle 
     =&\, -\frac{1}{8\kappa_{\hbox{\tiny {\it N}}}^2}\bigg[
 \frac{8 A_4 (\tau,\phi ) B_0 (\tau,\phi ) \Omega_0 (\tau,\phi )^2}{A_0 (\tau,\phi )}
 -\frac{4 \Omega_0 ^{(0,1)}(\tau,\phi }{A_0 (\tau,\phi )} \left(A_0 ^{(0,1)}(\tau,\phi ) \Omega_0 (\tau,\phi )+A_0 ^{(1,0)}(\tau,\phi )\right) 
 \nonumber \\
 &
 +\frac{2 A_0 ^{(0,1)}(\tau,\phi )}{A_0 (\tau,\phi ) B_0 (\tau,\phi )}\bigg(A_0 ^{(0,1)}(\tau,\phi )-\Omega_0 (\tau,\phi ) \Big(B_0 ^{(0,1)}(\tau,\phi ) \Omega_0 (\tau,\phi )+B_0 ^{(1,0)}(\tau,\phi )\Big)\bigg)
  \nonumber \\
 &
 -\frac{2 A_0 ^{(1,0)}(\tau,\phi )}{A_0 (\tau,\phi ) B_0 (\tau,\phi )}\left(B_0 ^{(0,1)}(\tau,\phi ) \Omega_0 (\tau,\phi )+B_0 ^{(1,0)}(\tau,\phi )\right)
 +8 \Big(A_4 (\tau,\phi )-2 B_0 (\tau,\phi ) \Omega_0 (\tau,\phi ) \Omega_4 (\tau,\phi )\Big)\
 \nonumber \\
 &
+ \frac{2}{B_0 (\tau,\phi )^2}\, \left(A_0 ^{(0,1)}(\tau,\phi ) B_0 ^{(0,1)}(\tau,\phi ) 
    -\left(B_0 ^{(0,1)}(\tau,\phi ) \Omega_0 (\tau,\phi )+B_0 ^{(1,0)}(\tau,\phi )\right)^2 \right)
  \nonumber \\
 &
 +\frac{4}{B_0 (\tau,\phi )}\bigg( B_0 ^{(2,0)}(\tau,\phi ) -A_0 ^{(0,2)}(\tau,\phi )+ \Omega_0 (\tau,\phi ) \left(3 B_0 ^{(0,1)}(\tau,\phi ) \Omega_0 ^{(0,1)}(\tau,\phi )+2 B_0 ^{(1,1)}(\tau,\phi )\right)  \bigg)
  \nonumber \\
 &
 +\frac{4}{B_0 (\tau,\phi )}\left(B_0 ^{(0,2)}(\tau,\phi ) \Omega_0 (\tau,\phi )^2+2 B_0 ^{(1,0)}(\tau,\phi ) \Omega_0 ^{(0,1)}(\tau,\phi )+B_0 ^{(0,1)}(\tau,\phi ) \Omega_0 ^{(1,0)}(\tau,\phi )\right)
   \nonumber \\
 &
 +8 \left(\Omega_0 ^{(0,1)}(\tau,\phi )^2+\Omega_0 (\tau,\phi ) \Omega_0 ^{(0,2)}(\tau,\phi )+\Omega_0 ^{(1,1)}(\tau,\phi )\right)
   \nonumber \\
 &
 +4\Big[ \Big(2-(\Delta_+-\Delta_-)^2 \Big) A_0 (\tau,\phi )+2 B_0 (\tau,\phi ) \Omega_0 (\tau,\phi )^2\Big] \Big(\alpha (\tau,\phi ) \beta^{\dagger}(\tau,\phi)+\alpha^{\dagger}(\tau,\phi) \beta (\tau,\phi )\Big)
 \bigg], \nonumber\\
\big\langle \mathcal{T}_{\tau \phi} \big\rangle =&
- \frac{1}{\kappa_{\hbox{\tiny {\it N}}}^2} \frac{B_0 (\tau,\phi )}{A_0 (\tau,\phi )} \bigg[    A_0 (\tau,\phi ) \Omega_4 (\tau,\phi )-A_4 (\tau,\phi ) \Omega_0 (\tau,\phi )
   \nonumber \\
 &
 -A_0 (\tau,\phi ) \Omega_0 (\tau,\phi ) \Big(\alpha (\tau,\phi ) \beta^{\dagger}(\tau,\phi)+\alpha^{\dagger}(\tau,\phi) \beta (\tau,\phi )\Big)
 \bigg],  \nonumber\\
\big\langle \mathcal{T}_{\phi\phi} \big\rangle =& 
- \frac{1}{\kappa_{\hbox{\tiny {\it N}}}^2} \frac{B_0 (\tau,\phi )}{A_0 (\tau,\phi )} \bigg[  
 A_4 (\tau,\phi ) +  A_0 (\tau,\phi ) \Big(\alpha (\tau,\phi ) \beta^{\dagger}(\tau,\phi)+\alpha^{\dagger}(\tau,\phi) \beta (\tau,\phi )\Big)
 \bigg] ;  
\end{align}
}
and  the scalar VEVs are
\begin{align}\label{HoloRen:ScalarVEVs}
    \big\langle  \mathcal{O}_{\Phi}^{\hbox{\tiny $(\Delta_-)$}}  \big\rangle = \frac{1}{2\kappa_{\hbox{\tiny {\it N}}}^2}\,(\Delta_+-\Delta_-) \alpha^{\dagger}(\tau,\phi)\,,\qquad 
    \big\langle  \mathcal{O}_{\Phi^{\dagger}}^{\hbox{\tiny $(\Delta_-)$}}  \big\rangle = \frac{1}{2\kappa_{\hbox{\tiny {\it N}}}^2}\,(\Delta_+-\Delta_-) \alpha(\tau,\phi)\,.
\end{align}

The trace of the holographic stress tensor is 
{\small
\begin{align}\label{HoloRen:TraceHoloVEV}
     \big\langle\mathcal{T} \big\rangle =&\: \gamma^{ij}_{\hbox{\tiny $(\partial)$}} \big\langle\mathcal{T}_{ij} \big\rangle
     \nonumber \\
 =&\:
     - \frac{1}{4\kappa_{\hbox{\tiny {\it N}}}^2}\,\frac{1}{A_0 (\tau,\phi ) B_0 (\tau,\phi )} \bigg[ 
   \frac{A_0 ^{(1,0)}(\tau,\phi )}{A_0 (\tau,\phi )} \left(B_0 ^{(0,1)}(\tau,\phi ) \Omega_0 (\tau,\phi )+B_0 ^{(1,0)}(\tau,\phi )+2 B_0 (\tau,\phi ) \Omega_0 ^{(0,1)}(\tau,\phi )\right)
   \nonumber \\
 &
 -\frac{A_0 ^{(0,1)}(\tau,\phi )}{A_0 (\tau,\phi )} \bigg(A_0 ^{(0,1)}(\tau,\phi )-\Omega_0 (\tau,\phi ) \left(B_0 ^{(0,1)}(\tau,\phi ) \Omega_0 (\tau,\phi )+B_0 ^{(1,0)}(\tau,\phi )+2 B_0 (\tau,\phi ) \Omega_0 ^{(0,1)}(\tau,\phi )\right)\bigg)
   \nonumber \\
 &
 -2 \left(2 B_0 ^{(1,0)}(\tau,\phi ) \Omega_0 ^{(0,1)}(\tau,\phi )+B_0 ^{(0,1)}(\tau,\phi ) \Omega_0 ^{(1,0)}(\tau,\phi )+B_0 ^{(0,2)}(\tau,\phi ) \Omega_0 (\tau,\phi )^2\right)
    \nonumber \\
 &
 +2 \left(A_0 ^{(0,2)}(\tau,\phi )-B_0 ^{(2,0)}(\tau,\phi ) \right)
 -2 \Omega_0 (\tau,\phi ) \left(3 B_0 ^{(0,1)}(\tau,\phi ) \Omega_0 ^{(0,1)}(\tau,\phi )+2 B_0 ^{(1,1)}(\tau,\phi )\right)
     \nonumber \\
 &
-4 B_0 (\tau,\phi ) \left(\Omega_0 ^{(0,1)}(\tau,\phi )^2+\Omega_0 (\tau,\phi ) \Omega_0 ^{(0,2)}(\tau,\phi )+\Omega_0 ^{(1,1)}(\tau,\phi )\right)
  \nonumber \\
 &
 +\frac{1}{B_0 (\tau,\phi )} \left(\left(B_0 ^{(0,1)}(\tau,\phi ) \Omega_0 (\tau,\phi )+B_0 ^{(1,0)}(\tau,\phi )\right)^2-A_0 ^{(0,1)}(\tau,\phi ) B_0 ^{(0,1)}(\tau,\phi )\right)
     \bigg]
        \nonumber \\
 &
     -\frac{1}{2\kappa_{\hbox{\tiny {\it N}}}^2} (\Delta_+-\Delta_-)^2 \bigg[  \alpha (\tau,\phi ) \beta^{\dagger}(\tau,\phi )+ \alpha^{\dagger}(\tau,\phi ) \beta (\tau,\phi )   \bigg]
     \nonumber \\
\equiv &\: -\frac{1}{2\kappa_{\hbox{\tiny {\it N}}}^2} \bigg[ R_{\hbox{\tiny $(\partial)$}}  
+(\Delta_+-\Delta_-)^2 \Big(  \alpha (\tau,\phi ) \beta^{\dagger}(\tau,\phi )+ \alpha^{\dagger}(\tau,\phi ) \beta (\tau,\phi ) \Big)   \bigg],
\end{align}
}
where $R_{\hbox{\tiny $(\partial)$}}$ is the Ricci scalar of the boundary metric \eqref{HoloRen:bdryMetric}.
Therefore, the stress tensor is not traceless and its trace is given by the gravitational conformal anomaly $\mathcal{A}$.\footnote{Recall that there is no conformal anomaly $\mathcal{A}_\Phi$ due to the scalar field since we are considering scalar masses such that  $\Delta_+-\Delta_-$ is not an integer \cite{Petkou:1999fv}.}

One can check that the holographic stress tensor is conserved, 
\begin{align}\label{HoloRen:ConservHoloVEV}
\nabla _{\hbox{\tiny $(\partial)$}}^i \left\langle \mathcal{T}_{ij}\right\rangle =0\,,
\end{align}
 where  $\nabla _{\hbox{\tiny $(\partial)$}}$ is the covariant derivative of the boundary metric \eqref{HoloRen:bdryMetric}.\footnote{To prove this one needs to expand \eqref{HoloRen:FG} to sufficient order in the FG coord $z$ and solve the associated equations of motion to find a relation that fixes $\partial_t A_4$ and  $\partial_{\phi} \Omega_4$ as a function of other free parameters and their derivatives.} In particular, this means that $ \left\langle \mathcal{T}_{ij}\right\rangle$ ultimately defines the energy and the angular momentum of the system that are conserved physical quantities. 
For that let $\Sigma_{\tau}$ be a spacelike hypersurface (of constant time $\tau$) with future-directed unit normal $\tau=\partial/\partial \tau$ ($\tau_i \tau^i=-1$). Then, given any Killing vector $\xi^i$, we can define the associated conserved charge
\begin{align}\label{HoloRen:conservedCharges}
    Q_{\xi}=\int_{\Sigma_{\tau}} \mathrm{d}\phi \sqrt{\sigma}  \left\langle \mathcal{T}_{ij}\right\rangle \tau^i \xi^j
\end{align}
where $\sigma_{ij}= \gamma_{ij}^{\hbox{\tiny $(\partial)$}} +\tau_i \tau_j$ is the induced metric on $\Sigma_{\tau}$.
 The conserved charges of interest are the energy $E$, for $\xi=\partial/\partial \tau$, and
the angular momentum $J$, for $\xi=\partial/\partial \phi$.

Note that if we had not included the finite contribution from $S_{\hbox{\tiny \!{\it ct},3}}$ then one would not have $\delta S_{\hbox{\tiny ren}}=0$ and therefore we would  end up with a ``holographic stress tensor'' that is not conserved due to a contribution proportional to  $\big\langle  \mathcal{O}_{\Phi} \big\rangle \nabla _{\hbox{\tiny $(\partial)$}}^i \alpha+\big\langle  \mathcal{O^{\dagger}}_{\Phi} \big\rangle \nabla _{\hbox{\tiny $(\partial)$}}^i \alpha^{\dagger}$. Consequently, the associated ``energy'' and ``angular momentum'' \eqref{HoloRen:conservedCharges} would not be conserved quantities and the first law would not be $\d E= T \dd S + \Omega_H \dd J$. It would instead contain an extra unphysical contribution proportional to $\big\langle  \mathcal{O}_{\Phi} \big\rangle \mathrm{d}\alpha+\big\langle  \mathcal{O^{\dagger}}_{\Phi} \big\rangle  \mathrm{d}\alpha^{\dagger}$ (see section~\ref{secA:NoetherFirstLaw}). Note that getting the conserved charges is not just a conceptual or academic exercise. If we do not get the physical $E,J$ we cannot compare correctly two solutions (\eg the hairy and BTZ black holes) at the same energy and angular momentum to find which one has higher entropy and thus dominates the microcanonical ensemble.

The general holographic renormalization procedure for an arbitrary asymptotically AdS$_3$ spacetime with a  massive neutral complex scalar field subject with Dirichlet boundary conditions for the metric and double-trace boundary conditions for the scalar field is now terminated. Onwards, we apply it to our case of interest. Namely, we want the conformal boundary (with metric $\mathrm{d}s^2_{\hbox{\tiny $(\partial)$}}$) to be the Einstein Static Universe and we choose to be in a frame that does not rotate at infinity.
 This implies that  
 \begin{equation}\label{HoloRen:confFrame}
 A_0(\tau,\phi)=1, \qquad B_0(\tau,\phi)=1, \qquad \Omega_0(\tau,\phi)=0\,.
 \end{equation}
We impose these as Dirichlet BCs on the metric. We further impose double-trace BCs on the scalar field $\beta(\tau,\phi)=\kappa \, \alpha(\tau,\phi)$ with $\alpha(\tau,\phi)=e^{-i\hat{\omega} \tau}e^{i m \phi} \alpha$  and $\beta(\tau,\phi)=e^{-i \hat{\omega} \tau}e^{i m \phi} \beta$. In these conditions, \eqref{HoloRen:conservedCharges} yields the energy and angular momentum as
\begin{align}\label{HoloRen:EnergyAngMom}
    & E\equiv Q_{\hbox{\tiny $\partial_{\tau}$}}= -\frac{1}{4G}\Big[ A_2+  \Big(2-(\Delta_+-\Delta_-)^2 \Big)\kappa\, |\alpha|^2  \Big], \nonumber\\  
    & J \equiv Q_{\hbox{\tiny $\partial_{\phi}$}}=\frac{L}{4 G}\,\Omega_2\,.
\end{align}
Note that \eqref{HoloRen:EnergyAngMom} give the energy for the double-trace BC ($\beta=\kappa \, \alpha\neq 0$) but also for Dirichlet BCs ($\alpha=0$ ie the standard quantization) or Neumann BC ($\beta=0$, ie the alternative quantization). 

Denote the coefficient of the leading term of the asymptotic expansion of $\Omega(R)$ by $ \hat{C}_{\Omega}$ and further note that the asymptotic expansion of $f(R)$ is $f \big|_{R\to \infty}\sim R^2+\cdots -C_f +\cdots$. Applying the Fefferman-Graham coordinate  transformation 
\begin{align}\label{HoloRen:FGgeneral_appendix}
  T=\tau\,, \qquad R=\frac{1}{z} -\frac{1}{2} \,\alpha^2  z^{1-(\Delta_+-\Delta_-)}-\frac{1}{2}\Big( A_2 +4\Delta_-\Delta_+ \alpha\beta\Big) \, z+\cdots
\end{align}
one then finds that $A_2=-(\frac{1}{2} C_f+4 \Delta_-\Delta_+ \, \alpha \,\beta)$ and $\Omega_2=\hat{C}_\Omega$. Finally, imposing the double-trace BC, $\beta=\kappa \, \alpha$, we find that \eqref{HoloRen:EnergyAngMom} can be written as 
\begin{align}\label{HoloRen:EnergyAngMomFinal}
    & E\equiv Q_{\hbox{\tiny $\partial_T$}}= \frac{1}{8G}\left(C_f + 4  \kappa\, |\alpha|^2  \right), \nonumber\\  
    & J \equiv Q_{\hbox{\tiny $\partial_{\phi}$}}=\frac{L}{4 G}\,\hat{C}_\Omega\,.
\end{align}
Thus, \eqref{HoloRen:EnergyAngMom} or \eqref{HoloRen:EnergyAngMomFinal} reduces to the expression \eqref{Mass:Def} given in the main text.

\section{Covariant Noether Charge formalism, Noether charges and First Law \label{secA:NoetherFirstLaw}}

\subsection{Covariant Noether Charges and First Law for  \texorpdfstring{AdS$_{d+1}$}{AdSd+1} hairy black holes \label{secA:NoetherFirstLawAdSd}} 

The starting point of the Noether charge formalism is the variation of the bulk action  $S_{\hbox{\tiny bulk}}\equiv S$ as given in \eqref{eqn:action}. Our results in this section are valid for any spacetime dimension $d+1\geq 3$ (not only for the $d=2$ case of the main text) so we will keep $d$ arbitrary in the discussion of this section. 
We follow closely the discussion presented by Wald and collaborators in \cite{Lee:1990nz,Wald:1993nt, Iyer:1994ys, Wald:1999wa} but our analysis applies to asymptotically AdS$_{d+1}$ spacetimes with a complex scalar field.

In the Noether formalism it is often convenient to write expressions in their dualized version as differential forms.\footnote{We use differential forms' relations in the appendix of \cite{Dias:2019wof}.} The Lagrangian is then viewed as a $(d+1)$-form rather than a scalar density and the action $S_{\hbox{\tiny bulk}}$ is equivalently written as 
\begin{equation} \label{Noether:Sform}
S_{\hbox{\tiny bulk}} = \frac{1}{2 \kappa_{\hbox{\tiny {\it N}}}^2} \int_{\mathcal{M}} \bigg( \star\, \mathcal{R} -2\Lambda \star\mathbb{I}-2 \mathrm{d}\Phi \wedge \star\mathrm{d}\Phi^{\dagger} -2 V(\Phi \Phi^{\dagger})  \star\mathbb{I} \bigg)\,,
\end{equation}
where $\star$ stands for the Hodge duality operation and the Ricci volume $(d+1)$-form is  $\mathcal{R}= R \star\mathbb{I}$ and $\star\mathbb{I}=\mathrm{Vol}_{(d+1)}$ is the  volume $(d+1)$-form on $\mathcal{M}$. $\kappa_{\hbox{\tiny {\it N}}}^2=8\pi G$ ($G$ being Newton's constant in $(d+1)$ dimensions), and the cosmological constant is given in terms of the AdS radius as  $\Lambda=-\frac{d(d-1)}{2L^2}$. In the main text, we are interested in the case where the scalar potential is $V(\Phi\Phi^{\dagger})=\mu^2 \Phi\Phi^{\dagger}$ but our analysis will apply to any potential $V$.

As usual, the first order variation of the action \eqref{Noether:Sform} leads to the classical equations of motion. This variation includes boundary terms which do not affect the equations of motion. However these boundary terms are fundamental for computing conserved Noether charges and the first law of thermodynamics of the system. In particular, we must consider all boundary terms, the gravitational one and all the matter field ones, to get the correct thermodynamic quantities and identities. This is not often not done in the literature, leading to ``versions of the first law'' that are not physical.  

Consider then the variation, $\delta S_{\hbox{\tiny bulk}}$, of the bulk action \eqref{Noether:Sform} with respect to  the field variations $\{ \delta g^{ab},\delta\Phi,\delta\Phi^{\dagger} \}$.  After integration by parts of $\delta S_{\hbox{\tiny bulk}}$ to eliminate covariant derivatives of the field variations $-$ that introduces boundary terms (\ie total derivatives)  $-$ one gets
\begin{align}\label{Noether:varS}
 \delta S_{\hbox{\tiny bulk}} = &\: \frac{1}{2 \kappa_{\hbox{\tiny {\it N}}}^2} \int_{\mathcal{M}} \!\! \mathrm{d}^{d+1}x\sqrt{-g}\left[ R_{ab}-\frac{1}{2} \,R \, g_{ab}+\Lambda g_{ab}- T^{\Phi}_{ab}\right] \delta g^{ab} + \delta S_g^{bdry}\,, 
 \\
   \,&+ \frac{1}{\kappa_{\hbox{\tiny {\it N}}}^2}\int _{\mathcal{M}} \!\!  \mathrm{d}^{d+1}x\sqrt{-g} \left[\nabla^2\Phi^{\dagger}-\frac{\partial V}{\partial \Phi }\right]\delta \Phi  
   + \frac{1}{\kappa_{\hbox{\tiny {\it N}}}^2}\int_{\mathcal{M}} \!\!  \mathrm{d}^{d+1}x\sqrt{-g} \left[\nabla^2\Phi -\frac{\partial V}{\partial \Phi^{\dagger}}\right]\delta \Phi^{\dagger}  + \delta S_{\Phi,\Phi^{\dagger}}^{bdry} \nonumber
\end{align}
where the scalar field energy momentum tensor is 
\begin{equation}\label{Noether:scalarTab}
 T^{\Phi}_{ab}= \nabla_a\Phi \nabla_b\Phi^{\dagger}+\nabla_a\Phi^{\dagger} \nabla_b\Phi 
-g_{ab}\Big( \nabla_c\Phi  \nabla^c\Phi +V \Big)\,.
\end{equation}
In \eqref{Noether:varS},  $ \delta S_g^{bdry}$ and  $\delta S_{\Phi,\Phi^{\dagger}}^{bdry}$  are  total divergence contributions (\ie a boundary terms\footnote{Recall the divergence theorem $\int _{\mathcal{M}}\mathrm{d}^{d+1}x\sqrt{-g}\,\nabla_{a }X^{ab}=\int _{\partial \mathcal{M}} \mathrm{d}^{d}x\sqrt{-h}\,n_{a } X^{ab}$ for $n_a$ the unit outward normal.}), given by 
\begin{subequations}\label{Noether:bdryTerms0}
\begin{align}
 \delta S_g^{bdry} =&\: \frac{1}{2 \kappa_{\hbox{\tiny {\it N}}}^2} \int_{\mathcal{M}} \!\! \mathrm{d}^{d+1}x \sqrt{-g} \,g^{ab}\,\delta R_{ab}=\frac{1}{2 \kappa_{\hbox{\tiny {\it N}}}^2} \int_{\mathcal{M}}  \!\! \mathrm{d}^{d+1}x \, \nabla^a \Big[\sqrt{-g}  \left( \nabla^b \delta g_{ab}-\nabla_a \delta g\right)\Big],
  \\
 \delta S_{\Phi,\Phi^{\dagger}}^{bdry}  =&\: \frac{1}{\kappa_{\hbox{\tiny {\it N}}}^2}\int_{\mathcal{M}}  \!\! \mathrm{d}^{d+1}x\, \nabla^c\Big[-\sqrt{-g}\Big((\nabla_c \Phi^{\dagger})\delta \Phi +(\nabla_c\Phi) \delta \Phi^{\dagger}\Big)\Big],  
\end{align}
\end{subequations}
where $\delta g=g^{ab}\delta g_{ab}$.
It will be convenient to use the differential form language and encode the information of these boundary terms into symplectic potential $d$-forms, respectively, as:
\begin{subequations}\label{Noether:bdryTerms}
\begin{align}
\theta(g,\delta g)=&\: \frac{1}{2 \kappa_{\hbox{\tiny {\it N}}}^2}  \star\Big[ \left( \nabla^b \delta g_{ab}-\nabla_a \delta g\right) \dd x^a \Big],  \label{Noether:bdryTermsG}  \\
\theta(\Phi,\delta \Phi; \Phi^{\dagger},\delta \Phi^{\dagger})=&\: - \frac{1}{\kappa_{\hbox{\tiny {\it N}}}^2} \Big( 
\star\dd\Phi^{\dagger} \wedge \delta\Phi +\star\dd\Phi \wedge \delta\Phi^{\dagger}    \Big)\,. \label{Noether:bdryTermsPhi}
\end{align}
\end{subequations}

Requiring that the variations of \eqref{Noether:varS} w.r.t. $\{ \delta g^{ab},\delta\Phi,\delta\Phi^{\dagger} \}$ vanish leads to the equations of motion of the system:\footnote{In the dualized version as diﬀerential forms, the equations of motion for the scalar read $\dd \star \dd\Phi-\frac{\partial V}{\partial \Phi }\,\star \mathbb{I}=0$ and $\dd \star \dd\Phi^{\dagger}-\frac{\partial V}{\partial \Phi^{\dagger} }\,\star \mathbb{I}=0$.}
\begin{subequations}
\label{Noether:EOM}
\begin{align}
&R_{ab} = 2\Lambda g_{ab} + \Big(\nabla_a\Phi \nabla_b\Phi^{\dagger}+\nabla_a\Phi^{\dagger} \nabla_b\Phi \Big)+2V g_{ab}  \,,
\label{Noether:EOMg}
\\
&\nabla^2\Phi -\frac{\partial V}{\partial \Phi^{\dagger}}=0\,, \qquad \nabla^2\Phi^{\dagger}-\frac{\partial V}{\partial \Phi }= 0\,, \label{Noether:EOMphi}
\end{align}%
\end{subequations}
where we contracted the graviton equation of motion in \eqref{Noether:varS}-\eqref{Noether:scalarTab} with the inverse metric to get the on-shell Ricci scalar $R=2\Lambda+2\nabla_a\Phi \nabla_b\Phi^{\dagger} +6 V$ and then inserted this quantity back into  \eqref{Noether:varS} to get the trace reversed equation of motion for the graviton \eqref{Noether:EOMg}.\footnote{The on-shell Ricci volume $(d+1)$-form is then $\mathcal{R}= R \star\mathbb{I}=\Big(2\Lambda+2\nabla_a\Phi \nabla_b\Phi^{\dagger} +6 V \Big) \star\mathbb{I}$.}

Using these relations on \eqref{Noether:Sform} one finds that the on-shell $(d+1)$-form Lagrangian reads
\begin{equation}\label{Noether:onshellL}
\mathcal{L}\big |_{on-shell} = \frac{2}{\kappa_{\hbox{\tiny {\it N}}}^2} \big( \Lambda+ V \big) \star \mathbb{I}.
\end{equation}

Given a diffeomorphism vector generator $\xi$ (for now, not necessarily a Killing vector field), we can construct the associated symplectic Noether current $d$-form \cite{Wald:1993nt, Iyer:1994ys, Wald:1999wa}:
 \begin{equation} \label{Noether:current}
J =\Theta\left(g,\mathcal{L}_\xi g\right) + \Theta\left(\Phi,\mathcal{L}_\xi \Phi; \Phi^{\dagger},\mathcal{L}_\xi \Phi^{\dagger}\right)- \ins_\xi \mathcal{L}\big |_{on-shell} \,,
\end{equation}
where $\Theta(\phi_i,\mathcal{L}_\xi \phi_i)\equiv \theta(\phi_i,\mathcal{L}_\xi \phi_i)$ for $\phi_i = \{g,\Phi,\Phi^{\dagger}\}$, \ie we make the replacements $\delta \phi_i\to \mathcal{L}_\xi \phi_i$ on the boundary terms \eqref{Noether:bdryTerms}. Here, $\mathcal{L}_\xi \phi_i$ is the  Lie derivative of the field $\phi_i$ along the diffeomorphism generator $\xi$. Also, $ \ins_\xi \mathcal{L}\big |_{on-shell} $ is the interior product  of $\xi$ with the $(d+1)$-form \eqref{Noether:onshellL}. 

It can be shown \cite{Lee:1990nz} that $\dd J=- E_i \,\mathcal{L}_\xi \phi_i$, where $E_i$ stands for the equations of motion (Einstein summation convention holds here) \cite{Wald:1993nt, Iyer:1994ys, Wald:1999wa}. Therefore, on-shell ($E_i=0$) the current is closed, \ie $\dd J=0$ for all $\xi$. It follows that there is a Noether charge $(d-1)$-form relative to $\xi^a$, call it $\widetilde{Q}_\xi$, locally constructed from $\{\xi, \phi_i\}$, such that on-shell one has $J=\dd \widetilde{Q}_\xi$, since in these conditions $\dd J=\dd^2\widetilde{Q}_\xi=0$ \cite{Wald:1993nt, Iyer:1994ys, Wald:1999wa}.\footnote{$\widetilde{Q}_\xi$ is defined uniquely up to the addition of a closed and thus exact form $\dd \psi$. The integral of $\widetilde{Q}_\xi$ over a closed surface $\Sigma$ is  the Noether charge of  $\Sigma$ relative to $\xi^a$ In \cite{Wald:1993nt, Iyer:1994ys, Wald:1999wa} $\widetilde{Q}_\xi$ is denoted simply as $Q$.} 

To evaluate \eqref{Noether:current}, and then find $\widetilde{Q}_\xi$, it is useful to recall the definition of Lie derivative of a scalar and of a torsion-free metric tensor (with a Levi-Civita connection)
\begin{subequations}\label{Noether:Cartan}
 \begin{align}
& \mathcal{L}_\xi \Phi=\xi^a\nabla_a \Phi =\ins_\xi d \Phi\,,  \label{Noether:CartanScalar}\\
& \mathcal{L}_\xi g_{ab}=2\nabla_{(a} \xi_{b)}\,,  \label{Noether:LieGrav} \\
&  \mathcal{L}_\xi A_{(p)} = \dd \left(\ins_\xi A_{(p)}\right)+ \ins_\xi \dd A_{(p)}  \label{Noether:LieForm}
\end{align}
\end{subequations}
where we have also given Cartan's formula for the Lie derivative of a $p$-form $A_{(p)}$ to be used later.\footnote{Identity \eqref{Noether:Cartan} for a scalar field is simply Cartan's formula \eqref{Noether:LieForm} for $p=0$.} 

Using these relations and the equations of motion \eqref{Noether:EOM}, one finds:\footnote{To get $\Theta\left(g,\mathcal{L}_\xi g\right)$ we further use the commutator relation $\left[\nabla_a,\nabla_b\right]\xi_c=R_{abcd}\xi^d$}   
 \begin{align}\label{Noether:ThetaG}
 \Theta\left(g,\mathcal{L}_\xi g\right) =&\: \star\Big( \frac{1}{\kappa_{\hbox{\tiny {\it N}}}^2}\left[\nabla^b \nabla_{(a}\xi_{b)} -g^{bc}\nabla_a\nabla_{(b}\nabla_{c)}\right]\dd x^a \Big)\nonumber\\
  =&\:\frac{1}{d!}\left[ \frac{1}{\kappa_{\hbox{\tiny {\it N}}}^2} \varepsilon_{a_1\cdots a_d a}\left( \nabla_b \nabla^{[b}\xi^{a]}+R^a_{\:\:b}\xi^b\right)\right]\dd x^{a_1}\wedge \cdots \wedge \dd x^{a_d}\,, 
  \\
  \Theta\left(\Phi,\mathcal{L}_\xi \Phi;\Phi^{\dagger},\mathcal{L}_\xi \Phi^{\dagger}\right) =&\:
  -\frac{1}{\kappa_{\hbox{\tiny {\it N}}}^2}\left( \star\mathrm{d}\Phi^{\dagger}\wedge \ins_\xi \mathrm{d}\Phi +\star \mathrm{d}\Phi\wedge \ins_\xi \mathrm{d}\Phi^{\dagger}\right)\nonumber\\
=&\:\frac{1}{d!}\left[ \frac{1}{\kappa_{\hbox{\tiny {\it N}}}^2} \varepsilon_{a_1\cdots a_d a}\left( -\nabla^a\Phi^{\dagger}\nabla_b\Phi -\nabla^a\Phi \nabla_b\Phi^{\dagger} \right)\xi^b \right]\dd x^{a_1}\wedge \cdots \wedge \dd x^{a_d},
\end{align}
and
\begin{equation}\label{Noether:ThetaLag}
- \ins_\xi \mathcal{L}\big |_{on-shell} = \frac{1}{9!} \left[ \frac{1}{\kappa_{\hbox{\tiny {\it N}}}^2} \varepsilon_{a_1\cdots a_d a}\left( 
-2\Lambda-2 V\right)\xi^b \right]\dd x^{a_1}\wedge \cdots \wedge \dd x^{a_d}.
\end{equation}
Note that most of the contributions in  \eqref{Noether:ThetaG}-\eqref{Noether:ThetaLag}  but one add-on to build the equation of motion for the graviton \eqref{Noether:EOMg} and thus will not contribute to the final current. The only  contribution that survives after using the graviton equation of motion  \eqref{Noether:EOMg} is the first term in \eqref{Noether:ThetaG} that can be rewritten as:
 \begin{eqnarray}\label{Noether:stardxi}
\frac{1}{9!}\left[ \frac{1}{\kappa_{\hbox{\tiny {\it N}}}^2} \varepsilon_{a_1\cdots a_9 a} \nabla_b \nabla^{[b}\xi^{a]} \right]\dd x^{a_1}\wedge \cdots \wedge \dd x^{a_9} =\frac{1}{2\kappa_{\hbox{\tiny {\it N}}}^2}\dd\star\dd\xi\,.
\end{eqnarray}
Thus, we finally conclude that the symplectic Noether current $d$-form \eqref{Noether:current} is indeed closed for all $\xi^a$ since it is given by 
\begin{eqnarray}\label{Noether:currentFinal1}
J=\dd \widetilde{Q}_\xi\,,
\end{eqnarray}
with the Noether $(d-1)$-form charge relative to $\xi^a$ being simply 
\begin{eqnarray}\label{Noether:chargeQxi}
\widetilde{Q}_\xi \equiv \frac{1}{2\kappa_{10}^2}\star \dd \xi  
\end{eqnarray}
and its integral over a closed surface $\Sigma$ is  the Noether $(d-1)$-form charge of $\Sigma$  relative to $\xi^a$.
Note that, consistent with the discussion below \eqref{Noether:current}, we find that on-shell the current $J$ is indeed conserved, $\dd J=\dd^2\widetilde{Q}_\xi=0$. 

Now we want to consider variation of the current \eqref{Noether:current} when we consider arbitrary variations  of the dynamical fields $\delta\phi_i=\{\delta g_{ab}, \delta \Phi, \delta \Phi^{\dagger} \}$ (by now, not necessarily a solution of the linearized equations of motion) and arbitrary solutions $\phi_i=\{g_{ab}, \Phi, \Phi^{\dagger} \}$ of \eqref{Noether:EOM}. This is given by 
 \begin{equation} \label{Noether:deltaCurrent0}
\delta J =\delta\Theta\left(g,\mathcal{L}_\xi g\right) + \delta\Theta\left(\Phi,\mathcal{L}_\xi \Phi; \Phi^{\dagger},\mathcal{L}_\xi \Phi^{\dagger}\right)- \ins_\xi \delta\mathcal{L} |_{on-shell}\,,
\end{equation}
where we take the arbitrary diffeomorphism vector generator $\xi$ to be held fixed in this variation.
From \eqref{Noether:varS}-\eqref{Noether:bdryTerms},  the variation of the interior product of the  $(d+1)$-form Lagrangian is 
\begin{align}\label{Noether:deltaL}
\ins_\xi \delta \mathcal{L} |_{on-shell} = &\: \ins_\xi \Big[ (\cdots)\delta g^{ab}+ (\cdots)\delta \Phi + (\cdots)\delta \Phi^{\dagger} \Big]+ \ins_\xi \mathrm{d}\theta\left(g,\delta g\right) +  \ins_\xi\mathrm{d}\theta\left(\Phi,\delta\Phi; \Phi^{\dagger},\delta \Phi^{\dagger}\right) \nonumber\\
 = &\:  \mathcal{L}_\xi \theta\left(g,\delta g\right)+ \mathcal{L}_\xi \theta\left(\Phi,\delta\Phi; \Phi^{\dagger},\delta \Phi^{\dagger}\right)
   - \dd \ins_\xi \theta\left(g,\delta g\right)  - \dd  \ins_\xi \theta\left(\Phi,\delta\Phi; \Phi^{\dagger},\delta \Phi^{\dagger}\right) 
\end{align}
where, to get the second line, we used the fact that the $(\cdots)$ vanish on-shell $-$ because they are given by the equations of motion  \eqref{Noether:EOM} $-$ and Cartan's formula \eqref{Noether:LieForm}. Plugging this into \eqref{Noether:deltaCurrent0} yields
 \begin{align} \label{Noether:deltaCurrent1}
\delta J =&\: \Big[ \delta\Theta\left(g,\mathcal{L}_\xi g\right)-\mathcal{L}_\xi \theta\left(g,\delta g\right) \Big]
 + \Big[ \delta\Theta\left(\Phi,\mathcal{L}_\xi \Phi; \Phi^{\dagger},\mathcal{L}_\xi \Phi^{\dagger}\right) -\mathcal{L}_\xi \theta\left(\Phi,\delta\Phi; \Phi^{\dagger},\delta \Phi^{\dagger}\right) \Big] 
 \nonumber \\
 &\: +  \dd \ins_\xi \theta\left(g,\delta g\right)  + \dd  \ins_\xi \theta\left(\Phi,\delta\Phi; \Phi^{\dagger},\delta \Phi^{\dagger}\right) 
\,,
\end{align}
Since $\theta$ is covariant, $\mathcal{L}_\xi \theta(\phi_i,\delta\phi_i)$ is the same as the variation induced in $\theta$ by the field variation $\hat{\delta}\phi_i = \mathcal{L}_\xi$. Thus $ \delta\Theta\left(\phi_i,\mathcal{L}_\xi \phi_i\right)-\mathcal{L}_\xi \theta\left(\phi_i,\delta \phi_i\right)=\mathit{w}(\phi_i,\delta \phi_i,\mathcal{L}_\xi \phi_i)$ where  the presymplectic current $d$-form $\mathit{w}$ is defined in terms of the antisymmetrized variation of $\theta$, or two distinct linear perturbations  $\delta_1\phi$  $\delta_2\phi$,  as $\mathit{w}(\phi_i,\delta_1 \phi_i,\delta_2 \phi_i)= \delta_1\theta\left(\phi_i, \delta_2 \phi_i\right)-\delta_2 \theta\left(\phi_i,\delta_1 \phi_i\right)$  \cite{Lee:1990nz}. Thus, \eqref{Noether:deltaCurrent1} can be written simply as   
 \begin{equation} \label{Noether:deltaCurrent2}
\delta J = \mathit{w}\left(g,\delta g,\mathcal{L}_\xi g\right) +\mathit{w}\left(\Phi,\delta\Phi,\mathcal{L}_\xi \Phi; \Phi^{\dagger},\delta \Phi^{\dagger},\mathcal{L}_\xi \Phi^{\dagger}\right) 
+  \dd \ins_\xi \theta\left(g,\delta g\right)  + \dd  \ins_\xi \theta\left(\Phi,\delta\Phi; \Phi^{\dagger},\delta \Phi^{\dagger}\right) 
\,,
\end{equation}
One can now introduce a a closed $d$-dimensional submanifold {\it without} boundary and define the presymplectic $d$-form associated with $\Sigma$ as the integral of the  presymplectic current $\mathit{w}$ over $\Sigma$:  $\Omega_\Sigma= \int_\Sigma  \mathit{w}$. Then, integration of \eqref{Noether:deltaCurrent2} yields
 \begin{align} \label{Noether:deltaCurrent3}        
   \Omega_\Sigma\left(g,\delta g,\mathcal{L}_\xi g\right) +\Omega_\Sigma \left(\Phi,\delta\Phi,\mathcal{L}_\xi \Phi; \Phi^{\dagger},\delta \Phi^{\dagger},\mathcal{L}_\xi \Phi^{\dagger}\right)  =&\:      \int_\Sigma \delta J -  \int_\Sigma \dd \Big[  \ins_\xi \theta\left(g,\delta g\right)  +   \ins_\xi \theta\left(\Phi,\delta\Phi; \Phi^{\dagger},\delta \Phi^{\dagger}\right)   \Big] 
   \nonumber\\
\Leftrightarrow \qquad \delta H_{\xi}  =&\:      \int_\Sigma \delta J -  \int_\Sigma \dd \Big[  \ins_\xi \theta\left(g,\delta g\right)  +   \ins_\xi \theta\left(\Phi,\delta\Phi; \Phi^{\dagger},\delta \Phi^{\dagger}\right)   \Big]  \,,     
   \end{align}
   The left hand side (LHS) of this equation defines the variation, $\delta H_{\xi}$, of the Hamiltonian (if it exists) conjugate to the evolution by the vector field $\xi^a$ on $\Sigma$ \cite{Wald:1993nt, Iyer:1994ys, Wald:1999wa}.

So far, we have taken $\xi$ to be just a diffeomorphism generator vector field and we also left  $\delta\phi_i=\{\delta g_{ab}, \delta \Phi, \delta \Phi^{\dagger} \}$ arbitrary. However, onwards we require that $\delta\phi_i=\{\delta g_{ab}, \delta \Phi, \delta \Phi^{\dagger} \}$ are solutions of the linearized equations of motion and we further we take  $\xi$ to be a Killing vector field  \cite{Wald:1993nt, Iyer:1994ys, Wald:1999wa}.\footnote{Although of no interest in the system at hand, the analysis done here also applies if $\xi^a$ is `just an asymptotically Killing vector field  \cite{anderson1996asymptotic, barnich2002covariant, barnich2003boundary, barnich2008surface, Compere:2007vx, Compere:2007az, Chow:2013gba}. For completeness, it should be noted that on generic systems \eqref{Noether:deltaCurrent3}, and thus \eqref{Noether:VarCharge},  might also have a contribution, usually denoted as $\widetilde{Q}_{\delta \xi}$ if $\xi^a$ happens to depend on the solution parameters $m_k$ (this is certainly not our case since we only consider the Killing vector fields  $\xi=\partial_t$ and $\xi=\partial_{\phi}$ responsible for time or rotational symmetries, respectively).} The latter condition means that $\mathcal{L}_\xi \phi_i=0$ for all dynamical fields   $\phi_i=\{ g_{ab}, \Phi, \Phi^{\dagger} \}$; \eg $\xi=\partial_t$ and $\xi=\partial_{\phi}$ are the Killing vector fields responsible for time and rotational symmetries, respectively.} On the other hand, when $\delta\phi_i$ are solutions of the linearized equations of motion, one can replace  $\delta J$ by $\delta \dd\tilde{Q}_\xi= \dd \delta \tilde{Q}_\xi$ in \eqref{Noether:deltaCurrent2}. All together,  
\eqref{Noether:deltaCurrent3} reduces then to
 \begin{align} \label{Noether:deltaH}
\delta H_{\xi} =&\:   \int_\Sigma \dd \omega_\xi = \int_{\partial \Sigma} \omega_\xi \,,  
\end{align}
where to get the last relation we considered a limiting process whereby the integral is first taken over a a compact region $K$ of $\Sigma$ with boundary $\partial K$, we then apply Stokes' theorem  $ \int_K \dd \omega_\xi = \int_{\partial K} \omega_\xi$, and finally we let $K$ approach $\Sigma$ \cite{Wald:1999wa}.
In \eqref{Noether:deltaH},  $\omega_\xi$ is the $(d-1)$-form  variation (in the moduli space) of the charge $\tilde{Q}_\xi$ associated to $\xi$ defined in \eqref{Noether:chargeQxi}:
\begin{equation}\label{Noether:VarCharge}
\omega_\xi \equiv  \delta \widetilde{Q}_\xi - \ins_\xi \theta\left(g,\delta g\right) -\ins_\xi \theta\left(\Phi,\delta\Phi; \Phi^{\dagger},\delta \Phi^{\dagger}\right),
\end{equation}
with $\tilde{Q}_\xi$  defined in \eqref{Noether:chargeQxi} and the two boundary terms $\theta$  defined in \eqref{Noether:bdryTerms}. Solutions of  AdS Einstein theory coupled to a neutral complex scalar field typically depend on  one or more parameters $m_k$ (say, with $k=1,\cdots$; \eg the horizon radius, ...). Thus, the variation of  $\widetilde{Q}_\xi $ along this moduli space of solutions is given by $\delta \widetilde{Q}_\xi =  \partial_{m_k} \!\widetilde{Q}_\xi \,  \dd m_k $.

It follows from  \eqref{Noether:deltaH} that the  integration of \eqref{Noether:VarCharge} over the sphere $S^{(d-1)}_{\infty}$ $-$ \ie \eqref{Noether:deltaH} for $\Sigma\equiv S^{(d-1)}_{\infty}$  $-$ yields the variation of the canonical conserved charge conjugate to the Killing vector field $\xi$. Moreover, if and only if $-$ and for generic systems this is certainly not granted $-$ the system is manifestly integrable, \footnote{Formally, the system is integrable if there exists a $d$-form $\mathcal{B}$ such that $\delta\int_{S^{(d-1)}_{\infty}} \ins_\xi \mathcal{B} = \int_{S^{(d-1)}_{\infty}} \ins_\xi \theta$ for a sphere $S^{(d-1)}_{\infty}$ at spacial infinity. In practice, this is true if the integral of \eqref{Noether:VarCharge} over the solution parameters exists. This is the case for the system at hand so we do not discuss how, in some cases, we might still be able to compute the conserved charges of the system when integrability is not manifest. In  \cite{Dias:2019wof}, the reader can find a discussion of such cases in the context of 11-dimensional and IIB supergravities, that contain gauge/scalar fields in addition to the graviton, and we use asymptotic scale-invariances of the systems to find the conserved charges.} then a second integration, this time over the solution parameters ($\int dm_k$), yields the canonically conserved charge conjugate to the Killing vector field $\xi$. For the stationary Killing vector field $\xi=\partial_t$ this is the energy $\mathcal{E}$, while for the axisymmetric Killing vector field $\xi=\partial_{\phi}$ this is the angular momentum $\mathcal{J}$:
 \begin{subequations}  \label{Noether:canonicalEnergy}
 \begin{align} 
\mathcal{E} \equiv &\:  \int \dd m_k  \,\delta \mathcal{E} \equiv \int \dd m_k  \int_{S^{(d-1)}_{\infty}} \omega_{\xi}\big|_{\xi=\partial_t}\,, \\
\mathcal{J} \equiv &\:   \int \dd m_k \, \delta \mathcal{J} \equiv \int \dd m_k  \int_{S^{(d-1)}_{\infty}} \left(-\omega_{\xi}\big|_{\xi=\partial_{\phi}} \right)\,.
  \end{align}
 \end{subequations}
It can be formally demonstrated from first principles that the conserved Noether charges \eqref{Noether:canonicalEnergy} (a.k.a. covariant phase space charges) do agree with the holographic conserved charges \eqref{HoloRen:conservedCharges} (see e.g \cite{Papadimitriou:2005ii} for $AdS$ Einstein-Maxwell theory) and later we will check that this is indeed also the case for the AdS$_{d+1}$ Einstein theory coupled to a neutral complex scalar field described by \eqref{Noether:Sform}.
   
Now we can also derive the first law of thermodynamics for gravitational solutions of  AdS$_{d+1}$ Einstein theory coupled to a neutral complex scalar field.  This first law should be independent of the asymptotic boundary conditions imposed on the scalar field.
Again, we follow closely the discussion of \cite{Wald:1993nt, Iyer:1994ys, Wald:1999wa}.

 Let us start by assuming that the solution is a (rotating) black hole. We can then consider the Killing vector field 
 \begin{align}\label{Noether:HorizonKVF}
 \xi \equiv K=\partial_t+\Omega_H \partial_{\phi} \,,
 \end{align}   
 where $\partial_t$ has unit norm at infinity and  $\Omega_H$ is precisely  the angular velocity at the horizon in which case  $\xi$ is null at the horizon, $|\xi|_{\mathcal{H}}=0$, \ie the event horizon is a Killing horizon generated by \eqref{Noether:HorizonKVF}. That is, one has 
  $|\xi|^2 = -f(r)g(r) + r^2\left[\Omega(r)-\Omega_H\right]^2$, which vanishes at $r = r_+$ since $f(r_+) = 0$ (condition that introduces the location of the horizon) and $\Omega_H \equiv \Omega(r_+)$.\footnote{Here, we are implicitly  assuming that we are in a frame that does not rotate at the AdS boundary. If this is not the case, $\Omega_H$ in \eqref{Noether:HorizonKVF} should be interpreted as the difference between the angular velocities at the horizon and infinity. Also, for $d\geq 4$, asymptotically AdS$_{d+1}$ can have more than one rotational planes with associated axisymmetric Killing fields $\partial_{\phi_j}$ and horizon angular velocities $\Omega_H^{(j)}$. Our analysis covers such cases; we just need the replacement $\Omega_H \partial_{\phi} \to \sum_j \Omega_H^{(j)} \partial_{\phi_j} $ in \eqref{Noether:HorizonKVF} and there is an angular momentum $J_j$ conjugate to each  $\partial_{\phi_j}$.} 
 Since our $\xi$ is a symmetry of all the fields $\phi_i=\{g,\Phi,\Phi^{\dagger}\}$ one has $\mathcal{L}_\xi \phi_i=0$ and thus $\mathit{w}\left(\phi_i,\delta \phi_i,\mathcal{L}_\xi \phi_i\right)=0$ in \eqref{Noether:deltaCurrent2} \cite{Wald:1993nt, Iyer:1994ys, Wald:1999wa}. Moreover, if we assume that  $\delta\phi_i=\{\delta g_{ab}, \delta \Phi, \delta \Phi^{\dagger} \}$ are solutions of the linearized equations of motion, one can replace  $\delta J$ by $\delta \dd\tilde{Q}_\xi= \dd \delta \tilde{Q}_\xi$ in \eqref{Noether:deltaCurrent2}. Altogether, in these conditions \eqref{Noether:deltaCurrent2} reads $\dd \delta  \widetilde{Q}_\xi  -
 \dd \ins_\xi \theta\left(\phi_i,\delta \phi_i\right)=0 $ (Einstein summation implicit), \ie the $\omega_\xi $ defined in \eqref{Noether:VarCharge} is closed: $\dd \omega_\xi =0$.
 Let us now integrate this relation over a $d$-dimensional Cauchy surface, $\mathcal{C}$, with boundary $\delta\mathcal{C}$ that includes two components: the bifurcating Killing horizon  (call it $\Sigma_{\mathcal{H}}$) and the asymptotic AdS boundary (which is simply the sphere $S^{(d-1)}_{\infty}$ at infinity). Using Stokes' theorem this yields $0=\int_{\mathcal{C}} \dd \omega_\xi =\int_{\partial \mathcal{C}}  \omega_\xi =\int_{S^{(d-1)}_{\infty}} \omega_\xi-\int_{\Sigma_{\mathcal{H}}} \omega_\xi$, where we used the fact that the outward unit normal of $\mathcal{C}$ at the inner boundary has opposite sign to the one at the asymptotic boundary. That is, 
  \begin{align} \label{Noether:FirstLaw0}
 \int_{S^{(d-1)}_{\infty}} \omega_\xi = \int_{\Sigma_{\mathcal{H}}} \omega_\xi 
 \end{align}   
 and thus the integral of  $\omega_\xi$ is independent of the radial position. From \eqref{Noether:canonicalEnergy} and \eqref{Noether:HorizonKVF}, the LHS of \eqref{Noether:FirstLaw0} is simply $\delta E- \Omega_H \delta J$. On the other hand, on general grounds (as we show for our particular system below), it can be formally established that the RHS of \eqref{Noether:FirstLaw0} is $T_H \delta S_H$, where $T_H$ and $S_H$ are the temperature and entropy of the black hole horizon. Altogether, \eqref{Noether:FirstLaw0} with the horizon generator \eqref{Noether:HorizonKVF} yields the first law of thermodynamics for AdS$_{d+1}$ Einstein theory coupled to a neutral complex scalar field described by the action \eqref{Noether:Sform}, namely
   \begin{align} \label{Noether:FirstLaw}
\delta \mathcal{E}- \Omega_H \delta \mathcal{J} = T_H \delta S_H \qquad \implies \quad \dd \mathcal{E}- \Omega_H \dd \mathcal{J} = T_H \dd S_H \,. 
 \end{align}  
Very importantly, note that variations of scalar field contributions do not appear at all in \eqref{Noether:FirstLaw}, as they should not: for double‑trace deformations, $\alpha$ (and thus also $\beta=\kappa\,\alpha$) is a state parameter, behaving as a VEV‑type quantity whose value is dynamically determined rather than externally fixed. In some literature, the scalar field boundary contribution \eqref{Noether:bdryTermsPhi} is not taken into account. Consequently, the contribution $\ins_\xi \theta\left(\Phi,\delta\Phi; \Phi^{\dagger},\delta \Phi^{\dagger}\right)$ in \eqref{Noether:VarCharge} is also not taken into account. This has two ultimate consequences: the associated \eqref{Noether:canonicalEnergy} does not give the correct {\it conserved} energy and angular momentum of the system and the associated ``first law'' contains a term proportional to  $(\beta \mathrm{d}\alpha+\beta^{\dagger} \mathrm{d}\alpha^{\dagger})$ if double-trace boundary conditions are imposed on the scalar field. Such relations are not physical. 
 
 Let us now derive the first law of thermodynamics for the boson stars described by the field ans\"atze \eqref{HairyAnsatz}.
Let us start by assuming $m\neq 0$. Then, the only linear combination of $\partial_t$ and $\partial_{\phi}$ that leaves the scalar field \eqref{scalar_ansatz} invariant is \eqref{BS:KVF}, namely
  \begin{align}\label{NoetherBS:KVF}
 \xi= \xi^a\partial_a=\partial_t+\frac{\omega}{m} \,\partial_{\phi} \,.
 \end{align}   
A symmetry of the solution must leave both the matter and metric fields invariant and thus a boson star with $m>0$ only has a single Killing vector field given by the helical vector field \eqref{NoetherBS:KVF}. In particular, this means that it is not time independent neither axisymmetric but it is time-periodic. Nevertheless, the boson star metric \eqref{metric_ansatz} only depends on the radial coordinate because the energy-momentum tensor of the complex scalar field,  
$ 8 \pi G \, T_{\mu\nu} = \partial_\mu \Phi \partial_\nu\Phi^{\dagger} +\partial_\mu \Phi^{\dagger} \partial_\nu\Phi - g_{\mu\nu}\left(g^{\alpha\beta}\partial_\alpha\Phi\partial_\beta\Phi^{\dagger} + \mu^2 \Phi \Phi^{\dagger}\right)$ is independent of $t$ and $\phi$.
Given these considerations, it is convenient to use \eqref{NoetherBS:KVF} to discuss the first law for boson stars. This is because  
 $\xi$ given by \eqref{NoetherBS:KVF} is a symmetry of all the boson star fields $\phi_i=\{g,\Phi,\Phi^{\dagger}\}$, $\mathcal{L}_\xi \phi_i=0$, and thus $\mathit{w}\left(\phi_i,\delta \phi_i,\mathcal{L}_\xi \phi_i\right)=0$ in \eqref{Noether:deltaCurrent2} \cite{Wald:1993nt, Iyer:1994ys, Wald:1999wa}. Moreover, as for the black hole case, if we assume again that  $\delta\phi_i=\{\delta g_{ab}, \delta \Phi, \delta \Phi^{\dagger} \}$ are solutions of the linearized equations of motion, one can replace  $\delta J$ by $\delta \dd\tilde{Q}_\xi= \dd \delta \tilde{Q}_\xi$ in \eqref{Noether:deltaCurrent2}. Altogether, in these conditions \eqref{Noether:deltaCurrent2} reads $\dd \delta  \widetilde{Q}_\xi  -
 \dd \ins_\xi \theta\left(\phi_i,\delta \phi_i\right)=0 $ (Einstein summation implicit), \ie the $\omega_\xi $ defined in \eqref{Noether:VarCharge} is closed: $\dd \omega_\xi =0$.
Let us now integrate this relation over a $d$-dimensional Cauchy surface, $\mathcal{C}$, with boundary $\delta\mathcal{C}$ that this time includes a single component, namely the asymptotic AdS boundary $S^{(d-1)}_{\infty}$ (since the origin $r=0$ is not a boundary of $\mathcal{C}$). Using again Stokes' theorem, this time it yields $0=\int_{\mathcal{C}} \dd \omega_\xi =\int_{\partial \mathcal{C}}  \omega_\xi =\int_{S^{(d-1)}_{\infty}} \omega_\xi$. That is, 
  \begin{align} \label{NoetherBS:FirstLaw}
 \int_{S^{(d-1)}_{\infty}} \omega_\xi = 0\, \quad \Leftrightarrow \quad \delta \mathcal{E}- \frac{\omega}{m}  \delta \mathcal{J} = 0 \qquad \implies \quad \dd \mathcal{E}= \frac{\omega}{m} \, \dd \mathcal{J} 
 \end{align}   
where, to obtain the integral, we used \eqref{Noether:canonicalEnergy} with $\xi$ given by \eqref{NoetherBS:KVF}. This is the desired first law of thermodynamics for the non-axisymmetric boson stars of AdS$_{d+1}$ Einstein theory coupled to a neutral complex scalar field described by the action \eqref{Noether:Sform}.

On the other hand, on general grounds (as we show for $d=2$ below), it can be formally established that $\mathcal{J}=m N$ where $N$ is the conserved charge  (the number of scalar particles) $-$ see \eqref{N:Def} $-$  associated to the conserved current of the $U(1)$ global symmetry $\Phi\to e^{i\,\eta} \Phi$ \cite{Ruffini:1969qy,Schunck:1996he} (for recent reviews on boson stars see \cite{Liebling:2012fv,Shnir:2022lba}).  Thus, the first law of thermodynamics for the non-axisymmetric boson stars of AdS$_{d+1}$ Einstein theory coupled to a neutral complex scalar field can be written as 
  \begin{align} \label{NoetherBS:FirstLaw2}
\dd \mathcal{E}= \omega \,\dd N \,,
 \end{align}   
 which is valid also for axisymmetric ($m=0$) boson stars.
 
 \subsection{Application: Noether Charges and First Law for \texorpdfstring{AdS$_{3}$}{AdS3} hairy black holes \label{secA:NoetherFirstLawAdS3}} 
 
Summarizing, we have discussed how we can formally compute the conserved Noether charges and derive the first law of thermodynamics for an arbitrary solution of AdS$_{d+1}$ Einstein theory coupled to a neutral complex scalar field with potential $V(\Phi\Phi^{\dagger})$. Next, we explicitly compute these quantities for the most general solution $-$ be it a black hole or a soliton $-$ of this theory with spherical topology for the case $d=2$ (the case $d>2$ proceeds in a very similar way since it simply amounts to replace the $S^1$ by $S^{d-1}$) and $V(\Phi\Phi^{\dagger})=\mu^2 \Phi\Phi^{\dagger}$. For that, we introduce the coordinate chart $\{t,r,\phi
\}$ (where $\phi$ parametrizes the 1-sphere $S^1$) and we assume that $\partial_t$ and $\partial_{\phi}$ are stationary and axisymmetric Killing vector fields of the solution. 
 For a most general ansatz that describes asymptotically AdS$_3$ stationary solutions with a massive complex scalar field we can use \eqref{HairyAnsatz} (w.l.g. in the Schwarzschild gauge whereby $r$ is simply the areal radius, \ie  $g_{\phi\phi}=r^2$) which we rewrite here for convenience:
 \begin{subequations} \label{Noether:HairyAnsatz}
 \begin{align}
   \label{Noether:metric_ansatz} ds^2 &= -f(r)g(r)\mathrm{d}t^2 + \frac{\mathrm{d}r^2}{f(r)} + r^2 \Big(\mathrm{d}\phi- \Omega(r)\mathrm{d}t \Big)^2, \\
   \label{Noether:scalar_ansatz} \Phi &= e^{-i\omega t + i m \phi}\psi(r)\,,
\end{align}
 \end{subequations}
where $\phi \sim \phi + 2\pi$ and regularity requires that the azimuthal number $m$ is an integer. We choose the conformal frame to be  such that the boundary metric is the Einstein Static Universe (so, a frame that does not rotate at infinity and thus $\omega=m \Omega_H$ by the equations of motion evaluated at the horizon and its regularity).
The temperature, entropy and angular velocity of a black hole with event horizon at $r=r_+$ (defined as the locus of $f(r_+)=0$) are  
  \begin{subequations} \label{Noether:TSO}
 \begin{align}
  T_H = \frac{1}{4\pi}\sqrt{g(r_+)}f'(r_+)\,, \qquad S_H = \frac{\pi}{2 \,G} \,r_+\,, \qquad \Omega_H=\Omega(r_+).
  \end{align}
 \end{subequations}
Equipped with  \eqref{HairyAnsatz}, the Noether 1-form  charge \eqref{Noether:chargeQxi} relative to the Killing horizon generator, \ie $\xi$ as given by \eqref{Noether:HorizonKVF}, and the boundary 2-forms \eqref{Noether:bdryTerms}  are given by
 \begin{subequations}
 \begin{align}
 \widetilde{Q}_\xi =&\: \frac{1}{2\kappa_{\hbox{\tiny {\it N}}}^2} \frac{1}{\sqrt{g}} \bigg\{
 \bigg[  fg \bigg( 2(\Omega-\Omega_H) +  r\,\Omega' \Big( 1+r^2(\Omega-\Omega_H)\frac{\Omega}{f g}  \Big) \bigg) -r\, \Omega (f g)' \bigg] \dd t 
 \nonumber \\
 &\: \hspace{1.5cm} + r\Big[ (fg)' - r^2(\Omega- \Omega_H)\Omega' \Big] \dd\phi  \bigg\} \,;   
 \\
 \theta\left(g,\delta g\right) = &\:    \frac{1}{2\kappa_{\hbox{\tiny {\it N}}}^2}\, r\left[\delta\left(\frac{(fg)'}{\sqrt{g}}\right) - r^2 \frac{\Omega'}{\sqrt{g}}\delta\Omega +  \frac{\sqrt{g}}{r} \delta f \right]  \dd t \wedge \dd\phi 
  \\
 \theta\left(\Phi,\delta\Phi; \Phi^{\dagger},\delta \Phi^{\dagger}\right) =&\: 
 -\frac{1}{\kappa_{\hbox{\tiny {\it N}}}^2}\,   r f\sqrt{g} \left[ \left(\Phi^{\dagger}\right)' \delta\Phi +\Phi' \delta\Phi^{\dagger} \right] \dd t \wedge \dd\phi 
   \nonumber \\
 &\:   -\frac{1}{\kappa_{\hbox{\tiny {\it N}}}^2}\,  \frac{r}{f \sqrt{g}} \Big[ \left( \partial_t +\Omega   \partial_{\phi}\right)\Phi^{\dagger} \delta\Phi +c.c. \Big] \dd r \wedge \dd \phi 
  \nonumber \\
 &\:  + \frac{1}{\kappa_{\hbox{\tiny {\it N}}}^2}\, \frac{1}{r \sqrt{g} } \Big[  \Big( g \partial_{\phi} \Phi^{\dagger} - r^2\,\frac{\Omega}{f} \left( \partial_t +\Omega   \partial_{\phi}\right)\Phi^{\dagger}  \Big) \delta\Phi   +c.c. \Big] \dd t \wedge \dd r
\end{align}
\end{subequations}
where $c.c.$ stands for complex conjugate of the previous term.
Thus, the variation of $ \widetilde{Q}_\xi$ together with the interior product of the boundary 2-forms are given by the 1-forms:
\begin{subequations}\label{Noether:deltaQ_INTtheta}
\begin{align}
&  \hspace{0.5cm}  \delta   \widetilde{Q}_\xi  = \frac{1}{2\kappa_{\hbox{\tiny {\it N}}}^2}\, r\left[\delta\left(\frac{(fg)'}{\sqrt{g}}\right) - r^2 \frac{\Omega'}{\sqrt{g}}\delta\Omega - r^2\frac{(\Omega- \Omega_H)}{\sqrt{g}}\delta\Omega'+ r^2\frac{(\Omega- \Omega_H)\Omega'}{2g^{3/2}}\delta g \right] \dd\phi + (\cdots)\dd t \,, \label{Noether:deltaQ_form}
\\
&   \hspace{0.5cm}    \ins_\xi \theta\left(g,\delta g\right) = \frac{1}{2\kappa_{\hbox{\tiny {\it N}}}^2}\, r\left[\delta\left(\frac{(fg)'}{\sqrt{g}}\right) - r^2 \frac{\Omega'}{\sqrt{g}}\delta\Omega +  \frac{\sqrt{g}}{r} \delta f \right]  \Big(\dd\phi -\Omega_H \dd t  \Big)\,, \label{Noether:delta-INTthetaG_form}
    \\
&  \hspace{0.5cm}   \ins_\xi \theta\left(\Phi,\delta\Phi; \Phi^{\dagger},\delta \Phi^{\dagger}\right) = -\frac{1}{\kappa_{\hbox{\tiny {\it N}}}^2}\,   r f\sqrt{g} \left[ \left(\Phi^{\dagger}\right)' \delta\Phi +\Phi' \delta\Phi^{\dagger} \right] \dd\phi + (\cdots)\dd t + (\cdots)\dd r \,, \label{Noether:delta-INTthetaPhi_form}
\end{align}
\end{subequations}
where $(\cdots)$ describes either the components $\dd t$ or $\dd r$ of the associated 1-forms that will make no contribution when we integrate over a constant $t$ and $r$ hypersurface (as we do below) so we do not display them. 
Inserting \eqref{Noether:deltaQ_INTtheta} into \eqref{Noether:chargeQxi} it follows that the variation of the Noether charge \eqref{Noether:VarCharge} relative to $\xi=\partial_t+\Omega_H \partial_{\phi}$ in our hairy system is
\begin{align}\label{Noether:chargeomegaxiHairy} 
   \omega_{\hbox{\tiny $\xi \! = \! \partial_t \! + \! \Omega_H \partial_{\phi}$}} =&\:  \frac{1}{2\kappa_{\hbox{\tiny {\it N}}}^2}\left[
    r^3\,\frac{\Omega- \Omega_H}{\sqrt{g}} \Big( \frac{\Omega'}{2 g}\delta g -\delta\Omega' \Big)
    -\sqrt{g}\, \delta f 
    +  2 r f\sqrt{g} \Big( \left(\Phi^{\dagger}\right)' \delta\Phi +\Phi' \delta\Phi^{\dagger} \Big) \right]\dd\phi \nonumber\\
    &\: + (\cdots)\dd t + (\cdots)\dd r\,,
\end{align}
Note that the variation  $\omega_{\hbox{\tiny $\xi \! = \! \partial_t $}}$  relative to the Killing vector $\xi = \partial_t$ can be obtained from \eqref{Noether:chargeomegaxiHairy} by simply setting $\Omega_H =0$ while  the variation  $\omega_{\hbox{\tiny $\xi \! = \! \partial_{\phi} $}}$  relative to the Killing vector $\xi = \partial_{\phi}$ can effectively be obtained from \eqref{Noether:chargeomegaxiHairy} by simply taking its derivative w.r.t. $\Omega_H$.
Now consider the asymptotic expansion of the fields $-$ of \eqref{HairyAnsatz} about the Einstein Static Universe $-$ in fractional powers of $\frac{r_+}{r}$:
 \begin{align}    \label{Noether:asympExpansionfgOphi}
f |_{r\to\infty} \sim &\: \frac{r^2}{L^2}+ \cdots + \mathcal{F}_2(r_+) +\cdots , \qquad 
g |_{r\to\infty} \sim \: 1+\cdots , \qquad 
\Omega |_{r\to\infty} \sim \: \mathcal{W}_2(r_+)\,\frac{r_+^2}{r^2}+\cdots , \nonumber \\
\Phi |_{r\to\infty} \sim &\: \mathcal{A}(r_+) \left(\frac{r_+}{r}\right)^{\Delta_-}+\cdots +  \mathcal{B}(r_+) \left(\frac{r_+}{r}\right)^{\Delta_+}+\cdots 
  \end{align}
 where $\cdots$ represent terms that are a function of the independent UV free parameters $\{\mathcal{F}_2,\mathcal{W}_2,\mathcal{A},\mathcal{B}\}$.
Inserting this expansion in \eqref{Noether:chargeomegaxiHairy} and the latter in \eqref{Noether:canonicalEnergy}, we find that the conserved energy and angular momentum of a hairy black hole are
 \begin{subequations}  \label{Noether:canonicalEnergyHairy0}
 \begin{align} 
\mathcal{E}\equiv &\: \int \dd m_k  \int_{S^{(1)}_{\infty}} \omega_{\xi}\big|_{\xi=\partial_t} \nonumber \\
= &\: \frac{1}{2 \kappa_{N}^2 L^2} \int \dd r_+  \int_{0}^{2\pi}\!\dd\phi  \: r_+ \Big[ -r_+ \mathcal{F}_2'(r_+)-2 \mathcal{F}_2(r_+)+\frac{\sqrt{r_+}}{\sqrt{L}} 8 \kappa  L \mathcal{A}(r_+) \mathcal{A}'(r_+)+\frac{\sqrt{L}}{\sqrt{r_+}}  6 \kappa   \mathcal{A}(r_+)^2 \Big]  \nonumber \\
= &\: -\frac{\pi}{\kappa_{\hbox{\tiny {\it N}}}^2} \frac{r_+^2}{L^2}   \bigg[ (\mathcal{F}_2(r_+) - 4 \frac{\sqrt{L}}{\sqrt{r_+}} \, \kappa \, \mathcal{A}(r_+)^2  \bigg]
=-\frac{\pi}{\kappa_{\hbox{\tiny {\it N}}}^2} \frac{r_+^2}{L^2}   \Big[ \mathcal{F}_2(r_+)-4 \mathcal{A}(r_+) \mathcal{B}(r_+) \Big] \,, \\
\mathcal{J} \equiv &\: - \int \dd m_k  \int_{S^{(1)}_{\infty}} \,  \omega_{\xi}\big|_{\xi=\partial_{\phi}} =
 \frac{1}{\kappa_{N}^2} \int \dd r_+  \int_{0}^{2\pi}\!\dd\phi  \: r_+ \Big[r_+ \mathcal{W}_2'(r_+)+2 \mathcal{W}_2(r_+) \Big]
 \nonumber \\
= &\: \frac{2\pi}{\kappa_{\hbox{\tiny {\it N}}}^2}r_+^2\,\mathcal{W}_2
 \,. 
  \end{align}
 \end{subequations}
Above, we have used the double-trace BC, $\mathcal{B}=\frac{\sqrt{L}}{\sqrt{r_+}}\,\kappa \mathcal{A}$ and the integrations in $r_+$ were possible because the present system is integrable.

To compare the Noether charges with the holographic ones studied in the previous section, it is now convenient to (re)introduce the Fefferman-Graham coordinates $\{\tau,\phi,z\}$ studied in section~\ref{secA:HoloRen}. The required FG coordinate transformation is \eqref{HoloRen:FGgeneral_appendix}, \ie $t=L \tau$ and  
$r=\frac{L}{z} -\frac{L}{2} \frac{r_+^{2\Delta_-}}{L^{2\Delta_-}}  \mathcal{A}^2 z^{1-(\Delta_+-\Delta_-)}-\frac{L}{4}\frac{r_+^2}{L^2} \mathcal{F}_2 z+\cdots$, which rewrites the asymptotic behaviour of \eqref{HairyAnsatz} as
\eqref{HoloRen:FG}-\eqref{HoloRen:FGexp} with $A_0=1,B_0=1, \Omega_0=0$ (because our solutions are conformal to the Einstein Static Universe with no rotation at infinity).
The expansion coefficients $\{\mathcal{F}_2,\mathcal{W}_2,\mathcal{A},\mathcal{B}\}$, when rewritten it terms of the FG expansion coefficients $\{A_2,\Omega_2,\alpha,\beta\}$ of  \eqref{HoloRen:FGexp} read
   \begin{align} \label{Noether:coefFG}
\mathcal{F}_2=\frac{L^2}{r_+^2}\,2\Big( A_2 +4\Delta_-\Delta_+ \alpha\beta\Big),\qquad \mathcal{W}_2=\frac{L}{r_+^2}\,\Omega_2\,, \qquad \mathcal{A}=\frac{L^{\Delta_-}}{r_+^{\Delta_-}}\,\alpha\,,  \qquad \mathcal{B}=\frac{L^{\Delta_+}}{r_+^{\Delta_+}}\,\beta\,,
 \end{align}  
and thus \eqref{Noether:canonicalEnergyHairy0} can be written as a function of the FG coefficients as (after setting $\beta=\kappa \, \alpha$) as
 \begin{subequations}  \label{Noether:canonicalEnergyHairy}
 \begin{align} 
    & \mathcal{E} = -\frac{1}{4G}\Big[ A_2+  \Big(2-(\Delta_+-\Delta_-)^2 \Big)\kappa\, |\alpha|^2  \Big], \nonumber\\  
    & \mathcal{J}  = \frac{L}{4 G}\,\Omega_2\,.
  \end{align}
 \end{subequations}
That is to say, as expected, the Noether energy and angular momentum \eqref{Noether:canonicalEnergyHairy} agree with the conserved charges \eqref{HoloRen:EnergyAngMom} computed using holographic renormalization: $\mathcal{E}\equiv E$ and $\mathcal{J}\equiv J$.
As noted previously, note that \eqref{Noether:canonicalEnergyHairy} or \eqref{HoloRen:EnergyAngMom} give the energy for the double-trace BC ($\beta=\kappa \, \alpha\neq 0$) but also for Dirichlet BCs ($\alpha=0$ ie the standard quantization) or Neumann BC ($\beta=0$, ie the alternative quantization). 

We can now also explicitly check that the first law of the AdS$_3$ hairy black hole is \eqref{Noether:FirstLaw}.
We simply need to evaluate \eqref{Noether:FirstLaw0} for the variation of the Noether charge  relative to $\xi=\partial_t+\Omega_H \, \partial_{\phi}$ given in \eqref{Noether:chargeomegaxiHairy}:
 \begin{subequations}  \label{Noether:1stlawLHS-RHS}
 \begin{align} 
&  \int_{S^1_{\infty}} \omega_\xi \Big|_{\hbox{\tiny $\xi \! = \! \partial_t \! + \! \Omega_H \partial_{\phi}$}}
= \delta \mathcal{E}- \Omega_H \delta \mathcal{J} \,, \label{Noether:1stlawLHS} \\
 &  \int_{\Sigma_{\mathcal{H}}} \omega_\xi \Big|_{\hbox{\tiny $\xi \! = \! \partial_t \! + \! \Omega_H \partial_{\phi}$}}
 = - \frac{\pi}{\kappa_{\hbox{\tiny {\it N}}}^2}  \left[\sqrt{g(r)}\, \delta f(r)\right] \Big|_{r=r_+}
    = \frac{\sqrt{g(r_+)}f'(r_+)}{4 \pi} \delta\left(\frac{\pi r_+}{2G}\right) = T_H \delta S_H\,. \label{Noether:1stlawRHS} 
\end{align}
 \end{subequations}
In \eqref{Noether:1stlawLHS} we simply used the variation of \eqref{Noether:canonicalEnergyHairy0},
$\delta \mathcal{E} = \int_{S^{1}_{\infty}} \omega_{\xi}\big|_{\xi=\partial_t}$ and
$ \delta \mathcal{J}= -\int_{S^{1}_{\infty}} \omega_{\xi}\big|_{\xi=\partial_{\phi}}$ as well as $\omega_{\hbox{\tiny $\xi \! = \! \partial_t \! + \! \Omega_H \partial_{\phi}$}} = \omega_{\hbox{\tiny $\partial_t$}}+ \Omega_H \omega_{\hbox{\tiny $ \partial_{\phi}$}}$. In \eqref{Noether:1stlawRHS}, we used the fact that  all terms of \eqref{Noether:chargeomegaxiHairy} except one do vanish at the horizon since $f(r_+)=0$ and $\Omega(r_+)= \Omega_H$, as well as  the thermodynamic definitions \eqref{Noether:TSO}  and, finally,   the fact that the variation of $f$ in the phase space is $\delta f (r_+) = \frac{\partial f}{\partial r_+} \delta r_+= f'(r_+)\delta r_+$  (and similarly for the variation of $S_H$). Concluding, \eqref{Noether:FirstLaw0} and \eqref{Noether:1stlawLHS-RHS} indeed imply that the first law of black hole thermodynamics for the hairy AdS black holes is  \eqref{Noether:FirstLaw} independently of whether we impose Dirichlet, Neumann or double-trace BCs for the scalar field. 

For completeness, we can also use the equations of motion to prove that the angular momentum $J$ and $U(1)$ conserved charge $N$ of boson stars satisfy the relation $J=m\, N$ and thus the boson star first law \eqref{NoetherBS:FirstLaw} can indeed be rewritten as \eqref{NoetherBS:FirstLaw2}. For that, note that the equation of motion \eqref{eqn:Omegaeom} for $\hat{\Omega}$ can be rewritten as:
\begin{align}  \label{eqn:Omegaeom2} 
 \frac{d}{dR}\left(-\frac{R^3 \hat{\Omega}'(R)}{\sqrt{g(R)}}\right) = m\,\frac{4 R\,\psi(R)^2}{f(R)\sqrt{g(R)}} \Big(\hat{\omega} -m\hat{\Omega}(R)\Big).
\end{align}
Now consider a boson star and integrate \eqref{eqn:Omegaeom2} in the region $0\leq R <\infty$
\begin{align}  \label{eqn:Omegaeom3}
  \frac{L}{8 G}\int_0^{\infty}\frac{d}{dR}\left(-\frac{R^3 \hat{\Omega}'(R)}{\sqrt{g(R)}}\right) \dd R  &= m \, \frac{L}{8 G} \int_0^{\infty} \frac{4 R\,\psi(R)^2}{f(R)\sqrt{g(R)}} \Big(\hat{\omega} -m\hat{\Omega}(R)\Big)\dd R \nonumber\\
 \Leftrightarrow \quad  \lim_{R \to \infty }\left(-\frac{L}{8G}\frac{R^3 \Omega'(R)}{\sqrt{g(R)}}\right) & =  m \, \frac{L}{8 G} \int_0^{\infty} \frac{4 R\,\psi(R)^2}{f(R)\sqrt{g(R)}} \Big(\hat{\omega} -m\hat{\Omega}(R)\Big)\dd R 
 \nonumber\\
 \Leftrightarrow \hspace{3.37cm} \quad  J & = m\, N
\end{align}
where we used the boson star boundary conditions  \eqref{BCs:OriginBSm0} and \eqref{BCs:OriginBSm} at the origin to find that the lower integration limit of the LHS of \eqref{eqn:Omegaeom3} vanishes and the definition \eqref{N:Def} of the $U(1)$ conserved charge $N$.\footnote{We have also used 
\begin{align}\label{HoloRen:EnergyAngMomAlternative}
    E\equiv Q_{\hbox{\tiny $\partial_T$}} &= \lim_{R\to \infty}\frac{\sqrt{f(R)g(R)}}{16G}\left(\frac{2 \kappa |\alpha|^2}{R} - 4 \sqrt{f(R)} + R\big[3 \psi(R)+4R\big]^2\right) =  \frac{1}{8G}(C_f + 4 \kappa  |\alpha|^2), \nonumber \\
J \equiv Q_{\hbox{\tiny $\partial_{\phi}$}} & = -\lim_{R\to\infty} \frac{L}{8 G}\frac{R^3\hat{\Omega}'(R)}{\sqrt{g(R)}}=\frac{L}{4 G}\,\hat{C}_\Omega.
\end{align}
which, of course, agrees with \eqref{HoloRen:EnergyAngMomFinal}. 
} 

In particular,  \eqref{eqn:Omegaeom3} shows that the boson star first law \eqref{NoetherBS:FirstLaw} can also be written as \eqref{NoetherBS:FirstLaw2}, where the latter holds also for axisymmetric ($m=0$) boson stars that are necessarily static (\ie that have $J=0$). 

\clearpage
\bibliographystyle{JHEP}
\bibliography{refHairyBTZ}

\end{document}